\documentclass[11 pt]{article}

\usepackage{authblk} % author package
\usepackage{tabularx} % extra features for tabular environment
\usepackage{amsmath,amssymb}  % improve math presentation
\usepackage{mathtools} % includes \coloneqq

\usepackage{graphicx} % takes care of graphic including machinery
\usepackage{subcaption} % package for subfigure
\usepackage{float}
\usepackage[margin=1in,letterpaper]{geometry} % decreases margins
\usepackage{cite} % takes care of citations
\usepackage[final]{hyperref} % adds hyper links inside the generated pdf file
\usepackage{hyperref}
\hypersetup{colorlinks,allcolors=black}
\usepackage[nobiblatex]{xurl}
\usepackage{lscape} % package for landscape
\usepackage{booktabs} % package for tex professional table
\usepackage{bm} % bold math 

\usepackage{setspace} % set space for text, figures, and tables 
\usepackage{natbib}
\usepackage{bibentry}
\usepackage[nottoc,numbib]{tocbibind} % add Reference to table of contents
\usepackage{sectsty} % to use \sectionfont
\usepackage{titlesec}
\titleformat{\paragraph}{\normalfont\normalsize\bfseries}{\theparagraph}{1em}{}
\titlespacing*{\paragraph}{0pt}{3.25ex plus 1ex minus .2ex}{1.5ex plus .2ex}

\usepackage{tikz} % graph in latex
\usetikzlibrary{positioning,arrows.meta} %tikz diagram
\usepackage{pgfplots}
\pgfplotsset{compat=newest} %this setting allow node to go to the correct place
\usepgfplotslibrary{fillbetween}
\usepackage{siunitx} % to use mm in tikz
\usetikzlibrary{{shapes.misc}}
\usepackage{pdflscape} %pdf landscape page

\newenvironment{chapquote}[2][2em]
  {\setlength{\@tempdima}{#1}%
   \def\chapquote@author{#2}%
   \parshape 1 \@tempdima \dimexpr\textwidth-2\@tempdima\relax%
   \itshape}
  {\par\normalfont\hfill--\ \chapquote@author\hspace*{\@tempdima}\par\bigskip}
\makeatother

\newcommand{\comment}[1]{}

\newcommand{\lnb}[1]{%
  \ln\left(#1\right)%
}

\renewcommand{\baselinestretch}{1.5} %  scales the default interline space to 1.5 its default value. 

\usepackage{setspace} % set space for text, figures, and tables 
\begin{document}
%The legibility of text can be enhanced by separating paragraphs with an amount of white space that will vary according to some design aesthetic.
\setlength{\parskip}{3pt plus1pt minus1pt}

\setlength{\abovedisplayskip}{3pt} % the space above equations 
\setlength{\belowdisplayskip}{3pt} % the space below equations 

%--------------------------------------------------------
\begin{titlepage}
\title{\textbf{Testing Business Cycle Theories: Evidence \\
from the Great Recession}
}

\author[1]{\textbf{Bo Li}  \thanks{ \protect\linespread{1}\protect\selectfont {\footnotesize Bo Li is with the Department of Finance, Arizona State University (boli15@asu.edu).} } }
\date{This Version: March 4th, 2024\\
\textcolor{blue}{\href{https://www.boli-finance.com/research}{\textcolor{blue}{[Click here for the latest version]}}}}
\maketitle

\vspace{-4mm}

\begin{abstract}
\fontsize{11}{13}\selectfont
\noindent Empirical business cycle studies using cross-country data usually cannot achieve causal relationships while within-country studies mostly focus on the bust period. We provide the first causal investigation into the boom period of the 1999-2010 U.S. cross-metropolitan business cycle. Using a novel research design, we show that credit expansion in private-label mortgages causes a differentially stronger boom (2000-2006) and bust (2007-2010) cycle in the house-related industries in the high net-export-growth areas. Most importantly, our unique research design enables us to perform the most comprehensive tests on theories (hypotheses) regarding the business cycle. We show that the following theories (hypotheses) cannot explain the cause of the 1999-2010 U.S. business cycle: the speculative euphoria hypothesis, the real business cycle theory, the collateral-driven credit cycle theory, the business uncertainty theory, and the extrapolative expectation theory.

\vspace{4ex}

\noindent \textbf{Keywords}: business cycle, financial crisis, credit supply, private-label mortgages \\
\noindent  \textbf{JEL Classification}: E32, G01, G21 \\

\end{abstract}

\thispagestyle{empty}
\end{titlepage}
%------------------------------------------------------------------

%--------------------------------------------------------
\begin{titlepage}

\centering

\textbf{\LARGE \vspace{3ex} \\
Testing Business Cycle Theories: Evidence \\
\vspace{4mm}
from the Great Recession}

%\title{\textbf{Credit Expansion and Housing Cycle}}
%\author[1]{\textbf{Bo Li}  \thanks{ \protect\linespread{1}\protect\selectfont {\footnotesize Bo Li is with the Department of Finance, Arizona State University (boli15@asu.edu).} } }
%\date{Nov 16th, 2023}
%\maketitle

\vspace{16ex}

\begin{abstract}
\fontsize{11}{13}\selectfont
\noindent Empirical business cycle studies using cross-country data usually cannot achieve causal relationships while within-country studies mostly focus on the bust period. We provide the first causal investigation into the boom period of the 1999-2010 U.S. cross-metropolitan business cycle. Using a novel research design, we show that credit expansion in private-label mortgages causes a differentially stronger boom (2000-2006) and bust (2007-2010) cycle in the house-related industries in the high net-export-growth areas. Most importantly, our unique research design enables us to perform the most comprehensive tests on theories (hypotheses) regarding the business cycle. We show that the following theories (hypotheses) cannot explain the cause of the 1999-2010 U.S. business cycle: the speculative euphoria hypothesis, the real business cycle theory, the collateral-driven credit cycle theory, the business uncertainty theory, and the extrapolative expectation theory.

\vspace{4ex}

\noindent \textbf{Keywords}: business cycle, financial crisis, credit supply, private-label mortgages \\
\noindent  \textbf{JEL Classification}: E32, G01, G21 \\

\end{abstract}

\thispagestyle{empty}
\end{titlepage}
%------------------------------------------------------------------

\clearpage 
\thispagestyle{empty}
\renewcommand{\baselinestretch}{1.3}
\tableofcontents
\renewcommand{\baselinestretch}{1.3}
\thispagestyle{empty}

%----------------------------------------------------------------------
% section 2: Introduction

\clearpage 
\pagenumbering{arabic}
\setcounter{page}{1}

%--------------------------------------------------------------
%--------------------------------------------------------------
% This is the beginning of the entire section of the Introduction
%--------------------------------------------------------------
%--------------------------------------------------------------
\clearpage

\begin{chapquote}{Amir Sufi and Alan M. Taylor, \textit{Financial Crises: A Survey, 2022}}
\noindent ``Crises do not occur randomly, and, as a result, an understanding of financial crises requires an investigation into the booms that precede them.”
\end{chapquote}

\vspace{-3mm}
\section{Introduction}

Finding the cause of the 1999-2010 U.S. business cycle is vital because the U.S. ended up with the deepest recession since the Great Depression in the 1930s. In the bust period, the U.S. economy experienced widespread mortgage defaults \citep{mayer2009rise, keys2010did}, massive failures in the banking industry \citep{bernanke2023nobel}, large consumption drop \citep{mian2013household, kaplan2020non}, and huge unemployment rise \citep{hoynes2012suffers, mian2014explains}. Identifying the cause and its major mechanisms is crucial for understanding of the economic connections among credit, housing, banking, and employment. It is also helpful for the design of regulatory frameworks and macroeconomic policies that monitor the economy, avoid a similar recession, and intervene in an early stage to heal the whole economy.

In the literature, there are multiple theories (hypotheses) trying to explain the business cycle. Prevalent hypotheses include speculative euphoria hypothesis \citep{kindleberger1978manias,minsky1986stabilizingan}, real business cycle theory \citep{prescott1986theory}, the collateral-driven credit cycle theory \citep{kiyotaki1997credit}, the business uncertainty theory \citep{bloom2009impact}, extrapolative expectation theory \citep{eusepi2011expectations}, and credit-drive household demand hypothesis \citep{mian2009consequences, schularick2012credit}. Each theory (hypothesis) has found some pieces of supporting evidence. So far, however, we have not gathered enough micro-style causal evidence to distinguish which theory (hypothesis) captures the major origin of the business cycle. The empirical difficulties are twofold. First, cross-country empirical studies usually only achieve correlation because of large endogenous differences in economic development, institutions, and culture. Second, within-country studies mostly focus on the bust periods without digging into the boom period, partially because of the difficulty in finding long-term incentives that have persistent geographic divergence that potentially causes the divergence in credit expansion. Prevalent theories and ample empirical evidence, however, support geographic convergence within a country \citep{kim2004historical}.

To preview, by combining insights from regional economics and international economics, we design a causal framework to locate the long-term incentive for credit that has persistent geographic divergence. Our research design enables us to find ample micro-style causal evidence for the ``credit-driven housing-dominant view". Specifically, induced by net export growth, credit expansion in private-label mortgages (non-jumbo) causes the 1999-2010 U.S. business cycle. This business cycle is much stronger in the high net-export-growth areas and housing-related industries experienced an amplified boom and bust cycle (``housing industry channel"). Most importantly, our unique research design empowers us to conduct the most comprehensive tests on the relevance of prevalent theories (hypotheses) regarding business cycles. We find that other hypotheses cannot explain the origin of this business cycle. We will describe our research design, then major findings, and lastly, our contribution.

\noindent \textbf{Research Design} Empirical studies of the business cycle by cross-country data face several challenges in identifying the cause and its primary channel. First, credit supply and household demand can be driven endogenously by institutional differences. For example, countries with higher prior economic growth, better legal protection of creditors, and a stronger regulatory system against financial fraud could have higher household demand that attracts international credit. Alternatively, these countries could attract more international funds that reduce interest rates and induce higher household demand. Second, we need to achieve a consistent measure of household debt so that such a measure can be comparable in the cross-section. For the largest part of household debt, mortgages (around 70\% in the US), the average maturity is around 45 years in Sweden, 30 years in the USA, and 15 years in Germany \citep{bernstein2021mortgage}. This fact means the same debt-to-income ratio due to the same amount of mortgage means relatively higher payback pressure for households in Germany but much lower payback pressure for households in Sweden. In a similar way, differences in the availability and level of social welfare programs also present difficulties in achieving a consistent measure of household leverage across countries. The above two difficulties prevent many empirical studies from accomplishing causal evidence by cross-country study (e.g., \cite{mian2017household, Muller2023credit}).

Within-country studies also face several challenges, though they can avoid the above difficulties faced by cross-country studies. First, any identification strategy requires cross-section differences in credit expansion, which in turn needs the treatment variable to capture the incentives to the credit supply.\footnote{To be precise, the `treatment variable' here can refer to the instrumental variable in an IV strategy, the running variable in a regression discontinuity design, and the treatment variable in a difference-in-difference design. For example, \cite{di2017credit} uses the interaction between anti-predatory lending laws and 2004 OCC preemption as the treatment variable in a difference-in-difference setting.} This requirement means the treatment variable must have enough area coverage (US mainland), geographic variation, and time coverage (99-10). Second, the geographic variation must persist over the business boom period (1999-2005) since the credit supply expansion (in corporate debt or in home mortgages) is a long-term financial decision. In other words, the treatment variable must provide persistent incentives to induce stronger credit expansion in certain areas or certain industries than others throughout the entire boom period. Short-run shocks, such as weather (e.g., rain, extreme temperature, and wildfires), are inadequate. Third, beyond variable nature, shocks to the treatment variable are required to constitute a solid identification strategy. Fourth, the underlying economic theory (or story) shall explain why the treatment variable (incentive) can induce stronger credit expansion in the boom period rather than in other periods (such as the prior period). Such a theory (or story) requires a large framework consist of the incentives of different types of participants, the mortgage market structure, and the time-varying nature of them.

Our within-US research design addresses the above challenges in three parts. In the first part, we operationalize the key idea of ``economic base theory" \citep{tiebout1962community} and construct a treatment variable that captures the long-term incentive of credit expansion: metropolitan exposure to net export growth of manufacturing industries. The ``economic base" (tradable sector) refers to the economic activities that a local area provides for the areas outside, thus bringing wealth to the local area. Most of the wealth will be reused locally via a money multiplier effect. By this theory, the tradable sector growth is the key driving force for local economic growth in the long term. Therefore, credit expansion in mortgages would be stronger in areas with stronger growth in tradable sector due to higher residual value of a mortgage given default.\footnote{Please see the model by \cite{li2024credit} for more details.} Ideal measurement of the composite and growth of table sector requires census-style data consisting of the accounting data of all firms, which is unavailable in reality. Alternativly, following \cite{li2024credit}, we proxy the composite (share) with manufacturing employment data at the industry-by-metropolitan level. In addition, we employ the substantial time series change (shift) in net export growth in the U.S. as a proxy for the relative growth at the industry level. Aggregating these two proxies (shift and share) can give us a good measure of the relative growth of the local tradable sector across metropolitan areas. By this construction, net export growth has enough area coverage (U.S. mainland) and enough time coverage (92-09). In addition, its geographic variation comes from the fact that the related manufacturing industries tend to cluster in just a few locations. Further, the persistence feature of net export growth comes from two dimensions. At the industry level, both increasing return to scales at the industry level and comparative advantages across nations make trade patterns persistent over time. At the individual level, job reallocation across different industries or locations is very costly.

\comment{
In the second part, our research design draws analysis from the model by \cite{li2024credit} to illustrate why the net export growth incentivizes mortgage credit expansion in the boom period (1999-2006) but not in the prior period (1992-1999). \cite{li2024credit} builds an economic model that incorporates the U.S. mortgage market structure and a sharp legal distinction between government-sponsored enterprise mortgages (GSEMs) and private-label mortgages (PLMs). In essence, by the legal constraint of non-discrimination, GSEMs cannot consider differences in regional economic conditions when setting up mortgage rates \citep{hurst2016regional}, where we use net export growth to capture the main differences in regional economic conditions. In contrast, PLMs are free from this constraint. Li's model explains that net export growth cannot induce house price growth in the prior period (92-99) when GSEMs dominate the market with government implicit guarantees. However, net export growth can induce a substantial house price boom and hence business boom 1999-2006 due to the sharp credit expansion in PLMs, which is documented by \cite{justiniano2022mortgage}. 
}

In the second part, our research design employs the gravity model-based instrumental variable approach developed by \cite{feenstra2019us} as our identification strategy. \cite{feenstra2019us} develop their IVs from a general equilibrium model and their IVs can isolate the exogenous part of U.S. imports and exports. Conceptually, their IVs capture the exogenous parts of net export growth due to (1) increasing world demand reflected in US export growth, (2) increasing world supply reflected in US import growth, and (3) tariff changes. They also use high-dimensional fixed effects to remove the potentially endogenous parts: (1) US industry-by-year supply-side shocks in exports and (2) US industry-by-year demands-side shocks in imports, and (3) pre-determined bilateral distance between U.S. and partner countries. They construct IVs separately for exports and imports, and we combine them together as an IV for net export growth.

In the third part of our research design, we rely on the model by \cite{li2024credit} to explain (1) why our framework is consistent with the viewpoint that net export growth cannot induce credit expansion in the prior period (1991-1999) and (2) the intuition for the business cycle. 

\noindent \textbf{Intuition for the Business Cycle} The basic intuition of the cross-metro differential business cycle story in this paper can be illustrated in the following Figure (\ref{fig_ModelIntuition_Graphs}). Conceptually, U.S. metropolitan areas can be divided into high and low net-export growth areas. Higher net export growth leads to higher household income growth, higher employment growth, and higher population growth gradually in the high net-export-growth area than in the low net-export-growth area. These differences grant the mortgage borrowers in the high net-export-growth area two advantages: (1) higher foreclosure price of house given default \footnote{In the empirical literature, there is ample evidence that household income growth, employment growth,  and population growth (including migration) can push up housing demand and then housing price, especially in the long term \citep{olsen1987demand}.} and (2) higher income growth in the future that can be recoursed by lenders after default.

This paragraph explains why our framework is consistent with the viewpoint that net export growth cannot induce credit expansion in the prior period (1991-1999). A key legal requirement is that government-sponsored enterprise mortgages cannot consider regional economic conditions (growth) \citep{hurst2016regional} in setting up mortgage rates. But private-label mortgages can. Since securitization innovation (notably the Copula approach by \cite{li2000copula} \citep{salmon2012formula}), ``global saving glut" \citep{bernanke2005global, bernanke2007global}, mortgage market deregulation \citep{di2017credit, lewis2023creditor}, and political push \citep{mian2013household} all occurred after 1999, private-label mortgages maintained high mortgage rates between 1991 and 1999 and only had a small market share. On the contrary, government-sponsored enterprise mortgages dominated the mortgage market with low rates due to economies of scale and the government's implicit guarantee.\footnote{Estimates show that the spread between government-sponsored enterprise mortgages and otherwise similar jumbo loans (purchased by private issuers) are, on average, between 15-40 basis points between 1996 and 2006 (see \cite{sherlund2008jumbo} and its summary of the literature).} Therefore, Without aggregate credit expansion, even high net-export-growth areas cannot undergo a business cycle before 1999.

However, documented by \cite{justiniano2022mortgage}, there is a huge credit expansion in the private-label mortgages that starts in 2003 summer.\footnote{\cite{justiniano2022mortgage} argues that this rate drop likely reflects mispricing, as shown in the subsequent increasing default rate.} Using loan-level data and a regression model, they identify a sharp and persistent decline in the spread between private-label mortgages and 10-year treasury yield started in 2003 summer. Given such credit expansion, our story predicts that private-label mortgages choose to increase strongly in high net-export-growth areas because of the two above advantages in borrowers (higher foreclosure price and higher borrower income growth). This differentially stronger growth in private-label mortgages in high net-export-growth areas eventually results in a much stronger boom and bust cycle in the house-related employment. 

%------------------------------------
\begin{figure}[H]

\begin{center}

\resizebox{6in}{4in}{%
%%\resizebox{\textwidth}{!}{%
\includegraphics[width=6in, height=4in]{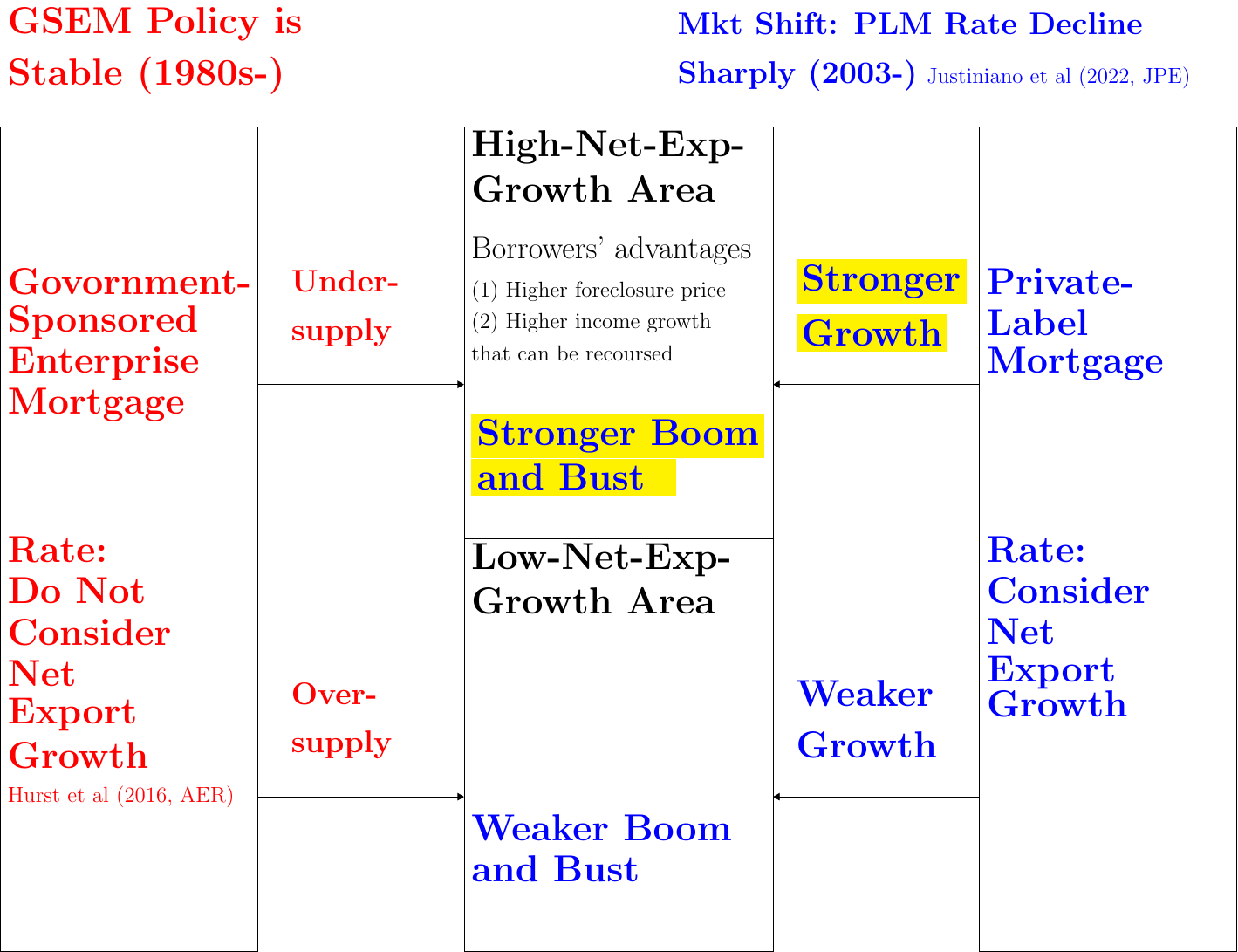}
} %end of resizebox

\end{center}

%---------------
% Figure setting: caption and label
%---------------
\caption{Model Intuition}
\label{fig_ModelIntuition_Graphs}

\end{figure}
%------------------------------------

\comment{
We explain the intuition in detail in the paragraph for the intuition of the business cycle below. A key legal requirement is that government-sponsored enterprise mortgages cannot consider regional economic conditions (growth) \citep{hurst2016regional} in setting up mortgage rates. But private-label mortgages can. Since securitization innovation (notably the Copula approach by \cite{li2000copula} \citep{salmon2012formula}), ``global saving glut" \citep{bernanke2005global, bernanke2007global}, mortgage market deregulation \citep{di2017credit, lewis2023creditor}, and political campaign \citep{mian2013household} all occurred after 1999, private-label mortgages maintained high mortgage rates between 1991 and 1999 and only had a small market share. On the contrary, government-sponsored enterprise mortgages dominated the mortgage market with low rates due to economies of scale and the government's implicit guarantee. Therefore, Without aggregate credit expansion, even high net-export-growth areas cannot undergo a business cycle.  
}

%--------------------------------------------------
%\subsection{Findings}
%--------------------------------------------------

\begin{figure}[H]

\begin{center}

\resizebox{5.4in}{!}{%
%\resizebox{\textwidth}{!}{%
\includegraphics[width=13.716cm, height=8.2cm]{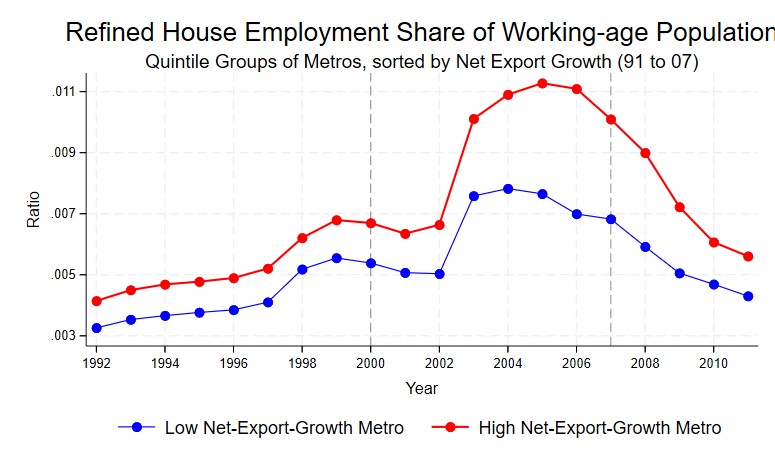}
} %end of resizebox

\end{center}

%---------------
% Figure setting: caption and label
%---------------
\caption{Refined House Employment Share of Working-age Population (1991-2011)}
\label{fig_RefineHouseEmpShr_92to11_Intro}

\end{figure}

\noindent \textbf{Findings} We illustrate our major findings in four parts. In the first part, we document two new empirical facts. There is a much stronger business (employment) cycle for house-related industries in the high net-export-growth metropolitan areas (HNEG areas) than in the low net-export-growth areas (LNEG areas) between 1999-2010. Figure (\ref{fig_RefineHouseEmpShr_92to11_Intro}) shows this stronger local house-related employment cycle. In the prior period (1992-2000), there is no difference in the dynamics of house-related employment share in the working-age population between the HNEG areas and the LNEG areas. However, in the boom period (2000-2006) characterized by excess credit supply in private-label mortgages \citep{justiniano2022mortgage}, the increase in house-related employment share is much stronger in the HNEG area (0.440\%) than one in the LNEG area (0.161\%). From 2007 to 2010, the drop is also stronger in the HNEG area (0.403\%) than in the LNEG area (0.214\%).

%------------------------------------
%------------------------------------

\begin{figure}[H]

\begin{center}

\resizebox{5.4in}{!}{%
%\resizebox{\textwidth}{!}{%
\includegraphics[width=13.716cm, height=8.2cm]{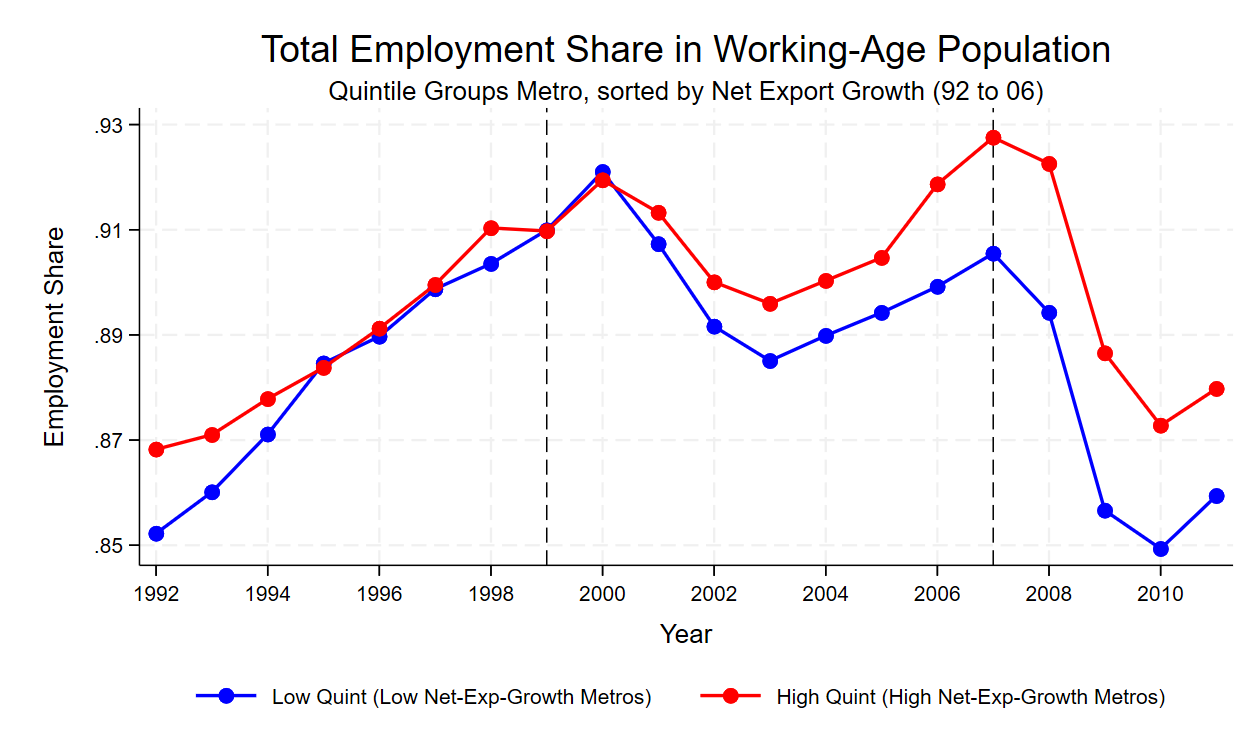}
} %end of resizebox

\end{center}

%---------------
% Figure setting: caption and label
%---------------
\caption{Total Employment Share (scaled by 1992 value) (1992-2011) across Metropolitan Areas}
\label{fig_TotEmpShr_92to11_Intro}

\end{figure}

In contrast, the total employment share of working-age population only experiences a differentially higher boom (2000-2006) the high net-export-growth metropolitan areas (HNEG areas) (-0.077\%) than in the low net-export-growth metropolitan areas (HNEG areas) (-2.184\%). for total employment. No stronger bust (2007-2010) for total employment is found in these areas (-5.477\% in the HNEG areas and -5.616\% in the LNEG areas). Figure (\ref{fig_TotEmpShr_92to11_Intro}) shows these trends. 

%------------------------------------
%------------------------------------

In the second part, using the instrumental variable by \cite{feenstra2019us}, we provide the first causal evidence that the credit expansion in the private-label mortgages (PLMs), rather than the government-sponsored enterprise mortgages (GSEMs), causes the employment boom and bust across metropolitan areas between 1999 and 2010. Intuitively, the gravity model-based instrumental variable (GIV) captures the exogenous parts of net export growth due to (1) rising world demand (supply) reflected in export (import) growth and (2) tariff changes while controlling for US industry-by-year supply (demand) shocks and pre-determined transportation cost. With this IV, we show that credit supply expansion in the private-label mortgages (PLMs) causes a much stronger cycle in house-related employment in the high net-export-growth areas than in the low net-export-growth areas. We term this result as ``housing industry channel", which emphasizes the much stronger employment cycle in residential construction industry, other supporting industries, and the mortgage industry.

Please note that ``causal evidence" means that we only find an incentive (net export growth) in the cross-section that induces credit expansion in private-label mortgages to be much stronger in the high net-export-growth areas. In contrast, We do not find the incentives that cause the aggregate credit expansion between 1999 and 2005. For the causes of the aggregate credit expansion, the literature have documented facts and evidence from financial innovation in securitization (notably the Copula formula by \cite{li2000copula} \citep{salmon2012formula}), international capital flow (``global saving glut" mainly 2003-2007 by \cite{bernanke2005global, bernanke2007global}), mortgage market deregulation (2004 preemption of national banks from state anti-predatory lending laws by the Office of the Comptroller of the Currency \citep{di2017credit}, 2005 Bankruptcy Abuse Prevention and Consumer Protection Act \citep{lewis2023creditor}), and political push (2002-2007 mortgage industry campaign contributions \citep{mian2013political}). Thus, our choice of 1999-2005 as the boom period tries to include all major events documented above and our choice also matches the mortgage boom period commonly used in the literature (see \cite{griffin2021drove} for a review).

\comment{
In the third part, we verify the “household demand channel” by showing that (1) the nontradable sector has a much stronger boom and bust than the tradable sector, and (2) inflation also experiences a boom and bust. Additionally, we use a placebo test to rule out the ``productivity capacity channel", which predicts that an increase in the banking loans to local corporate can cause a productivity increase, showing up in the rise of employment and establishments overall. Against such prediction, we show that bank loans to local small businesses do not increase as quickly as mortgage loans to households. Further, total business establishment rises slower than residential constructions. 
}

In the third part, we take advantage of empirical design and test alternative theories (hypotheses) regarding the business cycle. 
The first alternative one is the ``speculative euphoria hypothesis" by \cite{kindleberger1978manias, minsky1986stabilizingan}, which proposes that initial local economic growth can trigger bank to loose lending standard that fuels speculation by borrowers. The speculation eventually results in an unsustainable boom and ultimate crash. Due to the irrelevance of government-sponsored enterprise mortgages, we only distinguish speculation and credit expansion within the private-label mortgages (non-jumbo). we use the ``non-owner-occupied'' home purchase mortgages as a measure of speculation \citep{gao2020economic} and ``owner-occupied" ones as a measure of pure credit expansion. We use three tests to address the ``speculative euphoria hypothesis". Our first test shows that speculation (99-05) is largely caused by pure credit expansion. we use the residuals from the above regression as my measure of credit-independent speculation. Our second test shows that, compared to the dominant impact of pure credit expansion, credit-independent speculation only has a modest impact on the house employment cycle. Our third test shows that, without credit expansion, speculation does not respond to net export growth.

The second alternative theory is the famous ``real business cycle theory" by \cite{prescott1986theory},  which states that business boom and bust are mainly driven by shocks to the total productivity capacity of corporations. We provide three pieces of evidence against this theory. First, we show that net export growth causes a higher growth in the tradable employment in the high net-export-growth areas in both the boom and the bust periods, rather than only in the boom period. Second, commercial construction employment experiences neither a stronger boom nor a stronger bust in the high net-export-growth areas. Third, the debt (mainly home mortgages) in the household and nonprofit sector rather than the corporate sector experiences a boom and bust cycle.

We address two variants of the real business cycle theory: the ``natural disaster hypothesis" and ``technology shock hypothesis in construction sector". The ``natural disaster hypothesis" proposes that good whether conditions might drive the stronger boom and natural disaster might result in the stronger bust in the high net-export-growth areas. We present two pieces of opposing evidence. First, farm employment in the high net-export-growth areas experiences neither a stronger boom (00-06) nor a sharper bust (07-10). Second, manufacturing employment in these areas experiences a differentially stronger growth in both the boom (00-06) and the bust (07-10) periods. The ``technology shock hypothesis in the construction sector" argues that the technology advancement reduces the construction cost and other sectors increase their building demand. then the mortgage supply only overreacts to this shock and causes the boom and bust. We employ three pieces of opposing evidence to address this hypothesis. First, unlike the residential construction employment, the commercial construction employment experiences neither a stronger boom nor a stronger bust in the high net-export-growth areas (HNEG areas). Second, contrary to the trend in private-label mortgages (non-jumbo), government-sponsored enterprise mortgages experience neither a stronger boom (99-05) nor a stronger bust (05-08) in the HNEG areas. In other words, credit-qualified households do not increase their demand to the ``declining cost of housing" argued by this hypothesis. Third, We do not find a stronger boom (00-06) or a stronger bust (07-10) in manufacturing employment in the HNEG areas.

The third alternative theory is the ``collateral-driven credit cycle theory" developed by \cite{kiyotaki1997credit}. Their model highlights that a small shock to real estate can be amplified by a feedback loop via collateral and can ultimately result in large fluctuations. We offer two pieces of evidence against such a hypothesis. First, the ``collateral channel hypothesis" predicts that the house (collateral) price increase (decrease) shall be followed by mortgage amount increase (decrease). However, we show that private-label mortgages (non-jumbo) increases at the same pace as the house price between 1999 and 2005 and decreases before the house price crash between 2005 and 2010. Second, the hypothesis predicts that the debt-to-GDP ratio by corporations shall experience a much stronger boom and bust than households. However, we show that debt (mainly home mortgages) by household and nonprofit sector experiences a strong boom and bust cycle while no cycle can be found in debt by the business sector.

The fourth alternative theory is the ``The business uncertainty theory" \cite{bloom2009impact}, which emphasizes the uncertainty (that might rise from government policies) can cause firms to pause investment and hiring. We present two pieces of opposing evidence. First, we find commercial construction employment does not experience a differentially stronger drop during the bust period (07-10) in the high net-export-growth areas (HNEG areas). Additionally, tradable sector employment continues to enjoy differentially stronger expansion in the bust period (07-10) in the HNEG areas.

The fifth alternative theory, ``extrapolative expectation theory'' \citep{eusepi2011expectations}, states that local economic growth can drive extrapolation expectation in households that increases mortgage demand (and other consumption), which might overshoot and eventually lead to a stronger business cycle locally. Since government-sponsored enterprise mortgages (GSEMs) are still cheaper than non-jumbo private-label mortgages (PLMNJs) for credit-qualified households \citep{sherlund2008jumbo}, the above theory would predict that both GSEMs and PLMNJs show stronger growth in the high net-export-growth metropolitan areas. However, we only see such a trend in PLMNJs rather than in GSEMs. In addition, we use the evidence supporting exclusion restriction to help address the above ``extrapolative expectation theory''. We show that net export growth impacts house employment only via its impact on PLMNJs, rather than directly. Specifically, we show that in the prior period (91-99) without credit expansion net report growth did not increase house employment.

We conduct various tests to show that our main conclusion is robust. First, we consider the difference between states with and without the anti-predatory lending law (APL). In January 2004, the Office of the Comptroller of the Currency (OCC) preempted national banks from state-level anti-predatory lending laws \citep{di2017credit}. We find that preempted states experience neither a stronger boom nor a stronger bust in house-related employment. After controlling the state difference in APL, our main conclusion holds for all states. Second, we consider state-level differences in recourse law \citep{ghent2011recourse}. We do not find non-resource states experiencing a stronger boom or a stronger bust in house-related employment. After controlling for state-level differences in recourse law, our main conclusion is robust for all states. Third, I investigate state-level differences in the judicial requirement of foreclosure laws \citep{mian2015foreclosures}. Though non-judicial states experience a stronger boom and a stronger bust in house-related employment, our major conclusion still holds for all states. Fourth, our major conclusion still holds after controlling sand-state dummy. Fifth, our main conclusion is robust to the inclusion of state capital gain tax as a control.

%--------------------------------------------------
%\subsection{Contribution}
%--------------------------------------------------

\noindent \textbf{Contribution} We illustrate our contribution to the literature in four parts. First, we document two new empirical facts in the literature: (1) a stronger boom in total employment and (2) a stronger boom and a stronger bust in refined-house employment in the high net-export-growth metropolitan areas than in the low net-export-growth ones in the U.S. from 1999 to 2010.

In addition, our paper makes unique contributions to the literature on the causal impact of credit expansion on the U.S. 1999-2010 business cycle in five dimensions. First, our paper explains the differential business cycles across metropolitan areas (MSA) within the USA, a new dimension in the literature. In contrast, most papers on business cycle focus on the cross-country differences \cite{mian2017household, Muller2023credit} whereas some focus on differences across states \citep{di2017credit, choi2016sand, mian2015foreclosures}. Second, our unique method has three major advantages. The first advantage is that we capture the long-term incentive of credit expansion by building on the ``economic base theory”. This theory states that the tradable sector (for which we proxy it by net export growth) determines the local economy in the long term. The second advantage is that we can take advantage of the state-of-art instrumental variable by \cite{feenstra2019us} from International Economics for causal inference. The third advantage is that the IV approach by \cite{feenstra2019us} can cover the entire period of credit boom period (99-05) consisting of most major events in mortgage expansion. These events consist 2003-2007 ``global saving glut" by \cite{bernanke2005global, bernanke2007global}, financial innovation in securitization (particularly the Copula formula by \cite{li2000copula} \citep{salmon2012formula}),  mortgage market deregulation (2004 preemption by the Office of the Comptroller of the Currency \citep{di2017credit}, the 2005 Bankruptcy Abuse Prevention and Consumer Protection Act \citep{lewis2023creditor}), and political lobby (2002-2007 mortgage industry campaign contributions \citep{mian2013political}). In contrast, other causal studies mostly focus on one single event \citep{di2017credit,lewis2023creditor}. The above three advantages enable us to delve into the boom period preceding the bust period. In contrast, many previous papers focus on the bust period only \citep{mian2013household,mian2014explains}. Third and most importantly, we apply our method to test the relevance of almost all major theories (hypotheses) on the business cycle, including the speculative euphoria hypothesis, the real business cycle theory, the collateral-driven credit cycle theory, the business uncertainty theory, and the extrapolative expectation theory. No other paper can test as many theories and hypotheses as us. Our extensive tests add knowledge to the cause and the mechanism of the Great Depression and can advise future regulatory designs aimed at preventing it from happening again. Fourth, we also take advantage of a key legal constraint to distinguish the role of government-sponsored enterprise mortgages and private-label mortgages. Most other papers do not distinguish these two mortgages.\footnote{The only two exceptions that separate the role of government-sponsored enterprise mortgages and private-label mortgages are \cite{justiniano2022mortgage,mian2022credit}, though they do not show the irrelevance of government-sponsored enterprise mortgages to the differential business cycle across metropolitan areas.} Fifth, we show that housing-related employment is central to the business boom and bust cycle (the housing industry channel).

These five dimensions distinguish our paper from other papers in the literature. The most related paper is \cite{di2017credit}. They exploit the 2004 preemption of national banks (rather than state-chartered depository institutions and independent mortgage companies) from state antipredatory-lending laws (APL) by the Office of the Comptroller of the Currency (OCC). They find that, compared to non-APL states, preempted states experienced stronger growth in national banks’ credit expansion, house price, and nontradable employment in 2004-2006 but also stronger declines in these outcomes subsequently. Our paper differs from their paper in ******** angles. First, we explain the cross-metro differential business cycles while they explain the cross-state differential business cycles. Table (\ref{table_Robust.APLvsNone.RefineHouse.D00t06.D07t10}) shows that our cross-metro results are still very strong after controlling for the anti-predatory lending laws at the state level. Second, our empirical design builds on the “economic base theory” so that net export growth captures the long-term incentives of credit expansion in mortgages. One important advantage of our IV is that it covers the boom period (1991-2005) with almost all major events related to the mortgage market. In comparison, Their paper captures a single and perhaps temporary legal change as the incentive, thus ignoring technology innovation in securitization \citep{salmon2012formula}, 2005 Bankruptcy Abuse Prevention and Consumer Protection Act \citep{lewis2023creditor}, and other important changes happened or started before 2004 (such as the international capital inflow 2003-2007 termed as ``global saving glut" by \cite{bernanke2005global}). Third, they and we both provide causal evidence, but our IV strategy can do more. Specifically, our IV strategy can decompose the household speculation (non-owner-occupied private-label purchase mortgages) and show with three tests that credit expansion is a necessary condition for speculation. Fourth and most importantly, our research design and comprehensive data analyses allow us to go further and test the relevance of major alternative hypotheses. In contrast, their difference-in-difference research design can do little to test alternative hypotheses with data from different periods. Fifth, we take advantage of a key legal constraint to distinguish the roles of government-sponsored enterprise mortgages and private-label mortgages whereas they use the 2004 OCC's preemption to separate national banks from other mortgage institutions.

In spirit, our paper is similar to \cite{mian2020does}, which studies the state-level differential deregulation in the banking sector in the late 1970s and 1980s in the United States. They provide causal evidence that credit expansions amplify the business cycle via the household demand channel: credit expansions cause a differential higher rise in nontradable employment and nontradable goods. Different from their period, our paper focus on 1999-2010 US business cycle. Two other papers focus on the business downturn (2007-2009 or 2006-2009) and attribute the cause to the deterioration of house prices in household balance sheets. \cite{mian2013household} document that the deterioration of household balance sheets significantly predicts the sharp decline in U.S. consumption across areas between 2006 to 2009. \cite{mian2014explains} show that deterioration in household balance sheets predicts the sharp decline in U.S. employment across areas between 2007-2009 mainly via the ``household demand channel". \cite{favara2015credit} exploit the 1994-2005 state-level branching deregulation and show that early-deregulating states experience stronger growth in mortgage and house prices. However, their figure 4 shows that deregulation cannot explain well the house price growth from 2002 to 2005. Importantly, they do not study the period from 2005 to 2009.

Unlike our focus on the differential business cycle across metropolitan areas within a country, most empirical business cycle papers focus on cross-country differential business cycles. \cite{schularick2012credit} use 14 countries over the years 1870-2008 and show that credit growth is a powerful predictor of the financial crisis. \cite{mian2017household} use 30 countries from 1960 to 2012 and document that an increase in household debt predicts lower GDP growth and higher unemployment in the medium term. \cite{greenwood2022predictable} employ 42 countries from 1950 through 2016 and document that combination of rapid credit and asset price growth in the past three years in either the nonfinancial business or household sector is significantly associated with financial crisis in the next four years. \cite{Muller2023credit} use 117 countries between 1940 and 2014 to show that credit expansion to the nontradable sector is systematically associated with subsequent growth slowdown and financial crisis while credit expansion to the tradable sector can systematically predict sustainable output and productivity growth without a higher probability of financial crisis.

Third, we show that house-related industries dominate the business cycle. Previous studies do not study the comprehensive list of housing-related industries as we do, including the residential construction, other supporting industries, and mortgage banker industries. Thus we emphasize the central role of housing in the credit-induced business cycle.

%--------------------------------------------------------------
%--------------------------------------------------------------
% This is the end of the entire section 
%--------------------------------------------------------------
%--------------------------------------------------------------

%----------------------------------------------------------------------

%----------------------------------------------------------------------
% section 4: Research Desgin

%--------------------------------------------------------------------------------------
% new section
%--------------------------------------------------------------------------------------

%\clearpage

\section{Research Design}\label{sec:ResearchDesign}
In this section, we illustrate our empirical research design, which is similar to \cite{li2024credit}, that can provide causal evidence for the business cycle. First, we measure the fundamental incentive for mortgage credit by operationalizing the central ideal of the ``Economic Base theory" \citep{tiebout1962community}. Specifically, we use metropolitan exposure to net export growth of manufacturing industries (thereafter ``net export growth") as a proxy of the growth of local tradable sector, which by theory is the fundamental force of long-term economic growth. To overcome the endogeneity issue of OLS, we use an instrumental variable approach from International Economics by \cite{feenstra2019us} as our identification strategy.

\subsection{Operationalize the ``Economic Base Theory"}
In this subsection, we operationalize the key idea of ``economic base theory" by using metropolitan exposure to net export growth of manufacturing industries (thereafter ``net export growth") as a proxy of the growth of local tradable sector. This net export growth measure addresses the requirement of treatment variable in the introduction: enough area coverage (U.S. mainland), enough time coverage (91-09), enough geographic variation, and a feature of persistence.

Regional economics defines the ``economic base sector'' (tradable sector) as economic activities that a local area offers for the areas beyond its boundaries \citep{tiebout1962community, nijkamp1987regional}. The tradable sector brings wealth into the local area and majority of the money will be circulated locally via a multiplier effect through nontradable sector. By this theory, the economic base is the most important driving force for local economic growth in the long term \citep{nijkamp1987regional,thrall2002business,ling2013real}. Therefore, the tradable sector growth can predict demand-side factors that shape the long-term business growth, such as employment growth, household income growth, and population growth.\footnote{\cite{olsen1987demand} provides a survey on the demand factors of housing and business, including the three factors mentioned above.} By the above theory, credit expansion would be stronger in areas with stronger tradable sector growth due to at least two reasons: (1) higher foreclosure price of house and (2) higher income growth that can be recoursed by lenders given default.

Perfect measurement of local tradable sector growth requires census-style data to include the accounting data of all firms in the tradable sector, which is not available. Instead, following \cite{li2024credit}, we use the local employment composite as a proxy for the composite (share) of local tradable sector. Second, we use the time-series change in U.S. net export growth in manufacturing industries as a proxy for the relative growth (shift) at the industry level. Aggregating the above two proxies (share and shift) can achieve a good measure of the relative growth of the local treatable sector across metropolitan areas.\footnote{We acknowledge that the use of manufacturing does not include several other factors in the tradable sector: college town, retirement community, other tradable goods industries (e.g., natural resource), and tradable services (e.g., information technology and medical sector).} To account for the work commuting within local areas, we aggregate net export growth at the metropolitan level. 

We implement the above approach in two steps in data. First, in industry $g$ in year $t$, net export measure is defined as $ \text{NetExp}_{g,t}= \frac{Export_{g,t} - Import_{g,t}}{Y_{g,91}} $, where $Export_{g,t}$, $Import_{g,t}$, and $Y_{g,91}$ are US export, import and domestic production value in industry $g$ in year $t$ (or 1991), respectively. All terms are converted to 2007 US dollars. US domestic production in 1991 serves as the scaling factor. Year 1991 is chosen to avoid potential response of domestic production to trade in later periods.\footnote{The same choice of domestic production in 1991 as the denominator is taken by other papers like \cite{barrot2022import}.} Second, we use local employment composite (share) to aggregate net export growth at the metropolitan level across period: 
\begin{equation}\label{equ:NEG_m}
    \triangle_{t_{1},t_{2}}\text{NetExp}_{m} = \sum_{g} \big[  (L_{m,g,t_{0}}/L_{m.t_{0}}) * (\text{NetExp}_{g,t_{2}} - \text{NetExp}_{g,t_{1}}) \big]
\end{equation}
where $L_{m,g,t_{0}}$ and $L_{m.t_{0}}$ are the employment of industry $g$ and total employment, respectively, in metropolitan area $m$ in year $t_{0}$. I choose $t_{0} = t_{1}-1$ to make sure the employment share is pre-determined to the trade measure so that all changes in net export growth ($\triangle_{t_{1},t_{2}}\text{NetExp}_{m}$) is driven by changes in trade measures instead of employment composite. Year $t_{1}$ to $t_{2}$ is the period of interest.

The above net export growth satisfies the requirements of treatment variable. First, the net export growth has enough area coverage (U.S. mainland) and enough time coverage (91-09). Second, its geographic variation arises from the fact that the local tradable sector tends to cluster within several related industries due to economies of scale. Internal economies of scale increase the size of local manufacturing firms \citep{worldbank2009ch4} while external economies of scale attracts firms in the same and related industries to the clusters \citep{krugman2018internationalCh7,worldbank2009ch4}. Internal economies of scale include input purchase at a volume discount, fixed cost of plant operating, and learning in operation. External economies of scale consist specialization of suppliers, labor market pooling, and knowledge spillovers. Third, the long-term reduction in transportation costs and increased labor mobility in the last century strengthened the tradable sector clustering. Fourth, many global events in the 1980s and 1990s promoting international trade also added to the local industry clustering at the global level. These events include reforms and opening in large emerging countries, huge tariff reduction, the Dissolution of the Soviet Union in 1991, the establishment of North American Free Trade Agreement (NAFTA) in 1994, and the creation of World Trade Organization (WTO) in 1995. See \cite{worldtradereport2007} for a review. Fifth, the persistence feature of net export growth at the metropolitan level arises from three dimensions. At the industry level, both comparative advantages (due to technology level or natural endowment) and horizontal specialization (due to economies of scale) across nations make trade patterns persistent over time. At the local area level, clusters formation with huge costs is unlikely to change rapidly in a short period of time. At the individual level, human capital accumulation and job reallocation across industries are very difficult both locally and remotely.

\subsection{OLS and Its Bias}
In this subsection, we illustrate the potential bias from OLS regression specification. we begin with OLS specification that relates the employment growth to the growth of private-label mortgages (non-jumbo) (PLMNJ) at the county level:
\begin{equation}\label{eq:OLS_HP_PLMNJ}
\begin{split}
     \triangle_{00,06} TotalEmpShr_{c} = \beta * \triangle_{99,05} Ln(\text{Private-Mort}_{c}) + \gamma* Controls_{c}  + \epsilon_{c} 
\end{split}
\end{equation}

\noindent Here, the dependent variable $\triangle_{00,06} TotalEmpShr_{c}$ is the change of the total employment share in working-age population at county $c$ 00-06. The independent variable $\triangle_{99,05} Ln(PLMNJ_{c})$ is the growth rate of the dollar amount (07USD) of private-label mortgage (non-jumbo) (PLMNJ) at county $c$ 99-05.

\noindent \textbf{Omitted Variable Problems} 
Potential omitted variables could bias $\beta$. For example, the rapid net export growth and, hence, mortgage growth in 1999-2005 has been anticipated by employees in Silicon Valley so that house price and total employment grow before 1999 to reflect such expectation. In this case, $\beta$ could be biased downward because some of the effect of the mortgage shows up in total employment growth in earlier period, reducing the total employment growth 2000-2006.

\subsection{Gravity Model-based Instrumental Variable}\label{subsec:GIV_exports}
To overcome the endogenous concern regarding OLS specification, we use gravity model-based instrumental variable introduced by \cite{feenstra2019us} for net export growth. While they construct IV for exports and imports separately, we combine both as an IV for net exports. For illustration, we show the model of exports while leaving the model of imports in Appendix Section \ref{subsec:GIV_imports}. 

To form an instrumental variable for US exports at the industry-year level, \cite{feenstra2019us} builds on the key idea that eight other high-income countries' exports can instrument for the US exports since they both capture the world's rising demand. In addition, they incorporate tariff changes, which are commonly believed to be exogenous to foreign firms. Further, this method corrects for the supply shocks in the home country by employing a fixed effect to remove them.

To predict US export, the gravity model-based IV starts from a symmetric constant-elasticity equation by \cite{romalis2007nafta} for export: 
\begin{equation}{\label{eq:gravity_export}}
    \frac{X^{US,j}_{s,v,t}}{X^{i,j}_{s,v,t}} = \Bigg( \frac{w^{US}_{s,t}*d^{US,j}*\tau^{US,j}_{s,t}}{w^{i}_{s,t}*d^{i,j} * \tau^{i,j}_{s,t}} \Bigg) ^{1-\sigma}
\end{equation}
In the formula above, $X^{US,j}_{s,v,t}$ is US export to country $j$ in industry $s$ in product variant $v$ in year $t$. By the same notation, $X^{i,j}_{s,v,t}$ represents export from country $i$ to $j$. $w^{US}_{s,t}$ and $w^{i}_{s,t}$ are the relative marginal cost of production in the industry $s$ in the US and country $i$, respectively. $\tau^{US,j}_{s,t}$ and $\tau^{i,j}_{s,t}$ are the \textit{ad valorem} import tariff imposed by country $j$ on exports from the US and country $i$, respectively. $d^{US,j}$ and $d^{i,j}$ are the bilateral distance and other fixed trade costs from US to country $j$ and from country $i$ to country $j$, respectively. Lastly, $\sigma$ is the constant elasticity of substitution ($\sigma > 1$). 

The intuition of this gravity-style model is relatively straightforward. Competing with country $i$, US exports to country $j$ are decreasing in the ratio of bilateral distance, in the ratio of relative marginal cost, and in the ratio of \textit{ad valorem} total import tariff. 

Assume that there are $N^{i}_{s,t}$ identical product varieties exported by country $i$ to the country $j$ in the industry $s$ and year $t$, \cite{feenstra2019us} re-arranges this equation, multiply both sides with $N^{i}_{s,t}$, and sum over countries $i \neq US$:
\begin{equation*}
    X^{US,j}_{s,v,t}*\sum_{i\neq US} \big[ N^{i}_{s,t}(w^{i}_{s,t} d^{i,j})^{1-\sigma} \big] = (w^{US}_{s,t}d^{US,j}\tau^{US,j}_{s,t})^{1-\sigma} * \sum_{i\neq US} \big[N^{i}_{s,t}X^{i,j}_{s,v,t}  (\tau^{i,j}_{s,t})^{\sigma-1}\big]
\end{equation*}

As the above equation holds for any countries $i \neq US$, one can choose a set of countries that have similar economic conditions (so that they are close competitors of US exports) to make the prediction more accurate. \cite{feenstra2019us} use the eight high-income countries used by \cite{autor2013china}. 

Again, one can multiple $N^{US}_{s,t}$ (number of variants of products in US exports) on both sides and denote the sectoral exports $X^{US,j}_{s,t} \equiv X^{US,j}_{s,v,t}N^{US}_{s,t}$ and $X^{i,j}_{s,t} \equiv X^{i,j}_{s,v,t}N^{i}_{s,t}$, then one can get
\begin{equation*}
    X^{US,j}_{s,t}*\sum_{i\neq US} \big[ N^{i}_{s,t}*(w^{i}_{s,t} d^{i,j})^{1-\sigma} \big] = N^{US}_{s,t}*(w^{US}_{s,t}d^{US,j}\tau^{US,j}_{s,t})^{1-\sigma} * \sum_{i\neq US} \big[X^{i,j}_{s,t} * (\tau^{i,j}_{s,t})^{\sigma-1}\big]
\end{equation*}

With a few re-arrangements, one can get the formula for $ X^{US,j}_{s,t}$:
\begin{equation}
    X^{US,j}_{s,t} =   \frac{N^{i}_{s,t}*(w^{US}_{s,t}d^{US,j}\tau^{US,j}_{s,t})^{1-\sigma}}{\sum_{i\neq US} \big[ N^{i}_{s,t}*(w^{i}_{s,t} d^{i,j})^{1-\sigma} \big]}  * \bigg( \sum_{k\neq US} X^{k,j}_{s,t} \bigg)  * \sum_{i\neq US} \bigg[  \frac{ X^{i,j}_{s,t} }{\sum_{k\neq US} X^{k,j}_{s,t}}  * (\tau^{i,j}_{s,t})^{\sigma-1} \bigg]
\end{equation}
 
Note in the above formula, we multiply and divide by $\sum_{k\neq US} X^{k,j}_{s,t}$ to prepare for the following regression setup. Now we can take logs of the above equation and get the regression-style formula:
\vspace{-1mm}
\begin{equation} \label{eq:exp_gravityRegression}
\resizebox{0.92\textwidth}{!}{%
\begin{math}
\begin{aligned}
\lnb{X^{US,j}_{s,t}} & = \underbrace{\lnb{\sum_{k\neq US}X^{k,j}_{s,t}} }_{\text{Term 0}} +  \underbrace{\lnb{N^{US}_{s,t}(w^{US}_{s,t})^{1-\sigma}}}_{\text{Ind-Year FE: } \alpha^{US}_{s,t}} + \underbrace{(1-\sigma)\lnb{d^{US,j}}}_{\text{Importing-country FE: } \delta^{US,j}}  \\
& + \underbrace{ (1 - \sigma) \lnb{\tau^{US,j}_{s,t}} }_{\text{Term 1}} + \underbrace{ (\sigma-1) \lnb{ \Bigg\{  \sum_{i\neq US} \bigg[ \frac{ X^{i,j}_{s,t} }{\sum_{k\neq US} X^{k,j}_{s,t}} (\tau^{i,j}_{s,t})^{\sigma -1} \bigg] \Bigg\} ^{\frac{1}{\sigma-1}}  }}_{\text{Term 2:} (\sigma -1)\lnb{T^{j}_{s,t}} } + \epsilon^{j}_{s,t} \\ 
\end{aligned} 
\end{math}
} %end of \scalemath \resizebos
\end{equation}
Now, We can see that US exports to the country $j$ in the industry $s$ year $t$ can be divided into six terms. ``Term 0'' represents the exports from eight other high-income countries to the country $j$, which reflects the world demand. The second term $\alpha^{US}_{s,t}$, which represents the US supply shocks by industry and year, is potentially endogenous. We remove this term by the US industry-year fixed effects. The third term $\delta^{US,j}$ reflects the predetermined distance from the US to the destination market $j$ and all other industry- and year-invariant trade costs. We remove it by importing country fixed effects. ``Term 1" is the tariffs on US exports imposed by destination country $j$, which is out of control by US firms. I retain this term to capture the shocks from tariffs. ``Term 2" is the weighted average tariffs on non-US exports charged by destination country $j$, which is out of control of US firms. Intuitively, when this weighted average tariffs on non-US exports rise, country $j$ will import more from the US as substitutions. I retain this term to reflect this substitution effect. Lastly, the term $\epsilon^{j}_{s,t} = - \lnb{ \sum_{k\neq US} [ N^{i}_{s,t}(w^{i}_{s,t}d^{i,j})^{1-\sigma} ] } $ is unobserved and remains in the regression error term. 

After the above regression, we can construct predicted US exports that are isolated from supply shocks and the predetermined distance:
\begin{equation} \label{eq:gravityPreUSExp}
\lnb{\widehat{X^{US,j}_{s,t}}} = \lnb{\sum_{k\neq US}X^{k,j}_{s,t}} + \hat{\beta_1} *\lnb{\tau^{US,j}_{s,t}} + \hat{\beta_2}* \lnb{T^{j}_{s,t}}
\end{equation}

\subsection{Data Implementation}\label{subsec:Data_Implementation}
There are four detailed steps when implementing the above procedures in data to get net export and its GIV at the metropolitan level across periods. First, we estimate Eq (\ref{eq:exp_gravityRegression}) at the 6-digit HS industry level (5673 industries) and derive predicted US export (isolated from supply shocks) by Eq (\ref{eq:gravityPreUSExp}). Second, we aggregate predicted US exports across importing countries and crosswalk the 6-digit HS code (5673 industries) to the 4-digit revised SIC system (392 manufacturing industries) by the crosswalk with weights by \cite{acemoglu2016import}. We end up with predicted US exports to the world at the industry $g$ year $t$ level. We perform this aggregation by $ \widehat{X^{US}_{g,t}} = \sum_{s\in g}\sum_{j} \widehat{ X^{US,j}_{s,t} }$. In a similar fashion, we get predicted US imports from the world $ \widehat{M^{US}_{g,t}}$. Third, we derive the gravity model-based instrumental variable for net export at the industry-year level 
\begin{equation}\label{equ:givNEP_gt}
\text{givNetExp}_{g,t}^{US}= \frac{\widehat{X^{US}_{g,t}} - \widehat{M^{US}_{g,t}} }{ Y_{g,91} }
\end{equation}
where $Y_{g,91}$ is US domestic production in year 1991. Fourth, we use employment data from County Business Pattern to aggregate $givNEP$ at the metropolitan level across periods. 
\begin{equation}\label{equ:delta_givNEP_m}
    \triangle_{t_{1},t_{2}}\text{givNetExp}_{m} = \sum_{g} \big[  (L_{m,g,t^{\prime}_{0}}/L_{m.t^{\prime}_{0}}) * (\text{givNetExp}_{g,t_{2}} - \text{givNetExp}_{g,t_{1}}) \big]
\end{equation}
where $L_{m,g,t^{\prime}_{0}}$ and $L_{m.t^{\prime}_{0}}$ are the employment counts of industry $g$ and total employment, respectively, in metropolitan area $m$ in year $t^{\prime}_{0}$. Following \cite{acemoglu2016import}, we choose $t^{\prime}_{0} = t_{1}-3$ to prevent potential covariance due to data error between the dependent variable and the independent variable.

\noindent \textbf{Relevance Condition} Conceptually, the gravity model-based IV captures the exogenous part of net export growth due to (1) increasing world demand reflected in net export growth by other eight high-income counties and (2) tariff changes, after removing the US industry-year supply shocks. This relevance condition of this GIV is satisfied because it starts from the general equilibrium model by \cite{romalis2007nafta} and is derived from specific decomposition by \cite{feenstra2019us}. We will test this condition in data by robust F-statistics \citep{kleibergen2006generalized} and efficient F-statistics \cite{olea2013robust}. 

\noindent \textbf{Exclusion Restriction} We have exclusion restrictions at two levels. The first level refers to the gravity model-based instruments by \cite{feenstra2019us}. They have already removed supply-side shocks via industry fixed-effect in predicted US exports and demand-side shocks via industry fixed-effect in predicted US imports. Thus, exclusion restriction holds for the gravity model-based IV. The second level refers to our regression specification Eq (\ref{eq:OLS_HP_PLMNJ}), where the exclusion restriction means that net export growth can only affect house prices via private-label mortgages. Section \ref{sec:ExclusionRestriction} provides evidence supporting such claim. 
 
As with all instrumental variable estimates, our 2SLS estimates capture the local average treatment effects on compilers \citep{imbens1994identification}. In our setting, compilers are metropolitan-by-year observations that experience more industry-by-year US net export to the world following increases in gravity model-based predicted US net export.
 
%-------------------------------------------------------
%-------------------------------------------------------

%--------------------------------------------------------------------------------------
% end of the entire section 
%--------------------------------------------------------------------------------------

%----------------------------------------------------------------------

%----------------------------------------------------------------------
% section 3: Data Source

%--------------------------------------------------------------------------------------------------------------------------------------------------------------------------

%\clearpage
\section{Data Sources}\label{sec:data}
%--------------------------------------------------------------------------------------------------------------------------------------------------------------------------

I combine several datasets to study how credit expansion in mortgages causes a stronger business boom and bust in the high-net-export-growth areas (HNEG areas). The International trade and tariff data are new to the literature on business cycles, enabling me to use the gravity model-based instrumental variable \citep{feenstra2019us} as my identification strategy. 

%--------------------------------------------------------------------------------------------------------------------------------------------------------------------------
\subsection{Data for Net Export Growth}
%---------------------------------------------------------------------------------------------------------------------------

%---------------------------------------------------------------------------------------
\noindent \textbf{Trade Flow Data} International trade flow data are from the UN Comtrade Database.\footnote{The website of UN Comtrade Database is \url{https://comtrade.un.org/data/}.} This database contains bilateral imports and exports data for detailed products recorded under the six-digit Harmonized Commodity Description and Coding System (HS code). To convert trade value to 2007 USD dollar, I apply the Personal Consumption Expenditures Chain-type Price Index by Federal Reserve in St. Louis.\footnote{Federal Reserve in St. Louis provides Personal Consumption Expenditures Chain-type Price Index at \url{https://fred.stlouisfed.org/series/PCEPI}}. To crosswalk these trade data from a six-digit HS system to a four-digit SIC system, I use the crosswalk file and revised SIC system (392 manufacturing industries) in \cite{acemoglu2016import}.\footnote{This crosswalk file is also available from Prof. David Dorn's website: \url{https://www.ddorn.net/data.htm}. The further refined SIC system (392 manufacturing industries) and crosswalk file are available from \cite{acemoglu2016import}.}  \footnote{To make sure my calculation is correct, I calculate China's exports to the US and eight other high-income countries at the industry-year level from 1991 to 2007 and compare them to data provided by David Dorn's website. The trade data is within the section [D] Industry Trade Exposure via his data page \url{https://www.ddorn.net/data.htm}. Correlations between my calculation and his corresponding data are 0.9983 for China's export to the US and 0.9973 for China's export to eight other high-income countries.}

%---------------------------------------------------------------------------------------
\noindent \textbf{Tariff Data} Bilateral tariff schedule data at five-digit SITC product level between 1984 to 2011 is from \cite{feenstra2014international}.\footnote{The original data are collected from the TRAINS, IDB databases, and various other resources, with multiple cleaning steps and filling in missing values by other resources. The data work is described in Appendix C in \cite{feenstra2014international}.} To crosswalk tariff data from a five-digit SITC system to a six-digit HS system, I follow the methods in \cite{feenstra2019us}. Specifically, I first use crosswalk files from the Trade Statistics Branch (TSB) of the United Nations Statistics Division to convert the HS 2007 version to the HS 2002 version.\footnote{The crosswalks files between different HS versions are available from the UN Comtrade database: \url{https://unstats.un.org/unsd/trade/classifications/correspondence\%2Dtables.asp}.} Then I use a crosswalk from \cite{feenstra2005world} to match each six-digit HS code to one five-digit SITC2. When one six-digit HS code is matched to multiple SITC2 codes, I follow \cite{feenstra2019us} and use the one with the highest value share.

%---------------------------------------------------------------------------------------
\noindent \textbf{Manufacturing Production Data} I use the US 4-digit SIC manufacturing industry total domestic production (vship) in 1991 as the denominator to scale the trade value. Such data comes from NBER-Center for Economics Studies (NBER-CES) Manufacturing Industry Database. 1991 is the first year in analysis so production is unlikely to respond to trade change afterward.\footnote{This choice of scaling is also used by \cite{barrot2018import}.}

%---------------------------------------------------------------------------------------
\noindent \textbf{Manufacturing Employment Data} Employment data across detailed manufacturing industries at the county-by-year level in the U.S. comes from County Business Patterns (CBP) Database by U.S. Census.\footnote{County Business Patterns Database is available here: \url{https://www.census.gov/programs-surveys/cbp/data/datasets.html}.} Following the method by \cite{acemoglu2016import}, I use manufacturing employment data to aggregate net export growth and its IV at the metropolitan areas across periods. Refer to Section \ref{subsec:Data_Implementation} for implementation details.

%--------------------------------------------------------------------------------------------------------------------------------------------------------------------------
\subsection{Data for Mortgages and House Prices}
%--------------------------------------------------------------------------------------------------------------------------------------------------------------------------

%---------------------------------------------------------------------------------------
\noindent \textbf{Mortgage Data} Detailed loan-level mortgage data are from the Home Mortgage Disclosure Act (HMDA) database.\footnote{The Consumer Financial Protection Bureau (CFPB) provides 2007-2017 HMDA data"  \url{https://www.consumerfinance.gov/data\%2Dresearch/hmda/historic\%2Ddata/}. The Federal Financial Institutions Examination Council (FFIEC) maintains 2017-2021 HMDA data at \url{https://ffiec.cfpb.gov/data\%2Dpublication/2021}. CFPB offers links of 1990-2006 HMDA to the National Archives at \url{https://github.com/cfpb/HMDA_Data_Science_Kit/blob/master/hmda_data_links.md}.} In 1975, Congress enacted HMDA to improve public reporting of mortgage loans. Any financial institution must report HMDA data to its regulator if it meets certain criteria, such as a threshold for assets and whether the institution has a home office or branch in a Metropolitan Statistical Area (MSA). This annual database contains information on loan applications, borrower demographics, lender identifiers, and loan specifics such as purpose, amount, and location. The HMDA database provides near-universal coverage of the mortgage market. \cite{avery20102008} confirm that in 2008, commercial banks filing HMDA carried 93\% of the total mortgage dollars outstanding on commercial bank portfolios.\footnote{Although lenders with offices only in non-metropolitan areas are not required to file HMDA, 83.2\% of the population lived in metropolitan areas in 2006 \citep{dell2012credit}.}

We use the following filtering criteria for HMDA data. First, we keep originated loans and delete applications that are denied, withdrawn, or not accepted. Second, for loan types, we keep conventional and Federal Housing Administration-insured (FHA-insured) loans and delete loans insured by the Veterans Administration, Farm Service Agency, or Rural Housing Service. Third, for loan purposes, we mainly use home purchase mortgages for most empirical tests and add refinancing mortgages in some robustness tests. Fourth, for occupancy types, we keep both non-owner-occupied and owner-occupied loans and treat ``not applicable" as owner-occupied.\footnote{Based on the HMDA manual (\url{https://www.ffiec.gov/hmda/pdf/1998guide.pdf}), this ``not applicable" occupancy very likely refers to a multifamily dwelling where the borrower lives in. In terms of numbers, this ``not applicable" occupancy is only around 3.5\% of the number of non-owner-occupied loans and 0.59\% of the number of owner-occupied loans as of 2007.}

We use the HMDA database to construct loan volume (number and dollar amount) at the county-by-year level for government-sponsored enterprise mortgages (GSEM) and private-label mortgages (PLM). Based on the HMDA examination procedures, an institution is required to report the type of entities that purchase the loans that are originated (or purchased) and then sold in the same calendar year.\footnote{See ``Home Mortgage Disclosure Act
Examination Procedures" at \url{https://www.federalreserve.gov/boarddocs/caletters/2009/0910/09\%2D10_attachment.pdf}. These procedures imply that the HMDA can potentially under-estimate the mortgages that are sold as GSEM and PLM since the mortgages originated near the end of the calendar year need some time to be sold. Nonetheless, this potential underestimation can only bias my results to zero.} I follow the classification of PLMs and GSEMs by \cite{mian2022credit}.\footnote{\cite{mian2022credit} group five categories as PLMs when a mortgage is sold: (1) to a commercial bank, savings bank, or savings affiliation affiliate, (2) into private securitization, (3) to an affiliate institution, (4) to a life insurance company, credit union, mortgage bank, or finance company, and (5) to other types of purchasers.}

%---------------------------------------------------------------------------------------
\noindent \textbf{Conforming Loan Limits Data} I obtain conforming loan limits by year and county from Federal Housing Finance Agency.\footnote{Before and in 2007, conforming loan limits are set only at the national level: \url{https://www.fhfa.gov/AboutUs/Policies/Documents/Conforming\%2DLoan\%2DLimits/loanlimitshistory07.pdf}. From 2008 onward, conforming loan limits are set by year and by county: \url{https://www.fhfa.gov/DataTools/Downloads/Pages/Conforming\%2DLoan\%2DLimit.aspx}. } Conforming loan limits are, in general, different for 1-unit, 2-unit, 3-unit, and 4-unit dwellings in each year (and county). Since the HMDA data does not include information on the number of units in a home between 1991 and 2009, We use the 1-unit conforming loan limit for all mortgages. Thus, our conservative measure of non-jumbo mortgages can help avoid a potentially upward bias in results.

%---------------------------------------------------------------------------------------
\noindent \textbf{Consistent Counties Covered by HMDA} Following the suggestion by \cite{avery2007opportunities}, we infer counties that are consistently covered by HMDA based on the scope of metropolitan areas defined and updated by the U.S. Office of Management and Budget. Detailed steps are the same in \cite{li2024credit}. Consequently, We get 800 ``HMDA consistent counties after 1996" and 712 ``HMDA consistent counties after 1990".

%---------------------------------------------------------------------------------------
\noindent \textbf{U.S. House Price Data} We obtain the U.S. house price index data based on repeated sales at the county and ZIP levels from the Federal Housing Finance Agency.\footnote{The data are available at \url{https://www.fhfa.gov/DataTools/Downloads}.FHFA working paper \cite{bogin2019missing} describes the construction of the index and tests its accuracy via various methods. }

%---------------------------------------------------------------------------------------
\noindent \textbf{Merge House Price and Mortgage Data} For both figure and regression analysis regarding house prices, I require that the counties are covered by both house price data and mortgage data. The merged data set contains fewer counties compared to mortgage data since house price data covers fewer counties.

%--------------------------------------------------------------------------------------------------------------------------------------------------------------------------
\subsection{Data for Employment and Business}
%--------------------------------------------------------------------------------------------------------------------------------------------------------------------------
%---------------------------------------------------------------------------------------
\noindent \textbf{Aggregate Debts for Sectors} I obtain annual debt data for households, business (corporate and non-corporate), and government (federal and local) from the Federal Reserve.\footnote{The debt data is available here: \url{https://www.federalreserve.gov/releases/z1/dataviz/z1/nonfinancial_debt/table/}.} Such data also contains subcategories for household debt: mortgages, consumer credit, and other liability.

%---------------------------------------------------------------------------------------
\noindent \textbf{BEA Employment Data} I obtain annual employment data at the county level from the U.S. Bureau of Economic Analysis (BEA) for analysis on total employment because such data include both (1) wage and salary employment and (2) proprietor employment (self-employment).\footnote{The BEA employment data is available at \url{https://apps.bea.gov/regional/downloadzip.cfm}. In the category ``Personal Income (State and Local)", "CAEMP25S" contains data from 1969 to 2000, while "CAEMP25N" contains data from 2001 and onward.} This coverage is better than County Business Pattern employment data, which does not contain self-employment not working in commercial establishments.

%---------------------------------------------------------------------------------------
\noindent \textbf{CBP Employment Data} I get employment data in house-related and other industries from the County Business Pattern (CBP) database in the U.S. Census.\footnote{County Business Patterns Database is available at: \url{https://www.census.gov/programs-surveys/cbp/data/datasets.html}.} To derive the accurate number from ranges reported in CBP, I obtain the carefully imputed CBP data from \cite{eckert2020imputing}.\footnote{\cite{eckert2020imputing} provide final data, code, and detailed documentation of their methodology in imputing CBP data at \url{https://fpeckert.me/cbp/}.} For industry classification, I follow \cite{goukasian2010reaction} and \cite{mian2014explains}. Please see detailed information in Table (\ref{table_EmploymentIndustryClassification}).

%---------------------------------------------------------------------------------------
\noindent \textbf{New Residential Unit Permits} I get county-by-year new residential unit permit data from the U.S. Census.\footnote{New residential unit permits data is available at: \url{https://www.census.gov/construction/bps/index.html}.} To avoid reduced sample size due to missing observations because of non-survey years for some counties, I use the Census-imputed permit data.

%--------------------------------------------------------------------------------------------------------------------------------------------------------------------------
\subsection{Local Economic Conditions}
%--------------------------------------------------------------------------------------------------------------------------------------------------------------------------

%---------------------------------------------------------------------------------------
\noindent \textbf{IRS Household Income Data} I obtain household income data at the county-by-year level from the U.S. Internal Revenue Service (IRS).\footnote{For 1989 to 2018, the data is available at \url{https://www.irs.gov/statistics/soi\%2Dtax\%2Dstats\%2Dcounty\%2Ddata}.} The average household income at the county level is the adjusted gross income divided by the number of returns (households). The income is adjusted to the 2007 USD by the Personal Consumption Expenditures Chain-type Price Index (PCEPI) from the Federal Reserve Bank of St. Louis.

%---------------------------------------------------------------------------------------
\noindent \textbf{Local Control Variables} I obtain Control variables at county and ZIP code levels from U.S. Decennial Census Summary Files. Control variables in 1989 at the county level are from 1990 (March) Census Summary File 1C and 3C, whereas control variables at the ZIP level are from Summary File 3B.\footnote{The 1990 U.S. Census Summary File 1 is available at \url{https://www.census.gov/data/datasets/1990/dec/summary-file-1.html} and Summary File 3 is available at \url{https://www.census.gov/data/datasets/1990/dec/summary-file-3.html}. } Control variables in 1999 at the county level and the ZIP level are from both 2000 (March) Census Summary File 1 and 3.\footnote{The 2000 U.S. Census Files are available at: \url{https://www.census.gov/programs-surveys/decennial-census/guidance/2000.html}}

%--------------------------------------------------------------------------------------------------------------------------------------------------------------------------
\subsection{Counties Severely Affected by 2005 Hurricanes}\label{subsec:2005Hurricanes}
%--------------------------------------------------------------------------------------------------------------------------------------------------------------------------
Following \cite{li2024credit}, we delete twelve ``deeply affected counties by 2005 Hurricanes'' since they experienced unusual growth in mortgages due to hurricane damage and subsequent government subsidies.\footnote{I try my best to present the most robust results. Since outliers only largely affect results in regression but not the illustration in figures, I include these twelve counties in the figures but remove them from regressions and summary statistics.} In 2005, three Category 5 hurricanes (Katrina, Rita, and Wilma) caused enormous fatalities and damage (estimated \$125 billion).\footnote{These ``deeply affected counties'' include Monroe County (FL, 12087), Cameron Parish (LA, 22023), Jefferson Parish (LA, 22051), Orleans Parish (LA, 22071), Plaquemines Parish (LA, 22075), St. Bernard Parish (LA, 22087), St. Tammany Parish (LA, 22103), Vermilion Parish (LA, 22113), Hancock County (MS, 28045), Harrison County (MS, 28047), Jackson County (MS, 28059), Stone County (MS, 28131). }

\subsection{Summary Statistics and Figures}
Summary statistics of key variables are reported in Table (\ref{table_SumStat1}) and (\ref{table_SumStat2}), separated into different periods. In the prior period, starting with 712 ``HMDA consistent counties after 1990”, we remove seven counties due to heavy impact of the 2005 hurricanes described in section (\ref{subsec:2005Hurricanes}), ending up with 705 counties. Due to data availability, we have less number of observations for refined house employment and housing supply elasticity. In the boom period and bust period, beginning with 800 ``HMDA consistent counties after 1996”, we remove eight counties due to heavy impact of the 2005 hurricanes, resulting in 792 counties. Again, due to data availability, we have less number of observations for many variables of employment growth and housing supply elasticity. 

First, we can see that the total employment and refined house employment share experienced a clear boom and bust cycle. For example, while the mean of annualized change across counties is only 0.029\% in the prior period (1992-2000), it is 0.042\% in the boom period (1999-2005) and -0.089\% in the bust period (2007-2009). 

Second, the government-sponsored enterprise mortgages (GSEM) play an less important role in the boom period: its mean annualzed growth rate across counties is 14.950\% in the prior period but only 4.378\% in the boom period. 

Third, for the private-label mortgages (non-jumbo) (PLMNJ), it is interesting that the mean of growth rates across counties is similar between the prior period (17.072\%) and boom period (16.978\%). The similar mean values result from the use of different base values.\footnote{Specifically, the 1991 PLMNJ dollar amount is used for the growth rate between 1991 and 1999 while the 1999 PLMNJ dollar mount is used for the growth rate between 1999 and 2005.} The similar mean growth rates disguise the much larger increase in absolute dollar amount in private-label mortgages in the boom period . Figure (\ref{fig_GSEMvsPLMNJ_91t11_combine}) shows the differential huge increase in dollar amount in PLMNJ in the boom period, because time series values are scaled by 1991 dollar value only. Figure (\ref{fig_GSEMvsPLMNJ_91t11_combine}) and (\ref{fig_RefineHouseEmpShr_92To11}) hint that the impact of private-label mortgages (non-jumbo) shall be much larger than the government-sponsored enterprise mortgages in the boom and bust cycle of refined house employment since the former ones do show a differentially stronger increase in the high-net-export metropolitan areas.

Fourth, the net export growth rates have negative mean in the boom period (with a mean of -0.218\%  and standard deviation of 0.20\%) and in the prior period (with a mean of -0.18\%  and standard deviation of 0.17\%). This pattern is also present in the gravity model-based net export growth rates. 

We include control variables only in the starting year in each period in the county $c$, which avoids any impact from net export growth during that period. Control variables are employed to neutralize factors that may affect credit expansion. Our basic controls include average household income, the number of households, and the fraction of the labor force in the county $c$. Our housing controls include the number of house units, house vacancy rate, housing supply elasticity \citep{saiz2010geographic}, and fraction of renters in the occupied house units. Our demographic controls include the percentage of the white population, the fraction of population holding a Bachelor's degree and above, and the count of immigrants entering the U.S. between 1990 and 1999. We do not control local industry employment share, which is partially and jointly determined by my left-hand-side industry-level employment. The joint determinant relationship means that local industry employment shares are very likely ``bad controls''.

%----------------------------------------------------------------------

%----------------------------------------------------------------------
% section 5: Empirical Results

%------------------------------------------------------------
%------------------------------------------------------------
%\clearpage
%------------------------------------------------------------
%------------------------------------------------------------
\section{Empirical Results: Supporting Credit Expansion}
Our main empirical tests provide direct evidence that credit expansion in private-label mortgages (PLMs) instead of government-sponsored enterprise mortgages (GSEMs) cause the 1999-2010 boom and bust in the business boom and bust. To compare these two categories of mortgages under the same criteria, we use the conforming loan limits to get the non-jumbo category of private-label mortgage (PLMNJ) and remove the jumbo ones. we will focus on the comparison between private-label mortgages (non-jumbo) (PLMNJs) and the government-sponsored enterprise mortgages (GSEMs).\footnote{First, jumbo loan borrowers are usually not low-income households so credit expansion study shall focus on non-jumbo loans. Second, the jumbo category of private-label mortgages is much smaller in number when compared to non-jumbo ones. Third, including jumbo ones only strengthens my results on private-label mortgages.}

Based on the intuition described in the instruction, we predict that when the funding cost declines private-label mortgages (non-jumbo) expand more in the high net-export-growth areas in response to higher net export growth.

%------------------------------------------------------------
%------------------------------------------------------------
%------------------------------------------------------------
\subsection{Total Employment Boom (00-06) and Bust (07-10)}\label{subsec:causal_employment_inflation_boom_bust}
%------------------------------------------------------------
%------------------------------------------------------------
%------------------------------------------------------------

%------------------------------------------------------------
%------------------------------------------------------------
\textbf{Total Employment Boom (00-06)}
%------------------------------------------------------------
%------------------------------------------------------------

Let us first focus on my main hypothesis: total employment boom (00-06) as a result of credit expansion in response to net export growth across metropolitan areas. First, I expect private-label mortgages (non-jumbo) would experience stronger growth (99-05) in the high-net-export-growth metropolitan areas. Consequently, the above credit expansion causes total employment to grow stronger in the high-net-export-growth metropolitan areas.

\noindent \textbf{Hypothesis 1} In cross-section, growth in private-label mortgages (non-jumbo) (PLMNJs) in the boom period (1999-2005) causes the total employment boom (2000-2006).

To test the total employment boom, I perform the following regression. 
\begin{equation}\label{equ:TotEmpShre_reg_PLMNJ}
    \triangle_{00,06} TotalEmpShr_{c} = \beta * \triangle_{99,05} Ln(PLMNJ_{c}) + \gamma* \bm{Controls_{c}} + \epsilon_{c}
\end{equation}
The left-hand-side dependent variable $\triangle_{00,06} TotalEmpShr_{c}$ is the change of the total employment share in working-age population at county $c$ 2000-2006, and the key independent variable $\triangle_{99,05} Ln(PLMNJ_{c})$ is the growth rate of the dollar amount of private-label mortgage (non-jumbo) (PLMNJ) at county $c$ 1999-2005. $Controls_{c}$ indicates control variables at county $c$ in 1999. I use the gravity model-based instrument variable  $\triangle_{99,05}\text{givNetExp}_{m}$ for net export growth $\triangle_{99,05} Ln(PLMNJ_{c})$.

\comment{
We include control variables at county $c$ only in 1999, which prevents any impact from net export growth during the periods. Control variables are used to neutralize factors that may affect credit expansion for reasons unrelated to the main hypothesis. My basic controls include the number of households, average household income, and the fraction of the labor force at the county $c$. My housing controls include the number of house units, housing supply elasticity \citep{saiz2010geographic}, house vacancy rate, and fraction of renters in the occupied house units. My demographic controls include the fraction of population holding a Bachelor's degree and above, the percentage of the white population, and the count of immigrants entering the U.S. between 1990 and 1999. My industry controls include the ratio of the population that is in the art, entertainment, and recreation industries, that is in the health industries, that is in the tradable service industries, and that is college students. The industry controls capture the phenomena of retirement towns, medical centers, and college towns and the effect of the tradable service sector. 
}

Since employment growth is usually lagged, we choose the period for employment growth (00-06) to be a year later than the period for credit expansion (99-05). To prevent excessive influence from outliers, total employment share is winsorized at 0.5\% and 99.5\% level. Each regression uses analytical weight as the natural logarithm of the number of house units in the starting year 1999. Logarithm instead of the absolute number of house units is chosen to guard results from being dominated by a few super-populous counties. To take into account that households might commute to work across counties within a metropolitan area, I measure net export growth in the metropolitan area.\footnote{According to US Census, ``the general concept of a metropolitan statistical area is that of a core area containing a substantial population nucleus, together with adjacent communities having a high degree of economic and social integration with that core." (\url{https://www.census.gov/programs-surveys/metro-micro/about.html})} Furthermore, I cluster standard errors at the metropolitan area (CBSA 2003 code) level.

Table (\ref{table_BEA.TotEmpShr.D00t06.PLMNJ.D99t05.4Reg}) reports OLS, reduced-form, second-stage, and first-stage results in panel A, B, C, and D, respectively.  First, panel A shows the impact of PLMNJ growth (99-05, annualized) on total employment growth (00-06, annualized) are positive but insignificant. This result is consistent with our prediction that some of the growth of PLMNJ is expected. Reduced-form estimates in panel B show a significant and positive effect, since the gravity model-based instrument captures the exogenous and unexpected part of net export growth. The first-stage estimates in panel D are positive and significant at 1\% level. For the specification with full controls, clustered Kleibergen-Paap F-statistic \citep{kleibergen2006generalized} is 10.48, and the Montiel Olea-Pflueger Efficient F-Statistic \citep{olea2013robust} is 10.47, both of which are larger than 10. Thus, my estimates are very unlikely biased by a weak instrumental variable. Panel C reports 2SLS estimates for equation (\ref{equ:TotEmpShre_reg_PLMNJ}). Like reduced-form estimates, the 2SLS estimates are statistically significant and quite stable across various specifications.

Let us turn our attention to economic meaning in 2SLS with full controls in Table (\ref{table_BEA.TotEmpShr.D00t06.PLMNJ.D99t05.2SLS}). According to column (5) with all controls, one standard deviation in cross-sectional difference in annualized PLMNJ growth (99-05) results in $8.057\% \times 0.094 = 0.757\%$ difference in annualized total employment share growth (00-06) across metropolitan areas, translating into $4.540\%$ difference from 2000 to 2006. One standard deviation in cross-section difference in annualized total employment share growth (00-06) is $0.950\%$, translating into $5.701\%$ from 2000 to 2006. The two results mean that, one standard deviation in PLMNJ growth (99-05) can explain $4.540\% / 5.701\% = 79.71\%$ of one standard deviation in total employment share growth (00-06). 

The literature has shown that housing supply elasticity \citep{saiz2010geographic} is a very important determinant of mortgage growth. Our first-stage regression in Table (\ref{table_BEA.TotEmpShr.D00t06.PLMNJ.D99t05.2SLS}) verifies the above finding with coefficient being $-0.012$ and significant at 1\% level. However, the insignificant 2SLS estimate $0.001$ shows that housing supply elasticity does not impact the total employment growth directly after controlling for private-label mortgages. We provide a potential explanation for this result. Recall that lower elasticity in the local housing market means insufficient land supply for new buildings, thus attracting mortgage growth given the higher expected housing price in the near future. But lower elasticity also means that insufficient building supply cannot provide enough cheap establishment buildings for business expansion. Thus the total effect of housing supply elasticity on the total employment growth could be near zero after controlling for private-label mortgages.

We also show that the employment boom is robust to the alternative measures of total employment. In Appendix Table (\ref{table_RobustTotEmpShr.D00t06.PLMNJ.D99t05.4Reg}), we report OLS, reduced-form, first stage, and second stage results for three measures of alternative total employment in our context: wage and salary, nonfarm, and nonfarm private employment. Wage and salary employees are likely more impacted by business cycle than self-employed workers. Nonfarm employees are likely more influenced by business cycle than farm employees. Nonfarm private employee are nonfarm and non-government employees who are more vulnerable to business cycle than nonfarm government workers. Again, to guard results from excessive influence by outliers, alternative measures are all winsorized at 0.5\% and 99.5\% level. The reduced-form and 2SLS estimates together show that the employment boom results are very strong for the above three alternative measures, with significant levels all at 1\%. Specifically, the 2SLS coefficients ($0.081$, $0.099$, and $0.111$) are also very close to the the coefficient ($0.094$) for the total employment share growth.

%------------------------------------------------------------
%------------------------------------------------------------
\textbf{Total Employment Bust (07-10)}
%------------------------------------------------------------
%------------------------------------------------------------
The employment bust (07-10) is quite different from the employment boom (00-06) and is not stronger in the high net-export-growth metropolitan areas. Conceptually, this happens because, in the boom period, the high net-export-growth areas experience higher growth in tradable sector, housing-related sector, and nontradable sector. However, in the bust period, the high net-export-growth areas experience higher growth in tradable sector but higher drop in housing-related sector and nontradable sector. Thus, we do not find the total employment bust to have a differential trend in the high net-export-growth areas. The results in the rest of paper will make the above claim clear.

%------------------------------------------------------------
%------------------------------------------------------------
%------------------------------------------------------------
\subsection{The Housing Industry Channel}\label{subsec:HousingIndustryChannel}
%------------------------------------------------------------
%------------------------------------------------------------
%------------------------------------------------------------
Since the credit expansion in mortgages seem to dominant the 1999-2010 U.S. business cycle, I propose a ``housing industry channel", which states that the credit expansion in mortgages results in employment boom and bust primarily through the house-related industries. I will show causal evidence for this channel in this section. In this subsection \ref{subsec:HousingIndustryChannel}, to make regression coefficient more visible with digit, I multiple the employment share by 100 for all dependent variables.

\subsubsection{House-Related Employment}
This section provides causal evidence that growth in private-label mortgages (non-jumbo) 1999-2005 causes the house-related industries employment boom (2000-2006) and bust (2007-2010). For house-related industries, I use the definition by \cite{goukasian2010reaction}. The complete definition of all sub categories and finer SIC industries are provided in Table (\ref{table_EmploymentIndustryClassification}). I use crosswalk files from US census to crosswalk these 1987 SIC codes to 2007 NAICS codes. Then I process house employment data from the computed county business pattern data by \cite{eckert2020imputing}.\footnote{Fabian Eckert provides their computed county business pattern data at \url{https://fpeckert.me/cbp/}.}

Since the primary driving force is the private-label residential mortgages, not all house-related industries experience the same boom and bust pattern. In particular, commercial construction employment is likely not impacted as significantly as residential construction employment. In addition, real estate brokerage and management may not be impacted significantly because the mortgage boom may only trigger a shift of renters to homeowners while the entire real estate brokerage and management demand do not change. Due to these reasons, I define ``refined house-related industries" that include residential construction, supporting industries, and mortgage banks and brokers. I expect the boom and bust trend is stronger for ``refined house-related industries" but may be weaker for other house-related industries defined by \cite{goukasian2010reaction}.

%------------------------------------------------------------
%------------------------------------------------------------
\noindent \textbf{Refined House-Related Employment Boom (00-06) and Bust (07-10)}
%------------------------------------------------------------
%------------------------------------------------------------
I test the refined house-related employment boom and bust in the following single regression 
\begin{equation}\label{eq:RefineHouseEmpShrBoomBustonPLMNJ}
\resizebox{0.92\textwidth}{!}{$
\begin{aligned}
\triangle_{00,06} \& \triangle_{07,10} RefinedHouseEmpShr_{c} & = \beta_{00,06} * \triangle_{99,05} Ln(PLMNJ_{c}) \times Dum_{00,06} + \beta_{07,10} * \triangle_{99,05} Ln(PLMNJ_{c}) \times Dum_{07,10} \\
& + \gamma_{00,06}* \bm{Controls_{c}} \times Dum_{00,06} + \gamma_{07,09}* \bm{Controls_{c}} \times Dum_{07,10} + \epsilon_{period, c}
\end{aligned}
$} %end of \resizebox
\end{equation}
Controls, weight, and standard errors are the same as Eq(\ref{equ:TotEmpShre_reg_PLMNJ}). 

Table (\ref{table_RefineHouse.D00t06vsD07t10.PLMNJ.4Reg}) reports OLS, reduced-form, second stage, and the first stage of the stacked regression of refined house-related employment growth in the boom period (00-06) and the bust (07-10) periods. First, the OLS coefficients in panel A shows that the impact of PLMNJ growth (99-05) are significantly positive in the boom period (00-06) and significantly negative in the bust period (07-10), both of which are at 1\% level. The similar trend applies to reduced-form estimates in panel B and 2SLS estimates in panel C. Importantly, the coefficient magnitudes in 2SLS are around four times of OLS results in the specification with full controls in column (5), much lower than the average nine in the top three finance journals \citep{jiang2017have}. The smaller coefficients in the OLS regressions are consistent with our prediction in the research design that part of net export growth is likely expected by local workers and mortgage managers. First-stage estimates in panel D show the stable and strong positive correlation between the PLMNJ growth and gravity model-based IV for Net Export Growth (GIV-NEG), with large enough first-stage F-Statistics (clustered Kleibergen-Paap F-statistic is 12.05, and the Montiel Olea-Pflueger Efficient F-Statistic is 12.09). Therefore, my results are free from weak IV concerns. Let us turn our attention to the coefficient equality test of impact of PLMNJ growth in the boom and bust periods in the Table (\ref{table_RefineHouse.D00t06vsD07t10.PLMNJ.2SLS.wide}). For all specifications through columns (1)-(4), the chi-square statistics are large, and p-values are below 0.03, meaning the two coefficients are statistically different. To sum up, induced by exogenous net export growth, the growth of private-label mortgages (non-jumbo) (99-05) causes refined house-related employment to experience a larger boom (00-06) and a larger bust (07-10) in the high net-export-growth areas than the low net-export-growth areas.

In terms of economic meaning, one standard deviation in annualized PLMNJ growth (99-05) in cross section causes annualized refined house employment share to rise $8.137\% \times 0.382 / 100 = 0.031\%$ in boom (00-06) and to drop $8.137\% \times -0.696 / 100 = - 0.057\%$ in bust (07-10). Since the lengths of the boom and bust periods are different, I need to consider the time horizon: a longer period of credit expansion results in a shorter period of bust period. One standard deviation in six-year PLMNJ growth (99-05) causes refined house employment share to rise $0.031\% \times 6 = 0.186\%$ 2000-2006 and to drop $ 0.057\% \times 3 = 0.170\%$ 2007-2010. For annualized refined house employment share, one standard deviation is $0.049\%$ in boom (00-06) and $0.085\%$ in bust (07-10), translating into $0.049\% \times 6 = 0.294\%$ 2000-2006 and $0.085\% \times 3 = 0.256\%$ 2007-2010. The two results mean that one standard deviation in six-year PLMNJ growth (99-05) can explain $0.186\% / 0.294\% = 63.53\%$ of one standard deviation in six-year refined house employment boom (00-06) and $0.170\% / 0.256\% = 66.24\%$ of one standard deviation in three-year refined house employment bust (07-10). To sum up, six-year PLMNJ growth can explain $63.53\%$ refined house employment growth 2000-2006 and $66.24\%$ refined house employment drop 2007-2010. 

However, the 2SLS estimates in column (4) in Table (\ref{table_RefineHouse.D00t06vsD07t10.PLMNJ.2SLS.wide}) show that housing supply elasticity does not impact the refined house employment boom or bust directly after controlling other factors, particularly the private-label mortgages. Similar to our explanation above, although lower elasticity attracts higher mortgage growth, lower elasticity also reduces cheap establishment buildings for business expansion. Thus the total effect of housing supply elasticity on the refined house employment growth could be near zero after controlling for private-label mortgages.

%------------------------------------------------------------
%------------------------------------------------------------
\subsubsection{Residential Construction Employment Boom and Bust}\label{susubsec:ResiConstEmp}
%------------------------------------------------------------
%------------------------------------------------------------
Since the growth of private-label mortgages (99-05) is the key driving, I expect that residential construction employment experience both stronger boom (00-06) and stronger bust (07-10) in the high net-export-growth areas. However, there is no such trend in commercial construction employment. This subsection provides causal evidence for the above two predictions. 

I compare the residential construction employment boom and bust in the same specification in Equation (\ref{eq:RefineHouseEmpShrBoomBustonPLMNJ}) except that the dependent variable is the annualized residential construction employment growth in both boom (00-06) and bust (07-10) periods. Controls, weight, and standard errors are the same. Table (\ref{table_ResiConst.D00t06vsD07t10.PLMNJ.4Reg}) reports OLS, reduced-form, second stage, and the first stage of the stacked regression of residential construction employment growth in the boom period (00-06) and the bust (07-10) periods. First, panel A shows that the OLS coefficients of PLMNJ growth (99-05) are significantly positive in the boom period (00-06) and significantly negative in the bust period (07-10), both of which are at the 1\% level. Likewise, the reduced-form estimates in panel B and 2SLS estimates in panel C are all significant in the same directions. Importantly, the coefficient magnitudes in 2SLS are around four to five times of OLS results in specification with full controls in column (4), much lower than the average nine in the top three finance journals \citep{jiang2017have}. The smaller coefficients in the OLS regressions are consistent with our prediction that part of net export growth is likely expected by local employees and mortgage managers. First-stage estimates in panel D show a very robust and positive correlation between the PLMNJ growth and gravity model-based IV for Net Export Growth (GIV-NEG), with large enough first-stage F-Statistic (clustered Kleibergen-Paap F-statistic is 11.75, and the Montiel Olea-Pflueger Efficient F-Statistic is 11.78 in column (4)). Therefore, my results are free from weak IV concerns. For the coefficient equality test of impact of PLMNJ growth in the boom and bust periods in the Table (\ref{table_ResiConst.D00t06vsD07t10.PLMNJ.2SLS.wide}), the chi-square statistics are large, and p-values are below 0.04, meaning the two coefficients are statistically different. To sum up, induced by net export growth across metropolitan areas, the growth of private-label mortgages (non-jumbo) (99-05) causes residential construction employment to grow substantially in the boom period (00-06) but to drop significantly in the bust period (07-10).

In terms of economic meaning, one standard deviation in annualized PLMNJ growth (99-05) in cross-section causes annualized residential construction employment share to rise $8.137\% \times 0.171 / 100 = 0.014\%$ in boom (00-06) and to drop $8.137\% \times -0.261 / 100 = - 0.021\%$ in bust (07-10). Since the lengths of the boom and bust periods are different, I need to consider the time horizon: a longer period of credit expansion results in a shorter period of bust period. One standard deviation in six-year PLMNJ growth (99-05) can causes residential construction employment share to rise $0.014\% \times 6 = 0.084\%$ 2000-2006 and to drop $ 0.021\% \times 3 = 0.064\%$ 2007-2010. For annualized residential construction employment share, one standard deviation is $0.024\%$ in boom (00-06) and $0.085\%$ in bust (07-10), translating into $0.024\% \times 6 = 0.146\%$ 2000-2006 and $0.036\% \times 3 = 0.109\%$ 2007-2010. The two results mean that one standard deviation in six-year PLMNJ growth (99-05) can explain $0.083\% / 0.146\% = 57.34\%$ of one standard deviation in six-year residential construction employment boom (00-06) and $0.064\% / 0.109\% = 58.30\%$ of one standard deviation in three-year residential construction employment bust (07-10). To sum up, six-year PLMNJ growth can explain $57.34\%$ residential construction employment growth 2000-2006 and $58.30\%$ residential construction employment drop 2007-2010.

However, the 2SLS estimates in column (4) in Table (\ref{table_ResiConst.D00t06vsD07t10.PLMNJ.2SLS.wide}) show that housing supply elasticity does not impact the refined house employment boom or bust directly after controlling other factors, particularly the private-label mortgages. Although lower elasticity attracts higher mortgage growth, lower elasticity also reduces cheap establishment buildings for business expansion. Thus the total effect of housing supply elasticity on the refined house employment growth could be near zero after controlling for private-label mortgages.

\comment{
To address the concern that residential construction employment boom and bust may reflect the technology advancement in the general construction sector and mortgage growth over-react to such technology advancement, I perform the same test on the commercial construction employment in Appendix Table (\ref{table_ComConst.D00t06vsD07t10.PLMNJ.4Reg}). The insignificant results in OLS, reduced-form, and second-stage regressions help me exclude the possibility that technology advance in construction induces mortgages that eventually trigger the business cycle. 
}

\subsubsection{Other House Employment Boom and Bust}
%------------------------------------------------------------
%------------------------------------------------------------
\cite{goukasian2010reaction} also provide a list of sub-industries that support the house industries, for which I call ``other house industries". I expect that other house industries employment experience both a stronger boom (00-06) and a stronger bust (07-10) in the high net-export-growth areas. This subsection provides causal evidence for this prediction. 

I compare the other house industries employment boom and bust in a regression that is the same as Equation (\ref{eq:RefineHouseEmpShrBoomBustonPLMNJ}) except that the dependent variable is other house industries employment growth. Table (\ref{table_Other.D00t06vsD07t10.PLMNJ.4Reg}) reports OLS, reduced-form, second stage, and the first stage of the stacked regression of other house industries employment growth in the boom period (00-06) and the bust (07-10) periods. First, panel A shows that the OLS coefficients of PLMNJ growth (99-05) are significantly positive in the boom period (00-06) and significantly negative in the bust period (07-10), both of which are at 1\% level. Similar trends apply to reduced-form estimates in panel B and 2SLS estimates in panel C. Importantly, the coefficient magnitudes in 2SLS are between four to five times of OLS results across specifications, much lower than the average nine in the top three finance journals \citep{jiang2017have}. The relatively smaller coefficients in the OLS regressions are consistent with our prediction that part of net export growth is expected by local employees and mortgage bankers. First-stage estimates in panel D show a robust and positive correlation between the PLMNJ and gravity model-based IV for Net Export Growth (GIV-NEG), with large enough first-stage F-Statistic (clustered Kleibergen-Paap F-statistic is 11.75, and the Montiel Olea-Pflueger Efficient F-Statistic is 11.78 in column (4) with full controls). Therefore, my results are free from weak IV concerns. For the coefficient equality test of impact of PLMNJ growth (99-05) in the boom and bust periods in the Table (\ref{table_Other.D00t06vsD07t10.PLMNJ.2SLS.wide}), the chi-square statistics are large, and p-values are below 0.03, meaning the two coefficients are statistically different. To sum up, induced by net export growth across metropolitan areas, the growth of private-label mortgages (non-jumbo) (99-05) causes other house industry employment to grow substantially in the boom period (00-06) but to drop significantly in the bust period (07-10). Again, the 2SLS estimates in column (4) in Table (\ref{table_Other.D00t06vsD07t10.PLMNJ.2SLS.wide}) show that housing supply elasticity does not impact the other house employment boom or bust directly after controlling other factors, particularly the private-label mortgages.

In terms of economic meaning in the cross-section, one standard deviation in annualized PLMNJ growth (99-05) causes annualized other house employment share to rise $8.137\% \times 0.189 / 100 = 0.015\%$ in boom (00-06) and to drop $8.137\% \times -0.350 / 100 = - 0.028\%$ in bust (07-10). Considering horizon length, one standard deviation in six-year PLMNJ growth (99-05) causes other house employment share to rise $0.015\% \times 6 = 0.092\%$ 2000-2006 and to drop $ 0.028\% \times 3 = 0.085\%$ 2007-2010. For annualized other house employment share, one standard deviation is $0.028\%$ in boom (00-06) and $0.085\%$ in bust (07-10), translating into $0.028\% \times 6 = 0.174\%$ 2000-2006 and $0.047\% \times 3 = 0.142\%$ 2007-2010. The two results mean that one standard deviation in six-year PLMNJ growth (99-05) can explain $0.092\% / 0.174\% = 53.05\%$ of one standard deviation in six-year other house employment boom (00-06) and $0.085\% / 0.142\% = 60.10\%$ of one standard deviation in three-year other house employment bust (07-10). To sum up, six-year PLMNJ growth can explain $53.05\%$ other house employment growth 2000-2006 and $60.10\%$ other house employment drop 2007-2010.

\subsubsection{Mortgage Employment Boom and Bust}
%------------------------------------------------------------
%------------------------------------------------------------
I expect that mortgage industry employment can experience both a stronger boom (00-06) and a stronger bust (07-10) in the high net-export-growth areas. This subsection, however, provides evidence that the data noises in the County Business Pattern likely make the boom and bust in the mortgage industry employment not robust to controls. Further, I use private-label mortgage (non-jumbo) boom (99-05) and bust (05-08) to show evidence that the mortgage industry indeed experiences a stronger boom (99-05) and a stronger bust (05-08) in the high net-export-growth areas. 

First, I compare the mortgage industry employment boom and bust in a regression that is the same as Equation (\ref{eq:RefineHouseEmpShrBoomBustonPLMNJ}) except that the dependent variable is mortgage industry employment growth. Table (\ref{table_Mortgage.D00t06vsD07t10.PLMNJ.4Reg}) reports OLS, reduced-form, second stage, and the first stage of the stacked regression of mortgage industry employment growth in the boom period (00-06) and the bust (07-10) periods. Only for specifications without controls, reduced-form, first-stage, and second-stage regressions are all significant. Thus, my analysis only provides suggestive but not robust evidence that the PLMNJ growth likely triggers the mortgage industry employment boom and bust across metropolitan areas. 

Mortgage industry employment boom and bust give me a good chance to show that the data noises in the County Business Pattern likely make my results insignificant since the Home Mortgage Disclosure Act provides universal coverage of mortgages in metropolitan areas. The mortgage dollar amount is a better measure of the mortgage industry business boom and bust because such measure covers both intensive margin (productivity change given fixed employment) and extensive margin (business change associated with employment change). I focus on private-label mortgages because the legal constraint rules that only private-label mortgages (instead of government-sponsored enterprise mortgages) can consider regional economic conditions. Given the credit expansion during the 2000s \citep{justiniano2022mortgage}, I expect net export growth to induce private-label mortgages (non-jumbo) to show a stronger boom (99-05) in the high net-export-growth areas, which eventually leads to house prices stronger boom and bust in these areas. Eventually, the stronger house price bust leads to a stronger bust (05-08) in private-label mortgages (non-jumbo) in the same high net-export-growth areas. 

I compare the dollar amount of private-label mortgages (non-jumbo) boom and bust in a regression that is the same as Equation (\ref{eq:RefineHouseEmpShrBoomBustonPLMNJ}) except that the dependent variable is the growth of private-label mortgages (non-jumbo). Table (\ref{table_PLMNJ.D99t05vsD05t08.4Reg}) reports OLS, reduced-form, second stage, and the first stage of the stacked regression of growth of private-label mortgages (non-jumbo) in the boom period (00-06) and the bust (07-10) periods. First, reduced-form estimates in panel B shows that the coefficients of net export growth (99-05) are significantly positive in the boom period (00-06) and significantly negative in the bust period (07-10). The same trend applies to the 2SLS estimates in panel C. Importantly, the coefficient magnitudes in 2SLS are between three to four times of OLS results across specifications, much lower than the average nine in the top three finance journals \citep{jiang2017have}. First-stage estimates in panel D show the stable and strong positive correlation between the PLMNJ and gravity model-based IV for Net Export Growth (GIV-NEG), with large enough first-stage F-Statistics (clustered Kleibergen-Paap F-statistic is 17.78, and the Montiel Olea-Pflueger Efficient F-Statistic is 17.44). Therefore, my results are free from weak IV concerns. Let us turn our attention to the coefficient equality test of net export growth in the boom and bust periods in Table (\ref{table_PLMNJ.D99t05vsD05t08.2SLS.wide}). For all specifications through columns (1)-(4), the chi-square statistics are large, and p-values are below 0.01, meaning the two coefficients are statistically different. To sum up, induced by net export growth across metropolitan areas, private-label mortgage (non-jumbo) businesses experience a stronger boom (99-05) and a stronger bust (07-10) in the high net-export-growth areas.

In terms of economic meaning, one standard deviation in annualized net export growth (99-05) in cross section causes annualized PLMNJ growth to rise $0.200\% \times 10.851  = 2.174\%$ in boom (00-06) and to drop $0.200\% \times -12.438  = - 0.249\%$ in bust (07-10). Since the lengths of the boom and bust periods are different, I need to consider the time horizon: a longer period of credit expansion results in a shorter period of bust period. One standard deviation in six-year net export growth (99-05) causes PLMNJ growth to rise $2.174\% \times 6 = 13.045\%$ 2000-2006 and to drop $0.249\% \times 3 = 7.477\%$ 2007-2010. For annualized PLMNJ growth, one standard deviation is $8.137\%$ in boom (00-06) and $14.430\%$ in bust (07-10), translating into $8.137\% \times 6 = 48.820\%$ 2000-2006 and $14.430\% \times 3 = 43.291\%$ 2007-2010. The two results mean that one standard deviation in six-year net export growth (99-05) can explain $13.045\% / 48.820\% = 26.72\%$ of one standard deviation in six-year PLMNJ boom (00-06) and $7.4770\% / 43.291\% = 17.27\%$ of one standard deviation in three-year PLMNJ bust (07-10). To sum up, six-year net export growth can explain $26.72\%$ PLMNJ growth 2000-2006 and $17.27\%$ PLMNJ drop 2007-2010.

%################################################################################
%################################################################################
%################################################################################
%################################################################################
%------------------------------------------------------------
%------------------------------------------------------------
%------------------------------------------------------------
%\section{Robustness}\label{sec:Robustness}
%------------------------------------------------------------
%------------------------------------------------------------
%------------------------------------------------------------

\subsection{Housing Industry Channel: Additional Evidence}
In addition to the above evidence of employment growth in three closely related industries, I provide additional evidence to support the housing industry channel: evidence of total construction employment and residential unit permits.

\subsubsection{BEA Construction Employment Boom and Bust}
%------------------------------------------------------------
%------------------------------------------------------------
Since the growth of private-label mortgages (99-05) is the key driving force of the 1999-2009 business cycle, I expect that general construction employment experience both a stronger boom (00-06) and a stronger bust (07-10) in the high net-export-growth areas. This subsection provides causal evidence for this prediction. 

To test BEA total construction employment boom and bust, I perform the same regressions in Equation  (\ref{eq:RefineHouseEmpShrBoomBustonPLMNJ}) except that the dependent variable is the growth of total construction employment. Table (\ref{table_BEA.ConstEmpShr.D00t06vsD07t10.PLMNJ.D99t05.4Reg}) reports OLS, reduced-form, second stage, and the first stage of the stacked regression of BEA construction employment growth in the boom period (00-06) and the bust (07-10) periods. First, panel A shows that the OLS coefficients of PLMNJ growth (99-05) are significantly positive in the boom period (00-06) and significantly negative in the bust period (07-10). The same trend applies to reduced-form estimates in panel B and 2SLS estimates in panel C. Importantly, the coefficient magnitudes in 2SLS are around three to four times of OLS results in specification with full controls in column (5), much lower than the average nine in the top three finance journals \citep{jiang2017have}. The smaller coefficients in the OLS regressions are consistent with our prediction in the research design that part of net export growth is likely expected by local employees and mortgage bankers. First-stage estimates in panel D show the stable and strong positive correlation between the PLMNJ and gravity model-based IV for Net Export Growth (GIV-NEG), with large enough first-stage F-Statistics (clustered Kleibergen-Paap F-statistic is 10.48, and the Montiel Olea-Pflueger Efficient F-Statistic is 10.47). Therefore, my results are free from weak IV concerns. As for the coefficient equality test of the impact of PLMNJ growth in the boom and bust periods in Table (\ref{table_BEA.ConstEmpShr.D00t06vsD07t10.PLMNJ.D99t05.2SLS.wide}), the chi-square statistics are large, and p-values are below 0.05, meaning the two coefficients are statistically different. Again, the coefficients of housing supply elasticity are insignificant in both periods. To sum up, induced by net export growth, the growth of private-label mortgages (non-jumbo) (99-05) causes construction employment to experience a stronger boom (00-06) and a stronger bust (07-10) in the high net-export-growth areas.

In terms of economic meaning, one standard deviation in annualized PLMNJ growth (99-05) in cross-section causes annualized construction employment share to rise $8.057\% \times 0.015  = 0.121\%$ in boom (00-06) and to drop $8.057\% \times -0.022 = - 0.177\%$ in bust (07-10). Since the lengths of the boom and bust periods are different, I need to consider the time horizon: a longer period of credit expansion results in a shorter period of bust period. One standard deviation in six-year PLMNJ growth (99-05) causes construction employment share to rise $0.121\% \times 6 = 0.725\%$ 2000-2006 and to drop $ 0.177\% \times 3 = 0.532\%$ 2007-2010. For annualized construction employment share, one standard deviation is $0.216\%$ in boom (00-06) and $0.378\%$ in bust (07-10), translating into $0.216\% \times 6 = 1.295\%$ 2000-2006 and $0.378\% \times 3 = 1.134\%$ 2007-2010. The two results mean that one standard deviation in six-year PLMNJ growth (99-05) can explain $0.725\% / 1.295\% = 55.99\%$ of one standard deviation in six-year construction employment boom (00-06) and $0.532\% / 1.134\% = 46.90\%$ of one standard deviation in three-year construction employment bust (07-10). To sum up, six-year PLMNJ growth can explain $55.99\%$ construction employment growth 2000-2006 and $46.90\%$ construction employment drop 2007-2010.

\subsubsection{Residential unit Permit Boom and Bust}

In the previous section \ref{susubsec:ResiConstEmp}, I provide evidence of residential employment to support the housing industry channel. A natural prediction is that business planning on future residential units also experience a boom and bust triggered by growth in private-label mortgages (non-jumbo) in response to net export growth. This subsection provides causal evidence for such a prediction.

To test residential unit permit boom and bust, I perform the same regressions in Equation  (\ref{eq:RefineHouseEmpShrBoomBustonPLMNJ}) except that the dependent variable is the growth of permit value (imputed by the U.S. Census) growth.\footnote{U.S. Census inputs some of the unit permits for counties that are not required in certain years. Using the imputation results by the U.S. Census avoids zero values of non-reporting counties in certain years.} Table (\ref{table_PermitValue.D99t05vsD05t09.PLMNJ.4Reg}) reports OLS, reduced-form, second stage, and the first stage of the stacked regression of residential unit permit value growth in the boom period (99-05) and the bust (05-09) periods. Due to its forward-looking nature, residential building plans experienced earlier boom (99-05) and earlier bust (05-09) ahead of employment boom (00-06) and bust (07-10). First, OLS coefficients in panel A show that the impact of PLMNJ growth (99-05) is significantly positive in the boom period (99-05) and significantly negative in the bust period (05-09). The same trend applies to reduced-form estimates in panel B and 2SLS estimates in panel C. Importantly, the coefficient magnitudes in 2SLS are around two to four times of OLS results in specification with full controls in column (5), much lower than the average nine in the top three finance journals \citep{jiang2017have}. The smaller coefficients in the OLS regressions are consistent with our prediction that part of net export growth is likely expected by building companies. First-stage estimates in panel D show a robust and strong positive correlation between the PLMNJ and gravity model-based IV for Net Export Growth (GIV-NEG), with large enough first-stage F-Statistics (clustered Kleibergen-Paap F-statistic is 11.74, and the Montiel Olea-Pflueger Efficient F-Statistic is 11.80 in column (4)). Therefore, my results are free from weak IV concerns. For the coefficient equality test of the impact of PLMNJ growth in the boom and bust periods in Table (\ref{table_BEA.ConstEmpShr.D00t06vsD07t10.PLMNJ.D99t05.2SLS.wide}), the chi-square statistics are large, and p-values are below 0.01, meaning the boom and bust coefficients are statistically different. Again, the coefficients of housing supply elasticity are insignificant in both periods. To sum up, induced by net export growth, the growth of private-label mortgages (non-jumbo) (99-05) causes residential building applications to experience a stronger boom (99-05) and a stronger bust (05-09) in the high net-export-growth areas.

In terms of economic meaning, one standard deviation in annualized PLMNJ growth (99-05) in cross-section causes annualized residential units permit value to rise $8.137\% \times 0.713  = 5.801\%$ in boom (00-06) and to drop $8.137\% \times -1.648 = - 13.409\%$ in bust (07-10). Since the lengths of the boom and bust periods are different, I need to consider the time horizon: a longer period of credit expansion results in a shorter period of bust period. One standard deviation in six-year PLMNJ growth (99-05) causes residential units permit value to rise $5.801\% \times 6 = 34.808\%$ 2000-2006 and to drop $ 13.409\% \times 3 = 40.228\%$ 2007-2010. For annualized residential units permit value growth, one standard deviation is $9.448\%$ in boom (00-06) and $16.906\%$ in bust (07-10), translating into $9.448\% \times 6 = 56.688\%$ 2000-2006 and $16.906\% \times 3 = 50.717\%$ 2007-2010. The two results mean that one standard deviation in six-year PLMNJ growth (99-05) can explain $34.808\% / 56.688\% = 61.40\%$ of one standard deviation in six-year residential units permit value boom (00-06) and $40.228\% / 50.717\% = 79.32\%$ of one standard deviation in three-year residential units permit value bust (07-10). To sum up, six-year PLMNJ growth can explain $61.40\%$ residential units permit value growth 2000-2006 and $79.32\%$ residential units permit value drop 2007-2010.

%################################################################################
%################################################################################
% This is the end of the entire section (tex file)
%################################################################################
%################################################################################
%----------------------------------------------------------------------

%----------------------------------------------------------------------
% section 6: Empirical.AlternativeHypotheses.tex

%------------------------------------------------------------
%------------------------------------------------------------
\section{Empirical: Tests for Alternative Hypotheses}\label{subsec:Empirical.Alternative}
%------------------------------------------------------------
%------------------------------------------------------------
In this section, we perform the most comprehensive tests on the business cycle theories. We show that the following theories (hypotheses) cannot explain the origin of the 1999-2010 U.S. business cycle:

%------------------------------------------------------------
%------------------------------------------------------------
\subsection{Speculation Euphoria Hypothesis}
%------------------------------------------------------------
%------------------------------------------------------------
In this section, we perform three tests to show that credit expansion is a necessary condition for speculation and speculation plays a much less important role than credit expansion. This section is designed to address the concern from the ``speculative euphoria hypothesis"  that, even though credit expansion happens in the first place, speculation by borrowers may also play a central role by taking advantage of the credit expansion \citep{kindleberger1978manias,minsky1986stabilizingan}. Since government-sponsored enterprise mortgages (GSEMs) do not respond to net export growth (see Table (\ref{table_GSEM.D99t05vsD05t08.4Reg})) and cannot explain the cross-metro refined house employment cycle (see Table (\ref{table_RefineHouse.D00t06vsD07t10.GSEM.4Reg})), we only distinguish speculation and credit expansion within the private-label mortgages (non-jumbo). We use the ``non-owner-occupied'' home purchase mortgages to measure speculation as \citep{gao2020economic}. In contrast, we use ``owner-occupied" home purchase mortgages as a measure of pure credit expansion.

To preview empirical tests, my first test shows that the cross-metropolitan variation in speculation (99-05) is largely caused by pure credit expansion. We use the residuals from the above regression as my measure of credit-independent speculation, which is the part of growth rate of the speculation (non-owner-occupied) that cannot be explained by the growth rate of pure credit expansion (owner-occupied). My second test shows that, compared to the dominant role of pure credit expansion, credit-independent speculation plays a much less important role in explaining the refined house employment cycle. Specifically, credit expansion explains 5.37 times the amount of refined house employment growth explained by speculation . My third test focuses on the prior period (91-99) where there is large difference in net export growth across metropolitan areas but no credit expansion at aggregate level. We show that, without credit expansion, speculation does not respond to the divergence of local economic growth, proxied by net export growth.

Before going into regression analysis, it is useful to see the general trend and magnitude of pure credit expansion vs. speculation across time in Figure (\ref{fig_CreditExpansion_vs_Sepculation}). First, in the boom period (99-05), the pure credit expansion (owner-occupied home purchase) in absolute dollar amount is much higher than speculation (non-owner-occupied home purchase). For example, in the peak year of 2005 in the top-quintile metropolitan areas based on net export growth, pure credit expansion is $67.0$ Billion (07USD) while the speculation is only $13.9$ Billion (07USD). Growth of both time series are stronger in the high net-export-growth areas. Such observations are consistent with  my conclusion for tests 1 \& 2 discussed above. In addition, in the prior period, there is not much difference in speculation (non-owner-occupied home purchase) between the high and low net-export-growth quintile groups of metropolitan areas. This observation is consistent with the conclusion in my third test above. Please note that for ease of calculation and comparison, all dependent and independent variables will not be annualized in this section.

%------------------------------------------------------------
\subsubsection{Credit Expansion Causes Speculation}
%------------------------------------------------------------

In this section, we show that cross-metropolitan variation in speculation is largely caused by pure credit expansion. We use the following 2SLS specification: 
\begin{equation}\label{eq:PLMNJNonOwn_on_PLMNJOwn}
\resizebox{0.9\textwidth}{!}{$
\triangle_{99,05} Ln(PLMNJ\_NonOwn_{c}) = \underbrace{\beta * \triangle_{99,05} Ln(PLMNJ\_Own_{c})}_{\text{Credit-Induced Speculation}} + \underbrace{\gamma* \bm{Controls_{c}} + \epsilon_{c}}_{\text{Credit-Independent Speculation}}
$} %end of \resizebox
\end{equation} 
The left-hand-side dependent variable $\triangle_{99,05} Ln(PLMNJ\_NonOwn_{c})$ is the growth rate of the dollar amount (07USD) of non-owner-occupied private-label mortgages (non-jumbo) at county $c$ 99-05 and the key independent variable $\triangle_{99,05} Ln(PLMNJ\_Own_{c})$ is the growth rate of the dollar amount (07USD) of owner-occupied private-label mortgages (non-jumbo) at county $c$ 99-05. We use the gravity model-based instrumental variable ($\triangle_{99,05}\text{givNetExp}_{m}$) as IV for $\triangle_{99,05} Ln(PLMNJ\_Own_{c})$. Controls, weights, and standard errors are the same as Eq (\ref{eq:RefineHouseEmpShrBoomBustonPLMNJ}).

Table (\ref{table_Speculation.PLMNJNonOwn.D99t05.PLMNJOwn.4Reg}) reports OLS, reduced-form, second stage, and the first stage of the results. First, panel A shows the positive and significant impact of owner-occupied private-label mortgages (non-jumbo) (PLMNJ) (99-05, An) as credit expansion on the growth of non-owner-occupied PLMNJ as speculation. The coefficients are quite stable across specifications and are significant at 1\%. A similar trend applies to the reduced-form estimates in panel B. First-stage estimates in panel D show that the strong positive correlation between the GIV and PLMNJ (owner-occupied) growth is quite stable across various specifications. In column (4) with all control variables, the first-stage clustered Kleibergen-Paap F-statistic is 11.00. And the Montiel Olea-Pflueger Efficient F-statistics is 11.04. Thus, it is very unlikely that my estimates are biased by weak instruments. The 2SLS estimates in panel C are statistically significant at a one percent level and quite stable across various specifications. The 2SLS estimates are very close to OLS estimates throughout various specifications, where column (4) indicate a ratio of $1.148/1.121=1.024$, much lower than nine. In terms of economic meaning in 2SLS with full controls, one standard deviation of 6-year PLMNJ (owner-occupied, credit expansion) growth can cause $46.868 * 1.148= 53.805\%$ increase in 6-year PLMNJ (non-owner-occupied, speculation) growth, which is $53.805/87.310 = 561.625\%$ of one standard deviation of PLMNJ (non-owner-occupied) growth. Therefore, we can conclude that cross-metropolitan variation in speculation is largely caused by by credit expansion.

%-------------------------------------------------------
\subsubsection{Credit-Independent Speculation vs Pure Credit Expansion}
%-------------------------------------------------------
In this section, we use the decomposition from the first test to get credit-independent speculation. Then we show that, compared to the dominant role of pure credit expansion measured by owner-occupied private-label mortgages (non-jumbo), credit-independent speculation can only explain a tiny portion of refined house employment cycle.

To be specific, we use the following regression specification:
\begin{equation}\label{eq:HPIBoom_on_PureCredit_vs_Speculation}
\resizebox{0.92\textwidth}{!}{$
\begin{aligned}
\triangle_{00,06} \& \triangle_{07,10} RefinedHouseEmpShr_{c} & = \beta_{00,06} * \triangle_{99,05} Ln(PLMNJ\_Own_{c}) \times Dum_{00,06} + \beta_{07,10} * \triangle_{99,05} Ln(PLMNJ\_Own_{c}) \times Dum_{07,10} \\
& \theta_{00,06}* \text{Credit-Independent Speculation (c, 99-05)} \times Dum_{00,06} + \theta_{07,10}* \text{Credit-Independent Speculation (99-05)} \times Dum_{07,10} \\
& + \gamma_{00,06}* \bm{Controls_{c}} \times Dum_{00,06} + \gamma_{07,09}* \bm{Controls_{c}} \times Dum_{07,10} + \epsilon_{period, c}
\end{aligned}
$} %end of \resizebox
\end{equation}
The first key independent variable $\triangle_{99,05} Ln(PLMNJ\_Own_{c})$ is the growth rate of the dollar amount (07USD) of owner-occupied private-label mortgage (non-jumbo) (PLMNJ) as pure credit expansion at county $c$ 99-05. This variable is instrumented by the gravity model-based instrumental variable ($\triangle_{99,05}\text{givNetExp}_{m}$). The second key independent variable $\text{Credit-Independent Speculation}$ is derived from the regression Eq (\ref{eq:PLMNJNonOwn_on_PLMNJOwn}), which is the part of growth rate of the speculation (non-owner-occupied) that cannot be explained by the growth rate of pure credit expansion (owner-occupied).

Table (\ref{table_Speculation.RefHou.D00t06vsD0710.OwnCredit_vs_Speculation.4Reg}) reports OLS, reduced-form, second stage, and the first stage of the results. First, panel A shows the OLS estimates of pure credit expansion and credit-independent speculation are both significant on the boom and bust of refined house employment. Panel B reports the reduced-form estimates and shows that gravity-model-based net export growth explains the refined house employment cycle. First-stage estimates in panel D show that the strong positive correlation between the GIV of net export growth and the growth rate of pure credit expansion. In column (4) with all control variables, the first-stage clustered Kleibergen-Paap F-statistic is 11.17. And the Montiel Olea-Pflueger Efficient F-statistics is 11.19. Thus, it is very unlikely that my estimates are biased by weak instruments. The 2SLS estimates in panel C are statistically significant and quite stable across various specifications. In term of the economic meaning in the boom period (00-06), one standard deviation of six-year pure credit expansion (owner-occupied) can cause  $46.868\% * 0.397/100 = 0.186\%$ six-year refined house employment boom (00-06), translating into $0.186\%/0.294\%  =63.386\%$ of one standard deviation of  the latter. In comparison, one standard deviation of six-year growth of credit-independent speculation can cause $69.331\% * 0.050/100 = 0.347\%$ six-year refined house employment boom (00-06), translating into $0.347\%/0.294\%  =11.810\%$ of one standard deviation of the latter. In the bust period (07-10), one standard deviation of six-year pure credit expansion can cause  $46.868\% * 0.373/100 = 0.175\%$ three-year refined house employment bust (07-10), translating into $0.175\%/0.256\%  =68.160\%$ of the one standard deviation of the latter. In comparison, one standard deviation of six-year growth of credit-independent speculation can cause $69.331\% * 0.042/100 = 0.029\%$ three-year refined house employment bust (07-10), translating into $0.029\%/0.256\%  =11.353\%$ of one standard deviation of the latter. To sum up, the explanation power of the pure credit expansion is 5.37 times of the credit-independent speculation in the boom (00-06) and is 6 times in the bust (07-10)

Therefore, we conclude that pure credit expansion can explain the majority of the refined house employment cycle, while credit-independent speculation can only explain a much smaller part. Results in test one and two together show that home consumption demand (``owner-occupied") induced by credit expansion plays a much more important role than speculation (``non-owner-occupied"). This conclusion is quite different from the view by \cite{minsky1986stabilizingan, kindleberger1978manias}.\footnote{This conclusion only refers to the first mortgage market. Though out of the scope of this paper, it might be the case that speculation by investors plays an important role in the secondary mortgage market.}

%-------------------------------------------------------
\subsubsection{Speculation Cannot Grow Without Credit Expansion}
%-------------------------------------------------------
In this subsection, using the prior period (1991-1999), we show that speculation cannot respond to local net export growth without credit expansion. Recall that in the test in Eq (\ref{eq:PLMNJNonOwn_on_PLMNJOwn}), we have shown that net export growth causes the credit expansion that largely explains the growth the speculation in the boom period (1999-2005). To further show that credit expansion is a necessary condition for speculation, we focus on the prior period (91-99), in which there is no aggregate credit expansion, and show that even the net export growth cannot cause credit expansion. The financial friction rises from the fact that most U.S. mortgages use 20\% ratio of down payment and borrow the rest 80\% from lenders. Speculation with a mortgage is a highly-leveraged and risky behavior, with a monthly payment requirement and a high liquidity need at the sale. Without credit expansion characterized by low mortgage rate, the high cost in mortgage rate deters most potential speculation with a mortgage. 

I perform the test by the following regression:
\begin{equation}\label{eq:SpeculationPrior_on_NEG}
\resizebox{0.9\textwidth}{!}{$
\begin{aligned}
\triangle_{91,99} \& \triangle_{99,05} Ln(PLMNJ\_NonOwn_{c}) & =  \beta_{91,99} * \triangle_{91,99} \text{NetExp}_{m} \times Dum_{91,99} + \beta_{99,05} * \triangle_{99,05} \text{NetExp}_{m} \times Dum_{99,05} \\
& + \gamma_{91,99}* \bm{Controls_{c}} \times Dum_{91,99} + \gamma_{99,05}* \bm{Controls_{c}} \times Dum_{99,05}  + \epsilon_{period, c}
\end{aligned}
$} %end of \resizebox
\end{equation}
The left-hand-side dependent variable $\triangle_{91,99} \& \triangle_{99,05} Ln(PLMNJ\_NonOwn_{c})$ is the stacked growth rate of the dollar amount (07USD) of non-owner-occupied private-label mortgages (non-jumbo) (PLMNJ) as a measure of speculation at county $c$ 91-99 and 99-05, respectively. The key independent variable, net export growth, is instrumented by the gravity model-based IV in two periods.

Table (\ref{table_Speculation.PLMNJNonOwn.D91t99vsD99t05.4Reg}) reports OLS, reduced-form, second stage, and the first stage of the results. First, OLS estimates in panel A shows the net export growth is positively correlated with the speculation in the boom period (99-05) but not in the prior period (91-99). The same pattern shows up in the reduced-form estimates in panel B. First-stage estimates in panel D and E show the strong positive correlation between the net export growth and its GIV in both periods. In column (5) with all control variables, the first-stage clustered Kleibergen-Paap F-statistic is 74.02 in the prior (91-99) and 17.51 in the boom (99-05). And the Montiel Olea-Pflueger Efficient F-statistics is 77.63 and 17.06. Thus, my estimates are unlikely biased by weak instruments. The 2SLS estimates in panel C show that net export growth causes speculation in the boom period (99-05) but not in the prior period (91-99). This empirical result is more consistent with the view that net export growth can cause speculation only if there is aggregate credit expansion. Taken together, my three tests are more consistent with the view that credit expansion is a necessary condition for speculation and credit expansion plays a much larger role than speculation. Thus, my three tests help address the concern from ``speculative euphoria hypothesis" by \cite{kindleberger1978manias,minsky1986stabilizingan}.

%------------------------------------------------------------
%------------------------------------------------------------
\subsection{Real Business Cycle Theory}
%------------------------------------------------------------
%------------------------------------------------------------
Real business cycle theory \citep{prescott1986theory} predicts that business cycles are mainly driven by shocks to the total productivity capacity of corporations. In our study, there is a possibility that the stronger positive technology shocks in the boom and stronger negative technology shocks in the bust to corporations in the high net-export-growth areas mainly drive the employment cycle. To address this concern, we show three pieces of evidence against the predictions by the real business cycle hypothesis. First, we show that net export growth causes differential higher growth in the tradable employment in both the boom and the bust in the high net-export-growth areas. Second, commercial construction employment experiences neither a stronger growth in the boom nor a stronger drop in the bust in the high net-export-growth areas. Third, the debt (mainly home mortgages) in the household and nonprofit sector instead of corporate sector experiences a substantial boom and bust.

First, if the real business cycle hypothesis mainly explains the employment cycle, it implies that the tradable sector employment shall experience differentially higher growth in the boom period (99-05) and differentially stronger drop in the bust period (07-10) in the high net-export-growth areas. However, reduced-form and 2SLS estimates in Table (\ref{table_Tradable.D00t06vsD07t10.PLMNJ.4Reg}) show that tradable sector employment in the high net-export-growth areas continues to show a differentially stronger growth in the bust period (07-10). Thus the real business cycle hypothesis cannot explain at least the bust period for tradable employment growth. 

Second, since technology shocks trigger business boom and bust, the real business cycle hypothesis implies that commercial real estate construction shall experience a differentially higher growth in the boom period (99-05) and a differentially sharper drop in the bust period (07-10) in the high net-export-growth areas. In sharp contrast, OLS, reduced-form, and 2SLS estimates in Table (\ref{table_ComConst.D00t06vsD07t10.PLMNJ.4Reg}) show that commercial construction employment experiences neither a differentially higher growth in the boom period (99-05) nor a sharper drop in the bust period (07-10) in the high net-export-growth areas. Thus the real business cycle hypothesis cannot explain either the boom or the bust period for commercial construction employment. 

Third, the real business cycle hypothesis predicts that technology shocks mainly function via the corporate side, meaning corporate shall experience balance sheet (asset and debt) boom and bust over the business cycle. However, subfigure (a) in Figure (\ref{fig_DebtToGDPRatio_HHBusiGov_HHSubCategories}) shows that it is primarily the household and nonprofit sector, rather than the corporate sector, that experiences the debt boom and bust over the business cycle. In addition, subfigure (b) in Figure (\ref{fig_DebtToGDPRatio_HHBusiGov_HHSubCategories}) shows that home mortgages experience a huge boom and bust while consumer credit and other debt seem to have no boom and bust at all.

%------------------------------------------------------------
%------------------------------------------------------------
\subsection{Two Variants in Spirit of Real Business Cycle Theory}
%------------------------------------------------------------
%------------------------------------------------------------

This subsection provides evidence to address the concern from two variants in spirit of the real business cycle theory: natural disaster hypothesis and technology shock hypothesis in construction sector. 
%------------------------------------------------------------
%------------------------------------------------------------
\subsubsection{Natural Disaster Hypothesis}
%------------------------------------------------------------

Natural disasters could have a considerable impact on local economy. One possibility is that better weather condition drives the stronger economic boom and natural disasters trigger the stronger bust in the high net-export-growth areas. To address this concern, we show two pieces of evidence against the above predictions from farm and manufacture employment. First, we show that farm employment in the high net-export-growth areas does not experience a stronger boom (00-06) or a sharper bust (07-10). Second, manufacturing employment in the high net-export-growth areas experiences a stronger growth than the low net-export-growth areas in both the boom (00-06) and the bust (07-10) periods.

First, if the natural disaster hypothesis mainly explains the employment cycle, it implies that farm employment shall experience stronger boom and stronger bust in the high net-export-growth areas because farm production is heavily dependent on local weather. In contrast, OLS, reduced-form and 2SLS estimates in Table (\ref{table_BEA.Farm.D00t06vsD07t10.PLMNJ.4Reg}) show that farm employment in the high net-export-growth areas experience neither stronger boom (99-05) nor stronger bust (07-10). Thus the natural disaster hypothesis cannot explain boom and bust for farm employment growth.

Second, natural disaster hypothesis implies that manufacture employment in the high net-export-growth areas shall experience sharper boom and sharper bust because manufacture production also requires stable weather that does not harm plants or delay transportation of materials and products. However, reduced-form and 2SLS estimates in Table (\ref{table_BEA.Manufacture.D00t06vsD07t10.PLMNJ.4Reg}) show that manufacture employment experience a stronger growth in both boom (99-05) and bust (07-10) periods in the high net-export-growth areas. Thus the natural disaster hypothesis cannot explain the bust period for manufacture employment growth.

%------------------------------------------------------------
%------------------------------------------------------------
\subsubsection{Technology Shock Hypothesis in Construction Sector}
%------------------------------------------------------------
%------------------------------------------------------------

The boom and bust in house-related employment may reflect the technological shock (advancement) in the general construction sector. Such technological shock reduces the construction cost and other sectors increase their building demand accordingly. Then the mortgage sector overreacts to such technological advancement, thus causing the presidential employment boom and later dramatic bust in industries that are closely related to construction. We employ three tests to address this hypothesis. 

The technology shock hypothesis predicts that, like the residential construction employment, the commercial construction employment shall experience a stronger boom and a stronger bust in the high net-export-growth areas (HNEG areas). In addition, technology shock hypothesis implies that, similar to the private-label mortgages (non-jumbo), government-sponsored enterprise mortgages shall experience a stronger boom and a stronger bust in the HNEG areas. This trend shall happen because credit-qualified households increase their housing demand to the declining cost of housing, either for their first house or for shifting to a larger house. Third, this hypothesis may indicate that manufacture sector can potentially overreact to the declining cost of manufacturing plants, experiencing a stronger boom and a stronger bust in the high net-export-growth areas. 

However, the above three predictions are not verified by empirical tests. First, OLS, reduced-form, and second-stage estimates in Table (\ref{table_ComConst.D00t06vsD07t10.PLMNJ.4Reg}) show that commercial construction employment experiences neither a stronger boom (00-06) nor a stronger bust (07-10) in the high net-export-growth areas. Second, Table (\ref{table_GSEM.D99t05vsD05t08.4Reg}) show that government-sponsored enterprise mortgages experience neither differentially stronger boom (00-06) nor differentially stronger bust (07-10) in the high net-export-growth areas. Third, Table (\ref{table_BEA.Manufacture.D00t06vsD07t10.PLMNJ.4Reg}) shows that manufacturing employment enjoy differentially higher growth in both the boom (00-06) and bust (07-10) periods in the high net-export-growth areas. No evidence supports a differential stronger bust in the high net-export-growth areas. To sum up, the above three tests help me exclude the possibility that technology advancement in construction induces mortgages over-expansion that eventually triggers the business cycle.

%------------------------------------------------------------
%------------------------------------------------------------
\subsection{The Collateral-Driven Credit Cycle Theory}
%------------------------------------------------------------
%------------------------------------------------------------

\cite{kiyotaki1997credit} develop a theory that emphasizes the collateral channel through which a small shock can generate ultimate large fluctuations in business activities. Their main idea can be roughly summarized in the following figure (\ref{fig_Kiyotaki&Moore1997_CollateralChannel}). When a small positive (negative) shock hits the asset price of collateral that determines the borrowing capacity of the representative firm, the firm increases (decreases) the amount of debt from the bank. The increased (decreased) debt results in higher (lower) productivity and output that ultimately increases (decreases) the asset price. This feedback loop reinforces the initial effect of the positive (negative) shock to asset price and ultimately results in a much larger positive (negative) effect on the whole economy.

%------------------------------------
\begin{figure}[H]
\vspace{-2mm}
\begin{center}

\resizebox{\textwidth}{!}{%
%\resizebox{\textwidth}{!}{%
\includegraphics[width=14cm, height=4cm]{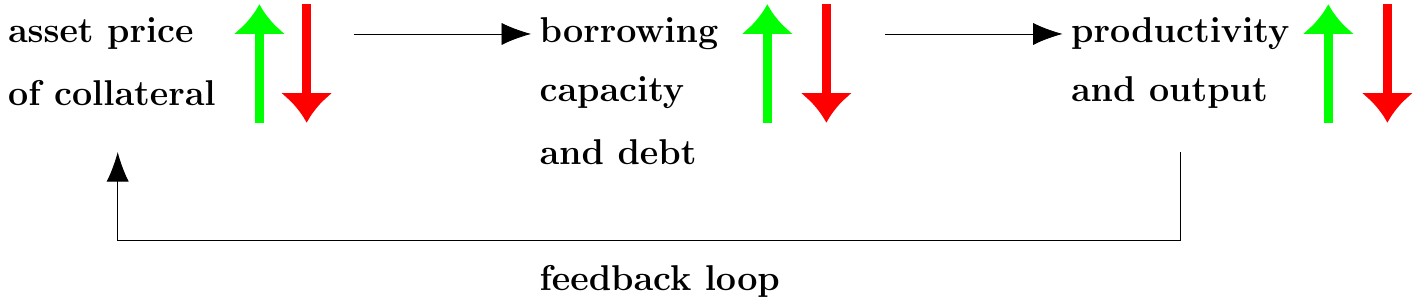}
} %end of resizebox

\end{center}
\vspace{-4mm}
%---------------
% Figure setting: caption and label
%---------------
\caption{The Collateral Channel by Kiyotaki and Moore (1997)}
\label{fig_Kiyotaki&Moore1997_CollateralChannel}

\end{figure}
%------------------------------------

The above collateral channel gives two predictions regarding the business cycle. First, the house (collateral) price shall increase (decrease) before the mortgage amount increases (decreases). Second, the debt-to-GDP ratio by corporations shall experience a much stronger boom and bust than households. 

We provide evidence against both predictions by the above collateral channel, thus indicating the collateral channel is at least not the major channel driving the 1999-2010 U.S. business cycle. For the first prediction, Figure (\ref{fig_PLMNJ_vs_HousePriceIndex}) shows that at aggregate level the credit expansion, measured by private-label mortgages (non-jumbo), increases at the same pace as the house price between 1999 and 2005, rather than following the house price appreciation. In addition, from 2005 to 2010, credit expansion drops dramatically (between 2005 and 2008) ahead of house price crash (between 2007 to 2010), instead of following the house price crash. The above two aggregate trends clearly argue against the first prediction by the collateral channel. 

For the second prediction, subfigure (a) in Figure (\ref{fig_DebtToGDPRatio_HHBusiGov_HHSubCategories}) shows the debt-to-GDP ratio by non-financial corporations do not change very much while the debt-to-GDP ratio by household and nonprofit organizations experiences a huge boom between 1999 to 2007.\footnote{The debt-to-GDP by non-financial non-corporations (mainly small businesses) does experience a modest boom between 1999 to 2007. This result seems to be consistent with the view in \cite{mian2020does} that the credit boom usually results in nontradable sector growth. This result is also consistent with the view in \cite{adelino2015house} that the house price appreciation increases small business employment via the collateral channel.} After the onset of the Great Depression, US government debt increases dramatically to save the economy via a set of programs. A few programs relieve household debt and mortgage credit dries up between 2006 to 2011, resulting in graduate drop in debt-to-GDP ratio of household and nonprofit organizations. Subfigure (b) shows that the home mortgage is the largest part of debt (by level) by household and nonprofit organizations and home mortgage experiences a much larger boom than consumer credit and other liabilities (by growth rate) between 1999 and 2005. Subfigure (a) and (b) together emphasize the dominant role of household home mortgages during the 1999-2010 business cycle, thus against the collateral-driven credit cycle theory that mainly emphasize the corporation sector.

%------------------------------------------------------------
%------------------------------------------------------------
\subsection{The Business Uncertainty Theory}
%------------------------------------------------------------
%------------------------------------------------------------

The amplified boom and bust in house-related employment in the high net-export-growth metropolitan areas may reflect the increased business uncertainty in these areas \citep{bloom2009impact}. The rising business uncertainty in the bust period can be caused by local government policies or other factors, thus causing firms to pause their investment and hiring temporarily. This section provides two pieces of empirical evidence against this business uncertainty theory. 

Facing the increases business uncertainty in the high net-export-growth metropolitan areas in the bust period (07-10), commercial construction sector and tradable sector would immediately pause some investment and hiring when most firms pause their investments and hiring. However, Table (\ref{table_ComConst.D00t06vsD07t10.PLMNJ.4Reg}) shows that commercial construction employment does not experience a differentially stronger drop in the high net-export-growth areas (HNEG areas) during the bust period (07-10). In addition, Table (\ref{table_Tradable.D00t06vsD07t10.PLMNJ.4Reg}) shows that tradable sector continues to enjoy differentially stronger expansion in the HNEG areas in the bust period (07-10). These two pieces evidence help address the potential concern of business uncertainty theory.

%------------------------------------------------------------
%------------------------------------------------------------
\subsection{Extrapolative Expectation Theory}\label{sec:ExtrapolativeExpectationTheory}
%------------------------------------------------------------
%------------------------------------------------------------

This subsection is designed to address the concern that in the boom period (99-05) the differential higher growth in private-label mortgages (non-jumbo) in the high net-export-growth areas is due to extrapolative expectation \cite{eusepi2011expectations} driven by net export growth, either directly or indirectly. Mortgage demand (and other consumption) might overshoot and eventually result in a stronger business cycle in the high net-export-growth areas. We provide two pieces of evidence against this ``extrapolative expectation theory''.

%------------------------------------------------------------
%------------------------------------------------------------
\subsubsection{Irrelevance of Government Mortgage}\label{sec:ExtrapolativeExpectationTheory_PLMNJ_vs_GSEM}
%------------------------------------------------------------
%------------------------------------------------------------

In this subsection, we show empirical evidence that the growth of GSEMs is irrelevant to refined house employment growth (our focus) in boom (99-05) and bust (07-10) periods. The evidence of the irrelevance of GSEMs can help argue against the ``extrapolative expectation theory''. Since government-sponsored enterprise mortgages (GSEMs) are still cheaper than corresponding non-jumbo private-label mortgages (PLMNJs) for credit-qualified households \citep{sherlund2008jumbo}, extrapolation expectation-driven demand predicts both GSEMs and PLMNJs show stronger growth in high net-export-growth metropolitan areas. However, in the following empirical test, we only see such a trend in PLMNJs instead of in GSEMs. Therefore, such empirical evidence helps argue against the demand view.

In addition, the irrelevance of GSEMs can also address the concern that in the boom period (99-05) with credit expansion in private-label mortgages, a ``peer effect'' between neighbors may induce credit-qualified households to use GSEMs for house consumption. If such a peer effect were to exist, ignoring the impact of GSEMs may overestimate the impact of private-label mortgages (non-jumbo). We argue that the supply-side mortgage rates of government-sponsored enterprise mortgages cannot respond to net export growth because a legal constraint rules that government-sponsored enterprise mortgages (GSEMs) cannot consider regional economic conditions in setting up mortgage rates \citep{hurst2016regional}.

Table (\ref{table_RefineHouse.D00t06vsD07t10.GSEM.4Reg}) reports OLS, reduced-form, second stage, and the first stage of the stacked regression of refined house employment growth in the boom (00-06) and bust (07-10) periods. First, panel A shows that the OLS coefficients of GSEM growth (99-05) are generally insignificantly in both the boom (00-06) and bust (07-10) periods. The only exception is the bust period without any control. Unlike OLS, the reduced-form estimates in panel B are all significantly positive in the boom period (99-05) and negative in the bust period  (07-10). 2SLS estimates in panel C are insignificant except for a marginally negative coefficient in column (2). First-stage estimates in panel D show no correlation between the GSEM growth and gravity model-based IV for Net Export Growth (GIV-NEG), with very small first-stage F-Statistics (clustered Kleibergen-Paap F-statistic is 2.131, and the Montiel Olea-Pflueger Efficient F-Statistic is 2.149). These results reflect the fact that exogenous net export growth causes a stronger boom and bust in refined house employment through channels other than GSEMs. Thus, our results on GSEMs can help address the concern of the ``extrapolative expectation theory" and the ``peer effect".

%------------------------------------------------------------
%------------------------------------------------------------
\subsubsection{Exclusion Restriction: Prior (91-99) vs. Boom (99-05)}\label{sec:ExclusionRestriction}
%------------------------------------------------------------
%------------------------------------------------------------

This subsection provides evidence supporting the exclusion restriction of our IV approach: net export growth impacts refined house employment only via its impact on PLMNJs. This subsection is designed to address the concern that net export growth directly causes stronger refined house employment growth in the high net-export-growth areas via the ``extrapolative expectation", making our IV strategy potentially overestimate the impact of PLMNJ growth. 

Our test focuses on prior period (91-99) when there is no credit expansion and the funding cost of private-label mortgages (non-jumbo) (PLMNJs) is relatively higher than the boom period (99-05). In the prior period, two factors prevent PLMNJs from expanding more in the high net-export-growth areas: (1) government-sponsored enterprise mortgages (GSEMs) dominate the markets because of their low rate resulting from implicit government insurance \citep{sherlund2008jumbo}, and (2) GSEMs can not consider differences in regional economic conditions \citep{hurst2016regional}. Thus, the prior period can potentially provide evidence supporting the exclusion restriction: net export growth impacts refined house employment only via its impact on PLMNJ growth. Put differently, we can potentially show that when net export growth did not cause higher PLMNJ growth in the high net-export-growth areas in the prior period (91-99), it did not cause differentially higher growth in refined house employment (00-06) in these areas. Please note that employment measure is one year lag than mortgages because labor adjustment is usually lagged. We choose a long period from 1991 to 1999 because of two events: (1) North American Free Trade Agreement enforced in 1994 and (2) World Trade Organization started in 1995. These two events make the prior period an ideal setting for a placebo test: there is strong divergence in net export growth across metropolitan areas but no credit expansion at the aggregate level.

In the first step, we test the relationship between refined house employment growth and net export growth by the following stacked regression specification for 2SLS
\begin{equation}\label{eq:RefineHouseEmpShr_PriorVSBoom}
\resizebox{0.92\textwidth}{!}{$
\begin{aligned}
\triangle_{92,00} \& \triangle_{00,06} RefineHouseEmpShr_{c} & = \beta_{92,00} * \triangle_{91,99} \text{NetExp}_{m} \times Dum_{92,00} + \beta_{00,06} * \triangle_{99,05} \text{NetExp}_{m} \times Dum_{00,06} \\
& + \gamma_{92,00}* \bm{Controls_{c}} \times Dum_{92,00} + \gamma_{00,06}* \bm{Controls_{c}} \times Dum_{00,06}  + \alpha_{92,00} + \alpha_{00,06} + \epsilon_{period, c}
\end{aligned}
$} %end of \resizebox
\end{equation}
$Controls_{c}$ indicates control variables at county $c$ in the period start year (either 1991 or 1999). We use the gravity model-based instrumental variable $\triangle_{91,99}\text{givNetExp}_{m}$ and $\triangle_{99,05}\text{givNetExp}_{m}$ as IVs for $\triangle_{91,99}\text{NetExp}_{m}$ and $\triangle_{99,05}\text{NetExp}_{m}$. To reduce the impact of outliers, we winsorize the dependent variable at the 3\% and 97\% level in each period.

Table (\ref{table_RefineHouse.D92t00vsD00t06.NEG.D91t99vsD99t05.4Reg}) reports OLS, reduced-form, second stage, and first stage of the stacked regression of refined house employment growth in the prior (91-99) and boom (99-05) periods.\footnote{Please note that we require counties to be consistently covered by the HMDA database since 1991. Because of the smaller inclusion of metropolitan areas and the availability of the housing price index in the early years, my sample size is reduced to $705 \times 2 = 1410$ counties in columns (1)-(2) and further reduced to $727 \times 2 = 1254$ in column (3)-(4) due to data availability of housing supply elasticity.} First,  panel B shows that the reduced-form estimates of net export growth in the boom period (00-06) are positive and statistically significant at a one percent level. However, the reduced-form estimates of net export growth in the prior period (91-99) are insignificant. The same trend applies to the 2SLS estimates in panel C. First-stage estimates in panels D and E show that the strong positive correlation between the net export growth and its gravity model-based instrumental variable is quite stable for both periods. First-stage F tests are significant for all specifications for both periods. With full controls, clustered Kleibergen-Paap F-statistic is 74.77 in the prior period (91-99) and 17.44 in the boom period (99-05). And the corresponding Montiel Olea-Pflueger Efficient F-Statistic is 78.55 and 16.98. My 2SLS estimates in the boom period is between two to four times of the OLS estimates, much lower than the average nine \citep{jiang2017have} Thus, my estimates are very unlikely biased by a weak instrumental variable. For the coefficient equality test in panel C, chi-square statistics are large, and the p-value is 0.046 with full controls, meaning that the two coefficients are statistically unequal. Taken together, the results from the above tables verify that refined house employment growth between 1991 and 1999 is not affected by net export growth. 

In the second step, we test the relationship between growth in private-label mortgages (non-jumbo) (PLMNJ) and net export growth in both periods. The regression test is the same as Equation (\ref{eq:RefineHouseEmpShr_PriorVSBoom}) except that the dependent variable is the growth in private-lable mortgages (non-jumbo) (PLMNJ) in the prior (91-99) and boom (99-05) periods. \footnote{Please note that we require counties to be consistently covered by the HMDA database since 1991. Because of the smaller inclusion of metropolitan areas in the early years, my sample size is reduced to $701 \times 2 = 1402$ counties in columns (1)-(2) and further reduced to $623 \times 2 = 1246$ counties in column (3)-(4) due to data availability of housing supply elasticity.}\footnote{In both periods, we drop four outliers, since these three outliers can make the coefficient of net export growth negative and significant in the prior period (91-99). Even though the negative and significant coefficient does not violate my conclusion, we drop the four outliers to show the results for most observations. In the Appendix Section \ref{subsec:App_EmpCreditExpansion}, we will compare results with and without the four outliers.} Table (\ref{table_PLMNJ.D91t99vsD99t05.NEG.D91t99vsD99t05.4Reg}) reports OLS, reduced-form, second stage, and first stage of the stacked regression of PLMNJ growth in the prior period (91-99) and boom period (99-05). First, panel A shows that the OLS coefficients of net export growth (99-05) are positive and statistically significant at a one percent level and are stable across specifications. However, the OLS estimates for net export growth (91-99) are insignificant. The same trend applies to reduced-form estimates in panel B and 2SLS estimates in panel C. First-stage estimates in panel D show that the strong positive correlation between the net export growth and its gravity model-based instrumental variable is quite stable for both periods. First-stage F tests are very large and relative stable across specifications. In the boom period, the 2SLS estimates are only around twice of the OLS estimates, much lower than the average nine \citep{jiang2017have}. Thus, my estimates are very unlikely biased by a weak instrumental variable. 
For the coefficient equality test of net export growth in prior and boom periods in 2SLS in panel C, the chi-square statistics are large, and p-values are below 0.01, meaning the two coefficients are statistically different. Taken together, results from the above table verify that PLMNJ, between 1991 and 1999, is not affected by net export growth.

Taking results from the above two tests, when net export growth does not increase PLMNJ growth in the prior period (91-99), it does not increase refined house employment growth (00-06). However, when net export growth does increase PLMNJ growth in the boom period (99-05), it does increase refined house employment growth (00-06). These results provide evidence supporting the exclusion restriction of the IV strategy: net export growth can only impact refined house employment growth via private-label mortgage (non-jumbo) growth. These results help address the concern that net export growth directly causes stronger refined house employment growth in the high net-export-growth areas.

%----------------------------------------------------------------------

%----------------------------------------------------------------------
% section 7: Empirical.Robustness.tex

%------------------------------------------------------------
%------------------------------------------------------------
\section{Empirical: Robustness}\label{subsec:Empirical.Robustness}
%------------------------------------------------------------
%------------------------------------------------------------

%------------------------------------------------------------
%------------------------------------------------------------
%\subsection{Robustness: State-Level Heterogeneity}\label{subsec:Empirical.Robustness.StateDifference}
%------------------------------------------------------------
%------------------------------------------------------------

In this subsection, we perform robustness tests to support the main conclusion that, induced by net export growth, credit expansion in private-label mortgages (non-jumbo) causes the 1999-2009 U.S. house employment boom (00-06) and bust (07-10) across metropolitan areas. We show that this main conclusion is robust to state-level differences in anti-predatory lending laws \citep{di2017credit}, recourse laws \cite{ghent2011recourse}, judicial requirement in foreclosure \cite{mian2015foreclosures}, difference between sand states and other states \cite{choi2016sand}, and state capital gain tax \cite{gao2020economic}.

%------------------------------------------------------------
\subsection{Anti-Predatory-Lending States vs. Other States}
%------------------------------------------------------------

In this subsection, we show that our main results hold after accounting for the state-level difference in the anti-predatory lending law. According to \cite{di2017credit}, prior to 2004, a dozen of states had already implemented anti-predatory laws to protect mortgage borrowers. However, on January 7th, 2004, the Office of the Comptroller of the Currency (OCC) preempted national banks (instead of state-chartered depository institutions and independent mortgage companies) from state anti-predatory lending law (APL law). They show that such deregulation results in the national bank’s credit expansion (relative to state-regulated institutions), house price growth, and nontradable employment rise in 2004-2006, but a sharp decline subsequently in these states relative to other states. Appendix Table 1 from \cite{di2017credit} summarizes the list of APL states before 2004. Since HMDA only contains annual data, we restrict my APL states to the states that implemented APL at least half a year before 2004, resulting in eleven APL states.\footnote{According to \cite{di2017credit}, these eleven APL states (including DC) are California, Connecticut, District of Columbia, Georgia, Maryland, Michigan, Minnesota, New York, North Carolina, Texas, West Virginia.}

Our stacked 2SLS regression is 
\begin{equation}\label{eq:RefineHouseEmp.BoomBustonPLMNJ_APLvsNone}
\resizebox{0.92\textwidth}{!}{$
\begin{aligned}
\triangle_{00,06} \& \triangle_{07,10} RefinedHouseEmpShr_{c} & = \beta_{Boom} * \triangle_{99,05} Ln(PLMNJ_{c}) \times Dum_{00,06} + \beta_{Bust} * \triangle_{99,05} Ln(PLMNJ_{c}) \times Dum_{07,10} \\
 & + \beta_{APL, Boom} * \triangle_{99,05} Ln(PLMNJ_{c}) \times Dum_{00,06} \times Dum_{APL} + \beta_{APL, Bust} * \triangle_{99,05} Ln(PLMNJ_{c}) \times Dum_{07,10} \times Dum_{APL} \\
 & + \gamma_{Boom} * \bm{Controls_{c}} \times Dum_{00,06} + \gamma_{Bust} * \bm{Controls_{c}} \times Dum_{07,10} + \epsilon_{c}
\end{aligned}
$} %end of \resizebox
\end{equation}
Controls, weight, and standard errors are the same as Eq(\ref{eq:RefineHouseEmpShrBoomBustonPLMNJ}).

Please note that in either boom or bust period, we have two endogenous variables here. $\triangle_{99,05} Ln(PLMNJ_{c})$ is instrumented by $\triangle_{99,05}\text{givNetExp}_{m}$ and $\triangle_{99,05} Ln(PLMNJ_{c}) \times Dum_{APL}$ is instrumented by $\triangle_{99,05}\text{givNetExp}_{m} \times Dum_{APL}$. For each of the two separate first-stage regression F-tests, we report Sanderson-Windmeijer robust (clustered) F-statistics \citep{sanderson2016weak}. The SW F-statistics is 11.22 for PLMNJ growth and is 15.41 for the interaction between PLMNJ growth and a dummy of APL-states in column (4), meaning each F-stage regression is significant and each instrument is strong for its endogenous variable. To evaluate the overall strength of two instruments, we report the p-value of robust (clustered) Kleibergen-Paap test statistic calculated by \citep{windmeijer2021testing}. This p-value is 0.0065 in column (4), meaning the two separate instruments are jointly strong for the two endogenous variables.

Table (\ref{table_Robust.APLvsNone.RefineHouse.D00t06.D07t10}) reports the above 2SLS results and shows two key conclusions. First, after controlling the potential differential trend in APL-states, for all metropolitan areas, the growth of private-label mortgages (non-jumbo) 1999-2005 leads to a stronger refined house employment boom (00-06) and a stronger bust (07-10) in the high net-export-growth areas. Second, compared to non-APL states, APL-states did not experience a differentially stronger refined house employment boom (00-06) or a differentially stronger bust (07-10) in the high net-export-growth metropolitan areas.

%------------------------------------------------------------
\subsection{Non-Recourse vs. Recourse States}
%------------------------------------------------------------

In this subsection, we show that the main conclusion is robust to the state-level difference in the mortgage recourse law. \cite{ghent2011recourse} document that, in 39 states of the U.S., mortgages are recourse loans.\footnote{39 recourse states can be found in Table 1 in \cite{ghent2011recourse}. 11 non-recourse states are Alaska, Arizona, California, Iowa, Minnesota, Montana, North Carolina, North Dakota, Oregon, Washington, and Wisconsin.} In such states, lenders could go after the borrower's additional assets to recover the mortgage loss not covered by the foreclosure sale through obtaining a deficiency judgment. They find that in recourse states, borrowers are less responsive to negative equity, and defaults are more likely to happen via a lender-friendly procedure. We worry that the non-recourse law may discourage lenders from increasing mortgage credit and induce households to be more willing to default when facing falling house prices. 

Regression specification is the same as Eq (\ref{eq:RefineHouseEmp.BoomBustonPLMNJ_APLvsNone}) except that the dummy variable is for non-recourse states. Table (\ref{table_Robust.NRCvsRC.RefineHouse.D00t06.D07t10}) reports the above 2SLS results and shows two key conclusions. First, after controlling the potentially different trend in the non-recourse states, for metropolitan areas in all states, the growth of private-label mortgages (non-jumbo) 1999-2005 leads to a stronger refined house employment boom (00-06) and a stronger bust (07-10) in the high net-export-growth metropolitan areas. Second, compared to recourse states, non-recourse states did not experience a stronger refined house employment bust (07-10) or a stronger bust (07-10). The SW F-statistics and p-value of robust (clustered) Kleibergen-Paap test statistics all show that instruments are separately and jointly strong for the two endogenous variables.

%------------------------------------------------------------
\subsection{Non-Judicial vs Judicial States}
%------------------------------------------------------------

In this subsection, we show that our main conclusion is robust to the differences between the non-judicial the judicial states. In addition, growth in private-label mortgages (non-jumbo) led to a stronger refined house employment boom (00-06) and a stronger bust in non-judicial states. \cite{mian2015foreclosures} shows that, in 20 states of the U.S., foreclosures of a delinquent property need judicial judgement.\footnote{20 judicial states can be found in Figure 2 in \cite{mian2015foreclosures} and \url{https://www.realtytrac.com/real-estate-guides/foreclosure-laws/}. The twenty judicial states are Connecticut, Delaware, Florida, Illinois, Indiana, Kansas, Kentucky, Louisiana, Maine, Maryland, Massachusetts, Nebraska, New Jersey, New Mexico, New York, North Dakota, Ohio, Pennsylvania, South Carolina, and Vermont.} In these states, in order to sell a delinquent property through foreclosure, lenders are required by law to file a notice with a judge providing evidence of the delinquency and get court approval. In contrast, in non-judicial states, the foreclosure process is much easier and does not need court approval. For more details, see \cite{mian2015foreclosures}. \cite{mian2015foreclosures} find that lenders in non-judicial states are twice as likely to foreclose on delinquent property. We worry that non-judicial law may encourage lenders to expand mortgages more aggressive, resulting in a stronger refined house employment boom (00-06) and a stronger bust (07-10) subsequently

Regression specification is the same as Eq (\ref{eq:RefineHouseEmp.BoomBustonPLMNJ_APLvsNone}) except that dummy variable is for non-judicial states. Table (\ref{table_Robust.NJDvsJD.RefineHouse.D00t06.D07t10}) reports the above 2SLS results and shows two key conclusions. First, after controlling the potentially different trend in non-recourse states, for all metropolitan areas, the growth of private-label mortgages (non-jumbo) 1999-2005 led to both a stronger boom (00-06) and a stronger bust (07-10) in refined house employment in the high net-export-growth areas. Second, compared to judicial states, non-judicial states experienced both a stronger boom (00-06) and a stronger bust (07-10) in refined house employment, caused by the growth of private-label mortgages (non-jumbo) 1999-2005. This evidence is consistent with the notion that non-judicial requirement ease the process of foreclosure \citep{mian2015foreclosures}. The SW F-statistics and p-value of robust (clustered) Kleibergen-Paap test statistics all show that instruments are both separately and jointly strong for the two endogenous variables.

%------------------------------------------------------------
\subsection{Sand vs Other States}
%------------------------------------------------------------

In this subsection, we show that our main conclusion is robust to the difference between the sand and non-sand states. In addition, growth in private-label mortgages (non-jumbo) leads to both a stronger boom and a stronger bust in the refined house employment in sand states. Many studies highlight that sand states (Arizona, California, Florida, and Nevada) experienced phenomenal housing cycles in comparison to the rest of the United States \citep{choi2016sand}. We expect that, in my setting, sand states may experience a stronger cycle in the refined house employment.

For regression specification, we add interaction of dummy variable for sand states and period dummy to Eq (\ref{eq:RefineHouseEmpShrBoomBustonPLMNJ}). We do not use interaction of three terms because limited number of metropolitan counties (only 73 metro counties in sand states) present a weak IV concern. Since here, my major focus is the differential housing boom and bust of sand states, a dummy variable can serve this purpose. In order to study the different impact from the growth of private-label mortgages, more granular data is necessary.

Table (\ref{table_Robust.SandvsNone.RefineHouse.D00t06.D07t10}) reports the above 2SLS results and shows two key conclusions. First, compared to other states, sand states experience a stronger refined house employment boom (00-06) and a stronger bust (07-10). Second, after controlling for the differential trend in the sand states, we only use the within-sand-states and within-other-states differences across metropolitan areas. Since the cross-group difference between sand states and other states is removed by the interaction terms of sand dummy and period dummy, my 2SLS shall be interpreted as evidence strongly supporting my main conclusion: induced by net export growth, private-label mortgage causes both a stronger boom (00-06) and a stronger bust (07-10) in the refined house employment across all metropolitan areas. Meanwhile, the reduced cross-metro variation due to sand-state dummy reduces the F-statistics: the kleibergen-Paap (2006) robust (clustered) statistics is 7.964 and Montiel Olea-Pflueger (2013) efficient statistics is 7.967. The reduced cross-metro variation might also result in the weaker but still marginal significant coefficients in the column (4) with full controls.

%------------------------------------------------------------
\subsection{State Capital Gain Tax}

In this section, we show that my major conclusion is robust to the inclusion of state capital gain tax as a control variable. \cite{gao2020economic} find that speculation measured by non-owner-occupied purchase mortgages are discouraged by the state capital gain tax and such speculation contribute to the housing boom and bust. Based on their findings, we expect state capital gain discourage housing boom and bust. 

For regression specification, we add interaction of state capital gain tax rate and period dummy to Eq (\ref{eq:RefineHouseEmpShrBoomBustonPLMNJ}). Table (\ref{table_Robust.StCapGainTax.RefineHouse.D00t06.D07t10}) reports the above 2SLS results and shows two key conclusions. First, after controlling for state capital gain tax rate, my major conclusion holds. Second, state capital gain tax rate does not contribute to the refined house employment boom (00-06) but strengthens the bust (07-10). This result partially contradicts with the discouraging effect of capital gain tax on housing speculation \citep{gao2020economic}. My result could be partially explained by the owner-occupied housing exclusion by state capital gain tax, which encourages owner-occupied home purchase. According to Tax Foundation\footnote{The Tax Foundation Report is here \url{https://taxfoundation.org/research/all/federal/capital-gains-taxes/}}, individual homeowner may exclude up to \$250,000 (\$500,000 for a couple) of capital gain if the homeowner had lived in the housing unit for at least two of the previous five years. The exemption can be taken only once every two years and applies at both the federal and state level. Thus my result may sum the two opposing effect of state capital gain on home purchase.

%----------------------------------------------------------------------

%----------------------------------------------------------------------
% section 8: Empirical Results: Double Difference

%\input{Empirical.DoubleDifference}
%----------------------------------------------------------------------

%----------------------------------------------------------------------
% section 9: Conclusion

%----------------------------------------------------------------------------
%\clearpage

\section{Conclusion}
The U.S. business boom and bust in the 2000s are unprecedented, resulting in the deepest recession since the Great Depression. This Great Recession has renewed the debate on the causes and mechanism of business cycles. Recent cross-country studies on business cycles only achieve association while within-country studies mainly focus on the bust periods. 

Using a new research design that builds on the ``economic base theory" by \cite{tiebout1956pure} and a novel instrumental approach by \cite{feenstra2019us}, we investigate into the boom period of the 1999-2010 U.S. cross-metro differential business cycle. We provide causal evidence that credit expansion in the private-label mortgages causes the differentially stronger boom (1999-2005) and bust (2007-2010) cycle in house-related industries in the high net-export-growth metropolitan areas than in the low net-export-growth areas. Most importantly, our unique research design enables us to test the relevance of almost all major hypothesis (theories) on the business cycles. We provide multiple pieces of evidence against each of the alternative theories (hypotheses): the speculative euphoria hypothesis, the real business cycle theory (including two variants: the natural disaster hypothesis and the technology shock in the construction sector), the collateral-driven credit cycle theory, the business uncertainty theory, and the extrapolative expectation theory. Further, our major conclusion is robust to the state-level differences in the anti-predatory lending law, the recourse law, the judicial requirement of foreclosure law, the classification of sand states, and the state-level capital gain tax rates.

%----------------------------------------------------------------------

\ifx\undefined\BySame
\newcommand{\BySame}{\leavevmode\rule[.5ex]{3em}{.5pt}\ }
\fi
\ifx\undefined\textsc
\newcommand{\textsc}[1]{{\sc #1}}
\newcommand{\emph}[1]{{\em #1\/}}
\let\tmpsmall\small
\renewcommand{\small}{\tmpsmall\sc}
\fi

%----------------------------------------------------------------------
% section 9: Figures and Tables

%------------------------------------------------------------
%------------------------------------------------------------
\pagebreak
%------------------------------------------------------------
%------------------------------------------------------------
\section{Figures and Tables}

%------------------------------------------------------------
% figure 0: fig_USMetroCty_NetExpGrowth

%------------------------------------------------------------
% fig_ZIP_MortGrowth_cbQuint_zipHalf&Quint

%------------------------------------
\begin{figure}[h!] 
    \centering
    \begin{subfigure}[t]{0.9\textwidth}
        \centering
        \includegraphics[height=7.5cm]{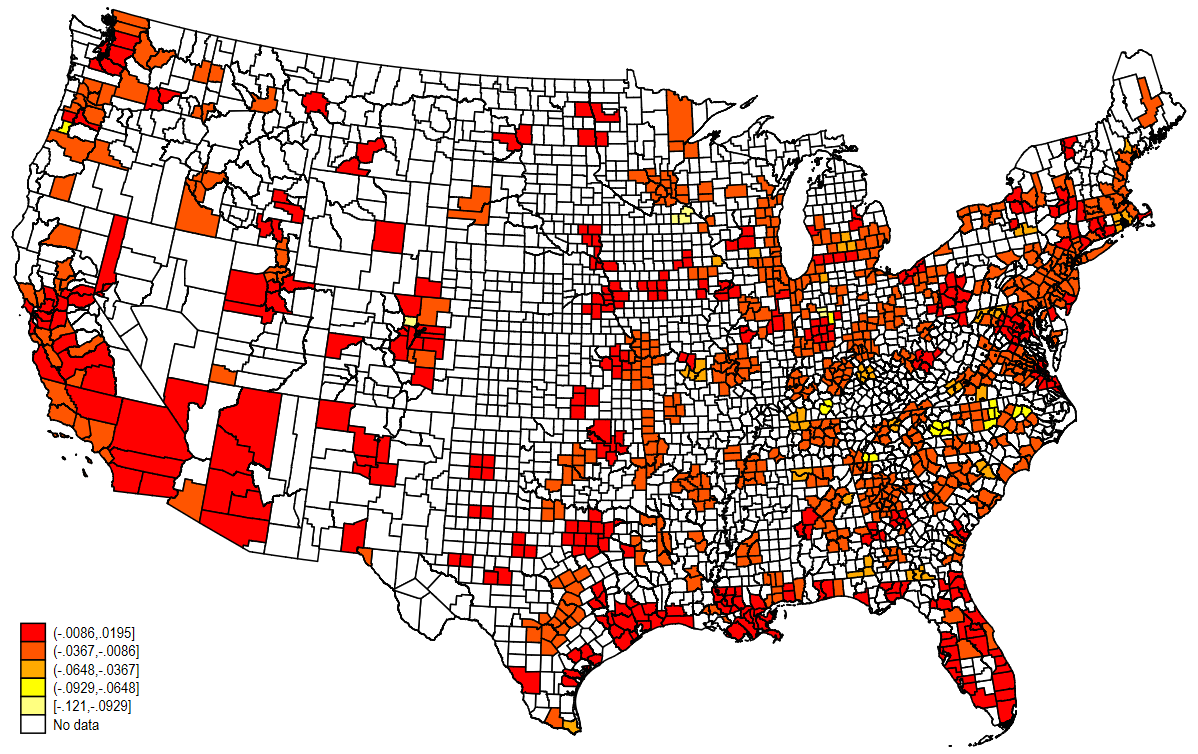}
        \caption{Net Export Growth (1991-1999)}
    \end{subfigure}%
    \hfill 
    \begin{subfigure}[t]{0.9\textwidth}
        \centering
        \includegraphics[height=7.5cm]{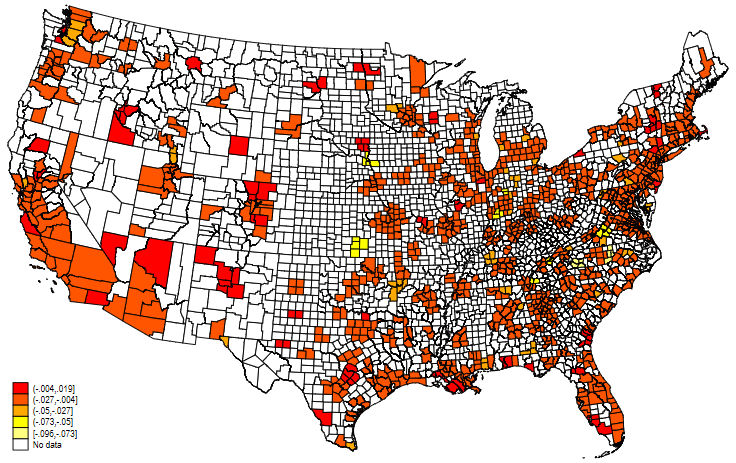}
        \caption{Net Export Growth (1999-2005)}
    \end{subfigure}
    \caption{\textbf{U.S. Mainland Metropolitan Heat Map of Net Export Growth: Two Periods}  \smallskip  \newline 
    {\footnotesize This figure displays the U.S. mainland metropolitan heat map of net export growth measure in two periods: subfigure (a) is for 1991-1999, and subfigure (b) is for 1999-2005. The net export growth measure at the metropolitan level for a period is defined in equation (\ref{equ:NEG_m}). In the above figures, each small area with a boundary is a county. Counties with white color are non-metropolitan areas in the 2003 CBSA version (1085 counties). Metropolitan counties are painted with colors ranging from yellow (for low net export growth) to red (for high net export growth) in five categories. 
        } %end of small font
    } % end of caption
    \label{fig_ZIP_MortGrowth_cbQuint_zipHalf&Quint}
    
\end{figure} 
%------------------------------------

%------------------------------------------------------------
%------------------------------------------------------------
% figure 1: fig_TotEmpShr_92To11

%------------------------------------------------------------
% figure 1: fig_TotEmpShr_92To11

\begin{figure}[h!] 
    \centering
    \includegraphics[width=16cm, height=12cm]{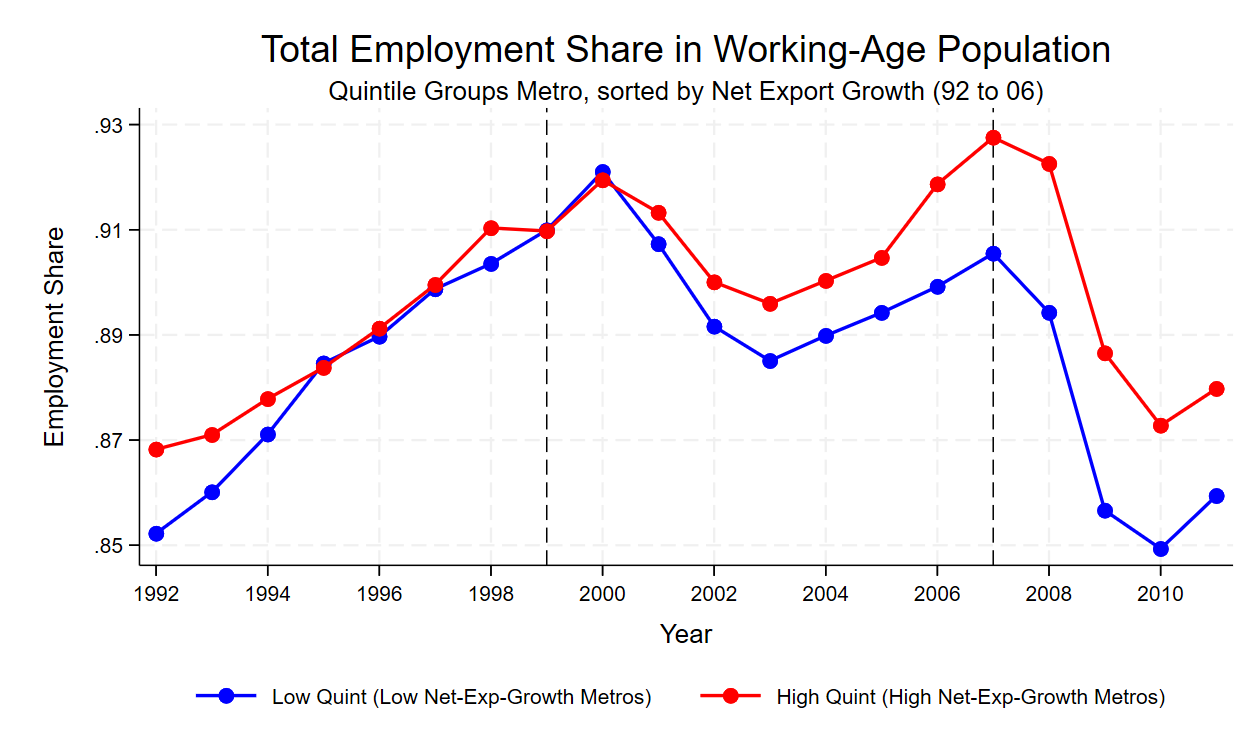}
    \caption{\textbf{Total Employment Share (92-11) in Metro Areas, Top vs. Bottom Quintile of Net Export Growth (91 to 07)} \smallskip \newline 
    {\footnotesize This figure displays the time series of weighted average total employment share in the working-age population for top and bottom quintile groups of Metropolitan Areas (MSAs) from 1992 to 2011. For the entire period, the quintile groups are sorted by net export growth (1991 to 2007) at the metropolitan level (CBSA code 2003 version). The sample includes 299 metropolitan areas (687 counties) that are consistently covered by the Employment sample between 1992 and 2011. These 687 out of HMDA consistent 712 counties have non-missing employment data from 1992 to 2011 from the Bureau of Economics Analysis in the U.S. Department of Commerce. The bottom quintile group comprises 60 metros (91 counties), and the top quintile group comprises 59 metros (104 counties) throughout the entire period. The total employment share is weighted by working-age population (age between 15 and 64) in counties within each group in each year. The red line represents the top quintile group, while the blue line represents the bottom quintile group.
    } %end of small font
    } % end of caption
    \label{fig_TotEmpShr_92To11}
    % note that \label is given after \caption.
\end{figure}

\pagebreak
%------------------------------------------------------------
%------------------------------------------------------------
% figure 2: fig_RefineHouseEmpShr_92To11

%------------------------------------------------------------
% figure 1: fig_RefineHouseEmpShr_92To11

\begin{figure}[h!] 
    \centering
    \includegraphics[width=16cm, height=12cm]{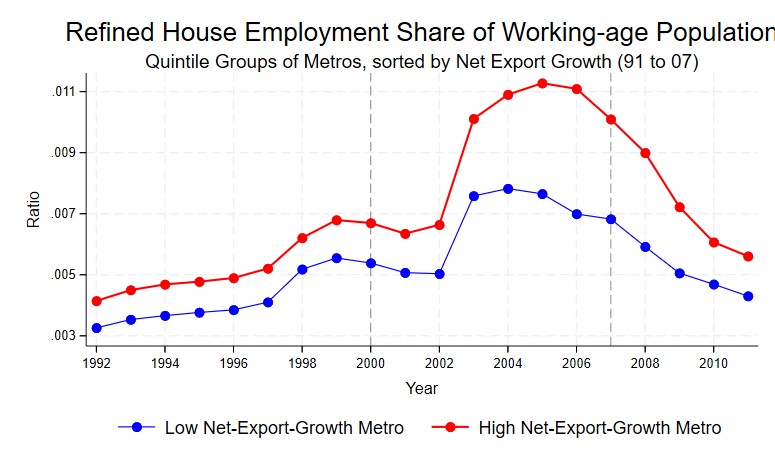}
    \caption{\textbf{Refined House Employment Share (92-11) in Metro Areas, Top vs. Bottom Quintile of Net Export Growth (91 to 07)} \smallskip \newline 
    {\footnotesize This figure displays the time series of weighted average refined house employment share in the working-age population for top and bottom quintile groups of Metropolitan Areas (MSAs) from 1992 to 2011. For the entire period, the quintile groups are sorted by net export growth (1991 to 2007) at the metropolitan level (CBSA code 2003 version). The sample includes 299 metropolitan areas (687 counties) that are consistently covered by the Employment sample between 1992 and 2011. These 687 out of HMDA consistent 712 counties have non-missing employment data from 1992 to 2011 from the Bureau of Economics Analysis in the U.S. Department of Commerce. The bottom quintile group comprises 60 metros (91 counties), and the top quintile group comprises 59 metros (104 counties) throughout the entire period. The refined house employment share is weighted by working-age population (age between 15 and 64) in counties within each group in each year. The red line represents the top quintile group, while the blue line represents the bottom quintile group.
    } %end of small font
    } % end of caption
    \label{fig_RefineHouseEmpShr_92To11}
    % note that \label is given after \caption.
\end{figure}

%\pagebreak
%------------------------------------------------------------
%------------------------------------------------------------
% figure 3: fig_GSEMvsPLMNJ_91t11_combine

%------------------------------------------------------------
% figure 2: fig_GSEMvsPLMNJ_91t11_combine

\begin{figure}[h!] 
    \centering
    \includegraphics[width=16cm, height=12cm]{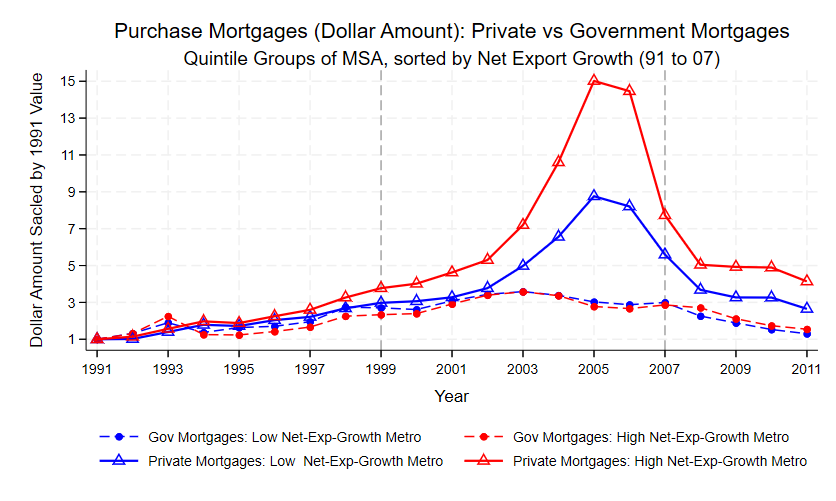}
    \caption{\textbf{Mortgage Growth (91-11) across Metropolitan Areas: GSEM vs. PLMNJ, High vs. Low Quintile Sorted by Net Export Growth (91-07).}  \smallskip \newline 
    {\footnotesize This figure displays the time series of the weighted-average dollar amount of Government-Sponsored Enterprise Mortgages (GSEM) (in dash lines with dots) and Private-Label Mortgages (Non-Jumbo) (PLMNJ) (in solid lines with triangles) for high and low quintile groups of metropolitan statistical areas (MSA) from 1991 to 2011. Both types of mortgages only include purchase loans. For the entire period, the quintile groups are sorted by net export growth (1991 to 2007) at the MSA level. The whole sample includes 301 MSA (712 counties) that are consistently covered by the HMDA sample after 1990 due to the smaller coverage of metropolitan areas in the early years. The low quintile group comprises 61 MSA (94 counties), and the high quintile group comprises 60 MSA (112 counties) throughout the entire period. The number of loans is weighted by the county-level housing units within each group in each year. Throughout the entire period, the time series of the weighted-average number of loans of each group are divided by their 1991 values, ensuring that both groups start at a value of 1 in 1991. The red lines represent the high quintile group, while the blue lines represent the low quintile group.
        } %end of small font
    } % end of caption
    \label{fig_GSEMvsPLMNJ_91t11_combine}
    % note that \label is given after \caption.
\end{figure}

%\pagebreak
%------------------------------------------------------------
%------------------------------------------------------------
% figure 3: fig_CreditExpansion_vs_Sepculation

%------------------------------------------------------------
% figure 3: fig_CreditExpansion_vs_Sepculation

\begin{figure}[h!] 
    \centering
    \includegraphics[width=16cm, height=12cm]{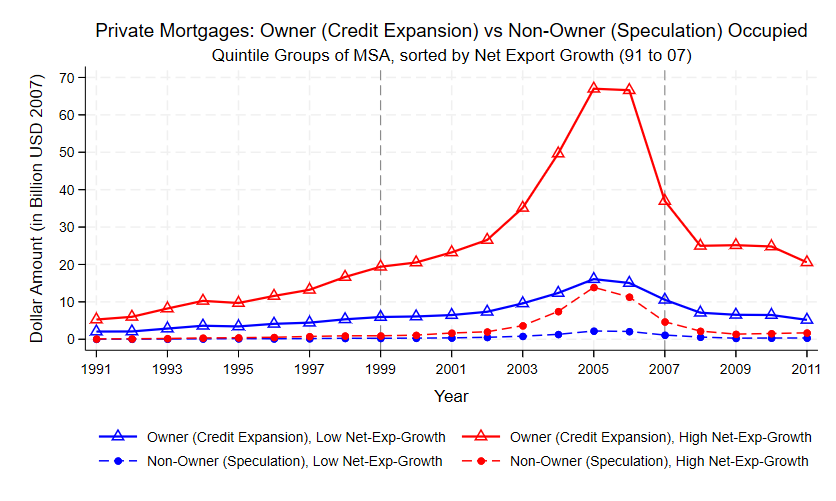}
    \caption{\textbf{Credit Expansion vs. Speculation (91-11) across Metropolitan Areas: High vs. Low Quintile Sorted by Net Export Growth (91-07).}  \smallskip \newline 
    {\footnotesize This figure displays the time series of dollar amount of ``owner-occupied" (credit expansion) home purchase private-label mortgages (in solid lines with triangles) and ``non-owner-occupied" (speculation) home purchase private-label mortgages (in dash lines with dots) from 1991 to 2011 for high and low quintile groups of metropolitan statistical areas (MSA). For the entire period, the quintile groups are sorted by net export growth (1991 to 2007) at the MSA level. The whole sample includes 301 MSA (712 counties) that are consistently covered by the HMDA sample after 1990 due to the smaller coverage of metropolitan areas in the early years. The low quintile group comprises 61 MSA (94 counties), and the high quintile group comprises 60 MSA (112 counties) throughout the entire period. The dollar amount is adjusted to the 2007 USD by the Personal Consumption Expenditures Chain-type Price Index (PCEPI) from Federal Reserve Bank of St. Louis. The red lines represent the high quintile group, while the blue lines represent the low quintile group.
        } %end of small font
    } % end of caption
    \label{fig_CreditExpansion_vs_Sepculation}
    % note that \label is given after \caption.
\end{figure}

\pagebreak
%------------------------------------------------------------
%------------------------------------------------------------
% figure 4: fig_PLMNJ_vs_HousePriceIndex

%------------------------------------------------------------
% fig_PLMNJ_vs_HousePriceIndex

\begin{figure}[h!] 
    \centering
    \includegraphics[width=16cm, height=12cm]{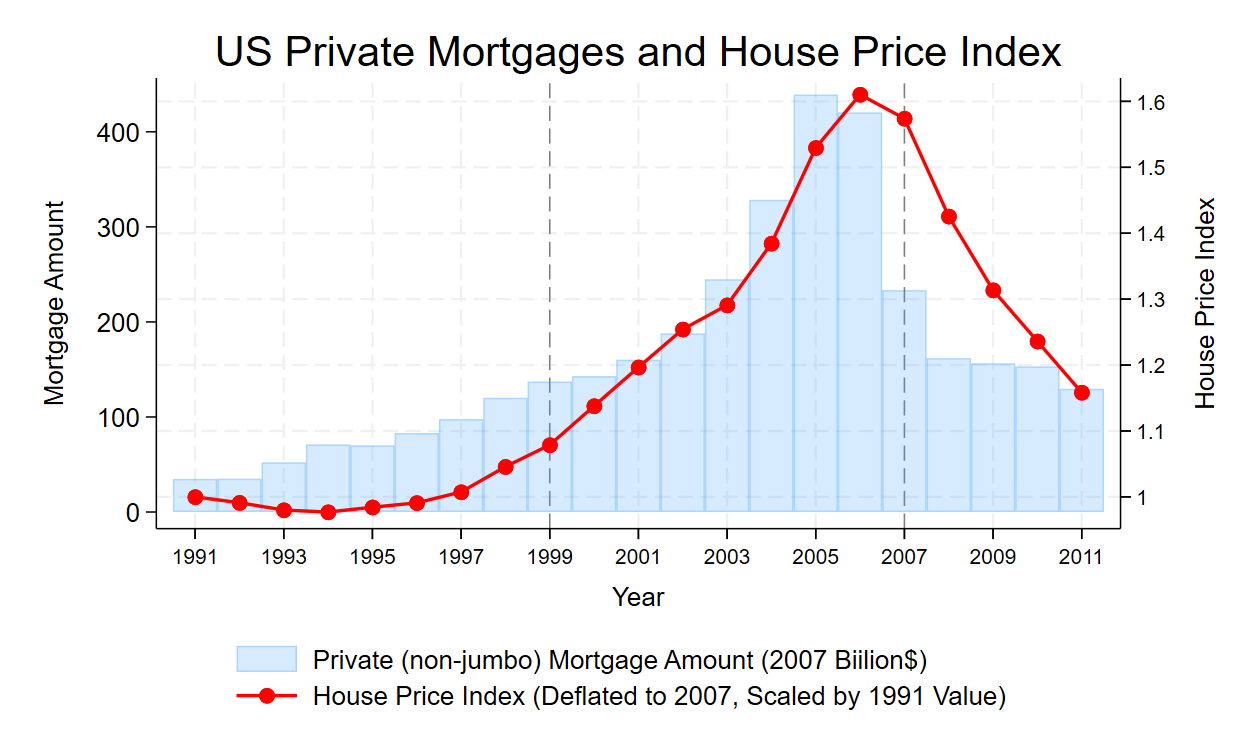}
    \caption{\textbf{Credit Expansion vs. House Price Index in USA (1991-2011).}  \smallskip \newline 
    {\footnotesize This figure displays the time series of dollar amount of private-label mortgages (non-jumbo) (in blue bars) and house price index (in red line) in USA from 1991 to 2011. The whole sample includes 301 MSA (679 counties) that are consistently covered by the HMDA sample after 1990 and by the house price index from Federal Housing Financing Agency (FHFA). The dollar amount of mortgages in billions is deflated to the 2007 USD by the Personal Consumption Expenditures Chain-type Price Index (PCEPI) from Federal Reserve Bank of St. Louis. The housing price index is first deflated by PCEPI to 2007 and further scaled by its 1991 value.
        } %end of small font
    } % end of caption
    \label{fig_PLMNJ_vs_HousePriceIndex}
    % note that \label is given after \caption.
\end{figure}

%\pagebreak
%------------------------------------------------------------
%------------------------------------------------------------
% figure 5: fig_DebtToGDPRatio_HHBusiGov_HHSubCategories

%------------------------------------------------------------
% fig_DebtToGDPRatio_HHBusiGov_HHSubCategories

%------------------------------------
\begin{figure}[h!] 
    \centering
    \begin{subfigure}[t]{0.9\textwidth}
        \centering
        \includegraphics[height=7.5cm]{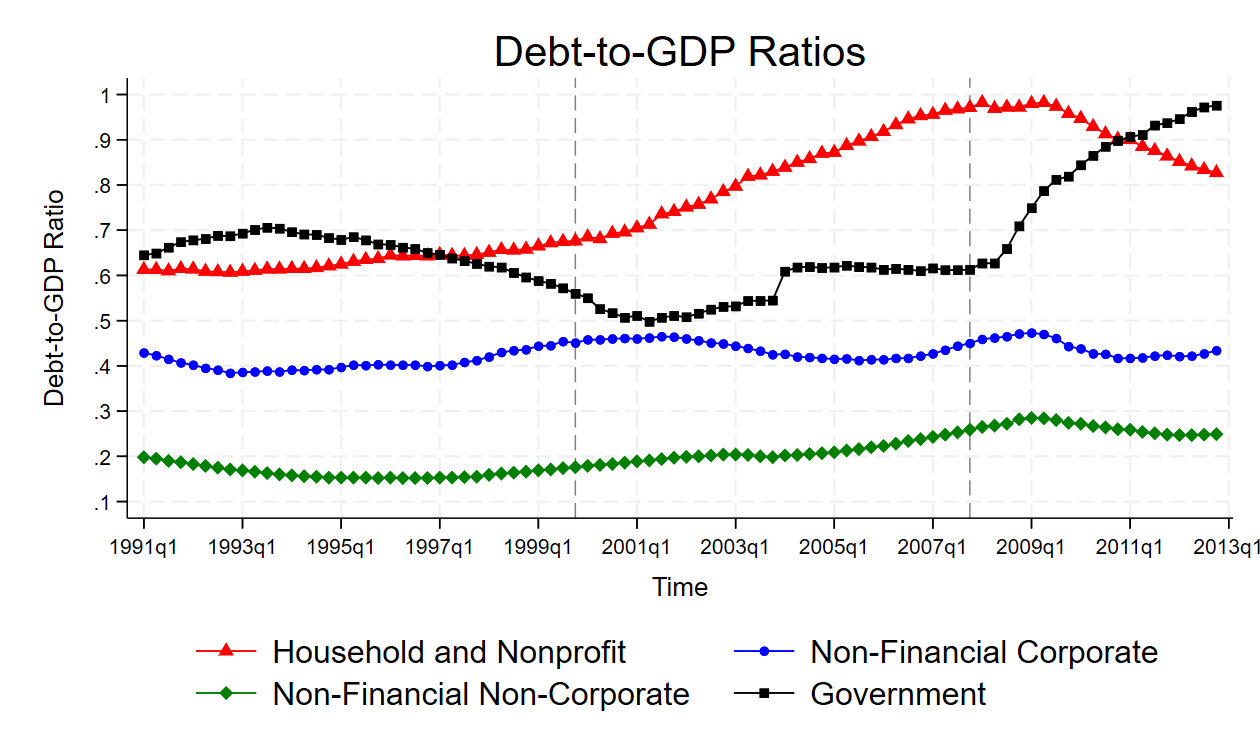}
        \caption{Households, Business, and Government}
    \end{subfigure}%
    \hfill 
    \begin{subfigure}[t]{0.9\textwidth}
        \centering
        \includegraphics[height=7.5cm]{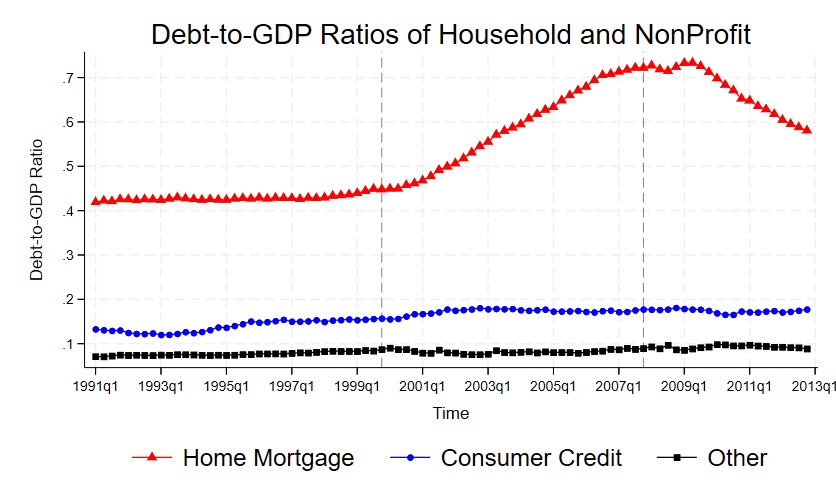}
        \caption{Households: Mortgages, Consumer Credit, and Other Liability}
    \end{subfigure}
    \caption{\textbf{Debt-to-GDP Ratios: Households Subcategories, Business, and Government}  \smallskip  \newline 
    {\footnotesize This figure displays the time series of debt-to-GDP ratios of households and nonprofit organizations, business organizations, and government in the USA from 1991 to 2012. Subfigure (a) includes debt-to-GDP ratios of three groups: households and nonprofit organizations, business organizations, and government. Subfigure (b) includes debt-to-GDP ratios of three sub-categories within the households and nonprofit organizations: mortgages, consumer credit and other liabilities. 
        } %end of small font
    } % end of caption
    \label{fig_DebtToGDPRatio_HHBusiGov_HHSubCategories}
    
\end{figure} 
%------------------------------------

%\pagebreak 
%-----------------------------------------------------------------
%%%%%%%%%%%%%%%%%%%%%%%%%%%%%%%%%%%%
% fig_ZIP_MortGrowth_cbQuint_zipHalf&Quint
%%%%%%%%%%%%%%%%%%%%%%%%%%%%%%%%%%%%
%\input{Figure/fig_ZIP_MortGrowth_cbQuint_zipHalf&Quint}

%\pagebreak 
%-----------------------------------------------------------------
%%%%%%%%%%%%%%%%%%%%%%%%%%%%%%%%%%%%
% fig_ZIP_HPIGrowth_cbQuint_zipHalf&Quint
%%%%%%%%%%%%%%%%%%%%%%%%%%%%%%%%%%%%
%\input{Figure/fig_ZIP_HPIGrowth_cbQuint_zipHalf&Quint.tex}

\pagebreak
%------------------------------------------------------------
% table 0: table_EmploymentIndustryClassification

%---------------------------------------------------------------

%%%%%%%%%%%%%%%%%%%%%%%%%%%%%%%%%%%%%%%%%%%%%%%%
% table_EmploymentIndustryClassification
%%%%%%%%%%%%%%%%%%%%%%%%%%%%%%%%%%%%%%%%%%%%%%%%

\noindent 

\begin{table}[h!]
\centering
\caption{
\textbf{Employment Industry Classification} \smallskip \newline
{\scriptsize
This table reports industry classification for employment data.
} % end of small font size
} % end of caption
\label{table_EmploymentIndustryClassification}
\resizebox{\columnwidth}{!}{%
\begin{tabular}{lllll}
\toprule 
\addlinespace
\multicolumn{5}{l}{\textbf{Panel A: Classification in Bureau of Economic Analysis Employment Data}}                                                                                                                                                                                          \\
\multicolumn{3}{l}{}                                                                            & \textbf{BEA File Code} (1969-2000)                & \textbf{BEA File Code} (2001-)         \\
\multicolumn{3}{l}{Total Employment}                                                            & 10                                                                               & 10                                                                           \\
\multicolumn{3}{l}{Wage and Salary Employment}                                                  & 20                                                                               & 20                                                                           \\
\multicolumn{3}{l}{Nonfarm Employment}                                                          & 80                                                                               & 80                                                                           \\
\multicolumn{3}{l}{Private Nonfarm Employment}                                                  & 90                                                                               & 90                                                                           \\
\multicolumn{3}{l}{Construction Employment}                                                     & 300                                                                              & 400                                                                          \\
\multicolumn{3}{l}{Farm Employment}                                                             & 70                                                                               & 70                                                                           \\
\multicolumn{3}{l}{Manufacture Employment}                                                      & 400                                                                              & 500                                                                          \\

\addlinespace
\midrule 
\addlinespace
\multicolumn{5}{l}{\begin{tabular}[c]{@{}l@{}}\textbf{Panel B: Industry Classification by \cite{goukasian2010reaction} in County} \\ \textbf{Business Pattern Data}\end{tabular}}                                                                                                                            \\
\textbf{1987 SIC Code}     & \multicolumn{4}{c}{\textbf{Industry Description}}                                                                                                                                                          \\
\multicolumn{1}{c}{}                                          & \multicolumn{4}{c}{Residential Construction}                                                                                                                                                      \\
1521                                                          & \multicolumn{4}{l}{General contractors—single-family houses}                                                                                                                                      \\
1522                                                          & \multicolumn{4}{l}{General contractors—residential buildings, other than single-family}                                                                                                           \\
1531                                                          & \multicolumn{4}{l}{Operative builders}                                                                                                                                                            \\
6552                                                          & \multicolumn{4}{l}{Land subdividers and developers, except cemeteries}                                                                                                                            \\
                                                              & \multicolumn{4}{c}{Other Employment}                                                                                                                                                              \\
1741                                                          & \multicolumn{4}{l}{Masonry, stone setting and other stone work}                                                                                                                                   \\
1771                                                          & \multicolumn{4}{l}{Concrete work}                                                                                                                                                                 \\
1791                                                          & \multicolumn{4}{l}{Structural steel erection}                                                                                                                                                     \\
1742                                                          & \multicolumn{4}{l}{Plastering, drywall, acoustical, and insulation work}                                                                                                                          \\
1761                                                          & \multicolumn{4}{l}{Roofing, siding, and sheet metal work}                                                                                                                                         \\
1731                                                          & \multicolumn{4}{l}{Electrical work}                                                                                                                                                               \\
                                                              & \multicolumn{4}{c}{Mortgage Employment}                                                                                                                                                           \\
6162                                                          & \multicolumn{4}{l}{Mortgage bankers and loan correspondents}                                                                                                                                      \\
                                                              & \multicolumn{4}{c}{Commercial Construction}                                                                                                                                                       \\
1522                                                          & \multicolumn{4}{l}{General contractors—residential buildings, other than single-family}                                                                                                           \\
1531                                                          & \multicolumn{4}{l}{Operative builders}                                                                                                                                                            \\
1541                                                          & \multicolumn{4}{l}{General contractors—industrial buildings and warehouses}                                                                                                                       \\
1542                                                          & \multicolumn{4}{l}{\begin{tabular}[c]{@{}l@{}}General contractors—nonresidential buildings, other than industrial \\ buildings and warehouses\end{tabular}}                                       \\
1611                                                          & \multicolumn{4}{l}{Highway and street construction, except elevated highways}                                                                                                                     \\
1622                                                          & \multicolumn{4}{l}{Bridge, tunnel and elevated highway construction}                                                                                                                              \\
1623                                                          & \multicolumn{4}{l}{Water, sewer, pipeline and communications and power line construction}                                                                                                         \\
1629                                                          & \multicolumn{4}{l}{Heavy construction, not elsewhere classified}                                                                                                                                  \\
8741                                                          & \multicolumn{4}{l}{Management services}                                                                                                                                                           \\
        & \multicolumn{4}{l}{\begin{tabular}[c]{@{}l@{}}\textbf{Goukasian and Majbouri (2010) House Employment} includes  \\ all industries in Table 1 in  \cite{goukasian2010reaction}\end{tabular}} \\
        & \multicolumn{4}{l}{\begin{tabular}[c]{@{}l@{}}Our \textbf{Refined House Employment} includes Residential \\ Construction, Other Employment, and Mortgage Employment\end{tabular}}                          \\
\addlinespace
\midrule 
\addlinespace

\multicolumn{5}{l}{\begin{tabular}[c]{@{}l@{}}\textbf{Panel C: Industry Classification by  \cite{mian2014explains} in County Business} \\ \textbf{Pattern Data}\end{tabular}}                                                                                                                                  \\

Tradable Employment & \multicolumn{4}{l}{Please refer to Appendix Table 1 in  \cite{mian2014explains}}                                                                                                 
\end{tabular}

} % end of resize box

\end{table}

\pagebreak
%------------------------------------------------------------
% table 0: table_SumStat1

%---------------------------------------------------------------

%%%%%%%%%%%%%%%%%%%%%%%%%%%%%%%%%%%%%%%%%%%%%%%%
% table_SumStat1
%%%%%%%%%%%%%%%%%%%%%%%%%%%%%%%%%%%%%%%%%%%%%%%%

\noindent 

\begin{table}[h!]
\centering
\caption{
\textbf{Summary Statistics} \smallskip \newline
{\scriptsize
This table reports summary statistics of variables used in regressions, separated into different periods.
} % end of small font size
} % end of caption
\label{table_SumStat1}
\resizebox{\columnwidth}{!}{%
\begin{tabular}{l*{6}{c}}
\toprule
Variable Names                     
            &\multicolumn{1}{c}{Count}&\multicolumn{1}{c}{Mean}&\multicolumn{1}{c}{SD}&\multicolumn{1}{c}{P25}&\multicolumn{1}{c}{P50}&\multicolumn{1}{c}{P75}\\     
\midrule 

\multicolumn{7}{l}{\textbf{Panel A. Boom Period}} \\
\addlinespace
Refined House Employment Growth (00-06, An)&         790&     0.00042&     0.00061&     0.00009&     0.00032&     0.00062\\ 
\addlinespace
Residential Construction Employment Growth (00-06, An)&         782&     0.00007&     0.00024&    -0.00004&     0.00003&     0.00012\\ 
\addlinespace
Other Employment Growth (00-06, An)&         787&     0.00030&     0.00042&     0.00006&     0.00021&     0.00044\\ 
\addlinespace
Mortgage Employment Growth (00-06, An)&         622&     0.00007&     0.00029&    -0.00001&     0.00003&     0.00010\\ 
\addlinespace
Commercial Construction Employment Growth (00-06, An)&         783&     0.00000&     0.00064&    -0.00010&     0.00001&     0.00014\\ 
\addlinespace
Tradable Employment Growth (00-06, An)&         792&    -0.00299&     0.00434&    -0.00458&    -0.00234&    -0.00078\\ 
\addlinespace
Goukasian and Majbouri (2010) House Emp Growth (00-06, An)&         791&     0.00097&     0.00119&     0.00034&     0.00080&     0.00153\\ 
\addlinespace
Total Employment Growth (00-06, An)&         765&    -0.00025&     0.00984&    -0.00566&    -0.00041&     0.00535\\ 
\addlinespace
Wage and Salary Employment Growth (00-06, An)&         765&    -0.00301&     0.00838&    -0.00768&    -0.00273&     0.00153\\ 
\addlinespace
Nonfarm Employment Growth (00-06, An)&         765&     0.00053&     0.00980&    -0.00485&     0.00055&     0.00589\\ 
\addlinespace
Private Nonfarm Employment Growth (00-06, An)&         765&     0.00077&     0.00904&    -0.00419&     0.00076&     0.00604\\  
\addlinespace
Construction Employment Growth (00-06, An)&         765&     0.00081&     0.00233&    -0.00011&     0.00079&     0.00183\\   
\addlinespace
Farm Employment Growth (00-06, An)&         765&    -0.00078&     0.00108&    -0.00099&    -0.00042&    -0.00014\\ 
\addlinespace
Manufacture Employment Growth (00-06, An)&         765&    -0.00416&     0.00497&    -0.00625&    -0.00361&    -0.00157\\ 
\addlinespace
PLMNJ Growth (07USD, 99-05, An)&         792&     0.16977&     0.08137&     0.11596&     0.16494&     0.21644\\ 
\addlinespace
PLMNJ (Non-Owner-Occupied) Growth (07USD, 99-05, An)&         705&     0.35274&     0.17428&     0.24616&     0.33980&     0.43666\\ 
\addlinespace
GSEM Growth (07USD, 99-05, An)&         792&     0.04378&     0.07115&     0.00156&     0.03893&     0.08807\\ 
\addlinespace
Net Export Growth (07USD, 99-05, An)&         792&    -0.00218&     0.00200&    -0.00264&    -0.00175&    -0.00120\\ 
\addlinespace
GIV Net Export Growth (07USD, 99-05, An)&         792&    -0.00101&     0.00107&    -0.00126&    -0.00084&    -0.00044\\ 
\addlinespace
Ln(Num of households, 99)&         792&    10.86569&     1.18574&    10.06677&    10.81489&    11.61388\\ 
\addlinespace
Ln(household Income, 99)&         792&    10.71970&     0.23687&    10.56255&    10.69462&    10.84019\\ 
\addlinespace
Ratio of Labor Force (1999)&         792&     0.65363&     0.05522&     0.62177&     0.65627&     0.69181\\ 
\addlinespace
Ln(Num of House Units, 99)&         792&    10.90055&     1.15434&    10.09358&    10.86442&    11.62624\\ 
\addlinespace
Housing supply elasticity&         701&     2.41944&     1.21671&     1.60555&     2.28972&     3.00291\\ 
\addlinespace
House Vacancy Rate (1999)&         792&     0.07906&     0.04943&     0.05213&     0.06867&     0.09194\\ 
\addlinespace
Ratio of Renters (1999)&         792&     0.29243&     0.09767&     0.22221&     0.27948&     0.34642\\ 
\addlinespace
Ratio of Bachelor Educated (1999)&         792&     0.20581&     0.08499&     0.14217&     0.19054&     0.25171\\ 
\addlinespace
Ratio of White Race (1999)&         792&     0.82145&     0.14330&     0.74142&     0.86038&     0.93570\\ 
\addlinespace
Ratio of Immigration (90-00)&         792&     0.02549&     0.02910&     0.00658&     0.01478&     0.03254\\ 
\addlinespace

\bottomrule
\end{tabular}

} % end of resize box

\end{table}

\pagebreak
%------------------------------------------------------------
% table 0: table_SumStat2

%---------------------------------------------------------------

%%%%%%%%%%%%%%%%%%%%%%%%%%%%%%%%%%%%%%%%%%%%%%%%
% table_SumStat2
%%%%%%%%%%%%%%%%%%%%%%%%%%%%%%%%%%%%%%%%%%%%%%%%

\noindent 

\begin{table}[h!]
\centering
\caption{
\textbf{Summary Statistics (Continued)} \smallskip \newline
{\scriptsize
This table reports summary statistics of variables used in regressions, separated into different periods.
} % end of small font size
} % end of caption
\label{table_SumStat2}
\resizebox{\columnwidth}{!}{%
\begin{tabular}{l*{6}{c}}
\toprule
Variable Names                     
            &\multicolumn{1}{c}{Count}&\multicolumn{1}{c}{Mean}&\multicolumn{1}{c}{SD}&\multicolumn{1}{c}{P25}&\multicolumn{1}{c}{P50}&\multicolumn{1}{c}{P75}\\     
\midrule 

\multicolumn{7}{l}{\textbf{Panel B . Bust Period}} \\
\addlinespace
Refined House Employment Growth (07-10, An)&         790&    -0.00089&     0.00107&    -0.00118&    -0.00069&    -0.00032\\ 
\addlinespace
Residential Construction Employment Growth (07-10, An)&         752&    -0.00022&     0.00036&    -0.00029&    -0.00015&    -0.00005\\ 
\addlinespace
Other Employment Growth (07-10, An)&         785&    -0.00051&     0.00081&    -0.00071&    -0.00039&    -0.00014\\ 
\addlinespace
Mortgage Employment Growth (07-10, An)&         689&    -0.00018&     0.00045&    -0.00023&    -0.00012&    -0.00004\\ 
\addlinespace
Commercial Construction Employment Growth (07-10, An)&         765&    -0.00015&     0.00070&    -0.00027&    -0.00007&     0.00006\\ 
\addlinespace
Tradable Employment Growth (07-10, An)&         792&    -0.00367&     0.00456&    -0.00540&    -0.00277&    -0.00115\\ 
\addlinespace
Goukasian and Majbouri (2010) House Emp Growth (07-10, An)&         791&    -0.00129&     0.00198&    -0.00206&    -0.00116&    -0.00030\\ 
\addlinespace
Total Employment Growth (07-10, An)&         765&    -0.01534&     0.01259&    -0.02162&    -0.01467&    -0.00861\\ 
\addlinespace
Wage and Salary Employment Growth (07-10, An)&         765&    -0.01477&     0.01216&    -0.02079&    -0.01391&    -0.00828\\ 
\addlinespace
Nonfarm Employment Growth (07-10, An)&         765&    -0.01517&     0.01253&    -0.02137&    -0.01476&    -0.00857\\ 
\addlinespace
Private Nonfarm Employment Growth (07-10, An)&         765&    -0.01529&     0.00999&    -0.02089&    -0.01455&    -0.00924\\ 
\addlinespace
Construction Employment Growth (07-10, An)&         765&    -0.00422&     0.00400&    -0.00588&    -0.00398&    -0.00254\\ 
\addlinespace
Farm Employment Growth (07-10, An)&         765&    -0.00018&     0.00066&    -0.00031&    -0.00008&     0.00001\\ 
\addlinespace
Manufacture Employment Growth (07-10, An)&         765&    -0.00449&     0.00497&    -0.00584&    -0.00342&    -0.00185\\ 
\addlinespace

\midrule
\multicolumn{7}{l}{\textbf{Panel C. Prior Period}} \\
\addlinespace
Refined House Employment Share (92-00, An)&         704&     0.00029&     0.00037&     0.00011&     0.00025&     0.00042\\ 
\addlinespace
PLMNJ Growth (07USD, 91-99, An)&         705&     0.17072&     0.08447&     0.11737&     0.16381&     0.22433\\ 
\addlinespace
PLMNJ (Non-Owner-Occupied) Growth (07USD, 91-99, An)&         705&     0.42898&     0.27037&     0.25359&     0.35455&     0.62461\\ 
\addlinespace
GSEM Growth (07USD, 91-99, An)&         705&     0.14950&     0.08763&     0.09098&     0.13970&     0.19043\\ 
\addlinespace
Net Export Growth (91-99, An)&         705&    -0.00175&     0.00172&    -0.00214&    -0.00134&    -0.00086\\ 
\addlinespace
GIV Net Export Growth (91-99, An)&         705&    -0.00039&     0.00099&    -0.00061&    -0.00026&    -0.00005\\ 
\addlinespace
Ln(Num of households, 91)&         705&    10.86665&     1.16236&    10.01891&    10.81158&    11.59256\\ 
\addlinespace
Ln(household Income, 91)&         705&    10.36543&     0.18413&    10.24665&    10.34772&    10.46248\\ 
\addlinespace
Ratio of Labor Force (1989)&         705&     0.66942&     0.05583&     0.63573&     0.67125&     0.70527\\ 
\addlinespace
Ln(Num of House Units, 91)&         705&    10.89443&     1.14511&    10.07762&    10.82760&    11.62445\\ 
\addlinespace
Housing supply elasticity&         627&     2.39402&     1.20581&     1.56184&     2.25943&     3.00165\\ 
\addlinespace
House Vacancy Rate (1989)&         705&     0.08274&     0.04649&     0.05494&     0.07114&     0.09611\\ 
\addlinespace
Ratio of Renters (1989)&         705&     0.31581&     0.09767&     0.24819&     0.30608&     0.36902\\ 
\addlinespace
Ratio of Bachelor Educated (1989)&         705&     0.19067&     0.08208&     0.12982&     0.17538&     0.23648\\ 
\addlinespace
Ratio of White Race (1989)&         705&     0.85856&     0.12814&     0.78412&     0.89830&     0.95911\\ 
\addlinespace
Ratio of Immigration (80-90)&         705&     0.01648&     0.02479&     0.00303&     0.00762&     0.01723\\
\addlinespace

\bottomrule
\end{tabular}

} % end of resize box

\end{table}

%---------------------------------------------------------------
%---------------------------------------------------------------
% Empirical: Main Tests 1
% Total Employment Growth in Boom (00-06)
%---------------------------------------------------------------
%---------------------------------------------------------------

\pagebreak
%-----------------------------------------------------------------

%%%%%%%%%%%%%%%%%%%%%%%%%%%%%%%%%%%%
% table_BEA.TotEmpShr.D00t06.PLMNJ.D99t05.2SLS
%%%%%%%%%%%%%%%%%%%%%%%%%%%%%%%%%%%%

%-----------------------------------------------------------------

%%%%%%%%%%%%%%%%%%%%%%%%%%%%%%%%%%%%
% table_BEA.TotEmpShr.D00t06.PLMNJ.D99t05.2SLS
%%%%%%%%%%%%%%%%%%%%%%%%%%%%%%%%%%%%

\noindent 

\begin{table}[h!]
\centering
\caption{
\textbf{2SLS Regression of Total Employment Growth in Boom Period (00-06) on PLMNJ Growth (99-05)} \smallskip \newline
{\footnotesize 
This table reports the first-stage and the second-stage results of 2SLS regression $\triangle_{00,06} TotalEmpShr_{c} = \beta * \triangle_{99,05} Ln(PLMNJ_{c})  + \gamma* \bm{Controls_{c}} + \alpha + \epsilon_{c}$. The left-hand-side dependent variable $\triangle_{00,06} TotalEmpShr_{c}$ is the change of the total employment share in working-age population at county $c$ 00-06. To prevent excessive influence from outliers, total employment share is winsorized at 0.5\% and 99.5\% level. The key independent variable $\triangle_{99,05} Ln(PLMNJ_{c})$ is the growth rate of the dollar amount (07USD) of private-label mortgage (non-jumbo) (PLMNJ) at county $c$ 99-05. $Controls_{c}$ indicates control variables at county $c$ in 1999. We use the gravity model-based instrumental variable ($\triangle_{99,05}\text{givNetExp}_{m}$) as IV for $\triangle_{99,05} Ln(PLMNJ_{c})$. For the first-stage F-test, we report Kleibergen-Paap (2006) robust (clustered) statistics and Montiel Olea-Pflueger (2013) efficient statistics. Regression is weighted by the natural logarithm of housing units in 1999. Standard errors are clustered at the CBSA level. ***, **, and * indicate significance at the 1\%, 5\%, and 10\% levels, respectively.
} % end of small font size
} % end of caption
\label{table_BEA.TotEmpShr.D00t06.PLMNJ.D99t05.2SLS}

\resizebox{0.90\columnwidth}{!}{%

\begin{tabular}{l*{5}{c}}
\toprule
        &\multicolumn{1}{p{3cm}}{\centering PLMNJ Growth \\ (99-05, An)}  &\multicolumn{4}{c}{\centering Total Employment Growth (00-06, An)} \\
            \cmidrule{2-6} 
            &\multicolumn{1}{c}{(1)}&\multicolumn{1}{c}{(2)}&\multicolumn{1}{c}{(3)}&\multicolumn{1}{c}{(4)}&\multicolumn{1}{c}{(5)}\\
            
\midrule
GIV Net Export Growth (99-05, An)&   12.555***&            &            &            &            \\
               &  (3.878)   &            &            &            &            \\
\addlinespace
PLMNJ Growth (99-05, An)&            &    0.074***&    0.078***&    0.080***&    0.094** \\
               &            &  (0.026)   &  (0.029)   &  (0.031)   &  (0.037)   \\
\addlinespace
Ln(Num of households, 99)&    0.015   &            &   -0.002***&   -0.021***&   -0.018** \\
               &  (0.035)   &            &  (0.001)   &  (0.007)   &  (0.008)   \\
\addlinespace
Ln(household Income, 99)&    0.010   &            &    0.005*  &    0.007***&   -0.007   \\
               &  (0.036)   &            &  (0.003)   &  (0.003)   &  (0.005)   \\
\addlinespace
Ratio of Labor Force (1999)&    0.145   &            &    0.002   &    0.004   &   -0.005   \\
               &  (0.100)   &            &  (0.012)   &  (0.016)   &  (0.017)   \\
\addlinespace
Ln(Num of House Units, 99)&   -0.001   &            &            &    0.020***&    0.017** \\
               &  (0.035)   &            &            &  (0.007)   &  (0.008)   \\
\addlinespace
Housing supply elasticity&   -0.012***&            &            &    0.000   &    0.001   \\
               &  (0.003)   &            &            &  (0.001)   &  (0.001)   \\
\addlinespace
House Vacancy Rate (1999)&    0.230***&            &            &   -0.026   &   -0.033*  \\
               &  (0.084)   &            &            &  (0.016)   &  (0.018)   \\
\addlinespace
Ratio of Renters (1999)&   -0.084   &            &            &   -0.008   &   -0.025***\\
               &  (0.059)   &            &            &  (0.008)   &  (0.009)   \\
\addlinespace
Ratio of Bachelor Educated (1999)&   -0.170*  &            &            &            &    0.052***\\
               &  (0.090)   &            &            &            &  (0.015)   \\
\addlinespace
Ratio of White Race (1999)&    0.020   &            &            &            &   -0.005   \\
               &  (0.035)   &            &            &            &  (0.004)   \\
\addlinespace
Ratio of Immigration (90-00)&    0.163   &            &            &            &    0.019   \\
               &  (0.194)   &            &            &            &  (0.033)   \\
\addlinespace
Constant       &   -0.118   &   -0.013***&   -0.050** &   -0.081***&    0.083*  \\
               &  (0.367)   &  (0.004)   &  (0.023)   &  (0.025)   &  (0.047)   \\
\midrule
Obs            &      684   &      765   &      765   &      684   &      684   \\
R2-adj         &    0.120   &      &     &      &      \\
Cluster SE     &     CBSA   &     CBSA   &     CBSA   &     CBSA   &     CBSA   \\
Weight         & Ln(HU99)   & Ln(HU99)   & Ln(HU99)   & Ln(HU99)   & Ln(HU99)   \\
KP F-Stat      &            &    20.97   &    19.44   &    12.87   &    10.48   \\
MOP F-Stat     &            &    20.00   &    18.89   &    12.85   &    10.47   \\
\bottomrule

\end{tabular}

} % end of resize box

\end{table}

\pagebreak 
%---------------------------------------------------------------

%%%%%%%%%%%%%%%%%%%%%%%%%%%%%%%%%%%%%%%%%%%%%%%%
% table_BEA.TotEmpShr.D00t06.PLMNJ.D99t05.4Reg
%%%%%%%%%%%%%%%%%%%%%%%%%%%%%%%%%%%%%%%%%%%%%%%%

%---------------------------------------------------------------

%%%%%%%%%%%%%%%%%%%%%%%%%%%%%%%%%%%%%%%%%%%%%%%%
% table_BEA.TotEmpShr.D00t06.PLMNJ.D99t05.4Reg
%%%%%%%%%%%%%%%%%%%%%%%%%%%%%%%%%%%%%%%%%%%%%%%%

\noindent 

\begin{table}[h!]
\centering
\caption{
\textbf{Four Regressions of Total Employment Growth in Boom Period (00-06) on PLMNJ Growth (99-05)} \smallskip \newline
{\footnotesize 
This table reports OLS, reduced-form, first stage, and second stage results of 2SLS regression $\triangle_{00,06} TotalEmpShr_{c} = \beta * \triangle_{99,05} Ln(PLMNJ_{c}) + \gamma* \bm{Controls_{c}} + \alpha + \epsilon_{c}$. The left-hand-side dependent variable $\triangle_{00,06} TotalEmpShr_{c}$ is the change of the total employment share in working-age population at county $c$ 00-06. To prevent excessive influence from outliers, total employment share is winsorized at 0.5\% and 99.5\% level. The key independent variable $\triangle_{99,05} Ln(PLMNJ_{c})$ is the growth rate of the dollar amount (07USD) of private-label mortgage (non-jumbo) (PLMNJ) at county $c$ 99-05. $Controls_{c}$ indicates control variables at county $c$ in 1999. We use the gravity model-based instrumental variable ($\triangle_{99,05}\text{givNetExp}_{m}$) as IV for $\triangle_{99,05} Ln(PLMNJ_{c})$. For the first-stage F-test, we report kleibergen-Paap (2006) robust (clustered) statistics and Montiel Olea-Pflueger (2013) efficient statistics. Each regression is weighted by the natural logarithm of housing units in 1999. Standard errors are clustered at the CBSA level. ***, **, and * indicate significance at the 1\%, 5\%, and 10\% levels, respectively.
} % end of small font size
} % end of caption
\label{table_BEA.TotEmpShr.D00t06.PLMNJ.D99t05.4Reg}

\resizebox{0.85\columnwidth}{!}{%

\begin{tabular}{l*{4}{c}}
\toprule
Dep Var (Panel A, B, and C)                     &\multicolumn{4}{c}{Total Employment Growth (00-06, An)} \\
            \cmidrule{2-5} 
            &\multicolumn{1}{c}{(1)}&\multicolumn{1}{c}{(2)}&\multicolumn{1}{c}{(3)}&\multicolumn{1}{c}{(4)}\\
            
\midrule
\multicolumn{5}{l}{\textbf{Panel A. OLS estimates}} \\
\addlinespace
PLMNJ Growth (99-05, An)&    0.006   &    0.007   &    0.004   &    0.005   \\
               &  (0.004)   &  (0.005)   &  (0.005)   &  (0.005)   \\
\addlinespace
R2-adj         &  0.00122   &   0.0187   &   0.0694   &   0.0880   \\
\addlinespace

\midrule
\multicolumn{5}{l}{\textbf{Panel B. Reduced-form estimates}} \\
\addlinespace
GIV Net Export Growth (99-05, An)&    1.171***&    1.172***&    1.024***&    1.181***\\
               &  (0.377)   &  (0.380)   &  (0.373)   &  (0.366)   \\
\addlinespace
R2-adj         &   0.0150   &   0.0315   &   0.0814   &    0.103   \\
\addlinespace

\midrule
\multicolumn{5}{l}{\textbf{Panel C . 2SLS estimates}} \\
\addlinespace
PLMNJ Growth (99-05, An)&    0.074***&    0.078***&    0.080***&    0.094** \\
               &  (0.026)   &  (0.029)   &  (0.031)   &  (0.037)   \\
\addlinespace

\addlinespace
\addlinespace

Dep Var (Panel D): &\multicolumn{4}{c}{PLMNJ Growth (99-05, An)} \\ 
\midrule 
\multicolumn{5}{l}{\textbf{Panel D . First-stage estimates}} \\
\addlinespace
GIV Net Export Growth (99-05, An)&   15.738***&   15.106***&   12.754***&   12.555***\\
               &  (3.437)   &  (3.426)   &  (3.555)   &  (3.878)   \\
\addlinespace
KP F-Stat      &    20.97   &    19.44   &    12.87   &    10.48   \\
MOP F-Stat     &    20.00   &    18.89   &    12.85   &    10.47   \\
\addlinespace
\midrule
\multicolumn{5}{l}{\textbf{Controls (for all Panels)}} \\
Basic Controls &            &  Y   &   Y    & Y        \\
Housing Controls &           &      & Y       & Y       \\
Demographic Controls &            &      &        &  Y        \\
\midrule              
Obs            &      765   &      765   &      684   &      684   \\
Cluster SE     &     CBSA   &     CBSA   &     CBSA   &     CBSA   \\
Weight         & Ln(HU99)   & Ln(HU99)   & Ln(HU99)   & Ln(HU99)   \\
\bottomrule

\end{tabular}

} % end of resize box

\end{table}

%---------------------------------------------------------------
%---------------------------------------------------------------
% Empirical: Main Tests 2.1 (1)
% Refined House Employment Growth in Boom (00-06) and Bust (07-10)
%---------------------------------------------------------------
%---------------------------------------------------------------

\pagebreak
%-----------------------------------------------------------------

%%%%%%%%%%%%%%%%%%%%%%%%%%%%%%%%%%%%
% table_RefineHouse.D00t06vsD07t10.PLMNJ.2SLS.wide
%%%%%%%%%%%%%%%%%%%%%%%%%%%%%%%%%%%%

%-----------------------------------------------------------------------
%%%%%%%%%%%%%%%%%%%%%%%%%%%%%%%%%%%%
% table_RefineHouse.D00t06vsD07t10.PLMNJ.2SLS.wide
%%%%%%%%%%%%%%%%%%%%%%%%%%%%%%%%%%%%

\noindent 

\begin{table}[h!]
\centering
\caption{
\textbf{2SLS Stacked Regression of Refined House Employment Growth in Boom (00-06) and Bust (07-10) Periods on PLMNJ Growth (99-05)} \smallskip \newline
{\scriptsize
This table reports 2SLS regression $\triangle_{00,06} \& \triangle_{07,10} RefinedHouseEmpShr_{c} = \beta_{00,06} * \triangle_{99,05} Ln(PLMNJ_{c}) \times Dum_{00,06} + \beta_{07,10} * \triangle_{99,05} Ln(PLMNJ_{c}) \times Dum_{07,10} + \gamma_{00,06}* \bm{Controls_{c}} \times Dum_{00,06} + \gamma_{07,10}* \bm{Controls_{c}} \times Dum_{07,10} + \epsilon_{period, c}$. The left-hand-side dependent variable $\triangle_{00,06} \& \triangle_{07,10} RefinedHouseEmpShr_{c}$ is the change of the refined house employment share in working-age population at county $c$ 00-06 and 07-10. To reduce the impact of outliers, the dependent variable is winsorized at 2\% and 98\% levels in each period. The key independent variable $\triangle_{99,05} Ln(PLMNJ_{c})$ is the growth rate of the dollar amount (07USD) of private-label mortgages (non-jumbo) at county $c$ 99-05. $Controls_{c}$ indicates control variables at county $c$ in the period start year 1999. We use the gravity model-based instrumental variable $\triangle_{99,05}\text{givNetExp}_{m}$ as the IV for $\triangle_{99,05}Ln(PLMNJ_{c})$. Regression is weighted by the natural logarithm of housing units in 1999.  For the first-stage F-test of two non-stacked samples, we report Kleibergen-Paap (2006) robust (clustered) statistics and Montiel Olea-Pflueger (2013) efficient statistics. Standard errors are clustered at the CBSA level. ***, **, and * indicate significance at the 1\%, 5\%, and 10\% levels, respectively.
\smallskip
} % end of small font size
} % end of caption
\label{table_RefineHouse.D00t06vsD07t10.PLMNJ.2SLS.wide}

\vspace{-2mm}

\resizebox{\columnwidth}{!}{%
\begin{tabular}{l*{8}{c}}
\toprule
\textbf{TSLS estimates}            &\multicolumn{8}{c}{Refined House Employment Growth (00-06 and 07-10, An)} \\
            \cmidrule{2-9} 
            &\multicolumn{2}{c}{(1)}&\multicolumn{2}{c}{(2)}&\multicolumn{2}{c}{(3)}&\multicolumn{2}{c}{(4)}\\
            
\midrule
PLMNJ Growth (07USD, 99-05, An) x Dum00t06&    0.378***&  (0.123)&    0.333***&  (0.125)&    0.331** &  (0.163)&    0.382** &  (0.170)\\ 
\addlinespace
PLMNJ Growth (07USD, 99-05, An) x Dum07t10&   -0.693***&  (0.195)&   -0.537***&  (0.193)&   -0.591** &  (0.264)&   -0.696** &  (0.309)\\ 
\addlinespace
Dum00t06       &   -0.023   &  (0.021)&   -0.432***&  (0.117)&   -0.419***&  (0.140)&   -0.265   &  (0.232)\\ 
\addlinespace
Dum07t10       &    0.030   &  (0.034)&    1.101***&  (0.182)&    1.151***&  (0.250)&    0.301   &  (0.372)\\ 
\addlinespace
Ln(Num of HH, 99) x Dum00t06&            &         &    0.000   &  (0.002)&   -0.019   &  (0.018)&   -0.039*  &  (0.021)\\ 
\addlinespace
Ln(Num of HH, 99) x Dum07t10&            &         &   -0.009***&  (0.003)&    0.038   &  (0.026)&    0.059*  &  (0.030)\\ 
\addlinespace
Ln(HH Income, 99) x Dum00t06&            &         &    0.042***&  (0.012)&    0.036** &  (0.014)&    0.023   &  (0.023)\\ 
\addlinespace
Ln(HH Income, 99) x Dum07t10&            &         &   -0.097***&  (0.020)&   -0.088***&  (0.025)&   -0.015   &  (0.037)\\ 
\addlinespace
Ratio of Labor Force (1999) x Dum00t06&            &         &   -0.057   &  (0.046)&   -0.007   &  (0.050)&    0.030   &  (0.054)\\ 
\addlinespace
Ratio of Labor Force (1999) x Dum07t10&            &         &    0.048   &  (0.077)&   -0.070   &  (0.080)&   -0.094   &  (0.091)\\ 
\addlinespace
Ln(Num of HU, 99) x Dum00t06&            &         &            &         &    0.021   &  (0.018)&    0.038*  &  (0.021)\\ 
\addlinespace
Ln(Num of HU, 99) x Dum07t10&            &         &            &         &   -0.052** &  (0.026)&   -0.066** &  (0.031)\\ 
\addlinespace
Housing supply elasticity x Dum00t06&            &         &            &         &    0.001   &  (0.003)&    0.002   &  (0.003)\\ 
\addlinespace
Housing supply elasticity x Dum07t10&            &         &            &         &   -0.003   &  (0.005)&   -0.004   &  (0.005)\\ 
\addlinespace
House Vacancy Rate (1999) x Dum00t06&            &         &            &         &    0.063   &  (0.067)&    0.029   &  (0.065)\\ 
\addlinespace
House Vacancy Rate (1999) x Dum07t10&            &         &            &         &   -0.086   &  (0.116)&   -0.018   &  (0.116)\\ 
\addlinespace
Ratio of Renters (1999) x Dum00t06&            &         &            &         &   -0.011   &  (0.032)&   -0.078** &  (0.040)\\ 
\addlinespace
Ratio of Renters (1999) x Dum07t10&            &         &            &         &    0.039   &  (0.046)&    0.207***&  (0.057)\\ 
\addlinespace
Ratio of Bachelor Educated (1999) x Dum00t06&            &         &            &         &            &         &   -0.000   &  (0.064)\\ 
\addlinespace
Ratio of Bachelor Educated (1999) x Dum07t10&            &         &            &         &            &         &   -0.206*  &  (0.107)\\ 
\addlinespace
Ratio of White Race (1999) x Dum00t06&            &         &            &         &            &         &   -0.011   &  (0.024)\\ 
\addlinespace
Ratio of White Race (1999) x Dum07t10&            &         &            &         &            &         &    0.047   &  (0.034)\\ 
\addlinespace
Ratio of Immigration (90-00) x Dum00t06&            &         &            &         &            &         &    0.418***&  (0.149)\\ 
\addlinespace
Ratio of Immigration (90-00) x Dum07t10&            &         &            &         &            &         &   -0.581***&  (0.217)\\ 
\addlinespace
\midrule
Obs            &     1580   &         &     1580   &         &     1398   &         &     1398   &         \\
Cluster SE     &     CBSA   &         &     CBSA   &         &     CBSA   &         &     CBSA   &         \\
Weight         & Ln(HU99)   &         & Ln(HU99)   &         & Ln(HU99)   &         & Ln(HU99)   &         \\
KP F-Stat (99-05, non-stack sample)       &    23.41   &         &    21.19   &         &    14.39   &         &    12.05   &         \\
MOP F-Stat (99-05, non-stack sample)  &    22.22   &         &    20.36   &         &    14.43   &         &    12.09   &         \\
CoefEqual\_Chi2 &   12.874   &         &    8.399   &         &    5.199   &         &    5.533   &         \\
CoefEqual\_PValue &    0.000   &         &    0.004   &         &    0.023   &         &    0.019   &         \\
\bottomrule

\end{tabular}

} % end of resize box

\end{table}

\pagebreak 
%---------------------------------------------------------------

%%%%%%%%%%%%%%%%%%%%%%%%%%%%%%%%%%%%%%%%%%%%%%%%
% table_RefineHouse.D00t06vsD07t10.PLMNJ.4Reg
%%%%%%%%%%%%%%%%%%%%%%%%%%%%%%%%%%%%%%%%%%%%%%%%

%---------------------------------------------------------------

%%%%%%%%%%%%%%%%%%%%%%%%%%%%%%%%%%%%%%%%%%%%%%%%
% table_RefineHouse.D00t06vsD07t10.PLMNJ.4Reg
%%%%%%%%%%%%%%%%%%%%%%%%%%%%%%%%%%%%%%%%%%%%%%%%

\noindent 

\begin{table}[h!]
\centering
\caption{
\textbf{Four Stacked Regressions of Refined House Employment Growth in Boom (00-06) and Bust (07-10) Periods on PLMNJ Growth (99-05)} \smallskip \newline
{\scriptsize
This table reports OLS, reduced-form, first stage, and second stages of stacked 2SLS regression $\triangle_{00,06} \& \triangle_{07,10} RefinedHouseEmpShr_{c} = \beta_{00,06} * \triangle_{99,05} Ln(PLMNJ_{c}) \times Dum_{00,06} + \beta_{07,10} * \triangle_{99,05} Ln(PLMNJ_{c}) \times Dum_{07,10} + \gamma_{00,06}* \bm{Controls_{c}} \times Dum_{00,06} + \gamma_{07,10}* \bm{Controls_{c}} \times Dum_{07,10} + \epsilon_{period, c}$. The left-hand-side dependent variable $\triangle_{00,06} \& \triangle_{07,10} RefinedHouseEmpShr_{c}$ is the change of the refined house employment share in working-age population at county $c$ 00-06 and 07-10. To reduce the impact of outliers, the dependent variable is winsorized at 2\% and 98\% levels in each period. The key independent variable $\triangle_{99,05} Ln(PLMNJ_{c})$ is the growth rate of the dollar amount (07USD) of private-label mortgages (non-jumbo) at county $c$ 99-05. $Controls_{c}$ indicates control variables at county $c$ in the period start year 1999. We use the gravity model-based instrumental variable $\triangle_{99,05}\text{givNetExp}_{m}$ as the IV for $\triangle_{99,05}Ln(PLMNJ_{c})$. Regression is weighted by the natural logarithm of housing units in 1999.  For the first-stage F-test of two non-stacked samples, we report Kleibergen-Paap (2006) robust (clustered) statistics and Montiel Olea-Pflueger (2013) efficient statistics. Standard errors are clustered at the CBSA level. ***, **, and * indicate significance at the 1\%, 5\%, and 10\% levels, respectively.
} % end of small font size
} % end of caption
\label{table_RefineHouse.D00t06vsD07t10.PLMNJ.4Reg}
\resizebox{0.95\columnwidth}{!}{%
\begin{tabular}{l*{4}{c}}
\toprule
Dep Var (Panel A, B, and C)                      &\multicolumn{4}{c}{Refined House Employment Growth (00-06 \& 07-10, An)} \\
            \cmidrule{2-5} 
            &\multicolumn{1}{c}{(1)}&\multicolumn{1}{c}{(2)}&\multicolumn{1}{c}{(3)}&\multicolumn{1}{c}{(4)}\\

\midrule
\multicolumn{5}{l}{\textbf{Panel A. OLS estimates}} \\
PLMNJ Growth (07USD, 99-05, An) x Dum00t06&    0.148***&    0.141***&    0.117***&    0.111***\\
               &  (0.027)   &  (0.026)   &  (0.027)   &  (0.025)   \\
\addlinespace
PLMNJ Growth (07USD, 99-05, An) x Dum07t10&   -0.256***&   -0.227***&   -0.174***&   -0.175***\\
\addlinespace
R2-adj         &    0.524   &    0.570   &    0.585   &    0.599   \\
\addlinespace

\midrule
\multicolumn{5}{l}{\textbf{Panel B. Reduced-form estimates}} \\
GIV Net Export Growth (99-05, An) x Dum00t06&    5.947***&    5.113***&    4.094** &    4.681***\\
               &  (1.716)   &  (1.676)   &  (1.735)   &  (1.700)   \\
\addlinespace
GIV Net Export Growth (99-05, An) x Dum07t10&  -10.902***&   -8.203***&   -7.256***&   -8.474***\\
               &  (2.380)   &  (2.372)   &  (2.434)   &  (2.518)   \\
\addlinespace
R2-adj         &    0.504   &    0.552   &    0.576   &    0.591   \\
\addlinespace

\midrule
\multicolumn{5}{l}{\textbf{Panel C . 2SLS estimates}} \\
\addlinespace
PLMNJ Growth (07USD, 99-05, An) x Dum00t06&    0.378***&    0.333***&    0.331** &    0.382** \\
               &  (0.123)   &  (0.125)   &  (0.163)   &  (0.170)   \\
\addlinespace
PLMNJ Growth (07USD, 99-05, An) x Dum07t10&   -0.693***&   -0.537***&   -0.591** &   -0.696** \\
               &  (0.195)   &  (0.193)   &  (0.264)   &  (0.309)   \\
               
\addlinespace
\addlinespace

Dep Var (Panel D): &\multicolumn{4}{c}{PLMNJ Growth (99-05, An)} \\ 
\midrule 
\multicolumn{5}{l}{\textbf{Panel D . First-stage estimates only for 99-05 (Non-stack sample)}} \\
\addlinespace
GIV NEG (99-05, An)&   15.753***&   15.278***&   12.243***&   12.134***\\
               &  (3.255)   &  (3.331)   &  (3.276)   &  (3.541)   \\
\addlinespace
KP F-Stat      &    23.41   &    21.19   &    14.39   &    12.05   \\
MOP F-Stat     &    22.22   &    20.36   &    14.43   &    12.09   \\
\addlinespace

\midrule
\multicolumn{5}{l}{\textbf{Controls (for all Panels)}} \\
DumPeriod  &    Y        &  Y   &   Y    & Y        \\
Basic Controls x DumPeriod &            &  Y   &   Y    & Y    \\
Housing Controls x DumPeriod &           &      & Y       & Y   \\
Demographic Controls x DumPeriod &            &      &      &  Y \\

\midrule              
Obs (Panel A, B, and C)          &     1580   &     1580   &     1398   &     1398   \\
Obs (Panel D)          &      790   &      790   &      699   &      699   \\
Cluster SE     &     CBSA   &     CBSA   &     CBSA   &     CBSA   \\
Weight         & Ln(HU99)   & Ln(HU99)   & Ln(HU99)   & Ln(HU99)   \\
\bottomrule
\end{tabular}

} % end of resize box

\end{table}

%---------------------------------------------------------------
%---------------------------------------------------------------
% Empirical: Main Tests 2.3
% Residential Construction Employment Growth in Boom (00-06) and Bust (07-10)
%---------------------------------------------------------------
%---------------------------------------------------------------

\pagebreak
%-----------------------------------------------------------------

%%%%%%%%%%%%%%%%%%%%%%%%%%%%%%%%%%%%
% table_ResiConst.D00t06vsD07t10.PLMNJ.2SLS.wide
%%%%%%%%%%%%%%%%%%%%%%%%%%%%%%%%%%%%

%-----------------------------------------------------------------------
%%%%%%%%%%%%%%%%%%%%%%%%%%%%%%%%%%%%
% table_ResiConst.D00t06vsD07t10.PLMNJ.2SLS.wide
%%%%%%%%%%%%%%%%%%%%%%%%%%%%%%%%%%%%

\noindent 

\begin{table}[h!]
\centering
\caption{
\textbf{2SLS Stacked Regression of Residential Construction Employment Growth in Boom (00-06) and Bust (07-10) Periods on PLMNJ Growth (99-05)} \smallskip \newline
{\scriptsize
This table reports 2SLS regression $\triangle_{00,06} \& \triangle_{07,10} ResiConstEmpShr_{c} = \beta_{00,06} * \triangle_{99,05} Ln(PLMNJ_{c}) \times Dum_{00,06} + \beta_{07,10} * \triangle_{99,05} Ln(PLMNJ_{c}) \times Dum_{07,10} + \gamma_{00,06}* \bm{Controls_{c}} \times Dum_{00,06} + \gamma_{07,10}* \bm{Controls_{c}} \times Dum_{07,10} + \epsilon_{period, c}$. The left-hand-side dependent variable $\triangle_{00,06} \& \triangle_{07,10} ResiConstEmpShr_{c}$ is the change of the residential construction employment share in working-age population at county $c$ 00-06 and 07-10. To make the coefficient more visible in the table, the residential employment share growth is multiplied by 100. The key independent variable $\triangle_{99,05} Ln(PLMNJ_{c})$ is the growth rate of the dollar amount (07USD) of private-label mortgages (non-jumbo) at county $c$ 99-05. $Controls_{c}$ indicates control variables at county $c$ in the period start year 1999. We use the gravity model-based instrumental variable $\triangle_{99,05}\text{givNetExp}_{m}$ as the IV for $\triangle_{99,05}Ln(PLMNJ_{c})$. Regression is weighted by the natural logarithm of housing units in 1999.  For the first-stage F-test of two non-stacked samples, we report Kleibergen-Paap (2006) robust (clustered) statistics and Montiel Olea-Pflueger (2013) efficient statistics. Standard errors are clustered at the CBSA level. ***, **, and * indicate significance at the 1\%, 5\%, and 10\% levels, respectively.
\smallskip
} % end of small font size
} % end of caption
\label{table_ResiConst.D00t06vsD07t10.PLMNJ.2SLS.wide}

\vspace{-2mm}

\resizebox{\columnwidth}{!}{%
\begin{tabular}{l*{8}{c}}
\toprule
\textbf{TSLS estimates}            &\multicolumn{8}{c}{Residential Construction Employment Growth (00-06 and 07-10, An)} \\
            \cmidrule{2-9} 
            &\multicolumn{2}{c}{(1)}&\multicolumn{2}{c}{(2)}&\multicolumn{2}{c}{(3)}&\multicolumn{2}{c}{(4)}\\
            
\midrule
PLMNJ Growth (99-05, An) x Dum00t06&    0.124** &  (0.056)&    0.144** &  (0.062)&    0.157*  &  (0.084)&    0.171** &  (0.087)\\
\addlinespace
PLMNJ Growth (99-05, An) x Dum07t10&   -0.231***&  (0.073)&   -0.232***&  (0.081)&   -0.241** &  (0.110)&   -0.261** &  (0.123)\\
\addlinespace
Dum00t06       &   -0.015   &  (0.010)&   -0.056   &  (0.062)&   -0.092   &  (0.077)&   -0.052   &  (0.128)\\ 
\addlinespace
Dum07t10       &    0.018   &  (0.013)&    0.274***&  (0.086)&    0.295** &  (0.117)&    0.095   &  (0.202)\\ 
\addlinespace
Ln(Num of HH, 99) x Dum00t06&            &         &   -0.005***&  (0.001)&    0.004   &  (0.010)&    0.001   &  (0.011)\\ 
\addlinespace
Ln(Num of HH, 99) x Dum07t10&            &         &    0.006***&  (0.002)&    0.021   &  (0.014)&    0.023   &  (0.015)\\ 
\addlinespace
Ln(HH Income, 99) x Dum00t06&            &         &    0.013*  &  (0.007)&    0.013   &  (0.008)&    0.009   &  (0.013)\\ 
\addlinespace
Ln(HH Income, 99) x Dum07t10&            &         &   -0.036***&  (0.011)&   -0.032***&  (0.012)&   -0.015   &  (0.020)\\ 
\addlinespace
Ratio of Labor Force (1999) x Dum00t06&            &         &   -0.077***&  (0.027)&   -0.047*  &  (0.028)&   -0.041   &  (0.028)\\ 
\addlinespace
Ratio of Labor Force (1999) x Dum07t10&            &         &    0.100** &  (0.043)&    0.052   &  (0.041)&    0.052   &  (0.042)\\ 
\addlinespace
Ln(Num of HU, 99) x Dum00t06&            &         &            &         &   -0.009   &  (0.010)&   -0.006   &  (0.011)\\ 
\addlinespace
Ln(Num of HU, 99) x Dum07t10&            &         &            &         &   -0.017   &  (0.014)&   -0.018   &  (0.015)\\ 
\addlinespace
Housing supply elasticity x Dum00t06&            &         &            &         &    0.000   &  (0.001)&    0.001   &  (0.001)\\ 
\addlinespace
Housing supply elasticity x Dum07t10&            &         &            &         &   -0.001   &  (0.002)&   -0.001   &  (0.002)\\ 
\addlinespace
House Vacancy Rate (1999) x Dum00t06&            &         &            &         &    0.103***&  (0.037)&    0.096***&  (0.037)\\ 
\addlinespace
House Vacancy Rate (1999) x Dum07t10&            &         &            &         &   -0.062   &  (0.058)&   -0.049   &  (0.058)\\ 
\addlinespace
Ratio of Renters (1999) x Dum00t06&            &         &            &         &    0.016   &  (0.015)&    0.004   &  (0.019)\\ 
\addlinespace
Ratio of Renters (1999) x Dum07t10&            &         &            &         &    0.008   &  (0.026)&    0.042   &  (0.028)\\ 
\addlinespace
Ratio of Bachelor Educated (1999) x Dum00t06&            &         &            &         &            &         &    0.001   &  (0.030)\\ 
\addlinespace
Ratio of Bachelor Educated (1999) x Dum07t10&            &         &            &         &            &         &   -0.055   &  (0.051)\\ 
\addlinespace
Ratio of White Race (1999) x Dum00t06&            &         &            &         &            &         &    0.003   &  (0.012)\\ 
\addlinespace
Ratio of White Race (1999) x Dum07t10&            &         &            &         &            &         &    0.014   &  (0.015)\\ 
\addlinespace
Ratio of Immigration (90-00) x Dum00t06&            &         &            &         &            &         &    0.104** &  (0.050)\\ 
\addlinespace
Ratio of Immigration (90-00) x Dum07t10&            &         &            &         &            &         &   -0.076   &  (0.085)\\ 
\addlinespace
\midrule
Obs            &     1534   &         &     1534   &         &     1355   &         &     1355   &         \\
Cluster SE     &     CBSA   &         &     CBSA   &         &     CBSA   &         &     CBSA   &         \\
Weight         & Ln(HU99)   &         & Ln(HU99)   &         & Ln(HU99)   &         & Ln(HU99)   &         \\
KP F-Stat (99-05, non-stack sample)       &    23.24   &         &    21.20   &         &    13.90   &         &    11.75   &         \\
MOP F-Stat (99-05, non-stack sample)  &    22.16   &         &    20.45   &         &    13.93   &         &    11.78   &         \\
CoefEqual\_Chi2 &    8.448   &         &    7.717   &         &    4.651   &         &    4.559   &         \\
CoefEqual\_PValue&    0.004   &         &    0.005   &         &    0.031   &         &    0.033   &         \\
\bottomrule

\end{tabular}

} % end of resize box

\end{table}

\pagebreak 
%---------------------------------------------------------------

%%%%%%%%%%%%%%%%%%%%%%%%%%%%%%%%%%%%%%%%%%%%%%%%
% table_ResiConst.D00t06vsD07t10.PLMNJ.4Reg
%%%%%%%%%%%%%%%%%%%%%%%%%%%%%%%%%%%%%%%%%%%%%%%%

%---------------------------------------------------------------

%%%%%%%%%%%%%%%%%%%%%%%%%%%%%%%%%%%%%%%%%%%%%%%%
% table_ResiConst.D00t06vsD07t10.PLMNJ.4Reg
%%%%%%%%%%%%%%%%%%%%%%%%%%%%%%%%%%%%%%%%%%%%%%%%

\noindent 

\begin{table}[h!]
\centering
\caption{
\textbf{Four Stacked Regressions of Residential Construction Employment Growth in Boom (00-06) and Bust (07-10) Periods on PLMNJ Growth (99-05)} \smallskip \newline
{\scriptsize
This table reports OLS, reduced-form, first stage, and second stages of stacked 2SLS regression $\triangle_{00,06} \& \triangle_{07,10} ResiConstEmpShr_{c} = \beta_{00,06} * \triangle_{99,05} Ln(PLMNJ_{c}) \times Dum_{00,06} + \beta_{07,10} * \triangle_{99,05} Ln(PLMNJ_{c}) \times Dum_{07,10} + \gamma_{00,06}* \bm{Controls_{c}} \times Dum_{00,06} + \gamma_{07,10}* \bm{Controls_{c}} \times Dum_{07,10} + \epsilon_{period, c}$. The left-hand-side dependent variable $\triangle_{00,06} \& \triangle_{07,10} ResiConstEmpShr_{c}$ is the change of the residential construction employment share in working-age population at county $c$ 00-06 and 07-10. To make the coefficient more visible in the table, the residential employment share growth is multiplied by 100. The key independent variable $\triangle_{99,05} Ln(PLMNJ_{c})$ is the growth rate of the dollar amount (07USD) of private-label mortgages (non-jumbo) at county $c$ 99-05. $Controls_{c}$ indicates control variables at county $c$ in the period start year 1999. We use the gravity model-based instrumental variable $\triangle_{99,05}\text{givNetExp}_{m}$ as the IV for $\triangle_{99,05}Ln(PLMNJ_{c})$. Regression is weighted by the natural logarithm of housing units in 1999.  For the first-stage F-test of two non-stacked samples, we report Kleibergen-Paap (2006) robust (clustered) statistics and Montiel Olea-Pflueger (2013) efficient statistics. Standard errors are clustered at the CBSA level. ***, **, and * indicate significance at the 1\%, 5\%, and 10\% levels, respectively.
} % end of small font size
} % end of caption
\label{table_ResiConst.D00t06vsD07t10.PLMNJ.4Reg}
\resizebox{\columnwidth}{!}{%
\begin{tabular}{l*{4}{c}}
\toprule
Dep Var (Panel A, B, and C)                      &\multicolumn{4}{c}{Residential Construction Emp Growth (00-06 \& 07-10, An)} \\
            \cmidrule{2-5} 
            &\multicolumn{1}{c}{(1)}&\multicolumn{1}{c}{(2)}&\multicolumn{1}{c}{(3)}&\multicolumn{1}{c}{(4)}\\

\midrule
\multicolumn{5}{l}{\textbf{Panel A. OLS estimates}} \\
PLMNJ Growth (99-05, An) x Dum00t06&    0.035***&    0.039***&    0.033***&    0.031***\\
               &  (0.009)   &  (0.009)   &  (0.010)   &  (0.009)   \\
\addlinespace
PLMNJ Growth (99-05, An) x Dum07t10&   -0.074***&   -0.076***&   -0.058***&   -0.058***\\
               &  (0.014)   &  (0.014)   &  (0.015)   &  (0.015)   \\
\addlinespace
R2-adj         &    0.238   &    0.271   &    0.316   &    0.319   \\
\addlinespace

\midrule
\multicolumn{5}{l}{\textbf{Panel B. Reduced-form estimates}} \\
GIV Net Exp Growth (99-05, An) x Dum00t06&    1.922** &    2.194***&    1.917** &    2.067***\\
               &  (0.748)   &  (0.776)   &  (0.761)   &  (0.721)   \\
\addlinespace
GIV Net Exp Growth (99-05, An) x Dum07t10&   -3.516***&   -3.451***&   -2.841***&   -3.055***\\
               &  (0.925)   &  (0.981)   &  (0.915)   &  (0.960)   \\
\addlinespace
R2-adj         &    0.228   &    0.260   &    0.310   &    0.315   \\
\addlinespace

\midrule
\multicolumn{5}{l}{\textbf{Panel C . 2SLS estimates}} \\
\addlinespace
PLMNJ Growth (99-05, An) x Dum00t06&    0.124** &    0.144** &    0.157*  &    0.171** \\
               &  (0.056)   &  (0.062)   &  (0.084)   &  (0.087)   \\
\addlinespace
PLMNJ Growth (99-05, An) x Dum07t10&   -0.231***&   -0.232***&   -0.241** &   -0.261** \\
               &  (0.073)   &  (0.081)   &  (0.110)   &  (0.123)   \\
               
\addlinespace
\addlinespace

Dep Var (Panel D): &\multicolumn{4}{c}{PLMNJ Growth (99-05, An)} \\ 
\midrule 
\multicolumn{5}{l}{\textbf{Panel D . First-stage estimates only for 99-05 (Non-stack sample)}} \\
\addlinespace
GIV Net Exp Growth (99-05, An)&   15.753***&   15.278***&   12.243***&   12.134***\\
               &  (3.255)   &  (3.331)   &  (3.276)   &  (3.541)   \\
\addlinespace
KP F-Stat      &    23.24   &    21.20   &    13.90   &    11.75   \\
MOP F-Stat     &    22.16   &    20.45   &    13.93   &    11.78   \\
\addlinespace

\midrule
\multicolumn{5}{l}{\textbf{Controls (for all Panels)}} \\
DumPeriod  &    Y        &  Y   &   Y    & Y        \\
Basic Controls x DumPeriod &            &  Y   &   Y    & Y    \\
Housing Controls x DumPeriod &           &      & Y       & Y   \\
Demographic Controls x DumPeriod &            &      &      &  Y \\

\midrule              
Obs (Panel A, B, and C)          &     1534   &     1534   &     1355   &     1355   \\
Obs (Panel D)          &      782   &      782   &      692   &      692   \\
Cluster SE     &     CBSA   &     CBSA   &     CBSA   &     CBSA   \\
Weight         & Ln(HU99)   & Ln(HU99)   & Ln(HU99)   & Ln(HU99)   \\
\bottomrule
\end{tabular}

} % end of resize box

\end{table}

%---------------------------------------------------------------
%---------------------------------------------------------------
% Empirical: Main Tests 2.4
% Other House Employment Growth in Boom (00-06) and Bust (07-10)
%---------------------------------------------------------------
%---------------------------------------------------------------

\pagebreak
%-----------------------------------------------------------------

%%%%%%%%%%%%%%%%%%%%%%%%%%%%%%%%%%%%
% table_Other.D00t06vsD07t10.PLMNJ.2SLS.wide
%%%%%%%%%%%%%%%%%%%%%%%%%%%%%%%%%%%%

%-----------------------------------------------------------------------
%%%%%%%%%%%%%%%%%%%%%%%%%%%%%%%%%%%%
% table_Other.D00t06vsD07t10.PLMNJ.2SLS.wide
%%%%%%%%%%%%%%%%%%%%%%%%%%%%%%%%%%%%

\noindent 

\begin{table}[h!]
\centering
\caption{
\textbf{2SLS Stacked Regression of Other House Employment Growth in Boom (00-06) and Bust (07-10) Periods on PLMNJ Growth (99-05)} \smallskip \newline
{\scriptsize
This table reports 2SLS regression $\triangle_{00,06} \& \triangle_{07,10} OtherHouseEmpShr_{c} = \beta_{00,06} * \triangle_{99,05} Ln(PLMNJ_{c}) \times Dum_{00,06} + \beta_{07,10} * \triangle_{99,05} Ln(PLMNJ_{c}) \times Dum_{07,10} + \gamma_{00,06}* \bm{Controls_{c}} \times Dum_{00,06} + \gamma_{07,10}* \bm{Controls_{c}} \times Dum_{07,10} + \epsilon_{period, c}$. The left-hand-side dependent variable $\triangle_{00,06} \& \triangle_{07,10} OtherHouseEmpShr_{c}$ is the change of the other house employment share in working-age population at county $c$ 00-06 and 07-10. To reduce the impact of outliers, the dependent variable is winsorized at 5\% and 95\% levels in each period. To make the coefficient more visible in the table, the other employment share growth is multiplied by 100. The key independent variable $\triangle_{99,05} Ln(PLMNJ_{c})$ is the growth rate of the dollar amount (07USD) of private-label mortgages (non-jumbo) at county $c$ 99-05. $Controls_{c}$ indicates control variables at county $c$ in the period start year 1999. We use the gravity model-based instrumental variable $\triangle_{99,05}\text{givNetExp}_{m}$ as the IV for $\triangle_{99,05}Ln(PLMNJ_{c})$. Regression is weighted by the natural logarithm of housing units in 1999.  For the first-stage F-test of two non-stacked samples, we report Kleibergen-Paap (2006) robust (clustered) statistics and Montiel Olea-Pflueger (2013) efficient statistics. Standard errors are clustered at the CBSA level. ***, **, and * indicate significance at the 1\%, 5\%, and 10\% levels, respectively.
\smallskip
} % end of small font size
} % end of caption
\label{table_Other.D00t06vsD07t10.PLMNJ.2SLS.wide}

\vspace{-2mm}

\resizebox{\columnwidth}{!}{%
\begin{tabular}{l*{8}{c}}
\toprule
\textbf{TSLS estimates}            &\multicolumn{8}{c}{Other House Employment Growth (00-06 and 07-10, An)} \\
            \cmidrule{2-9} 
            &\multicolumn{2}{c}{(1)}&\multicolumn{2}{c}{(2)}&\multicolumn{2}{c}{(3)}&\multicolumn{2}{c}{(4)}\\
            
\midrule
PLMNJ Growth (07USD, 99-05, An) x Dum00t06&    0.204***&  (0.068)&    0.175***&  (0.066)&    0.164** &  (0.083)&    0.189** &  (0.084)\\ 
\addlinespace
PLMNJ Growth (07USD, 99-05, An) x Dum07t10&   -0.358***&  (0.107)&   -0.284***&  (0.101)&   -0.305** &  (0.138)&   -0.350** &  (0.145)\\ 
\addlinespace
Dum00t06       &   -0.006   &  (0.011)&   -0.181***&  (0.070)&   -0.105   &  (0.082)&   -0.006   &  (0.134)\\ 
\addlinespace
Dum07t10       &    0.013   &  (0.018)&    0.390***&  (0.104)&    0.334** &  (0.137)&    0.119   &  (0.197)\\ 
\addlinespace
Ln(Num of HH, 99) x Dum00t06&            &         &    0.001   &  (0.001)&   -0.012   &  (0.012)&   -0.022*  &  (0.013)\\ 
\addlinespace
Ln(Num of HH, 99) x Dum07t10&            &         &   -0.006***&  (0.002)&    0.016   &  (0.017)&    0.035*  &  (0.019)\\ 
\addlinespace
Ln(HH Income, 99) x Dum00t06&            &         &    0.014*  &  (0.007)&    0.006   &  (0.008)&   -0.002   &  (0.013)\\ 
\addlinespace
Ln(HH Income, 99) x Dum07t10&            &         &   -0.028** &  (0.011)&   -0.020   &  (0.014)&   -0.003   &  (0.020)\\ 
\addlinespace
Ratio of Labor Force (1999) x Dum00t06&            &         &    0.026   &  (0.027)&    0.029   &  (0.028)&    0.048   &  (0.030)\\ 
\addlinespace
Ratio of Labor Force (1999) x Dum07t10&            &         &   -0.030   &  (0.042)&   -0.065   &  (0.044)&   -0.099** &  (0.047)\\ 
\addlinespace
Ln(Num of HU, 99) x Dum00t06&            &         &            &         &    0.015   &  (0.012)&    0.024*  &  (0.013)\\ 
\addlinespace
Ln(Num of HU, 99) x Dum07t10&            &         &            &         &   -0.024   &  (0.017)&   -0.040** &  (0.019)\\ 
\addlinespace
Housing supply elasticity x Dum00t06&            &         &            &         &   -0.000   &  (0.001)&    0.000   &  (0.002)\\ 
\addlinespace
Housing supply elasticity x Dum07t10&            &         &            &         &    0.000   &  (0.003)&   -0.001   &  (0.003)\\ 
\addlinespace
House Vacancy Rate (1999) x Dum00t06&            &         &            &         &   -0.031   &  (0.039)&   -0.049   &  (0.038)\\ 
\addlinespace
House Vacancy Rate (1999) x Dum07t10&            &         &            &         &   -0.002   &  (0.064)&    0.032   &  (0.063)\\ 
\addlinespace
Ratio of Renters (1999) x Dum00t06&            &         &            &         &   -0.029*  &  (0.017)&   -0.066***&  (0.021)\\ 
\addlinespace
Ratio of Renters (1999) x Dum07t10&            &         &            &         &    0.050** &  (0.026)&    0.126***&  (0.032)\\ 
\addlinespace
Ratio of Bachelor Educated (1999) x Dum00t06&            &         &            &         &            &         &    0.008   &  (0.037)\\ 
\addlinespace
Ratio of Bachelor Educated (1999) x Dum07t10&            &         &            &         &            &         &   -0.026   &  (0.049)\\ 
\addlinespace
Ratio of White Race (1999) x Dum00t06&            &         &            &         &            &         &   -0.010   &  (0.013)\\ 
\addlinespace
Ratio of White Race (1999) x Dum07t10&            &         &            &         &            &         &    0.026   &  (0.019)\\ 
\addlinespace
Ratio of Immigration (90-00) x Dum00t06&            &         &            &         &            &         &    0.192** &  (0.079)\\ 
\addlinespace
Ratio of Immigration (90-00) x Dum07t10&            &         &            &         &            &         &   -0.348***&  (0.119)\\ 
\addlinespace
\midrule
Obs            &     1572   &         &     1572   &         &     1390   &         &     1390   &         \\
Cluster SE     &     CBSA   &         &     CBSA   &         &     CBSA   &         &     CBSA   &         \\
Weight         & Ln(HU99)   &         & Ln(HU99)   &         & Ln(HU99)   &         & Ln(HU99)   &         \\
KP F-Stat (99-05, non-stack sample)       &    23.25   &         &    20.97   &         &    14.25   &         &    12.02   &         \\
MOP F-Stat (99-05, non-stack sample)  &    22.06   &         &    20.11   &         &    14.27   &         &    12.05   &         \\
CoefEqual\_Chi2 &   11.846   &         &    8.818   &         &    5.252   &         &    6.372   &         \\
CoefEqual\_PValue &    0.001   &         &    0.003   &         &    0.022   &         &    0.012   &         \\
\bottomrule

\end{tabular}

} % end of resize box

\end{table}

\pagebreak 
%---------------------------------------------------------------

%%%%%%%%%%%%%%%%%%%%%%%%%%%%%%%%%%%%%%%%%%%%%%%%
% table_Other.D00t06vsD07t10.PLMNJ.4Reg
%%%%%%%%%%%%%%%%%%%%%%%%%%%%%%%%%%%%%%%%%%%%%%%%

%---------------------------------------------------------------

%%%%%%%%%%%%%%%%%%%%%%%%%%%%%%%%%%%%%%%%%%%%%%%%
% table_Other.D00t06vsD07t10.PLMNJ.4Reg
%%%%%%%%%%%%%%%%%%%%%%%%%%%%%%%%%%%%%%%%%%%%%%%%

\noindent 

\begin{table}[h!]
\centering
\caption{
\textbf{Four Stacked Regressions of Other House Employment Growth in Boom (00-06) and Bust (07-10) Periods on PLMNJ Growth (99-05)} \smallskip \newline
{\scriptsize
This table reports OLS, reduced-form, first stage, and second stages of stacked 2SLS regression $\triangle_{00,06} \& \triangle_{07,10} OtherHouseEmpShr_{c} = \beta_{00,06} * \triangle_{99,05} Ln(PLMNJ_{c}) \times Dum_{00,06} + \beta_{07,10} * \triangle_{99,05} Ln(PLMNJ_{c}) \times Dum_{07,10} + \gamma_{00,06}* \bm{Controls_{c}} \times Dum_{00,06} + \gamma_{07,10}* \bm{Controls_{c}} \times Dum_{07,10} + \epsilon_{period, c}$. The left-hand-side dependent variable $\triangle_{00,06} \& \triangle_{07,10} OtherHouseEmpShr_{c}$ is the change of the other house employment share in working-age population at county $c$ 00-06 and 07-10. To reduce the impact of outliers, the dependent variable is winsorized at 5\% and 95\% levels in each period. To make the coefficient more visible in the table, the other employment share growth is multiplied by 100. The key independent variable $\triangle_{99,05} Ln(PLMNJ_{c})$ is the growth rate of the dollar amount (07USD) of private-label mortgages (non-jumbo) at county $c$ 99-05. $Controls_{c}$ indicates control variables at county $c$ in the period start year 1999. We use the gravity model-based instrumental variable $\triangle_{99,05}\text{givNetExp}_{m}$ as the IV for $\triangle_{99,05}Ln(PLMNJ_{c})$. Regression is weighted by the natural logarithm of housing units in 1999.  For the first-stage F-test of two non-stacked samples, we report Kleibergen-Paap (2006) robust (clustered) statistics and Montiel Olea-Pflueger (2013) efficient statistics. Standard errors are clustered at the CBSA level. ***, **, and * indicate significance at the 1\%, 5\%, and 10\% levels, respectively.
} % end of small font size
} % end of caption
\label{table_Other.D00t06vsD07t10.PLMNJ.4Reg}
\resizebox{0.95\columnwidth}{!}{%
\begin{tabular}{l*{4}{c}}
\toprule
Dep Var (Panel A, B, and C)                      &\multicolumn{4}{c}{Other House Employment Growth (00-06 \& 07-10, An)} \\
            \cmidrule{2-5} 
            &\multicolumn{1}{c}{(1)}&\multicolumn{1}{c}{(2)}&\multicolumn{1}{c}{(3)}&\multicolumn{1}{c}{(4)}\\

\midrule
\multicolumn{5}{l}{\textbf{Panel A. OLS estimates}} \\
PLMNJ Growth (07USD, 99-05, An) x Dum00t06&    0.090***&    0.086***&    0.076***&    0.074***\\
               &  (0.015)   &  (0.015)   &  (0.017)   &  (0.016)   \\
\addlinespace
PLMNJ Growth (07USD, 99-05, An) x Dum07t10&   -0.151***&   -0.137***&   -0.107***&   -0.104***\\
               &  (0.024)   &  (0.024)   &  (0.026)   &  (0.024)   \\
\addlinespace
R2-adj         &    0.534   &    0.564   &    0.574   &    0.587   \\
\addlinespace

\midrule
\multicolumn{5}{l}{\textbf{Panel B. Reduced-form estimates}} \\
GIV Net Export Growth (99-05, An) x Dum00t06&    3.203***&    2.672***&    2.025*  &    2.316** \\
               &  (1.036)   &  (0.962)   &  (1.044)   &  (1.035)   \\
\addlinespace
GIV Net Export Growth (99-05, An) x Dum07t10&   -5.659***&   -4.355***&   -3.781** &   -4.312***\\
               &  (1.399)   &  (1.321)   &  (1.461)   &  (1.423)   \\
\addlinespace
R2-adj         &    0.509   &    0.541   &    0.561   &    0.576   \\
\addlinespace

\midrule
\multicolumn{5}{l}{\textbf{Panel C . 2SLS estimates}} \\
\addlinespace
PLMNJ Growth (07USD, 99-05, An) x Dum00t06&    0.204***&    0.175***&    0.164** &    0.189** \\
               &  (0.068)   &  (0.066)   &  (0.083)   &  (0.084)   \\
\addlinespace
PLMNJ Growth (07USD, 99-05, An) x Dum07t10&   -0.358***&   -0.284***&   -0.305** &   -0.350** \\
               &  (0.107)   &  (0.101)   &  (0.138)   &  (0.145)   \\
               
\addlinespace
\addlinespace

Dep Var (Panel D): &\multicolumn{4}{c}{PLMNJ Growth (99-05, An)} \\ 
\midrule 
\multicolumn{5}{l}{\textbf{Panel D . First-stage estimates only for 99-05 (Non-stack sample)}} \\
\addlinespace
GIV NEG (99-05, An)&   15.753***&   15.278***&   12.243***&   12.134***\\
               &  (3.255)   &  (3.331)   &  (3.276)   &  (3.541)   \\
\addlinespace
KP F-Stat      &    23.25   &    20.97   &    14.25   &    12.02   \\
MOP F-Stat     &    22.06   &    20.11   &    14.27   &    12.05   \\
\addlinespace

\midrule
\multicolumn{5}{l}{\textbf{Controls (for all Panels)}} \\
DumPeriod  &    Y        &  Y   &   Y    & Y        \\
Basic Controls x DumPeriod &            &  Y   &   Y    & Y    \\
Housing Controls x DumPeriod &           &      & Y       & Y   \\
Demographic Controls x DumPeriod &            &      &      &  Y \\

\midrule              
Obs (Panel A, B, and C)         &     1572   &     1572   &     1390   &     1390   \\
Obs (Panel D)          &      787   &      787   &      696   &      696   \\
Cluster SE     &     CBSA   &     CBSA   &     CBSA   &     CBSA   \\
Weight         & Ln(HU99)   & Ln(HU99)   & Ln(HU99)   & Ln(HU99)   \\
\bottomrule
\end{tabular}

} % end of resize box

\end{table}

%---------------------------------------------------------------
%---------------------------------------------------------------
% Empirical: Main Tests 2.5 (1)
% Mortgage Employment Growth in Boom (00-06) and Bust (07-10)
%---------------------------------------------------------------
%---------------------------------------------------------------

\pagebreak
%-----------------------------------------------------------------

%%%%%%%%%%%%%%%%%%%%%%%%%%%%%%%%%%%%
% table_Mortgage.D00t06vsD07t10.PLMNJ.2SLS.wide
%%%%%%%%%%%%%%%%%%%%%%%%%%%%%%%%%%%%

%\input{Table_Purch/table_Mortgage.D00t06vsD07t10.PLMNJ.2SLS.wide}

\pagebreak 
%---------------------------------------------------------------

%%%%%%%%%%%%%%%%%%%%%%%%%%%%%%%%%%%%%%%%%%%%%%%%
% table_Mortgage.D00t06vsD07t10.PLMNJ.4Reg
%%%%%%%%%%%%%%%%%%%%%%%%%%%%%%%%%%%%%%%%%%%%%%%%

%---------------------------------------------------------------

%%%%%%%%%%%%%%%%%%%%%%%%%%%%%%%%%%%%%%%%%%%%%%%%
% table_Mortgage.D00t06vsD07t10.PLMNJ.4Reg
%%%%%%%%%%%%%%%%%%%%%%%%%%%%%%%%%%%%%%%%%%%%%%%%

\noindent 

\begin{table}[h!]
\centering
\caption{
\textbf{Four Stacked Regressions of Mortgage Employment Growth in Boom (00-06) and Bust (07-10) Periods on PLMNJ Growth (99-05)} \smallskip \newline
{\scriptsize
This table reports OLS, reduced-form, first stage, and second stages of stacked 2SLS regression $\triangle_{00,06} \& \triangle_{07,10} MortgageEmpShr_{c} = \beta_{00,06} * \triangle_{99,05} Ln(PLMNJ_{c}) \times Dum_{00,06} + \beta_{07,10} * \triangle_{99,05} Ln(PLMNJ_{c}) \times Dum_{07,10} + \gamma_{00,06}* \bm{Controls_{c}} \times Dum_{00,06} + \gamma_{07,10}* \bm{Controls_{c}} \times Dum_{07,10} + \epsilon_{period, c}$. The left-hand-side dependent variable $\triangle_{00,06} \& \triangle_{07,10} MortgageEmpShr_{c}$ is the change of the mortgage employment share in working-age population at county $c$ 00-06 and 07-10. To reduce the impact of outliers, the dependent variable is winsorized at 2\% and 98\% levels in each period. To make the coefficient more visible in the table, the mortgage employment share growth is multiplied by 100. The key independent variable $\triangle_{99,05} Ln(PLMNJ_{c})$ is the growth rate of the dollar amount (07USD) of private-label mortgages (non-jumbo) at county $c$ 99-05. $Controls_{c}$ indicates control variables at county $c$ in the period start year 1999. We use the gravity model-based instrumental variable $\triangle_{99,05}\text{givNetExp}_{m}$ as the IV for $\triangle_{99,05}Ln(PLMNJ_{c})$. Regression is weighted by the natural logarithm of housing units in 1999.  For the first-stage F-test of two non-stacked samples, we report Kleibergen-Paap (2006) robust (clustered) statistics and Montiel Olea-Pflueger (2013) efficient statistics. Standard errors are clustered at the CBSA level. ***, **, and * indicate significance at the 1\%, 5\%, and 10\% levels, respectively.
} % end of small font size
} % end of caption
\label{table_Mortgage.D00t06vsD07t10.PLMNJ.4Reg}
\resizebox{0.95\columnwidth}{!}{%
\begin{tabular}{l*{4}{c}}
\toprule
Dep Var (Panel A, B, and C)                      &\multicolumn{4}{c}{Mortgage Employment Growth (00-06 \& 07-10, An)} \\
            \cmidrule{2-5} 
            &\multicolumn{1}{c}{(1)}&\multicolumn{1}{c}{(2)}&\multicolumn{1}{c}{(3)}&\multicolumn{1}{c}{(4)}\\

\midrule
\multicolumn{5}{l}{\textbf{Panel A. OLS estimates}} \\
PLMNJ Growth (07USD, 99-05, An) x Dum00t06&    0.008   &    0.004   &    0.004   &    0.003   \\
               &  (0.009)   &  (0.007)   &  (0.008)   &  (0.008)   \\
\addlinespace
PLMNJ Growth (07USD, 99-05, An) x Dum07t10&   -0.001   &    0.007   &    0.007   &    0.004   \\
               &  (0.015)   &  (0.011)   &  (0.013)   &  (0.013)   \\
\addlinespace
R2-adj         &    0.272   &    0.386   &    0.394   &    0.399   \\
\addlinespace

\midrule
\multicolumn{5}{l}{\textbf{Panel B. Reduced-form estimates}} \\
GIV Net Export Growth (99-05, An) x Dum00t06&    1.128***&    0.435   &    0.301   &    0.480   \\
               &  (0.422)   &  (0.414)   &  (0.448)   &  (0.446)   \\
\addlinespace
GIV Net Export Growth (99-05, An) x Dum07t10&   -1.689** &   -0.544   &   -0.972   &   -1.297   \\
               &  (0.702)   &  (0.787)   &  (0.874)   &  (0.995)   \\
\addlinespace
R2-adj         &    0.275   &    0.386   &    0.394   &    0.400   \\
\addlinespace

\midrule
\multicolumn{5}{l}{\textbf{Panel C . 2SLS estimates}} \\
\addlinespace
PLMNJ Growth (07USD, 99-05, An) x Dum00t06&    0.074** &    0.029   &    0.025   &    0.039   \\
               &  (0.030)   &  (0.028)   &  (0.039)   &  (0.039)   \\
\addlinespace
PLMNJ Growth (07USD, 99-05, An) x Dum07t10&   -0.113** &   -0.037   &   -0.094   &   -0.125   \\
               &  (0.053)   &  (0.055)   &  (0.096)   &  (0.119)   \\
               
\addlinespace
\addlinespace

Dep Var (Panel D): &\multicolumn{4}{c}{PLMNJ Growth (99-05, An)} \\ 
\midrule 
\multicolumn{5}{l}{\textbf{Panel D . First-stage estimates only for 99-05 (Non-stack sample)}} \\
\addlinespace
GIV NEG (99-05, An)&   15.753***&   15.278***&   12.243***&   12.134***\\
               &  (3.255)   &  (3.331)   &  (3.276)   &  (3.541)   \\
\addlinespace
KP F-Stat      &    24.16   &    21.54   &    12.38   &    11.13   \\
MOP F-Stat     &    23.11   &    20.76   &    12.26   &    10.99   \\
\addlinespace

\midrule
\multicolumn{5}{l}{\textbf{Controls (for all Panels)}} \\
DumPeriod  &    Y        &  Y   &   Y    & Y        \\
Basic Controls x DumPeriod &            &  Y   &   Y    & Y    \\
Housing Controls x DumPeriod &           &      & Y       & Y   \\
Demographic Controls x DumPeriod &            &      &      &  Y \\

\midrule              
Obs (Panel A, B, and C)          &     1311   &     1311   &     1150   &     1150   \\
Obs (Panel D)          &      622   &      622   &      544   &      544   \\
Cluster SE     &     CBSA   &     CBSA   &     CBSA   &     CBSA   \\
Weight         & Ln(HU99)   & Ln(HU99)   & Ln(HU99)   & Ln(HU99)   \\
\bottomrule
\end{tabular}

} % end of resize box

\end{table}

%---------------------------------------------------------------
%---------------------------------------------------------------
% Empirical: Main Tests 2.5 (2)
% PLMNJ Growth in Boom (00-06) and Bust (07-10)
%---------------------------------------------------------------
%---------------------------------------------------------------

\pagebreak
%-----------------------------------------------------------------

%%%%%%%%%%%%%%%%%%%%%%%%%%%%%%%%%%%%
% table_PLMNJ.D99t05vsD05t08.2SLS.wide
%%%%%%%%%%%%%%%%%%%%%%%%%%%%%%%%%%%%

%----------------------------------------------------------------

%%%%%%%%%%%%%%%%%%%%%%%%%%%%%%%%%%%%
% table_PLMNJ.D99t05vsD05t08.2SLS.wide
%%%%%%%%%%%%%%%%%%%%%%%%%%%%%%%%%%%%

\noindent 

\begin{table}[h!]
\centering
\caption{
\textbf{2SLS Stacked Regression of PLMNJ Growth in Boom (99-05) and Bust (05-08) Periods on Net Export Growth (99-05)} \smallskip \newline
{\scriptsize 
This table reports 2SLS stacked regression $\triangle_{99,05} \& \triangle_{05,08} Ln(PLMNJ_{c}) = \beta_{99,05} * \triangle_{99,05} \text{NetExp}_{m} \times Dum_{99,05} + \beta_{05,08} * \triangle_{99,05} \text{NetExp}_{m} \times Dum_{05,08} + \gamma_{99,05}* \bm{Controls_{c}} \times Dum_{99,05} + \gamma_{05,08}* \bm{Controls_{c}} \times Dum_{05,08}  + \alpha_{99,05} + \alpha_{05,08} + \epsilon_{period, c}$. The left-hand-side dependent variable $\triangle_{99,05} \& \triangle_{05,08} Ln(PLMNJ_{c})$ is the stacked growth rate of the dollar amount (07USD) of private-label mortgages (non-jumbo) (PLMNJ) at county $c$ 99-05 and 05-08. The key independent variable $\triangle_{99,05} \text{NetExp}_{m}$ is the growth rate of net export at the metropolitan area (CBSA03 code) $m$ 99-05. $Controls_{c}$ indicates control variables at county $c$ in the period start year 1999. We use the gravity model-based instrumental variable $\triangle_{99,05}\text{givNetExp}_{m}$ as IVs for $\triangle_{99,05}\text{NetExp}_{m}$. Regression is weighted by the natural logarithm of housing units in the start year 1999. We report the statistics and p-values for the tests of coefficient equality between $\beta_{99,05}$ and $\beta_{05,08}$. For the first-stage F-test of two separate non-stack samples, we report kleibergen-Paap (2006) robust (clustered) statistics and Montiel Olea-Pflueger (2013) efficient statistics. Standard errors are clustered at the CBSA level. ***, **, and * indicate significance at the 1\%, 5\%, and 10\% levels, respectively.
} % end of small font size
} % end of caption
\label{table_PLMNJ.D99t05vsD05t08.2SLS.wide}

\resizebox{0.95\columnwidth}{!}{%

\begin{tabular}{l*{8}{c}}
\toprule
\textbf{TSLS estimates}            &\multicolumn{8}{c}{PLMNJ Growth (99-05 and 05-08, An)} \\
            \cmidrule{2-9} 
            &\multicolumn{2}{c}{(1)}&\multicolumn{2}{c}{(2)}&\multicolumn{2}{c}{(3)}&\multicolumn{2}{c}{(4)}\\
            
\midrule
Net Export Growth (99-05, An) x Dum99t05&   13.694***&  (3.547)&   13.365***&  (3.643)&   11.145***&  (3.254)&   10.851***&  (3.331)\\ 
\addlinespace
Net Export Growth (99-05, An) x Dum05t08&  -16.325***&  (6.194)&  -16.915***&  (6.457)&  -13.335** &  (5.782)&  -12.438** &  (5.669)\\ 
\addlinespace
Dummry 99-05   &    0.200***&  (0.009)&    0.181   &  (0.212)&    0.352   &  (0.236)&   -0.339   &  (0.371)\\ 
\addlinespace
Dummry 05-08   &   -0.320***&  (0.018)&   -1.010***&  (0.375)&   -1.711***&  (0.436)&   -0.022   &  (0.720)\\ 
\addlinespace
Ln(Num of HH, 99) x pd\_D99t05&            &         &    0.006** &  (0.003)&    0.021   &  (0.032)&    0.003   &  (0.031)\\ 
\addlinespace
Ln(Num of HH, 99) x pd\_D05t08&            &         &   -0.010   &  (0.006)&    0.083   &  (0.081)&    0.157*  &  (0.082)\\ 
\addlinespace
Ln(HH Income, 99) x pd\_D99t05&            &         &   -0.004   &  (0.022)&   -0.027   &  (0.025)&    0.035   &  (0.036)\\ 
\addlinespace
Ln(HH Income, 99) x pd\_D05t08&            &         &    0.077*  &  (0.041)&    0.161***&  (0.046)&   -0.001   &  (0.070)\\ 
\addlinespace
Ratio of Labor Force (1999) x pd\_D99t05&            &         &   -0.021   &  (0.107)&    0.081   &  (0.099)&    0.127   &  (0.102)\\ 
\addlinespace
Ratio of Labor Force (1999) x pd\_D05t08&            &         &   -0.043   &  (0.234)&   -0.355   &  (0.224)&   -0.534** &  (0.241)\\ 
\addlinespace
Ln(Num of HU, 99) x pd\_D99t05&            &         &            &         &   -0.010   &  (0.033)&    0.009   &  (0.032)\\ 
\addlinespace
Ln(Num of HU, 99) x pd\_D05t08&            &         &            &         &   -0.105   &  (0.081)&   -0.180** &  (0.082)\\ 
\addlinespace
Housing supply elasticity x pd\_D99t05&            &         &            &         &   -0.011***&  (0.003)&   -0.011***&  (0.003)\\ 
\addlinespace
Housing supply elasticity x pd\_D05t08&            &         &            &         &    0.029***&  (0.007)&    0.028***&  (0.007)\\ 
\addlinespace
House Vacancy Rate (1999) x pd\_D99t05&            &         &            &         &    0.179** &  (0.087)&    0.187** &  (0.084)\\ 
\addlinespace
House Vacancy Rate (1999) x pd\_D05t08&            &         &            &         &   -0.143   &  (0.184)&   -0.130   &  (0.181)\\ 
\addlinespace
Ratio of Renters (1999) x pd\_D99t05&            &         &            &         &   -0.107** &  (0.053)&   -0.054   &  (0.058)\\ 
\addlinespace
Ratio of Renters (1999) x pd\_D05t08&            &         &            &         &    0.316** &  (0.128)&    0.287*  &  (0.152)\\ 
\addlinespace
Ratio of Bachelor Educated (1999) x pd\_D99t05&            &         &            &         &            &         &   -0.239** &  (0.094)\\ 
\addlinespace
Ratio of Bachelor Educated (1999) x pd\_D05t08&            &         &            &         &            &         &    0.610***&  (0.176)\\ 
\addlinespace
Ratio of White Race (1999) x pd\_D99t05&            &         &            &         &            &         &    0.018   &  (0.035)\\ 
\addlinespace
Ratio of White Race (1999) x pd\_D05t08&            &         &            &         &            &         &    0.074   &  (0.075)\\ 
\addlinespace
Ratio of Immigration (90-00) x pd\_D99t05&            &         &            &         &            &         &    0.138   &  (0.197)\\ 
\addlinespace
Ratio of Immigration (90-00) x pd\_D05t08&            &         &            &         &            &         &   -0.407   &  (0.392)\\ 
\addlinespace
\midrule
Obs            &     1584   &         &     1584   &         &     1402   &         &     1402   &         \\
Cluster SE     &     CBSA   &         &     CBSA   &         &     CBSA   &         &     CBSA   &         \\
Weight         & Ln(HU99)   &         & Ln(HU99)   &         & Ln(HU99)   &         & Ln(HU99)   &         \\
KP F-Stat      &    23.36   &         &    23.11   &         &    16.44   &         &    17.78   &         \\
MOP F-Eff      &    22.91   &         &    22.63   &         &    16.08   &         &    17.44   &         \\
CoefEqual\_Chi2 &   11.266   &         &   10.375   &         &    9.183   &         &    8.386   &         \\
CoefEqual\_PValue &    0.001   &         &    0.001   &         &    0.002   &         &    0.004   &         \\
\bottomrule

\end{tabular}

} % end of resize box

\end{table}

\pagebreak 
%---------------------------------------------------------------

%%%%%%%%%%%%%%%%%%%%%%%%%%%%%%%%%%%%%%%%%%%%%%%%
% table_PLMNJ.D99t05vsD05t08.4Reg
%%%%%%%%%%%%%%%%%%%%%%%%%%%%%%%%%%%%%%%%%%%%%%%%

%---------------------------------------------------------------

%%%%%%%%%%%%%%%%%%%%%%%%%%%%%%%%%%%%%%%%%%%%%%%%
% table_PLMNJ.D99t05vsD05t08.4Reg
%%%%%%%%%%%%%%%%%%%%%%%%%%%%%%%%%%%%%%%%%%%%%%%%

\noindent 

\begin{table}[h!]
\centering
\caption{
\textbf{Four Stacked Regressions of PLMNJ Growth in Boom (99-05) and Bust (05-08) Periods on Net Export Growth (99-05)} \smallskip \newline
{\scriptsize
$\triangle_{99,05} \& \triangle_{05,08} Ln(PLMNJ_{c}) = \beta_{99,05} * \triangle_{99,05} \text{NetExp}_{m} \times Dum_{99,05} + \beta_{05,08} * \triangle_{99,05} \text{NetExp}_{m} \times Dum_{05,08} + \gamma_{99,05}* \bm{Controls_{c}} \times Dum_{99,05} + \gamma_{05,08}* \bm{Controls_{c}} \times Dum_{05,08}  + \alpha_{99,05} + \alpha_{05,08} + \epsilon_{period, c}$. The left-hand-side dependent variable $\triangle_{99,05} \& \triangle_{05,08} Ln(PLMNJ_{c})$ is the stacked growth rate of the dollar amount (07USD) of private-label mortgages (non-jumbo) (PLMNJ) at county $c$ 99-05 and 05-08. The key independent variable $\triangle_{99,05} \text{NetExp}_{m}$ is the growth rate of net export at the metropolitan area (CBSA03 code) $m$ 99-05. $Controls_{c}$ indicates control variables at county $c$ in the period start year 1999. We use the gravity model-based instrumental variable $\triangle_{99,05}\text{givNetExp}_{m}$ as IVs for $\triangle_{99,05}\text{NetExp}_{m}$. Regression is weighted by the natural logarithm of housing units in the start year 1999. For the first-stage F-test of two separate non-stack samples, we report kleibergen-Paap (2006) robust (clustered) statistics and Montiel Olea-Pflueger (2013) efficient statistics. Standard errors are clustered at the CBSA level. ***, **, and * indicate significance at the 1\%, 5\%, and 10\% levels, respectively.
} % end of small font size
} % end of caption
\label{table_PLMNJ.D99t05vsD05t08.4Reg}

\resizebox{0.95\columnwidth}{!}{%
\begin{tabular}{l*{4}{c}}
\toprule
Dep Var (Panel A, B, and C)                      &\multicolumn{4}{c}{PLMNJ Growth (99-05 \& 05-08, An)} \\
            \cmidrule{2-5} 
            &\multicolumn{1}{c}{(1)}&\multicolumn{1}{c}{(2)}&\multicolumn{1}{c}{(3)}&\multicolumn{1}{c}{(4)}\\

\midrule
\multicolumn{5}{l}{\textbf{Panel A. OLS estimates}} \\
Net Export Growth (99-05, An) x Dum99t05&    8.391***&    8.110***&    5.993***&    6.564***\\
               &  (1.552)   &  (1.575)   &  (1.627)   &  (1.729)   \\
\addlinespace
Net Export Growth (99-05, An) x Dum05t08&   -6.095** &   -6.377** &   -2.113   &   -3.519   \\
               &  (2.806)   &  (2.839)   &  (3.018)   &  (2.983)   \\
\addlinespace
R2-adj         &    0.803   &    0.804   &    0.829   &    0.833   \\
\addlinespace

\midrule
\multicolumn{5}{l}{\textbf{Panel B. Reduced-form estimates}} \\
GIV Net Export Growth (99-05, An) x Dum99t05&   15.753***&   15.278***&   12.243***&   12.134***\\
               &  (3.256)   &  (3.332)   &  (3.277)   &  (3.543)   \\
\addlinespace
GIV Net Export Growth (99-05, An) x Dum05t08&  -18.780***&  -19.336***&  -14.647***&  -13.909** \\
               &  (5.772)   &  (5.988)   &  (5.609)   &  (6.104)   \\
\addlinespace
R2-adj         &    0.804   &    0.806   &    0.831   &    0.834   \\
\addlinespace

\midrule
\multicolumn{5}{l}{\textbf{Panel C . 2SLS estimates}} \\
\addlinespace
Net Export Growth (99-05, An) x Dum99t05&   13.694***&   13.365***&   11.145***&   10.851***\\
               &  (3.547)   &  (3.643)   &  (3.254)   &  (3.331)   \\
\addlinespace
Net Export Growth (99-05, An) x Dum05t08&  -16.325***&  -16.915***&  -13.335** &  -12.438** \\
               &  (6.194)   &  (6.457)   &  (5.782)   &  (5.669)   \\
\addlinespace
\addlinespace

Dep Var (Panel D): &\multicolumn{4}{c}{Net Export Growth (99-05, An)} \\ 
\midrule 
\multicolumn{5}{l}{\textbf{Panel D . First-stage estimates only for 99-05 (non-stack sample)}} \\
\addlinespace
GIV Net Export Growth (99-05, An) x Dum99t05&    1.150***&    1.143***&    1.098***&    1.118***\\
               &  (0.238)   &  (0.238)   &  (0.271)   &  (0.265)   \\
\addlinespace
Kleibergen-Paap F-Stat      &    23.36   &    23.11   &    16.44   &    17.78   \\
MOP F-Eff &    22.91   &    22.63   &    16.08   &    17.44   \\

\addlinespace
\addlinespace

\midrule
\multicolumn{5}{l}{\textbf{Controls (for all Panels)}} \\
Period Dum &  Y   &   Y    & Y    &  Y \\
Basic Controls x Dum &            &  Y   &   Y    & Y        \\
Housing Controls x Dum &           &      & Y       & Y        \\
Demographic Controls x Dum &            &      &        &  Y     \\

\midrule              
Obs (Panel A, B, and C)         &     1584   &     1584   &     1402   &     1402   \\
Obs (Panel D)           &      792   &      792   &      701   &      701   \\
Cluster SE     &     CBSA   &     CBSA   &     CBSA   &     CBSA    \\
Weight         & Ln(HU99)   & Ln(HU99)   & Ln(HU99)   & Ln(HU99)   \\
\bottomrule
\end{tabular}

} % end of resize box

\end{table}

%%%%%%%%%%%%%%%%%%%%%%%%%%%%%%%%%%%%%%%%%%%%%%%%%%%%%%%%%%%%%%%%%%%%%%%%%%%
%%%%%%%%%%%%%%%%%%%%%%%%%%%%%%%%%%%%%%%%%%%%%%%%%%%%%%%%%%%%%%%%%%%%%%%%%%%
\pagebreak
% Housing Industry Channel: Other Evidence
% BEA construction employment and Residential Building Permits. 
%%%%%%%%%%%%%%%%%%%%%%%%%%%%%%%%%%%%%%%%%%%%%%%%%%%%%%%%%%%%%%%%%%%%%%%%%%%
%%%%%%%%%%%%%%%%%%%%%%%%%%%%%%%%%%%%%%%%%%%%%%%%%%%%%%%%%%%%%%%%%%%%%%%%%%%

%---------------------------------------------------------------
%---------------------------------------------------------------
% BEA Construction Employment Growth in Boom (00-06) and Bust (07-10)
%---------------------------------------------------------------
%---------------------------------------------------------------

\pagebreak
%-----------------------------------------------------------------

%%%%%%%%%%%%%%%%%%%%%%%%%%%%%%%%%%%%
% table_BEA.ConstEmpShr.D00t06vsD07t10.PLMNJ.D99t05.2SLS.wide
%%%%%%%%%%%%%%%%%%%%%%%%%%%%%%%%%%%%

%-----------------------------------------------------------------------
%%%%%%%%%%%%%%%%%%%%%%%%%%%%%%%%%%%%
% table_BEA.ConstEmpShr.D00t06vsD07t10.PLMNJ.D99t05.2SLS.wide
%%%%%%%%%%%%%%%%%%%%%%%%%%%%%%%%%%%%

\noindent 

\begin{table}[h!]
\centering
\caption{
\textbf{2SLS Stacked Regression of Construction Employment Growth in Boom (00-06) and Bust (07-10) Periods on PLMNJ Growth (99-05)} \smallskip \newline
{\scriptsize
This table reports 2SLS regression $\triangle_{00,06} \& \triangle_{07,10} ConstEmpShr_{c} = \beta_{00,06} * \triangle_{99,05} Ln(PLMNJ_{c}) \times Dum_{00,06} + \beta_{07,10} * \triangle_{99,05} Ln(PLMNJ_{c}) \times Dum_{07,10} + \gamma_{00,06}* \bm{Controls_{c}} \times Dum_{00,06} + \gamma_{07,10}* \bm{Controls_{c}} \times Dum_{07,10} + \epsilon_{period, c}$. The left-hand-side dependent variable $\triangle_{00,06} \& \triangle_{07,10} ConstEmpShr_{c}$ is the change of the construction employment share in working-age population at county $c$ 00-06 and 07-10. To reduce the impact of outliers, dependent variable is winsorized at 1\% and 99\% levels in each period. The key independent variable $\triangle_{99,05} Ln(PLMNJ_{c})$ is the growth rate of the dollar amount (07USD) of private-label mortgages (non-jumbo) at county $c$ 99-05. $Controls_{c}$ indicates control variables at county $c$ in the period start year 1999. We use the gravity model-based instrumental variable $\triangle_{99,05}\text{givNetExp}_{m}$ as the IV for $\triangle_{99,05}Ln(PLMNJ_{c})$. Regression is weighted by the natural logarithm of housing units in 1999.  For the first-stage F-test of two non-stacked samples, we report Kleibergen-Paap (2006) robust (clustered) statistics and Montiel Olea-Pflueger (2013) efficient statistics. Standard errors are clustered at the CBSA level. ***, **, and * indicate significance at the 1\%, 5\%, and 10\% levels, respectively.
\smallskip
} % end of small font size
} % end of caption
\label{table_BEA.ConstEmpShr.D00t06vsD07t10.PLMNJ.D99t05.2SLS.wide}

\vspace{-2mm}

\resizebox{\columnwidth}{!}{%
\begin{tabular}{l*{8}{c}}
\toprule
\textbf{TSLS estimates}            &\multicolumn{8}{c}{Construction Employment Growth (00-06 and 07-10, An)} \\
            \cmidrule{2-9} 
            &\multicolumn{2}{c}{(1)}&\multicolumn{2}{c}{(2)}&\multicolumn{2}{c}{(3)}&\multicolumn{2}{c}{(4)}\\
            
\midrule
PLMNJ Growth (07USD, 99-05, An) x Dum00t06&    0.015***&  (0.005)&    0.015***&  (0.005)&    0.013** &  (0.006)&    0.015** &  (0.006)\\ 
\addlinespace
PLMNJ Growth (07USD, 99-05, An) x Dum07t10&   -0.022***&  (0.008)&   -0.019** &  (0.008)&   -0.018*  &  (0.010)&   -0.022** &  (0.011)\\ 
\addlinespace
Dum00t06       &   -0.002*  &  (0.001)&   -0.012***&  (0.004)&   -0.014***&  (0.005)&    0.000   &  (0.008)\\ 
\addlinespace
Dum07t10       &   -0.000   &  (0.001)&    0.032***&  (0.008)&    0.026***&  (0.010)&   -0.000   &  (0.017)\\ 
\addlinespace
Ln(Num of HH, 99) x Dum00t06&            &         &   -0.000** &  (0.000)&   -0.003***&  (0.001)&   -0.003***&  (0.001)\\ 
\addlinespace
Ln(Num of HH, 99) x Dum07t10&            &         &   -0.000   &  (0.000)&   -0.001   &  (0.002)&   -0.000   &  (0.002)\\ 
\addlinespace
Ln(HH Income, 99) x Dum00t06&            &         &    0.001***&  (0.000)&    0.001***&  (0.000)&   -0.000   &  (0.001)\\ 
\addlinespace
Ln(HH Income, 99) x Dum07t10&            &         &   -0.003***&  (0.001)&   -0.002*  &  (0.001)&    0.000   &  (0.002)\\ 
\addlinespace
Ratio of Labor Force (1999) x Dum00t06&            &         &   -0.005***&  (0.002)&   -0.003   &  (0.002)&   -0.003   &  (0.002)\\ 
\addlinespace
Ratio of Labor Force (1999) x Dum07t10&            &         &   -0.000   &  (0.004)&   -0.006   &  (0.004)&   -0.007   &  (0.004)\\ 
\addlinespace
Ln(Num of HU, 99) x Dum00t06&            &         &            &         &    0.003***&  (0.001)&    0.003***&  (0.001)\\ 
\addlinespace
Ln(Num of HU, 99) x Dum07t10&            &         &            &         &    0.000   &  (0.002)&   -0.000   &  (0.002)\\ 
\addlinespace
Housing supply elasticity x Dum00t06&            &         &            &         &    0.000   &  (0.000)&    0.000   &  (0.000)\\ 
\addlinespace
Housing supply elasticity x Dum07t10&            &         &            &         &    0.000   &  (0.000)&   -0.000   &  (0.000)\\ 
\addlinespace
House Vacancy Rate (1999) x Dum00t06&            &         &            &         &    0.000   &  (0.003)&   -0.001   &  (0.003)\\ 
\addlinespace
House Vacancy Rate (1999) x Dum07t10&            &         &            &         &   -0.010   &  (0.007)&   -0.007   &  (0.006)\\ 
\addlinespace
Ratio of Renters (1999) x Dum00t06&            &         &            &         &   -0.003*  &  (0.002)&   -0.004** &  (0.002)\\ 
\addlinespace
Ratio of Renters (1999) x Dum07t10&            &         &            &         &    0.006** &  (0.003)&    0.011***&  (0.003)\\ 
\addlinespace
Ratio of Bachelor Educated (1999) x Dum00t06&            &         &            &         &            &         &    0.004   &  (0.002)\\ 
\addlinespace
Ratio of Bachelor Educated (1999) x Dum07t10&            &         &            &         &            &         &   -0.006   &  (0.004)\\ 
\addlinespace
Ratio of White Race (1999) x Dum00t06&            &         &            &         &            &         &    0.001   &  (0.001)\\ 
\addlinespace
Ratio of White Race (1999) x Dum07t10&            &         &            &         &            &         &    0.001   &  (0.002)\\ 
\addlinespace
Ratio of Immigration (90-00) x Dum00t06&            &         &            &         &            &         &    0.012***&  (0.004)\\ 
\addlinespace
Ratio of Immigration (90-00) x Dum07t10&            &         &            &         &            &         &   -0.021** &  (0.010)\\ 
\addlinespace
\midrule
Obs            &     1530   &         &     1530   &         &     1368   &         &     1368   &         \\
Cluster SE     &     CBSA   &         &     CBSA   &         &     CBSA   &         &     CBSA   &         \\
Weight         & Ln(HU99)   &         & Ln(HU99)   &         & Ln(HU99)   &         & Ln(HU99)   &         \\
KP F-Stat (99-05, non-stack sample)       &    20.97   &         &    19.44   &         &    12.87   &         &    10.48   &         \\
MOP F-Stat (99-05, non-stack sample)  &    20.00   &         &    18.89   &         &    12.85   &         &    10.47   &         \\
CoefEqual\_Chi2 &   10.192   &         &    8.385   &         &    4.154   &         &    5.548   &         \\
CoefEqual\_PValue &    0.001   &         &    0.004   &         &    0.042   &         &    0.019   &         \\
\bottomrule

\end{tabular}

} % end of resize box

\end{table}

\pagebreak 
%---------------------------------------------------------------

%%%%%%%%%%%%%%%%%%%%%%%%%%%%%%%%%%%%%%%%%%%%%%%%
% table_BEA.ConstEmpShr.D00t06vsD07t10.PLMNJ.D99t05.4Reg
%%%%%%%%%%%%%%%%%%%%%%%%%%%%%%%%%%%%%%%%%%%%%%%%

%---------------------------------------------------------------

%%%%%%%%%%%%%%%%%%%%%%%%%%%%%%%%%%%%%%%%%%%%%%%%
% table_BEA.ConstEmpShr.D00t06vsD07t10.PLMNJ.D99t05.4Reg
%%%%%%%%%%%%%%%%%%%%%%%%%%%%%%%%%%%%%%%%%%%%%%%%

\noindent 

\begin{table}[h!]
\centering
\caption{
\textbf{Four Stacked Regressions of Construction Employment Growth in Boom (00-06) and Bust (07-10) Periods on PLMNJ Growth (99-05)} \smallskip \newline
{\scriptsize
This table reports OLS, reduced-form, first stage, and second stages of stacked 2SLS regression $\triangle_{00,06} \& \triangle_{07,10} ConstEmpShr_{c} = \beta_{00,06} * \triangle_{99,05} Ln(PLMNJ_{c}) \times Dum_{00,06} + \beta_{07,10} * \triangle_{99,05} Ln(PLMNJ_{c}) \times Dum_{07,10} + \gamma_{00,06}* \bm{Controls_{c}} \times Dum_{00,06} + \gamma_{07,10}* \bm{Controls_{c}} \times Dum_{07,10} + \epsilon_{period, c}$. The left-hand-side dependent variable $\triangle_{00,06} \& \triangle_{07,10} ConstEmpShr_{c}$ is the change of the construction employment share in working-age population at county $c$ 00-06 and 07-10. To reduce the impact of outliers, dependent variable is winsorized at 1\% and 99\% levels in each period. The key independent variable $\triangle_{99,05} Ln(PLMNJ_{c})$ is the growth rate of the dollar amount (07USD) of private-label mortgages (non-jumbo) at county $c$ 99-05. $Controls_{c}$ indicates control variables at county $c$ in the period start year 1999. We use the gravity model-based instrumental variable $\triangle_{99,05}\text{givNetExp}_{m}$ as the IV for $\triangle_{99,05}Ln(PLMNJ_{c})$. Regression is weighted by the natural logarithm of housing units in 1999.  For the first-stage F-test of two non-stacked samples, we report Kleibergen-Paap (2006) robust (clustered) statistics and Montiel Olea-Pflueger (2013) efficient statistics. Standard errors are clustered at the CBSA level. ***, **, and * indicate significance at the 1\%, 5\%, and 10\% levels, respectively.
} % end of small font size
} % end of caption
\label{table_BEA.ConstEmpShr.D00t06vsD07t10.PLMNJ.D99t05.4Reg}
\resizebox{0.9\columnwidth}{!}{%
\begin{tabular}{l*{4}{c}}
\toprule
Dep Var (Panel A, B, and C)                      &\multicolumn{4}{c}{Construction Emp Growth (00-06 \& 07-10, An)} \\
            \cmidrule{2-5} 
            &\multicolumn{1}{c}{(1)}&\multicolumn{1}{c}{(2)}&\multicolumn{1}{c}{(3)}&\multicolumn{1}{c}{(4)}\\

\midrule
\multicolumn{5}{l}{\textbf{Panel A. OLS estimates}} \\
PLMNJ Growth (07USD, 99-05, An) x Dum00t06&    0.006***&    0.006***&    0.006***&    0.006***\\
               &  (0.001)   &  (0.001)   &  (0.001)   &  (0.001)   \\
\addlinespace
PLMNJ Growth (07USD, 99-05, An) x Dum07t10&   -0.011***&   -0.010***&   -0.008***&   -0.008***\\
               &  (0.002)   &  (0.002)   &  (0.002)   &  (0.002)   \\
\addlinespace
R2-adj         &    0.536   &    0.554   &    0.569   &    0.576   \\
\addlinespace

\midrule
\multicolumn{5}{l}{\textbf{Panel B. Reduced-form estimates}} \\
GIV Net Export Growth (99-06, An) x Dum00t06&    0.235***&    0.226***&    0.160*  &    0.187** \\
               &  (0.083)   &  (0.079)   &  (0.087)   &  (0.087)   \\
\addlinespace
GIV Net Export Growth (99-06, An) x Dum07t10&   -0.351***&   -0.289** &   -0.227   &   -0.273*  \\
               &  (0.133)   &  (0.122)   &  (0.150)   &  (0.152)   \\
\addlinespace
R2-adj         &    0.513   &    0.533   &    0.555   &    0.563   \\
\addlinespace

\midrule
\multicolumn{5}{l}{\textbf{Panel C . 2SLS estimates}} \\
\addlinespace
PLMNJ Growth (07USD, 99-05, An) x Dum00t06&    0.015***&    0.015***&    0.013** &    0.015** \\
               &  (0.005)   &  (0.005)   &  (0.006)   &  (0.006)   \\
\addlinespace
PLMNJ Growth (07USD, 99-05, An) x Dum07t10&   -0.022***&   -0.019** &   -0.018*  &   -0.022** \\
               &  (0.008)   &  (0.008)   &  (0.010)   &  (0.011)   \\
               
\addlinespace
\addlinespace

Dep Var (Panel D): &\multicolumn{4}{c}{PLMNJ Growth (99-05, An)} \\ 
\midrule 
\multicolumn{5}{l}{\textbf{Panel D . First-stage estimates only for 99-05 (Non-stack sample)}} \\
\addlinespace
GIV Net Export Growth (99-05, An)&   15.738***&   15.106***&   12.754***&   12.555***\\
               &  (3.437)   &  (3.426)   &  (3.555)   &  (3.878)   \\
\addlinespace
KP F-Stat      &    20.97   &    19.44   &    12.87   &    10.48   \\
MOP F-Stat     &    20.00   &    18.89   &    12.85   &    10.47   \\
\addlinespace

\midrule
\multicolumn{5}{l}{\textbf{Controls (for all Panels)}} \\
DumPeriod  &    Y        &  Y   &   Y    & Y        \\
Basic Controls x DumPeriod &            &  Y   &   Y    & Y    \\
Housing Controls x DumPeriod &           &      & Y       & Y   \\
Demographic Controls x DumPeriod &            &      &      &  Y \\

\midrule              
Obs (Panel A, B, and C)          &     1530   &     1530   &     1368   &     1368   \\
Obs (Panel D)          &      765   &      765   &      684   &      684   \\
Cluster SE     &     CBSA   &     CBSA   &     CBSA   &     CBSA   \\
Weight         & Ln(HU99)   & Ln(HU99)   & Ln(HU99)   & Ln(HU99)   \\
\bottomrule
\end{tabular}

} % end of resize box

\end{table}

%---------------------------------------------------------------
%---------------------------------------------------------------
% Residential Building Permit Value Growth in Boom (00-06) and Bust (07-10)
%---------------------------------------------------------------
%---------------------------------------------------------------

\pagebreak
%-----------------------------------------------------------------

%%%%%%%%%%%%%%%%%%%%%%%%%%%%%%%%%%%%
% table_PermitValue.D99t05vsD05t09.PLMNJ.2SLS.wide
%%%%%%%%%%%%%%%%%%%%%%%%%%%%%%%%%%%%

%-----------------------------------------------------------------------
%%%%%%%%%%%%%%%%%%%%%%%%%%%%%%%%%%%%
% table_PermitValue.D99t05vsD05t09.PLMNJ.2SLS.wide
%%%%%%%%%%%%%%%%%%%%%%%%%%%%%%%%%%%%

\noindent 

\begin{table}[h!]
\centering
\caption{
\textbf{2SLS Stacked Regression of Residential Units Permit Value Growth in Boom (99-05) and Bust (05-09) Periods on PLMNJ Growth (99-05)} \smallskip \newline
{\scriptsize
This table reports 2SLS regression $\triangle_{99,05} \& \triangle_{05,09} Ln(PermitValue_{c}) = \beta_{99,05} * \triangle_{99,05} Ln(PLMNJ_{c}) \times Dum_{99,05} + \beta_{05,09} * \triangle_{99,05} Ln(PLMNJ_{c}) \times Dum_{05,09} + \gamma_{99,05}* \bm{Controls_{c}} \times Dum_{99,05} + \gamma_{05,09}* \bm{Controls_{c}} \times Dum_{05,09} + \epsilon_{period, c}$. The left-hand-side dependent variable $\triangle_{99,05} \& \triangle_{05,09} Ln(PermitValue_{c})$ is the residential units permit value (by imputation by U.S. Census) growth at county $c$ 99-05 and 05-09. The key independent variable $\triangle_{99,05} Ln(PLMNJ_{c})$ is the growth rate of the dollar amount (07USD) of private-label mortgages (non-jumbo) at county $c$ 99-05. $Controls_{c}$ indicates control variables at county $c$ in the period start year 1999. We use the gravity model-based instrumental variable $\triangle_{99,05}\text{givNetExp}_{m}$ as the IV for $\triangle_{99,05}Ln(PLMNJ_{c})$. Regression is weighted by the natural logarithm of housing units in 1999.  For the first-stage F-test of two non-stacked samples, we report Kleibergen-Paap (2006) robust (clustered) statistics and Montiel Olea-Pflueger (2013) efficient statistics. Standard errors are clustered at the CBSA level. ***, **, and * indicate significance at the 1\%, 5\%, and 10\% levels, respectively.
\smallskip
} % end of small font size
} % end of caption
\label{table_PermitValue.D99t05vsD05t09.PLMNJ.2SLS.wide}

\vspace{-2mm}

\resizebox{\columnwidth}{!}{%
\begin{tabular}{l*{8}{c}}
\toprule
\textbf{TSLS estimates}            &\multicolumn{8}{c}{Residential Units Permit Value Growth (99-05 \& 05-09, An)} \\
            \cmidrule{2-9} 
            &\multicolumn{2}{c}{(1)}&\multicolumn{2}{c}{(2)}&\multicolumn{2}{c}{(3)}&\multicolumn{2}{c}{(4)}\\
            
\midrule
PLMNJ Growth (07USD, 99-05, An) x Dum99t05&    0.689***&  (0.184)&    0.827***&  (0.200)&    0.661***&  (0.252)&    0.713***&  (0.264)\\ 
\addlinespace
PLMNJ Growth (07USD, 99-05, An) x Dum05t09&   -1.588***&  (0.480)&   -1.462***&  (0.485)&   -1.660***&  (0.638)&   -1.648***&  (0.577)\\ 
\addlinespace
Dum99t05       &   -0.051   &  (0.032)&    0.723***&  (0.196)&    0.554***&  (0.187)&    0.977***&  (0.375)\\ 
\addlinespace
Dum05t09       &   -0.027   &  (0.078)&    1.096***&  (0.356)&    0.732*  &  (0.420)&    2.240***&  (0.753)\\ 
\addlinespace
Ln(Num of HH, 99) x Dum99t05&            &         &   -0.006   &  (0.005)&    0.002   &  (0.048)&   -0.026   &  (0.048)\\ 
\addlinespace
Ln(Num of HH, 99) x Dum05t09&            &         &    0.000   &  (0.006)&    0.031   &  (0.081)&    0.117   &  (0.082)\\ 
\addlinespace
Ln(HH Income, 99) x Dum99t05&            &         &   -0.056***&  (0.021)&   -0.040** &  (0.019)&   -0.071** &  (0.036)\\ 
\addlinespace
Ln(HH Income, 99) x Dum05t09&            &         &   -0.116***&  (0.037)&   -0.080** &  (0.040)&   -0.223***&  (0.069)\\ 
\addlinespace
Ratio of Labor Force (1999) x Dum99t05&            &         &   -0.208** &  (0.085)&   -0.159*  &  (0.094)&   -0.108   &  (0.089)\\ 
\addlinespace
Ratio of Labor Force (1999) x Dum05t09&            &         &    0.151   &  (0.193)&    0.116   &  (0.200)&   -0.089   &  (0.225)\\ 
\addlinespace
Ln(Num of HU, 99) x Dum99t05&            &         &            &         &   -0.013   &  (0.048)&    0.010   &  (0.047)\\ 
\addlinespace
Ln(Num of HU, 99) x Dum05t09&            &         &            &         &   -0.033   &  (0.082)&   -0.116   &  (0.082)\\ 
\addlinespace
Housing supply elasticity x Dum99t05&            &         &            &         &    0.000   &  (0.005)&    0.001   &  (0.005)\\ 
\addlinespace
Housing supply elasticity x Dum05t09&            &         &            &         &   -0.005   &  (0.010)&   -0.006   &  (0.010)\\ 
\addlinespace
House Vacancy Rate (1999) x Dum99t05&            &         &            &         &    0.159   &  (0.116)&    0.105   &  (0.121)\\ 
\addlinespace
House Vacancy Rate (1999) x Dum05t09&            &         &            &         &    0.288   &  (0.252)&    0.315   &  (0.246)\\ 
\addlinespace
Ratio of Renters (1999) x Dum99t05&            &         &            &         &    0.139*  &  (0.084)&   -0.002   &  (0.090)\\ 
\addlinespace
Ratio of Renters (1999) x Dum05t09&            &         &            &         &    0.123   &  (0.158)&    0.144   &  (0.180)\\ 
\addlinespace
Ratio of Bachelor Educated (1999) x Dum99t05&            &         &            &         &            &         &    0.081   &  (0.093)\\ 
\addlinespace
Ratio of Bachelor Educated (1999) x Dum05t09&            &         &            &         &            &         &    0.612***&  (0.206)\\ 
\addlinespace
Ratio of White Race (1999) x Dum99t05&            &         &            &         &            &         &   -0.080** &  (0.034)\\ 
\addlinespace
Ratio of White Race (1999) x Dum05t09&            &         &            &         &            &         &    0.042   &  (0.128)\\ 
\addlinespace
Ratio of Immigration (90-00) x Dum99t05&            &         &            &         &            &         &    0.393** &  (0.170)\\ 
\addlinespace
Ratio of Immigration (90-00) x Dum05t09&            &         &            &         &            &         &   -0.898   &  (0.585)\\ 
\addlinespace
\midrule
Obs            &     1582   &         &     1582   &         &     1401   &         &     1401   &         \\
Cluster SE     &     CBSA   &         &     CBSA   &         &     CBSA   &         &     CBSA   &         \\
Weight         & Ln(HU99)   &         & Ln(HU99)   &         & Ln(HU99)   &         & Ln(HU99)   &         \\
KP F-Stat (99-05, non-stack sample)       &    23.43   &         &    21.04   &         &    13.97   &         &    11.74   &         \\
MOP F-Stat (99-05, non-stack sample)  &    22.30   &         &    20.29   &         &    14.03   &         &    11.80   &         \\
CoefEqual\_Chi2 &   16.078   &         &   14.832   &         &    9.361   &         &   10.906   &         \\
CoefEqual\_PValue &    0.000   &         &    0.000   &         &    0.002   &         &    0.001   &         \\
\bottomrule

\end{tabular}

} % end of resize box

\end{table}

\pagebreak 
%---------------------------------------------------------------

%%%%%%%%%%%%%%%%%%%%%%%%%%%%%%%%%%%%%%%%%%%%%%%%
% table_PermitValue.D99t05vsD05t09.PLMNJ.4Reg
%%%%%%%%%%%%%%%%%%%%%%%%%%%%%%%%%%%%%%%%%%%%%%%%

%---------------------------------------------------------------

%%%%%%%%%%%%%%%%%%%%%%%%%%%%%%%%%%%%%%%%%%%%%%%%
% table_PermitValue.D99t05vsD05t09.PLMNJ.4Reg
%%%%%%%%%%%%%%%%%%%%%%%%%%%%%%%%%%%%%%%%%%%%%%%%

\noindent 

\begin{table}[h!]
\centering
\caption{
\textbf{Four Stacked Regressions of Residential Units Permit Value Growth in Boom (99-05) and Bust (05-09) Periods on PLMNJ Growth (99-05)} \smallskip \newline
{\scriptsize
This table reports OLS, reduced-form, first stage, and second stages of stacked 2SLS regression $\triangle_{99,05} \& \triangle_{05,09} Ln(PermitValue_{c}) = \beta_{99,05} * \triangle_{99,05} Ln(PLMNJ_{c}) \times Dum_{99,05} + \beta_{05,09} * \triangle_{99,05} Ln(PLMNJ_{c}) \times Dum_{05,09} + \gamma_{99,05}* \bm{Controls_{c}} \times Dum_{99,05} + \gamma_{05,09}* \bm{Controls_{c}} \times Dum_{05,09} + \epsilon_{period, c}$. The left-hand-side dependent variable $\triangle_{99,05} \& \triangle_{05,09} Ln(PermitValue_{c})$ is the residential units permit value (by imputation by U.S. Census) growth at county $c$ 99-05 and 05-09. The key independent variable $\triangle_{99,05} Ln(PLMNJ_{c})$ is the growth rate of the dollar amount (07USD) of private-label mortgages (non-jumbo) at county $c$ 99-05. $Controls_{c}$ indicates control variables at county $c$ in the period start year 1999. We use the gravity model-based instrumental variable $\triangle_{99,05}\text{givNetExp}_{m}$ as the IV for $\triangle_{99,05}Ln(PLMNJ_{c})$. Regression is weighted by the natural logarithm of housing units in 1999.  For the first-stage F-test of two non-stacked samples, we report Kleibergen-Paap (2006) robust (clustered) statistics and Montiel Olea-Pflueger (2013) efficient statistics. Standard errors are clustered at the CBSA level. ***, **, and * indicate significance at the 1\%, 5\%, and 10\% levels, respectively.
} % end of small font size
} % end of caption
\label{table_PermitValue.D99t05vsD05t09.PLMNJ.4Reg}
\resizebox{\columnwidth}{!}{%
\begin{tabular}{l*{4}{c}}
\toprule
Dep Var (Panel A, B, and C)                      &\multicolumn{4}{c}{Residential Units Permit Value Growth (99-05 \& 05-09, An)} \\
            \cmidrule{2-5} 
            &\multicolumn{1}{c}{(1)}&\multicolumn{1}{c}{(2)}&\multicolumn{1}{c}{(3)}&\multicolumn{1}{c}{(4)}\\

\midrule
\multicolumn{5}{l}{\textbf{Panel A. OLS estimates}} \\
PLMNJ Growth (07USD, 99-05, An) x Dum99t05&    0.393***&    0.404***&    0.414***&    0.415***\\
               &  (0.047)   &  (0.046)   &  (0.040)   &  (0.040)   \\
\addlinespace
PLMNJ Growth (07USD, 99-05, An) x Dum05t09&   -0.614***&   -0.582***&   -0.512***&   -0.473***\\
               &  (0.104)   &  (0.100)   &  (0.112)   &  (0.108)   \\
\addlinespace
R2-adj         &    0.745   &    0.754   &    0.768   &    0.779   \\
\addlinespace

\midrule
\multicolumn{5}{l}{\textbf{Panel B. Reduced-form estimates}} \\
GIV Net Export Growth (99-05, An) x Dum99t05&   10.853***&   12.641***&    8.094** &    8.647** \\
               &  (3.098)   &  (3.386)   &  (3.311)   &  (3.356)   \\
\addlinespace
GIV Net Export Growth (99-05, An) x Dum05t09&  -24.777***&  -22.025***&  -20.374***&  -20.050***\\
               &  (6.732)   &  (6.424)   &  (6.864)   &  (6.805)   \\
\addlinespace
R2-adj         &    0.724   &    0.735   &    0.753   &    0.766   \\
\addlinespace

\midrule
\multicolumn{5}{l}{\textbf{Panel C . 2SLS estimates}} \\
\addlinespace
PLMNJ Growth (07USD, 99-05, An) x Dum99t05&    0.689***&    0.827***&    0.661***&    0.713***\\
               &  (0.184)   &  (0.200)   &  (0.252)   &  (0.264)   \\
\addlinespace
PLMNJ Growth (07USD, 99-05, An) x Dum05t09&   -1.588***&   -1.462***&   -1.660***&   -1.648***\\
               &  (0.480)   &  (0.485)   &  (0.638)   &  (0.577)   \\
               
\addlinespace
\addlinespace

Dep Var (Panel D): &\multicolumn{4}{c}{PLMNJ Growth (99-05, An)} \\ 
\midrule 
\multicolumn{5}{l}{\textbf{Panel D . First-stage estimates only for 99-05 (Non-stack sample)}} \\
\addlinespace
GIV NEG (99-05, An)&   15.753***&   15.278***&   12.243***&   12.134***\\
               &  (3.255)   &  (3.331)   &  (3.276)   &  (3.541)   \\
\addlinespace
KP F-Stat      &    23.43   &    21.04   &    13.97   &    11.74   \\
MOP F-Stat     &    22.30   &    20.29   &    14.03   &    11.80   \\
\addlinespace

\midrule
\multicolumn{5}{l}{\textbf{Controls (for all Panels)}} \\
DumPeriod  &    Y        &  Y   &   Y    & Y        \\
Basic Controls x DumPeriod &            &  Y   &   Y    & Y    \\
Housing Controls x DumPeriod &           &      & Y       & Y   \\
Demographic Controls x DumPeriod &            &      &      &  Y \\

\midrule              
Obs (Panel A, B, and C)          &     1582   &     1582   &     1401   &     1401   \\
Obs (Panel D)          &      792   &      792   &      701   &      701   \\
Cluster SE     &     CBSA   &     CBSA   &     CBSA   &     CBSA   \\
Weight         & Ln(HU99)   & Ln(HU99)   & Ln(HU99)   & Ln(HU99)   \\
\bottomrule
\end{tabular}

} % end of resize box

\end{table}

%%%%%%%%%%%%%%%%%%%%%%%%%%%%%%%%%%%%%%%%%%%%%%%%%%%%%%%%%%%%%%%%%%%%%%%%%%%
%%%%%%%%%%%%%%%%%%%%%%%%%%%%%%%%%%%%%%%%%%%%%%%%%%%%%%%%%%%%%%%%%%%%%%%%%%%
\pagebreak
% Robustness Section
%%%%%%%%%%%%%%%%%%%%%%%%%%%%%%%%%%%%%%%%%%%%%%%%%%%%%%%%%%%%%%%%%%%%%%%%%%%
%%%%%%%%%%%%%%%%%%%%%%%%%%%%%%%%%%%%%%%%%%%%%%%%%%%%%%%%%%%%%%%%%%%%%%%%%%%

%---------------------------------------------------------------
%---------------------------------------------------------------
% Robustness 1. Relevance: GSEM vs PLMNJ
% Refined HOuse Employment Growth in Boom (00-06) and Bust (07-10)
%---------------------------------------------------------------
%---------------------------------------------------------------

%\pagebreak 
%---------------------------------------------------------------

%%%%%%%%%%%%%%%%%%%%%%%%%%%%%%%%%%%%%%%%%%%%%%%%
% table_RefineHouse.D00t06vsD07t10.GSEM.4Reg
%%%%%%%%%%%%%%%%%%%%%%%%%%%%%%%%%%%%%%%%%%%%%%%%

%---------------------------------------------------------------

%%%%%%%%%%%%%%%%%%%%%%%%%%%%%%%%%%%%%%%%%%%%%%%%
% table_RefineHouse.D00t06vsD07t10.GSEM.4Reg
%%%%%%%%%%%%%%%%%%%%%%%%%%%%%%%%%%%%%%%%%%%%%%%%

\noindent 

\begin{table}[h!]
\centering
\caption{
\textbf{Four Stacked Regressions of Refined House Employment Growth in Boom (00-06) and Bust (07-10) Periods on GSEM Growth (99-05)} \smallskip \newline
{\scriptsize
This table reports OLS, reduced-form, first stage, and second stages of stacked 2SLS regression $\triangle_{00,06} \& \triangle_{07,10} RefinedHouseEmpShr_{c} = \beta_{00,06} * \triangle_{99,05} Ln(GSEM_{c}) \times Dum_{00,06} + \beta_{07,10} * \triangle_{99,05} Ln(GSEM_{c}) \times Dum_{07,10} + \gamma_{00,06}* \bm{Controls_{c}} \times Dum_{00,06} + \gamma_{07,10}* \bm{Controls_{c}} \times Dum_{07,10} + \epsilon_{period, c}$. The left-hand-side dependent variable $\triangle_{00,06} \& \triangle_{07,10} RefinedHouseEmpShr_{c}$ is the change of the refined house employment share in working-age population at county $c$ 00-06 and 07-10. To reduce the impact of outliers, the dependent variable is winsoried at 2\% and 98\% levels in each period. To make the coefficient more visible in the table, the refined house employment share is multiplied by 100. The key independent variable $\triangle_{99,05} Ln(GSEM_{c})$ is the growth rate of the dollar amount (07USD) of government-sponsored enterprise mortgages at county $c$ 99-05. $Controls_{c}$ indicates control variables at county $c$ in the period start year 1999. We use the gravity model-based instrumental variable $\triangle_{99,05}\text{givNetExp}_{m}$ as the IV for $\triangle_{99,05}Ln(GSEM_{c})$. Regression is weighted by the natural logarithm of housing units in 1999.  For the first-stage F-test of two non-stacked samples, we report Kleibergen-Paap (2006) robust (clustered) statistics and Montiel Olea-Pflueger (2013) efficient statistics. Standard errors are clustered at the CBSA level. ***, **, and * indicate significance at the 1\%, 5\%, and 10\% levels, respectively.
} % end of small font size
} % end of caption
\label{table_RefineHouse.D00t06vsD07t10.GSEM.4Reg}
\resizebox{0.95\columnwidth}{!}{%
\begin{tabular}{l*{4}{c}}
\toprule
Dep Var (Panel A, B, and C)                      &\multicolumn{4}{c}{Refined House Employment Growth (00-06 \& 07-10, An)} \\
            \cmidrule{2-5} 
            &\multicolumn{1}{c}{(1)}&\multicolumn{1}{c}{(2)}&\multicolumn{1}{c}{(3)}&\multicolumn{1}{c}{(4)}\\

\midrule
\multicolumn{5}{l}{\textbf{Panel A. OLS estimates}} \\
GSEM Growth (07USD, 99-05, An) x Dum00t16&   -0.000   &    0.000   &    0.000   &    0.001   \\
               &  (0.000)   &  (0.000)   &  (0.000)   &  (0.000)   \\
\addlinespace
GSEM Growth (07USD, 99-05, An) x Dum07t10&    0.002***&    0.000   &    0.000   &   -0.000   \\
               &  (0.001)   &  (0.001)   &  (0.001)   &  (0.001)   \\
\addlinespace
R2-adj         &    0.476   &    0.518   &    0.545   &    0.559   \\
\addlinespace

\midrule
\multicolumn{5}{l}{\textbf{Panel B. Reduced-form estimates}} \\
GIV Net Export Growth (99-05, An) x Dum00t06&    0.060***&    0.052***&    0.040** &    0.046** \\
               &  (0.019)   &  (0.018)   &  (0.019)   &  (0.019)   \\
\addlinespace
GIV Net Export Growth (99-05, An) x Dum07t10&   -0.115***&   -0.087***&   -0.076***&   -0.088***\\
               &  (0.025)   &  (0.025)   &  (0.025)   &  (0.026)   \\
\addlinespace
R2-adj         &    0.474   &    0.523   &    0.549   &    0.564   \\
\addlinespace

\midrule
\multicolumn{5}{l}{\textbf{Panel C . 2SLS estimates}} \\
\addlinespace
GSEM Growth (07USD, 99-05, An) x Dum00t16&    0.028   &    0.010   &    0.011   &    0.013   \\
               &  (0.040)   &  (0.006)   &  (0.011)   &  (0.011)   \\
\addlinespace
GSEM Growth (07USD, 99-05, An) x Dum07t10&   -0.047   &   -0.016*  &   -0.021   &   -0.023   \\
               &  (0.061)   &  (0.009)   &  (0.016)   &  (0.017)   \\
               
\addlinespace
\addlinespace

Dep Var (Panel D): &\multicolumn{4}{c}{GSEM Growth (99-05, An)} \\ 
\midrule 
\multicolumn{5}{l}{\textbf{Panel D . First-stage estimates only for 99-05 (Non-stack sample)}} \\
\addlinespace
GIV NEG (99-05, An)&    2.173   &    5.288** &    3.373   &    3.522   \\
               &  (2.986)   &  (2.566)   &  (2.632)   &  (2.514)   \\
\addlinespace
KP F-Stat      &    0.538   &    4.368   &    1.884   &    2.131   \\
MOP F-Stat     &    0.539   &    4.317   &    1.882   &    2.149   \\
\addlinespace

\midrule
\multicolumn{5}{l}{\textbf{Controls (for all Panels)}} \\
DumPeriod  &    Y        &  Y   &   Y    & Y        \\
Basic Controls x DumPeriod &            &  Y   &   Y    & Y    \\
Housing Controls x DumPeriod &           &      & Y       & Y   \\
Demographic Controls x DumPeriod &            &      &      &  Y \\

\midrule              
Obs (Panel A, B, and C)          &     1580   &     1580   &     1398   &     1398   \\
Obs (Panel D)           &      790   &      790   &      699   &      699   \\
Cluster SE     &     CBSA   &     CBSA   &     CBSA   &     CBSA   \\
Weight         & Ln(HU99)   & Ln(HU99)   & Ln(HU99)   & Ln(HU99)   \\
\bottomrule
\end{tabular}

} % end of resize box

\end{table}

%---------------------------------------------------------------
%---------------------------------------------------------------
% Robustness 2. Exclusion Restriction: Prior vs Boom 
% (1) Refined HOuse Employment Growth and (2) PLMNJ Growth in Prior (92-00) and Boom (00-06) 
%---------------------------------------------------------------
%---------------------------------------------------------------

%\pagebreak 
%---------------------------------------------------------------

%%%%%%%%%%%%%%%%%%%%%%%%%%%%%%%%%%%%%%%%%%%%%%%%
% table_RefineHouse.D92t00vsD00t06.NEG.D91t99vsD99t05.4Reg
%%%%%%%%%%%%%%%%%%%%%%%%%%%%%%%%%%%%%%%%%%%%%%%%

%---------------------------------------------------------------

%%%%%%%%%%%%%%%%%%%%%%%%%%%%%%%%%%%%%%%%%%%%%%%%
% table_RefineHouse.D92t00vsD00t06.NEG.D91t99vsD99t05.4Reg
%%%%%%%%%%%%%%%%%%%%%%%%%%%%%%%%%%%%%%%%%%%%%%%%

\noindent 

\begin{table}[h!]
\centering
\caption{
\textbf{Four Stacked Regressions of Refined House Employment Growth on Net Export Growth in Prior (92-00) and Boom (00-06) Periods} \smallskip \newline
{\scriptsize
This table reports OLS, reduced-form, first stage and second stages of stacked 2SLS regression $\triangle_{92,00} \& \triangle_{00,06} RefineHouseEmpShr_{c} = \beta_{92,00} * \triangle_{91,99} \text{NetExp}_{m} \times Dum_{92,00} + \beta_{00,06} * \triangle_{99,05} \text{NetExp}_{m} \times Dum_{00,06} + \gamma_{92,00}* \bm{Controls_{c}} \times Dum_{92,00} + \gamma_{00,06}* \bm{Controls_{c}} \times Dum_{00,06}  + \alpha_{92,00} + \alpha_{00,06} + \epsilon_{period, c}$. The left-hand-side dependent variable $\triangle_{92,00} \& \triangle_{00,06} RefineHouseEmpShr_{c}$ is the stacked change of refined house employment growth at county $c$ 92-00 and 00-06. The key independent variable $\triangle_{91,99} \text{NetExp}_{m}$ and $\triangle_{99,05} \text{NetExp}_{m}$ are the growth rate of net export at the metropolitan area (CBSA03 code) $m$ 91-99 and 99-05, respectively. $Controls_{c}$ indicates control variables at county $c$ in the period start year (either 1991 or 1999). We use the gravity model-based instrumental variable $\triangle_{91,99}\text{givNetExp}_{m}$ and $\triangle_{99,05}\text{givNetExp}_{m}$ as IVs for $\triangle_{91,99}\text{NetExp}_{m}$ and $\triangle_{99,05}\text{NetExp}_{m}$. Each regression is weighted by the natural logarithm of housing units in the start year (either 1991 or 1999). For the first-stage F-test of two non-stacked samples, we report Kleibergen-Paap (2006) robust (clustered) statistics and Montiel Olea-Pflueger (2013) efficient statistics. Standard errors are clustered at the CBSA level. ***, **, and * indicate significance at the 1\%, 5\%, and 10\% levels, respectively.
} % end of small font size
} % end of caption
\label{table_RefineHouse.D92t00vsD00t06.NEG.D91t99vsD99t05.4Reg}
\resizebox{0.75\columnwidth}{!}{%
\begin{tabular}{l*{4}{c}}
\toprule
Dep Var (Panel A, B, and C)                      &\multicolumn{4}{c}{Refined House Employment Growth (92-00 \& 00-06, An)} \\
            \cmidrule{2-5} 
            &\multicolumn{1}{c}{(1)}&\multicolumn{1}{c}{(2)}&\multicolumn{1}{c}{(3)}&\multicolumn{1}{c}{(4)}\\

\midrule
\multicolumn{5}{l}{\textbf{Panel A. OLS estimates}} \\
Net Exp Growth (91-99, An) x Dum91t99&    0.015** &    0.009*  &    0.008   &    0.009   \\
               &  (0.007)   &  (0.005)   &  (0.006)   &  (0.006)   \\
\addlinespace
Net Exp Growth (99-05, An) x Dum99t05&    0.027***&    0.021** &    0.011   &    0.016*  \\
               &  (0.009)   &  (0.009)   &  (0.009)   &  (0.009)   \\
\addlinespace
R2-adj         &    0.481   &    0.520   &    0.535   &    0.553   \\
\addlinespace

\midrule
\multicolumn{5}{l}{\textbf{Panel B. Reduced-form estimates}} \\
GIV Net Exp Growth (91-99, An) x Dum91t99&    0.001   &    0.005   &    0.003   &    0.004   \\
               &  (0.008)   &  (0.008)   &  (0.009)   &  (0.009)   \\
\addlinespace
GIV Net Exp Growth (99-05, An) x Dum99t05&    0.059***&    0.049***&    0.046***&    0.055***\\
               &  (0.018)   &  (0.017)   &  (0.017)   &  (0.016)   \\
\addlinespace
R2-adj         &    0.482   &    0.520   &    0.538   &    0.556   \\
\addlinespace

\midrule
\multicolumn{5}{l}{\textbf{Panel C . 2SLS estimates}} \\
\addlinespace
Net Exp Growth (91-99, An) x Dum91t99 &    0.001   &    0.005   &    0.002   &    0.004   \\
               &  (0.007)   &  (0.007)   &  (0.008)   &  (0.008)   \\
\addlinespace
Net Exp Growth (99-05, An) x Dum99t05 &    0.050***&    0.042** &    0.041** &    0.048** \\
               &  (0.019)   &  (0.018)   &  (0.020)   &  (0.020)   \\
\addlinespace
CoefEqual\_Chi2 &    5.234   &    3.340   &    2.698   &    3.997   \\
CoefEqual\_PValue &    0.022   &    0.068   &    0.100   &    0.046   \\
\addlinespace
\addlinespace

Dep Var (Panel D): &\multicolumn{4}{c}{Net Export Growth (91-99, An)} \\ 
\midrule 
\multicolumn{5}{l}{\textbf{Panel D . First-stage estimates only for 91-99 (non-stack sample)}} \\
\addlinespace
GIV Net Exp Growth (91-99, An)&    1.152***&    1.152***&    1.089***&    1.085***\\
               &  (0.126)   &  (0.126)   &  (0.125)   &  (0.125)   \\
\addlinespace
KP F-Stat      &    83.18   &    83.77   &    75.92   &    74.77   \\
MOP F-Stat  &    84.87   &    85.48   &    79.82   &    78.55   \\

\addlinespace
\addlinespace

Dep Var (Panel E): &\multicolumn{4}{c}{Net Export Growth (99-05, An)} \\ 
\midrule 
\multicolumn{5}{l}{\textbf{Panel E . First-stage estimates only for 99-05 (non-stack sample)}} \\
\addlinespace
GIV Net Exp Growth (99-05, An)&    1.173***&    1.163***&    1.122***&    1.132***\\
               &  (0.232)   &  (0.232)   &  (0.271)   &  (0.271)   \\
\addlinespace
KP F-Stat      &    25.54   &    25.16   &    17.13   &    17.44   \\
MOP F-Stat  &    24.81   &    24.46   &    16.63   &    16.98   \\

\midrule
\multicolumn{5}{l}{\textbf{Controls (for all Panels)}} \\
DumPeriod &  Y   &   Y    & Y    &  Y    \\
Basic Controls x DumPeriod &            &  Y   &   Y    & Y       \\
Housing Controls x DumPeriod &           &      & Y       & Y       \\
Demographic Controls x DumPeriod &            &      &        &  Y       \\
\midrule              
Obs (Panel A, B, and C)         &     1410   &     1410   &     1254   &     1254   \\
Obs (Panel D and E)           &      705   &      705   &      627   &      627   \\
Cluster SE     &     CBSA   &     CBSA   &     CBSA   &     CBSA    \\
Weight         & {\scriptsize Ln(HU-Start)}   & {\scriptsize Ln(HU-Start)}   &{\scriptsize Ln(HU-Start)}   &{\scriptsize Ln(HU-Start)}      \\
\bottomrule
\end{tabular}

} % end of resize box

\end{table}

%\pagebreak 
%---------------------------------------------------------------

%%%%%%%%%%%%%%%%%%%%%%%%%%%%%%%%%%%%%%%%%%%%%%%%
% table_PLMNJ.D91t99vsD99t05.NEG.D91t99vsD99t05.4Reg
%%%%%%%%%%%%%%%%%%%%%%%%%%%%%%%%%%%%%%%%%%%%%%%%

%---------------------------------------------------------------

%%%%%%%%%%%%%%%%%%%%%%%%%%%%%%%%%%%%%%%%%%%%%%%%
% table_PLMNJ.D91t99vsD99t05.NEG.D91t99vsD99t05.4Reg
%%%%%%%%%%%%%%%%%%%%%%%%%%%%%%%%%%%%%%%%%%%%%%%%

\noindent 

\begin{table}[h!]
\centering
\caption{
\textbf{Four Stacked Regressions of PLMNJ Growth on Net Export Growth in Prior (91-99) and Boom (99-05) Period} \smallskip \newline
{\scriptsize
This table reports OLS, reduced-form, first stage and second stage of 2SLS stacked regression $\triangle_{91,99} \quad \&  \quad \triangle_{99,05}  Ln(PLMNJ_{c}) = \beta_{91,99} * \triangle_{91,99} \text{NetExp}_{m} \times Dum_{91,99} + \beta_{99,05} * \triangle_{99,05} \text{NetExp}_{m} \times Dum_{99,05} + \gamma_{91,99}* \bm{Controls_{c}} \times Dum_{91,99} + \gamma_{99,05}* \bm{Controls_{c}} \times Dum_{99,05}  + \alpha_{91,99} + \alpha_{99,05} + \epsilon_{period, c}$. The left-hand-side dependent variable $\triangle_{91,99} \& \triangle_{99,05} Ln(PLMNJ_{c})$ is the stacked growth rate of the dollar amount (07USD) of private-label mortgages (non-jumbo) (PLMNJ) at county $c$ 91-99 and 99-05. The key independent variable $\triangle_{91,99} \text{NetExp}_{m}$ and $\triangle_{99,05} \text{NetExp}_{m}$ are the growth rate of net export at the metropolitan area (CBSA03 code) $m$ 91-99 and 99-05, respectively. $Controls_{c}$ indicates control variables at county $c$ in the period start year either 1991 or 1999. We use the gravity model-based instrumental variable $\triangle_{91,99}\text{givNetExp}_{m}$ and $\triangle_{99,05}\text{givNetExp}_{m}$ as IVs for $\triangle_{91,99}\text{NetExp}_{m}$ and $\triangle_{99,05}\text{NetExp}_{m}$. Each regression is weighted by the natural logarithm of housing units in the start year (either 1991 or 1999). For the first-stage F-test of two non-stack samples, we report Kleibergen-Paap (2006) robust (clustered) statistics and Montiel Olea-Pflueger (2013) efficient statistics. We report Standard errors are clustered at the CBSA level. ***, **, and * indicate significance at the 1\%, 5\%, and 10\% levels, respectively.
} % end of small font size
} % end of caption
\label{table_PLMNJ.D91t99vsD99t05.NEG.D91t99vsD99t05.4Reg}
\resizebox{0.75\columnwidth}{!}{%
\begin{tabular}{l*{4}{c}}
\toprule
Dep Var (Panel A, B, and C)                      &\multicolumn{4}{c}{PLMNJ Growth (91-99 \& 99-05, annualized)} \\
            \cmidrule{2-5} 
            &\multicolumn{1}{c}{(1)}&\multicolumn{1}{c}{(2)}&\multicolumn{1}{c}{(3)}&\multicolumn{1}{c}{(4)}\\

\midrule
\multicolumn{5}{l}{\textbf{Panel A. OLS estimates}} \\
Net Export Growth (91-99, An) x Dum91t99&    4.119   &    4.540   &    1.462   &    0.679   \\
               &  (3.236)   &  (3.214)   &  (3.490)   &  (3.265)   \\
\addlinespace
Net Export Growth (99-05, An) x Dum99t05&    8.262***&    7.979***&    6.296***&    7.084***\\
               &  (1.782)   &  (1.766)   &  (1.961)   &  (1.964)   \\
\addlinespace
R2-adj         &    0.819   &    0.820   &    0.835   &    0.844   \\
\addlinespace

\midrule
\multicolumn{5}{l}{\textbf{Panel B. Reduced-form estimates}} \\
GIV Net Export Growth (91-99, An) x Dum91t99&   -1.982   &   -2.372   &   -7.752   &   -7.820*  \\
               &  (4.935)   &  (4.858)   &  (5.219)   &  (4.666)   \\
\addlinespace
GIV Net Export Growth (99-05, An) x Dum99t05&   16.958***&   16.297***&   14.198***&   14.389***\\
               &  (3.644)   &  (3.748)   &  (3.754)   &  (3.908)   \\
\addlinespace
R2-adj         &    0.819   &    0.820   &    0.836   &    0.845   \\
\addlinespace

\midrule
\multicolumn{5}{l}{\textbf{Panel C . 2SLS estimates}} \\
\addlinespace
Net Export Growth (91-99, An) x Dum91t99&   -1.730   &   -2.067   &   -7.400   &   -7.492   \\
               &  (4.373)   &  (4.312)   &  (5.492)   &  (4.837)   \\
\addlinespace
Net Export Growth (99-05, An) x Dum99t05&   14.753***&   14.311***&   13.116***&   13.361***\\
               &  (3.956)   &  (4.035)   &  (3.982)   &  (4.016)   \\
\addlinespace
CoefEqual\_Chi2 &    8.854   &    8.376   &    9.477   &    11.45   \\
CoefEqual\_PValue &    0.003   &    0.004   &    0.002   &    0.001   \\              
\addlinespace
\addlinespace

Dep Var (Panel D): &\multicolumn{4}{c}{Net Export Growth (91-99, An)} \\ 
\midrule 
\multicolumn{5}{l}{\textbf{Panel D . First-stage estimates only for 91-99 (non-stack sample)}} \\
\addlinespace
GIV Net Export Growth (91-99, An)&    1.146***&    1.148***&    1.048***&    1.044***\\
               &  (0.180)   &  (0.178)   &  (0.191)   &  (0.194)   \\
\addlinespace
KP F-Stat      &    40.61   &    41.60   &    30.02   &    29.03   \\
MOP F-Stat     &    40.79   &    41.91   &    30.76   &    29.74   \\

\addlinespace
\addlinespace

Dep Var (Panel E): &\multicolumn{4}{c}{Net Export Growth (99-05, An)} \\ 
\midrule 
\multicolumn{5}{l}{\textbf{Panel E . First-stage estimates only for 99-05 (non-stack sample)}} \\
\addlinespace
GIV Net Export Growth (99-05, An)&    1.149***&    1.139***&    1.082***&    1.077***\\
               &  (0.253)   &  (0.252)   &  (0.296)   &  (0.293)   \\
\addlinespace
KP F-Stat      &    20.66   &    20.42   &    13.41   &    13.52   \\
MOP F-Stat     &    19.92   &    19.70   &    12.94   &    13.05   \\

\midrule
\multicolumn{5}{l}{\textbf{Controls (for all Panels)}} \\
DumPeriod  &    Y        &  Y   &   Y    & Y        \\
Basic Controls x DumPeriod &            &  Y   &   Y    & Y        \\
Housing Controls x DumPeriod &           &      & Y       & Y        \\
Demographic Controls x DumPeriod &            &      &        &  Y     \\

\midrule              
Obs (Panel A, B, and C)   &     1402   &     1402   &     1246   &     1246   \\
Obs (Panel D and E)    &      701   &      701   &      623   &      623   \\
Cluster SE     &     CBSA   &     CBSA   &     CBSA   &     CBSA     \\
Weight         & {\scriptsize Ln(HU-Start)}   & {\scriptsize Ln(HU-Start)}   &{\scriptsize Ln(HU-Start)}   &{\scriptsize Ln(HU-Start)}   \\
\bottomrule
\end{tabular}

} % end of resize box

\end{table}

%---------------------------------------------------------------
%---------------------------------------------------------------
% Robustness: Evidence Against Speculation
%---------------------------------------------------------------
%---------------------------------------------------------------

\pagebreak 
%---------------------------------------------------------------

%%%%%%%%%%%%%%%%%%%%%%%%%%%%%%%%%%%%%%%%%%%%%%%%
% table_Speculation.PLMNJNonOwn.D99t05.PLMNJOwn.4Reg
%%%%%%%%%%%%%%%%%%%%%%%%%%%%%%%%%%%%%%%%%%%%%%%%

%---------------------------------------------------------------

%%%%%%%%%%%%%%%%%%%%%%%%%%%%%%%%%%%%
% table_Speculation.PLMNJNonOwn.D99t05.PLMNJOwn.4Reg
%%%%%%%%%%%%%%%%%%%%%%%%%%%%%%%%%%%%

\noindent 

\begin{table}[h!]
\centering
\caption{
\textbf{2SLS Regression of PLMNJ (Non-Owner-Occupied) Growth on PLMNJ (Owner-Occupied) Growth in Boom Period (99-05)} \smallskip \newline
{\footnotesize
This table reports OLS, reduced-form, first stage, and second stage results of 2SLS regression $\triangle_{99,05} Ln(PLMNJ\_NonOwn_{c}) = \beta * \triangle_{99,05} Ln(PLMNJ\_Own_{c}) + \gamma* \bm{Controls_{c}} + \alpha + \epsilon_{c}$. The left-hand-side dependent variable $\triangle_{99,05} Ln(PLMNJ\_NonOwn_{c})$ is the growth rate of the dollar amount of non-owner-occupied private-label mortgages (non-jumbo) at county $c$ 99-05 and the key independent variable $\triangle_{99,05} Ln(PLMNJ\_Own_{c})$ is the growth rate of the dollar amount of owner-occupied private-label mortgages (non-jumbo) at county $c$ 99-05. $Controls_{c}$ indicates control variables at county $c$ in 1999. We use the gravity model-based instrumental variable ($\triangle_{99,05}\text{givNetExp}_{m}$) as IV for $\triangle_{99,05} Ln(PLMNJ\_Own_{c})$. For the first-stage F-test, we report Kleibergen-Paap (2006) robust (clustered) statistics and Montiel Olea-Pflueger (2013) efficient statistics. Regression is weighted by the natural logarithm of housing units in 1999 as the weight for the housing market. Standard errors are clustered at the CBSA level. ***, **, and * indicate significance at the 1\%, 5\%, and 10\% levels, respectively.
} % end of small font size
} % end of caption
\label{table_Speculation.PLMNJNonOwn.D99t05.PLMNJOwn.4Reg}

\resizebox{\columnwidth}{!}{%

\begin{tabular}{l*{4}{c}}
\toprule
         &\multicolumn{4}{c}{PLMNJ (Non-Owner-Occupied) Growth (07USD, 99-05)} \\
            \cmidrule{2-5} 
            &\multicolumn{1}{c}{(1)}&\multicolumn{1}{c}{(2)}&\multicolumn{1}{c}{(3)}&\multicolumn{1}{c}{(4)}\\
            
\midrule
\multicolumn{5}{l}{\textbf{Panel A. OLS estimates}} \\
\addlinespace
PLMNJ (Owner-Occupied, Credit Expansion) Growth (99-05)&    1.134***&    1.125***&    1.098***&    1.121***\\
               &  (0.073)   &  (0.066)   &  (0.069)   &  (0.065)   \\
\addlinespace

R2-adj         &    0.377   &    0.407   &    0.404   &    0.434   \\
\addlinespace

\midrule
\multicolumn{5}{l}{\textbf{Panel B. Reduced-form estimates}} \\
\addlinespace
GIV Net Export Growth (99-05)&   17.499***&   15.974** &   13.093** &   13.283** \\
               &  (6.704)   &  (6.453)   &  (5.735)   &  (5.465)   \\
\addlinespace
R2-adj         &   0.0155   &   0.0501   &   0.0718   &   0.0903   \\
\addlinespace

\midrule
\multicolumn{5}{l}{\textbf{Panel C . 2SLS estimates}} \\
\addlinespace
PLMNJ (Owner-Occupied, Credit Expansion) Growth (99-05)&    1.177***&    1.106***&    1.115***&    1.148***\\
               &  (0.349)   &  (0.331)   &  (0.364)   &  (0.350)   \\
\addlinespace

\addlinespace
\addlinespace

Dep Var (Panel D): &\multicolumn{4}{c}{PLMNJ (Owner-Occupied) Growth (07USD, 99-05)} \\ 
\midrule 
\multicolumn{5}{l}{\textbf{Panel D . First-stage estimates}} \\
\addlinespace
GIV Net Export Growth (99-05)&   14.876***&   14.410***&   11.743***&   11.518***\\
               &  (3.096)   &  (3.139)   &  (3.178)   &  (3.501)   \\
               
\addlinespace
KP F-Stat      &    23.34   &    21.16   &    13.85   &    11.00   \\
MOP F-Stat      &    22.30   &    20.38   &    13.94   &    11.04   \\
\addlinespace
\midrule
\multicolumn{5}{l}{\textbf{Controls (for all Panels)}} \\
Basic Controls &            &  Y   &   Y    & Y     \\
Housing Controls &           &      & Y       & Y    \\
Demographic Controls &            &      &        &  Y   \\
\midrule              
Obs            &      774   &      774   &      683   &      683   \\
Cluster SE     &     CBSA   &     CBSA   &     CBSA   &     CBSA   \\
Weight         & Ln(HU99)   & Ln(HU99)   & Ln(HU99)   & Ln(HU99)   \\
\bottomrule

\end{tabular}

} % end of resize box

\end{table}

\pagebreak 
%-----------------------------------------------------------------------
%%%%%%%%%%%%%%%%%%%%%%%%%%%%%%%%%%%%
% table_Speculation.RefHou.D00t06vsD0710.OwnCredit_vs_Speculation.4Reg
%%%%%%%%%%%%%%%%%%%%%%%%%%%%%%%%%%%%

%-----------------------------------------------------------------

%%%%%%%%%%%%%%%%%%%%%%%%%%%%%%%%%%%%
% table_Speculation.RefHou.D00t06vsD0710.OwnCredit_vs_Speculation.4Reg
%%%%%%%%%%%%%%%%%%%%%%%%%%%%%%%%%%%%

\noindent 

\begin{table}[h!]
\centering
\caption{
\textbf{2SLS Regression of Refined House Employment Growth in Boom (00-06) and Bust (07-10) Periods on Credit Expansion and Speculation in Boom Period (99-05)} \smallskip \newline
{\footnotesize 
This table reports OLS, reduced-form, first stage, and second stages of stacked 2SLS regression $\triangle_{00,06} \& \triangle_{07,10} RefinedHouseEmpShr_{c} = \beta_{00,06} * \triangle_{99,05} Ln(PLMNJ\_Own_{c}) \times Dum_{00,06} + \beta_{07,10} * \triangle_{99,05} Ln(PLMNJ\_Own_{c}) \times Dum_{07,10} + \theta_{00,06}* \text{Credit-Independent Speculation (c, 99-05)} \times Dum_{00,06} + \theta_{07,10}* \text{Credit-Independent Speculation (99-05)} \times Dum_{07,10} +  \gamma_{00,06}* \bm{Controls_{c}} \times Dum_{00,06} + \gamma_{07,10}* \bm{Controls_{c}} \times Dum_{07,10} + \epsilon_{period, c}$. The left-hand-side dependent variable $\triangle_{00,06} \& \triangle_{07,10} RefinedHouseEmpShr_{c}$ is the change of the refined house employment share in working-age population at county $c$ 00-06 and 07-10. To reduce the impact of outliers, the dependent variable is winsorized at 2\% and 98\% levels in each period. As a proxy for pure credit expansion, $\triangle_{99,05} Ln(PLMNJ\_Own_{c})$ is the growth rate of the dollar amount (07USD) of private-label mortgages (non-jumbo) for owner-occupied homes at county $c$ 99-05. As a proxy for credit-independent speculation at county $c$ 99-05, \text{Credit-Independent Speculation (c, 99-05)}  is derived from the regression Eq (\ref{eq:PLMNJNonOwn_on_PLMNJOwn}), which is a part of growth rate of the non-owner-occupied private-label mortgage (non-jumbo) that cannot be explained by the growth rate of owner-occupied private-label mortgage (non-jumbo). $Controls_{c}$ indicates control variables at county $c$ in the period start year 1999. We use the gravity model-based instrumental variable $\triangle_{99,05}\text{givNetExp}_{m}$ as the IV for $\triangle_{99,05}Ln(PLMNJ\_Own_{c})$. Regression is weighted by the natural logarithm of housing units in 1999.  For the first-stage F-test of two non-stacked samples, we report Kleibergen-Paap (2006) robust (clustered) statistics and Montiel Olea-Pflueger (2013) efficient statistics. Standard errors are clustered at the CBSA level. ***, **, and * indicate significance at the 1\%, 5\%, and 10\% levels, respectively.
} % end of small font size
\smallskip 
} % end of caption
\label{table_Speculation.RefHou.D00t06vsD0710.OwnCredit_vs_Speculation.4Reg}

\resizebox{0.70\columnwidth}{!}{%

\begin{tabular}{l*{4}{c}}
\toprule
       &\multicolumn{4}{c}{House Price Growth (07USD, 99-05)} \\
       \cmidrule{2-5} 
        &\multicolumn{1}{c}{(1)}   &\multicolumn{1}{c}{(2)}   &\multicolumn{1}{c}{(3)}   &\multicolumn{1}{c}{(4)}    \\

\midrule
\multicolumn{5}{l}{\textbf{Panel A. OLS estimates}} \\
\addlinespace
PLMNJ Own Growth (Credit Expansion, 99-05) x Dum00t06&    0.136***&    0.130***&    0.107***&    0.100***\\
               &  (0.028)   &  (0.027)   &  (0.028)   &  (0.025)   \\
\addlinespace
PLMNJ Own Growth (Credit Expansion, 99-05) x Dum07t10&   -0.117***&   -0.103***&   -0.077***&   -0.078***\\
               &  (0.023)   &  (0.022)   &  (0.022)   &  (0.020)   \\
\addlinespace
PLMNJ Growth (Credit-Ind Speculation, 99-05) x Dum00t06&    0.053***&    0.058***&    0.049***&    0.045** \\
               &  (0.018)   &  (0.017)   &  (0.017)   &  (0.019)   \\
\addlinespace
PLMNJ Growth (Credit-Ind Speculation, 99-05) x Dum07t10&   -0.046***&   -0.044***&   -0.046***&   -0.037** \\
               &  (0.012)   &  (0.012)   &  (0.013)   &  (0.015)   \\
\addlinespace
R2-adj         &    0.501   &    0.537   &    0.552   &    0.567   \\
\addlinespace

\midrule
\multicolumn{5}{l}{\textbf{Panel B. Reduced-form estimates}} \\
\addlinespace
GIV Net Export Growth (99-05) x Dum00t06&    5.963***&    5.172***&    4.071** &    4.624***\\
               &  (1.687)   &  (1.664)   &  (1.730)   &  (1.679)   \\
\addlinespace
GIV Net Export Growth (99-05) x Dum07t10&   -5.543***&   -4.223***&   -3.733***&   -4.317***\\
               &  (1.170)   &  (1.179)   &  (1.223)   &  (1.261)   \\
\addlinespace
PLMNJ Growth (Credit-Ind Speculation, 99-05) x Dum00t06&    0.051***&    0.056***&    0.046***&    0.044** \\
               &  (0.019)   &  (0.018)   &  (0.017)   &  (0.019)   \\
\addlinespace
PLMNJ Growth (Credit-Ind Speculation, 99-05) x Dum07t10&   -0.045***&   -0.043***&   -0.044***&   -0.036** \\
               &  (0.012)   &  (0.012)   &  (0.014)   &  (0.015)   \\
\addlinespace
R2-adj         &    0.486   &    0.523   &    0.545   &    0.562   \\
\addlinespace

\midrule
\multicolumn{5}{l}{\textbf{Panel C . 2SLS estimates}} \\
\addlinespace
PLMNJ Own Growth (Credit Expansion, 99-05) x Dum00t06&    0.402***&    0.358***&    0.345** &    0.397** \\
               &  (0.134)   &  (0.137)   &  (0.176)   &  (0.184)   \\
\addlinespace
PLMNJ Own Growth (Credit Expansion, 99-05) x Dum07t10&   -0.373***&   -0.293***&   -0.319** &   -0.373** \\
               &  (0.108)   &  (0.107)   &  (0.144)   &  (0.170)   \\
\addlinespace
PLMNJ Growth (Credit-Ind Speculation, 99-05) x Dum00t06&    0.055***&    0.060***&    0.054***&    0.050** \\
               &  (0.018)   &  (0.017)   &  (0.018)   &  (0.020)   \\
\addlinespace
PLMNJ Growth (Credit-Ind Speculation, 99-05) x Dum07t10&   -0.048***&   -0.046***&   -0.052***&   -0.042** \\
               &  (0.015)   &  (0.015)   &  (0.017)   &  (0.018)   \\
\addlinespace 
\addlinespace

Dep Var (Panel D): &\multicolumn{4}{c}{PLMNJ Own Growth (Credit Expansion, 99-05)} \\ 
\midrule 
\multicolumn{5}{l}{\textbf{Panel D . First-stage estimates}} \\
\addlinespace
GIV Net Export Growth (99-05) x Dum00t06&   14.871***&   14.434***&   11.730***&   11.573***\\
               &  (3.083)   &  (3.153)   &  (3.163)   &  (3.487)   \\
\addlinespace
KP F-Stat      &    23.21   &    21.08   &    13.92   &    11.17   \\
MOP F-Stat      &    22.13   &    20.23   &    13.96   &    11.19   \\
\addlinespace
\midrule
\multicolumn{5}{l}{\textbf{Controls (for all Panels)}} \\
Basic Controls &            &  Y   &   Y    & Y     \\
Housing Controls &           &      & Y       & Y     \\
Demographic Controls &            &      &        &  Y    \\
\midrule
Obs (Panel A, B, \& C)  &     1546   &     1546   &     1364   &     1364   \\
Obs (Panel D)           &      773   &      773   &      682   &      682   \\
Cluster SE     &     CBSA   &     CBSA   &     CBSA   &     CBSA   \\
Weight         & Ln(HU99)   & Ln(HU99)   & Ln(HU99)   & Ln(HU99)   \\
\bottomrule

\end{tabular}

} % end of resize box

\end{table}

\pagebreak 
%---------------------------------------------------------------

%%%%%%%%%%%%%%%%%%%%%%%%%%%%%%%%%%%%%%%%%%%%%%%%
% table_Speculation.PLMNJNonOwn.D91t99vsD99t05.4Reg
%%%%%%%%%%%%%%%%%%%%%%%%%%%%%%%%%%%%%%%%%%%%%%%%
%----------------------------------------------------------------

%%%%%%%%%%%%%%%%%%%%%%%%%%%%%%%%%%%%
% table_Speculation.PLMNJNonOwn.D91t99vsD99t05.4Reg
%%%%%%%%%%%%%%%%%%%%%%%%%%%%%%%%%%%%

\noindent 

\begin{table}[h!]
\centering
\caption{
\textbf{2SLS Stacked Regression of Non-Owner-Occupied PLMNJ on Net Export Growth in Prior (91-99) and Boom (99-05) Periods} \smallskip \newline
{\scriptsize 
This table reports OLS, reduced-form, first stage, and second stage results of 2SLS regression $\triangle_{91,99} \& \triangle_{99,05} Ln(PLMNJ\_NonOwn_{c}) = \beta_{91,99} * \triangle_{91,99} \text{NetExp}_{m} \times Dum_{91,99} + \beta_{99,05} * \triangle_{99,05} \text{NetExp}_{m} \times Dum_{99,05} + \gamma_{91,99}* \bm{Controls_{c}} \times Dum_{91,99} + \gamma_{99,05}* \bm{Controls_{c}} \times Dum_{99,05}  + \epsilon_{period, c}$. The left-hand-side dependent variable $\triangle_{91,99} \& \triangle_{99,05} Ln(PLMNJ\_NonOwn_{c})$ is the stacked growth rate of the dollar amount (07USD) of non-owner-occupied private-label mortgages (non-jumbo) (PLMNJ) at county $c$ 91-99 and 99-05, respectively. The key independent variable $\triangle_{91,99} \text{NetExp}_{m}$ and $\triangle_{99,05} \text{NetExp}_{m}$ are the net export growth at the metropolitan area (CBSA03 code) $m$ 91-99 and 99-05, respectively. $Controls_{c}$ indicates control variables at county $c$ in the period start year, either 1991 or 1999. We use the gravity model-based instrumental variable $\triangle_{91,99}\text{givNetExp}_{m}$ and $\triangle_{99,05}\text{givNetExp}_{m}$ as IVs for $\triangle_{91,99}\text{NetExp}_{m}$ and $\triangle_{99,05}\text{NetExp}_{m}$. We report the statistics and p-values for the tests of coefficient equality between $\beta_{91,99}$ and $\beta_{99,05}$. For the first-stage F-test of two separate non-stack samples, we report kleibergen-Paap (2006) robust (clustered) statistics and Montiel Olea-Pflueger (2013) efficient statistics. Regression is weighted by the natural logarithm of housing units in the start year (either 1991 or 1999). Standard errors are clustered at the CBSA level. ***, **, and * indicate significance at the 1\%, 5\%, and 10\% levels, respectively.
} % end of small font size
} % end of caption
\label{table_Speculation.PLMNJNonOwn.D91t99vsD99t05.4Reg}

\resizebox{0.86\columnwidth}{!}{%

\begin{tabular}{l*{4}{c}}
\toprule
\textbf{TSLS estimates}            &\multicolumn{4}{c}{Private-label Mortgage (Non-Owner) Growth (91-99 or 99-05, An)} \\
            \cmidrule{2-5} 
            &\multicolumn{1}{c}{(1)}&\multicolumn{1}{c}{(2)}&\multicolumn{1}{c}{(3)}&\multicolumn{1}{c}{(4)}\\
            
\midrule
\multicolumn{5}{l}{\textbf{Panel A. OLS estimates}} \\
Net Export Growth (91-99, An) x Dum91t99&   -8.436   &   -3.039   &   -6.007   &   -7.890   \\
               &  (6.412)   &  (6.272)   &  (6.873)   &  (6.749)   \\
\addlinespace
Net Export Growth (99-05, An) x Dum99t05&    9.939***&   10.242***&    8.981** &    9.478***\\
               &  (3.764)   &  (3.509)   &  (3.722)   &  (3.642)   \\
\addlinespace
R2-adj         &    0.757   &    0.772   &    0.777   &    0.780   \\
\addlinespace

\midrule
\multicolumn{5}{l}{\textbf{Panel B. Reduced-form estimates}} \\
GIV Net Export Growth (91-99, An) x Dum91t99&   -3.337   &   -2.596   &   -3.723   &   -5.770   \\
               &  (9.855)   & (10.204)   & (10.714)   & (10.836)   \\
\addlinespace
GIV Net Export Growth (99-05, An) x Dum99t05&   18.950** &   18.629***&   18.129***&   18.251***\\
               &  (7.458)   &  (6.941)   &  (6.946)   &  (7.008)   \\
\addlinespace
R2-adj         &    0.757   &    0.771   &    0.777   &    0.779   \\
\addlinespace

\midrule
\multicolumn{5}{l}{\textbf{Panel C . 2SLS estimates}} \\
\addlinespace
Net Export Growth (91-99, An) x Dum91t99&   -2.894   &   -2.252   &   -3.419   &   -5.315   \\
               &  (8.522)   &  (8.812)   &  (9.707)   &  (9.797)   \\
\addlinespace
Net Export Growth (99-05, An) x Dum99t05&   16.120** &   15.982** &   16.129** &   16.100***\\
               &  (6.980)   &  (6.227)   &  (6.430)   &  (6.085)   \\
\addlinespace
\addlinespace

Dep Var (Panel D): &\multicolumn{4}{c}{Net Export Growth (91-99, An)} \\ 
\midrule 
\multicolumn{5}{l}{\textbf{Panel D . First-stage estimates only for 91-99 (non-stack sample)}} \\
\addlinespace
GIV Net Export Growth (91-99, An)&    1.153***&    1.153***&    1.089***&    1.086***\\
               &  (0.126)   &  (0.126)   &  (0.125)   &  (0.126)   \\
\addlinespace
KP F-Stat      &    83.14   &    83.86   &    75.38   &    74.02   \\
MOP F-Stat      &    84.81   &    85.57   &    79.14   &    77.63   \\

\addlinespace
\addlinespace

Dep Var (Panel E): &\multicolumn{4}{c}{Net Export Growth (99-05, An) } \\ 
\midrule 
\multicolumn{5}{l}{\textbf{Panel E . First-stage estimates only for 99-05 (non-stack sample)}} \\
\addlinespace
GIV Net Export Growth (99-05, An)&    1.176***&    1.166***&    1.124***&    1.134***\\
               &  (0.232)   &  (0.232)   &  (0.271)   &  (0.271)   \\
\addlinespace
KP F-Stat      &    25.67   &    25.27   &    17.22   &    17.51   \\
MOP F-Stat      &    24.94   &    24.56   &    16.72   &    17.06   \\

\midrule
\multicolumn{5}{l}{\textbf{Controls (DumPeriod for stacked sample and no DumPeriod for non-stacked sample)}} \\
DumPeriod  &    Y        &  Y   &   Y    & Y     \\
Basic Controls x DumPeriod &            &  Y   &   Y    & Y    \\
Housing Controls x DumPeriod &           &      & Y       & Y    \\
Demographic Controls x DumPeriod &            &      &        &  Y    \\

\midrule              
Obs (Panel A, B, and C)         &     1410   &     1410   &     1254   &     1254   \\
Obs (Panel D and E)           &      705   &      705   &      627   &      627   \\
Cluster SE     &     CBSA   &     CBSA   &     CBSA   &     CBSA   \\
Weight         & {\scriptsize Ln(HU-Start)}   & {\scriptsize Ln(HU-Start)}   &{\scriptsize Ln(HU-Start)}   &{\scriptsize Ln(HU-Start)}   \\
\bottomrule

\end{tabular}

} % end of resize box

\end{table}

%---------------------------------------------------------------
%---------------------------------------------------------------
% Robustness 3. Real Business Cycle Hypothesis
% A. Tradable Employment Growth in Boom (00-06) and Bust (07-10)
% B. Commericial Construction Employment
%---------------------------------------------------------------
%---------------------------------------------------------------

%\pagebreak 
%---------------------------------------------------------------

%%%%%%%%%%%%%%%%%%%%%%%%%%%%%%%%%%%%%%%%%%%%%%%%
% table_Tradable.D00t06vsD07t10.PLMNJ.4Reg
%%%%%%%%%%%%%%%%%%%%%%%%%%%%%%%%%%%%%%%%%%%%%%%%

%---------------------------------------------------------------

%%%%%%%%%%%%%%%%%%%%%%%%%%%%%%%%%%%%%%%%%%%%%%%%
% table_Tradable.D00t06vsD07t10.PLMNJ.4Reg
%%%%%%%%%%%%%%%%%%%%%%%%%%%%%%%%%%%%%%%%%%%%%%%%

\noindent 

\begin{table}[h!]
\centering
\caption{
\textbf{Four Stacked Regressions of Tradable Employment Growth in Boom (00-06) and Bust (07-10) Periods on PLMNJ Growth (99-05)} \smallskip \newline
{\scriptsize
This table reports OLS, reduced-form, first stage, and second stages of stacked 2SLS regression $\triangle_{00,06} \& \triangle_{07,10} TradableEmpShr_{c} = \beta_{00,06} * \triangle_{99,05} Ln(PLMNJ_{c}) \times Dum_{00,06} + \beta_{07,10} * \triangle_{99,05} Ln(PLMNJ_{c}) \times Dum_{07,10} + \gamma_{00,06}* \bm{Controls_{c}} \times Dum_{00,06} + \gamma_{07,10}* \bm{Controls_{c}} \times Dum_{07,10} + \epsilon_{period, c}$. The left-hand-side dependent variable $\triangle_{00,06} \& \triangle_{07,10} TradableEmpShr_{c}$ is the change of the Tradable employment share in working-age population at county $c$ 00-06 and 07-10. The key independent variable $\triangle_{99,05} Ln(PLMNJ_{c})$ is the growth rate of the dollar amount (07USD) of private-label mortgages (non-jumbo) at county $c$ 99-05. $Controls_{c}$ indicates control variables at county $c$ in the period start year 1999. We use the gravity model-based instrumental variable $\triangle_{99,05}\text{givNetExp}_{m}$ as the IV for $\triangle_{99,05}Ln(PLMNJ_{c})$. Regression is weighted by the natural logarithm of housing units in 1999.  For the first-stage F-test of two non-stacked samples, we report Kleibergen-Paap (2006) robust (clustered) statistics and Montiel Olea-Pflueger (2013) efficient statistics. Standard errors are clustered at the CBSA level. ***, **, and * indicate significance at the 1\%, 5\%, and 10\% levels, respectively.
} % end of small font size
} % end of caption
\label{table_Tradable.D00t06vsD07t10.PLMNJ.4Reg}
\resizebox{0.95\columnwidth}{!}{%
\begin{tabular}{l*{4}{c}}
\toprule
Dep Var (Panel A, B, and C)                      &\multicolumn{4}{c}{Tradable Employment Growth (00-06 \& 07-10, An)} \\
            \cmidrule{2-5} 
            &\multicolumn{1}{c}{(1)}&\multicolumn{1}{c}{(2)}&\multicolumn{1}{c}{(3)}&\multicolumn{1}{c}{(4)}\\

\midrule
\multicolumn{5}{l}{\textbf{Panel A. OLS estimates}} \\
PLMNJ Growth (07USD, 99-05, An) x Dum00t06&    0.004** &    0.004** &    0.002   &    0.003   \\
               &  (0.002)   &  (0.002)   &  (0.002)   &  (0.002)   \\
\addlinespace
PLMNJ Growth (07USD, 99-05, An) x Dum07t10&    0.007***&    0.007***&    0.005** &    0.006***\\
               &  (0.002)   &  (0.002)   &  (0.002)   &  (0.002)   \\
\addlinespace
R2-adj         &     1584   &     1584   &     1402   &     1402   \\
\addlinespace

\midrule
\multicolumn{5}{l}{\textbf{Panel B. Reduced-form estimates}} \\
GIV Net Export Growth (99-05, An) x Dum00t06&    0.955***&    0.959***&    0.814***&    0.871***\\
               &  (0.182)   &  (0.178)   &  (0.174)   &  (0.163)   \\
\addlinespace
GIV Net Export Growth (99-05, An) x Dum07t10&    0.905***&    0.937***&    0.859***&    0.941***\\
               &  (0.176)   &  (0.179)   &  (0.188)   &  (0.190)   \\
\addlinespace
R2-adj         &     1584   &     1584   &     1402   &     1402   \\
\addlinespace

\midrule
\multicolumn{5}{l}{\textbf{Panel C . 2SLS estimates}} \\
\addlinespace
PLMNJ Growth (07USD, 99-05, An) x Dum00t06&    0.061***&    0.063***&    0.066***&    0.072***\\
               &  (0.016)   &  (0.016)   &  (0.020)   &  (0.023)   \\
\addlinespace
PLMNJ Growth (07USD, 99-05, An) x Dum07t10&    0.057***&    0.061***&    0.070***&    0.078***\\
               &  (0.014)   &  (0.015)   &  (0.023)   &  (0.028)   \\
               
\addlinespace
\addlinespace

Dep Var (Panel D): &\multicolumn{4}{c}{PLMNJ Growth (99-05, An)} \\ 
\midrule 
\multicolumn{5}{l}{\textbf{Panel D . First-stage estimates only for 99-05 (Non-stack sample)}} \\
\addlinespace
GIV NEG (99-05, An)&   15.753***&   15.278***&   12.243***&   12.134***\\
               &  (3.255)   &  (3.331)   &  (3.276)   &  (3.541)   \\
\addlinespace
KP F-Stat      &    23.43   &    21.04   &    13.97   &    11.74   \\
MOP F-Stat     &    22.30   &    20.29   &    14.03   &    11.80   \\
\addlinespace

\midrule
\multicolumn{5}{l}{\textbf{Controls (for all Panels)}} \\
DumPeriod  &    Y        &  Y   &   Y    & Y        \\
Basic Controls x DumPeriod &            &  Y   &   Y    & Y    \\
Housing Controls x DumPeriod &           &      & Y       & Y   \\
Demographic Controls x DumPeriod &            &      &      &  Y \\

\midrule              
Obs (Panel A, B, and C)          &     1584   &     1584   &     1402   &     1402   \\
Obs (Panel D)          &      792   &      792   &      701   &      701   \\
Cluster SE     &     CBSA   &     CBSA   &     CBSA   &     CBSA   \\
Weight         & Ln(HU99)   & Ln(HU99)   & Ln(HU99)   & Ln(HU99)   \\
\bottomrule
\end{tabular}

} % end of resize box

\end{table}

\pagebreak 
%---------------------------------------------------------------

%%%%%%%%%%%%%%%%%%%%%%%%%%%%%%%%%%%%%%%%%%%%%%%%
% table_ComConst.D00t06vsD07t10.PLMNJ.4Reg
%%%%%%%%%%%%%%%%%%%%%%%%%%%%%%%%%%%%%%%%%%%%%%%%

%---------------------------------------------------------------

%%%%%%%%%%%%%%%%%%%%%%%%%%%%%%%%%%%%%%%%%%%%%%%%
% table_ComConst.D00t06vsD07t10.PLMNJ.4Reg
%%%%%%%%%%%%%%%%%%%%%%%%%%%%%%%%%%%%%%%%%%%%%%%%

\noindent 

\begin{table}[h!]
\centering
\caption{
\textbf{Four Stacked Regressions of Commercial Construction Employment Growth in Boom (00-06) and Bust (07-10) Periods on PLMNJ Growth (99-05)} \smallskip \newline
{\scriptsize
This table reports OLS, reduced-form, first stage, and second stages of stacked 2SLS regression $\triangle_{00,06} \& \triangle_{07,10} ComConstEmpShr_{c} = \beta_{00,06} * \triangle_{99,05} Ln(PLMNJ_{c}) \times Dum_{00,06} + \beta_{07,10} * \triangle_{99,05} Ln(PLMNJ_{c}) \times Dum_{07,10} + \gamma_{00,06}* \bm{Controls_{c}} \times Dum_{00,06} + \gamma_{07,10}* \bm{Controls_{c}} \times Dum_{07,10} + \epsilon_{period, c}$. The left-hand-side dependent variable $\triangle_{00,06} \& \triangle_{07,10} ComConstEmpShr_{c}$ is the change of the commercial construction employment share in working-age population at county $c$ 00-06 and 07-10. The key independent variable $\triangle_{99,05} Ln(PLMNJ_{c})$ is the growth rate of the dollar amount (07USD) of private-label mortgages (non-jumbo) at county $c$ 99-05. $Controls_{c}$ indicates control variables at county $c$ in the period start year 1999. We use the gravity model-based instrumental variable $\triangle_{99,05}\text{givNetExp}_{m}$ as the IV for $\triangle_{99,05}Ln(PLMNJ_{c})$. Regression is weighted by the natural logarithm of housing units in 1999.  For the first-stage F-test of two non-stacked samples, we report Kleibergen-Paap (2006) robust (clustered) statistics and Montiel Olea-Pflueger (2013) efficient statistics. Standard errors are clustered at the CBSA level. ***, **, and * indicate significance at the 1\%, 5\%, and 10\% levels, respectively.
} % end of small font size
} % end of caption
\label{table_ComConst.D00t06vsD07t10.PLMNJ.4Reg}
\resizebox{\columnwidth}{!}{%
\begin{tabular}{l*{4}{c}}
\toprule
Dep Var (Panel A, B, and C)                      &\multicolumn{4}{c}{Commercial Construction Emp Growth (00-06 \& 07-10, An)} \\
            \cmidrule{2-5} 
            &\multicolumn{1}{c}{(1)}&\multicolumn{1}{c}{(2)}&\multicolumn{1}{c}{(3)}&\multicolumn{1}{c}{(4)}\\

\midrule
\multicolumn{5}{l}{\textbf{Panel A. OLS estimates}} \\
PLMNJ Growth (99-05, An) x Dum00t06&    0.004   &    0.001   &   -0.007   &   -0.006   \\
               &  (0.017)   &  (0.018)   &  (0.021)   &  (0.021)   \\
\addlinespace
PLMNJ Growth (99-05, An) x Dum07t10&   -0.014   &   -0.015   &    0.001   &    0.002   \\
               &  (0.031)   &  (0.030)   &  (0.031)   &  (0.032)   \\
\addlinespace
R2-adj         &   0.0255   &   0.0372   &   0.0686   &   0.0665   \\
\addlinespace

\midrule
\multicolumn{5}{l}{\textbf{Panel B. Reduced-form estimates}} \\
GIV Net Exp Growth (99-05, An) x Dum00t06&   -1.221   &   -1.534   &   -2.846   &   -2.696   \\
               &  (2.612)   &  (2.698)   &  (2.713)   &  (2.637)   \\
\addlinespace
GIV Net Exp Growth (99-05, An) x Dum07t10&    0.112   &    0.741   &    0.206   &   -0.093   \\
               &  (1.870)   &  (2.001)   &  (2.267)   &  (2.271)   \\
\addlinespace
R2-adj         &   0.0255   &   0.0374   &   0.0697   &   0.0675   \\
\addlinespace

\midrule
\multicolumn{5}{l}{\textbf{Panel C . 2SLS estimates}} \\
\addlinespace
PLMNJ Growth (99-05, An) x Dum00t06&   -0.079   &   -0.101   &   -0.235   &   -0.223   \\
               &  (0.166)   &  (0.173)   &  (0.217)   &  (0.209)   \\
\addlinespace
PLMNJ Growth (99-05, An) x Dum07t10&    0.007   &    0.049   &    0.017   &   -0.008   \\
               &  (0.120)   &  (0.129)   &  (0.185)   &  (0.188)   \\
               
\addlinespace
\addlinespace

Dep Var (Panel D): &\multicolumn{4}{c}{PLMNJ Growth (99-05, An)} \\ 
\midrule 
\multicolumn{5}{l}{\textbf{Panel D . First-stage estimates only for 99-05 (Non-stack sample)}} \\
\addlinespace
GIV NEG (99-05, An)&   15.753***&   15.278***&   12.243***&   12.134***\\
               &  (3.255)   &  (3.331)   &  (3.276)   &  (3.541)   \\
\addlinespace
KP F-Stat      &    23.12   &    21.09   &    13.79   &    11.63   \\
MOP F-Stat     &    22.05   &    20.35   &    13.82   &    11.66   \\
\addlinespace

\midrule
\multicolumn{5}{l}{\textbf{Controls (for all Panels)}} \\
DumPeriod  &    Y        &  Y   &   Y    & Y        \\
Basic Controls x DumPeriod &            &  Y   &   Y    & Y    \\
Housing Controls x DumPeriod &           &      & Y       & Y   \\
Demographic Controls x DumPeriod &            &      &      &  Y \\

\midrule              
Obs (Panel A, B, and C)          &     1548   &     1548   &     1367   &     1367   \\
Obs (Panel D)          &      783   &      783   &      693   &      693   \\
Cluster SE     &     CBSA   &     CBSA   &     CBSA   &     CBSA   \\
Weight         & Ln(HU99)   & Ln(HU99)   & Ln(HU99)   & Ln(HU99)   \\
\bottomrule
\end{tabular}

} % end of resize box

\end{table}

%---------------------------------------------------------------
%---------------------------------------------------------------
% Robustness 4. Natural Disaster Hypothesis
% BEA Farm and Manufacture Employment Growth in Boom (00-06) and Bust (07-10)
%---------------------------------------------------------------
%---------------------------------------------------------------

%\pagebreak 
%---------------------------------------------------------------

%%%%%%%%%%%%%%%%%%%%%%%%%%%%%%%%%%%%%%%%%%%%%%%%
% table_BEA.Farm.D00t06vsD07t10.PLMNJ.4Reg
%%%%%%%%%%%%%%%%%%%%%%%%%%%%%%%%%%%%%%%%%%%%%%%%

%---------------------------------------------------------------

%%%%%%%%%%%%%%%%%%%%%%%%%%%%%%%%%%%%%%%%%%%%%%%%
% table_BEA.Farm.D00t06vsD07t10.PLMNJ.4Reg
%%%%%%%%%%%%%%%%%%%%%%%%%%%%%%%%%%%%%%%%%%%%%%%%

\noindent 

\begin{table}[h!]
\centering
\caption{
\textbf{Four Stacked Regressions of Farm Employment Growth in Boom (00-06) and Bust (07-10) Periods on PLMNJ Growth (99-05)} \smallskip \newline
{\scriptsize
This table reports OLS, reduced-form, first stage, and second stages of stacked 2SLS regression $\triangle_{00,06} \& \triangle_{07,10} FarmEmpShr_{c} = \beta_{00,06} * \triangle_{99,05} Ln(PLMNJ_{c}) \times Dum_{00,06} + \beta_{07,10} * \triangle_{99,05} Ln(PLMNJ_{c}) \times Dum_{07,10} + \gamma_{00,06}* \bm{Controls_{c}} \times Dum_{00,06} + \gamma_{07,10}* \bm{Controls_{c}} \times Dum_{07,10} + \epsilon_{period, c}$. The left-hand-side dependent variable $\triangle_{00,06} \& \triangle_{07,10} FarmEmpShr_{c}$ is the change of the Farm employment share in working-age population at county $c$ 00-06 and 07-10. The key independent variable $\triangle_{99,05} Ln(PLMNJ_{c})$ is the growth rate of the dollar amount (07USD) of private-label mortgages (non-jumbo) at county $c$ 99-05. $Controls_{c}$ indicates control variables at county $c$ in the period start year 1999. We use the gravity model-based instrumental variable $\triangle_{99,05}\text{givNetExp}_{m}$ as the IV for $\triangle_{99,05}Ln(PLMNJ_{c})$. Regression is weighted by the natural logarithm of housing units in 1999.  For the first-stage F-test of two non-stacked samples, we report Kleibergen-Paap (2006) robust (clustered) statistics and Montiel Olea-Pflueger (2013) efficient statistics. Standard errors are clustered at the CBSA level. ***, **, and * indicate significance at the 1\%, 5\%, and 10\% levels, respectively.
} % end of small font size
} % end of caption
\label{table_BEA.Farm.D00t06vsD07t10.PLMNJ.4Reg}
\resizebox{0.95\columnwidth}{!}{%
\begin{tabular}{l*{4}{c}}
\toprule
Dep Var (Panel A, B, and C)                      &\multicolumn{4}{c}{Farm Employment Growth (00-06 \& 07-10, An)} \\
            \cmidrule{2-5} 
            &\multicolumn{1}{c}{(1)}&\multicolumn{1}{c}{(2)}&\multicolumn{1}{c}{(3)}&\multicolumn{1}{c}{(4)}\\

\midrule
\multicolumn{5}{l}{\textbf{Panel A. OLS estimates}} \\
PLMNJ Growth (07USD, 99-05, An) x Dum00t06&   -0.120   &   -0.196***&   -0.150** &   -0.125** \\
               &  (0.077)   &  (0.072)   &  (0.069)   &  (0.061)   \\
\addlinespace
PLMNJ Growth (07USD, 99-05, An) x Dum07t10&    0.006   &   -0.008   &   -0.008   &   -0.005   \\
               &  (0.027)   &  (0.025)   &  (0.029)   &  (0.029)   \\
\addlinespace
R2-adj         &    0.317   &    0.438   &    0.458   &    0.506   \\
\addlinespace

\midrule
\multicolumn{5}{l}{\textbf{Panel B. Reduced-form estimates}} \\
GIV Net Export Growth (99-06, An) x Dum00t06&   -1.797   &   -4.876   &   -3.997   &   -5.478   \\
               &  (3.778)   &  (3.861)   &  (3.855)   &  (3.598)   \\
\addlinespace
GIV Net Export Growth (99-06, An) x Dum07t10&    2.058   &    1.422   &    1.506   &    1.618   \\
               &  (1.989)   &  (2.005)   &  (2.077)   &  (2.153)   \\
\addlinespace
R2-adj         &    0.312   &    0.425   &    0.452   &    0.502   \\
\addlinespace

\midrule
\multicolumn{5}{l}{\textbf{Panel C . 2SLS estimates}} \\
\addlinespace
PLMNJ Growth (07USD, 99-05, An) x Dum00t06&   -0.114   &   -0.323   &   -0.313   &   -0.436   \\
               &  (0.243)   &  (0.267)   &  (0.320)   &  (0.301)   \\
\addlinespace
PLMNJ Growth (07USD, 99-05, An) x Dum07t10&    0.131   &    0.094   &    0.118   &    0.129   \\
               &  (0.122)   &  (0.129)   &  (0.165)   &  (0.177)   \\
               
\addlinespace
\addlinespace

Dep Var (Panel D): &\multicolumn{4}{c}{PLMNJ Growth (99-05, An)} \\ 
\midrule 
\multicolumn{5}{l}{\textbf{Panel D . First-stage estimates only for 99-05 (Non-stack sample)}} \\
\addlinespace
GIV NEG (99-05, An)&   15.738***&   15.106***&   12.754***&   12.555***\\
               &  (3.437)   &  (3.426)   &  (3.555)   &  (3.878)   \\
\addlinespace
KP F-Stat      &    20.97   &    19.44   &    12.87   &    10.48   \\
MOP F-Stat     &    20.00   &    18.89   &    12.85   &    10.47   \\
\addlinespace

\midrule
\multicolumn{5}{l}{\textbf{Controls (for all Panels)}} \\
DumPeriod  &    Y        &  Y   &   Y    & Y        \\
Basic Controls x DumPeriod &            &  Y   &   Y    & Y    \\
Housing Controls x DumPeriod &           &      & Y       & Y   \\
Demographic Controls x DumPeriod &            &      &      &  Y \\

\midrule              
Obs (Panel A, B, and C)          &     1530   &     1530   &     1368   &     1368   \\
Obs (Panel D)          &      765   &      765   &      684   &      684   \\
Cluster SE     &     CBSA   &     CBSA   &     CBSA   &     CBSA   \\
Weight         & Ln(HU99)   & Ln(HU99)   & Ln(HU99)   & Ln(HU99)   \\
\bottomrule
\end{tabular}

} % end of resize box

\end{table}

\pagebreak 
%---------------------------------------------------------------

%%%%%%%%%%%%%%%%%%%%%%%%%%%%%%%%%%%%%%%%%%%%%%%%
% table_BEA.Manufacture.D00t06vsD07t10.PLMNJ.4Reg
%%%%%%%%%%%%%%%%%%%%%%%%%%%%%%%%%%%%%%%%%%%%%%%%

%---------------------------------------------------------------

%%%%%%%%%%%%%%%%%%%%%%%%%%%%%%%%%%%%%%%%%%%%%%%%
% table_BEA.Manufacture.D00t06vsD07t10.PLMNJ.4Reg
%%%%%%%%%%%%%%%%%%%%%%%%%%%%%%%%%%%%%%%%%%%%%%%%

\noindent 

\begin{table}[h!]
\centering
\caption{
\textbf{Four Stacked Regressions of Manufacture Employment Growth in Boom (00-06) and Bust (07-10) Periods on PLMNJ Growth (99-05)} \smallskip \newline
{\scriptsize
This table reports OLS, reduced-form, first stage, and second stages of stacked 2SLS regression $\triangle_{00,06} \& \triangle_{07,10} ManufactureEmpShr_{c} = \beta_{00,06} * \triangle_{99,05} Ln(PLMNJ_{c}) \times Dum_{00,06} + \beta_{07,10} * \triangle_{99,05} Ln(PLMNJ_{c}) \times Dum_{07,10} + \gamma_{00,06}* \bm{Controls_{c}} \times Dum_{00,06} + \gamma_{07,10}* \bm{Controls_{c}} \times Dum_{07,10} + \epsilon_{period, c}$. The left-hand-side dependent variable $\triangle_{00,06} \& \triangle_{07,10} ManufactureEmpShr_{c}$ is the change of the manufacture employment share in working-age population at county $c$ 00-06 and 07-10. The key independent variable $\triangle_{99,05} Ln(PLMNJ_{c})$ is the growth rate of the dollar amount (07USD) of private-label mortgages (non-jumbo) at county $c$ 99-05. $Controls_{c}$ indicates control variables at county $c$ in the period start year 1999. We use the gravity model-based instrumental variable $\triangle_{99,05}\text{givNetExp}_{m}$ as the IV for $\triangle_{99,05}Ln(PLMNJ_{c})$. Regression is weighted by the natural logarithm of housing units in 1999.  For the first-stage F-test of two non-stacked samples, we report Kleibergen-Paap (2006) robust (clustered) statistics and Montiel Olea-Pflueger (2013) efficient statistics. Standard errors are clustered at the CBSA level. ***, **, and * indicate significance at the 1\%, 5\%, and 10\% levels, respectively.
} % end of small font size
} % end of caption
\label{table_BEA.Manufacture.D00t06vsD07t10.PLMNJ.4Reg}
\resizebox{0.95\columnwidth}{!}{%
\begin{tabular}{l*{4}{c}}
\toprule
Dep Var (Panel A, B, and C)                      &\multicolumn{4}{c}{Manufacture Employment Growth (00-06 \& 07-10, An)} \\
            \cmidrule{2-5} 
            &\multicolumn{1}{c}{(1)}&\multicolumn{1}{c}{(2)}&\multicolumn{1}{c}{(3)}&\multicolumn{1}{c}{(4)}\\

\midrule
\multicolumn{5}{l}{\textbf{Panel A. OLS estimates}} \\
PLMNJ Growth (07USD, 99-05, An) x Dum00t06&    0.002   &    0.003   &    0.003   &    0.004   \\
               &  (0.003)   &  (0.003)   &  (0.003)   &  (0.003)   \\
\addlinespace
PLMNJ Growth (07USD, 99-05, An) x Dum07t10&    0.005** &    0.005** &    0.004   &    0.005*  \\
               &  (0.002)   &  (0.002)   &  (0.003)   &  (0.003)   \\
\addlinespace
R2-adj         &    0.454   &    0.466   &    0.474   &    0.489   \\
\addlinespace

\midrule
\multicolumn{5}{l}{\textbf{Panel B. Reduced-form estimates}} \\
GIV Net Export Growth (99-06, An) x Dum00t06&    1.015***&    1.084***&    1.035***&    1.147***\\
               &  (0.348)   &  (0.343)   &  (0.378)   &  (0.350)   \\
\addlinespace
GIV Net Export Growth (99-06, An) x Dum07t10&    0.983***&    0.965***&    0.890***&    0.971***\\
               &  (0.283)   &  (0.271)   &  (0.298)   &  (0.285)   \\
\addlinespace
R2-adj         &    0.477   &    0.490   &    0.495   &    0.513   \\
\addlinespace

\midrule
\multicolumn{5}{l}{\textbf{Panel C . 2SLS estimates}} \\
\addlinespace
PLMNJ Growth (07USD, 99-05, An) x Dum00t06&    0.064***&    0.072***&    0.081***&    0.091***\\
               &  (0.024)   &  (0.025)   &  (0.028)   &  (0.029)   \\
\addlinespace
PLMNJ Growth (07USD, 99-05, An) x Dum07t10&    0.062***&    0.064***&    0.070***&    0.077***\\
               &  (0.020)   &  (0.020)   &  (0.025)   &  (0.028)   \\
               
\addlinespace
\addlinespace

Dep Var (Panel D): &\multicolumn{4}{c}{PLMNJ Growth (99-05, An)} \\ 
\midrule 
\multicolumn{5}{l}{\textbf{Panel D . First-stage estimates only for 99-05 (Non-stack sample)}} \\
\addlinespace
GIV NEG (99-05, An)&   15.738***&   15.106***&   12.754***&   12.555***\\
               &  (3.437)   &  (3.426)   &  (3.555)   &  (3.878)   \\
\addlinespace
KP F-Stat      &    20.97   &    19.44   &    12.87   &    10.48   \\
MOP F-Stat     &    20.00   &    18.89   &    12.85   &    10.47   \\
\addlinespace

\midrule
\multicolumn{5}{l}{\textbf{Controls (for all Panels)}} \\
DumPeriod  &    Y        &  Y   &   Y    & Y        \\
Basic Controls x DumPeriod &            &  Y   &   Y    & Y    \\
Housing Controls x DumPeriod &           &      & Y       & Y   \\
Demographic Controls x DumPeriod &            &      &      &  Y \\

\midrule              
Obs (Panel A, B, and C)          &     1530   &     1530   &     1368   &     1368   \\
Obs (Panel D)          &      765   &      765   &      684   &      684   \\
Cluster SE     &     CBSA   &     CBSA   &     CBSA   &     CBSA   \\
Weight         & Ln(HU99)   & Ln(HU99)   & Ln(HU99)   & Ln(HU99)   \\
\bottomrule
\end{tabular}

} % end of resize box

\end{table}

%---------------------------------------------------------------
%---------------------------------------------------------------
% Robustness 5. Technology Shock Hypothesis in Construction Sector
% A. Placebo: 100 x GM10 CBP Commerical Construction (above)
% B. Government-Sponsored Enterprise Mortgages Boom and Bust 
% C. BEA.Manufacture Employment (above)
%---------------------------------------------------------------
%---------------------------------------------------------------

\pagebreak 
%---------------------------------------------------------------

%%%%%%%%%%%%%%%%%%%%%%%%%%%%%%%%%%%%%%%%%%%%%%%%
% table_GSEM.D99t05vsD05t08.4Reg
%%%%%%%%%%%%%%%%%%%%%%%%%%%%%%%%%%%%%%%%%%%%%%%%

%---------------------------------------------------------------

%%%%%%%%%%%%%%%%%%%%%%%%%%%%%%%%%%%%%%%%%%%%%%%%
% table_GSEM.D99t05vsD05t08.4Reg
%%%%%%%%%%%%%%%%%%%%%%%%%%%%%%%%%%%%%%%%%%%%%%%%

\noindent 

\begin{table}[h!]
\centering
\caption{
\textbf{Four Stacked Regressions of GSEM Growth in Boom (99-05) and Bust (05-08) Periods on Net Export Growth (99-05)} \smallskip \newline
{\scriptsize
This table reports OLS, reduced-form, first stage and second stages of 2SLS stacked regression $\triangle_{99,05} \&  \triangle_{05,08}  Ln(GSEM_{c}) = \beta_{99,05} * \triangle_{99,05} \text{NetExp}_{m} \times Dum_{99,05} + \beta_{05,08} * \triangle_{99,05} \text{NetExp}_{m} \times Dum_{05,08} + \gamma_{99,05}* \bm{Controls_{c}} \times Dum_{99,05} + \gamma_{05,08}* \bm{Controls_{c}} \times Dum_{05,08}  + \alpha_{99,05} + \alpha_{05,08} + \epsilon_{period, c}$. The left-hand-side dependent variable $\triangle_{99,05} \& \triangle_{05,08} Ln(GSEM_{c})$ is the stacked growth rate of the dollar amount of government-sponsored enterprise mortgages (GSEMs) at county $c$ 99-05 and 05-08. The key independent variable $\triangle_{99,05} \text{NetExp}_{m} $ is the growth rate of net export at the metropolitan area (CBSA03 code) $m$ 99-05. $Controls_{c}$ indicates control variables at county $c$ in the period start year 1999. We use the gravity model-based instrumental variable $\triangle_{99,05}\text{givNetExp}_{m} \times Dum_{99,05} $ and $\triangle_{99,05}\text{givNetExp}_{m} \times Dum_{05,08} $ as IVs for $\triangle_{99,05}\text{NetExp}_{m} \times Dum_{99,05}$ and $\triangle_{99,05}\text{NetExp}_{m} \times Dum_{05,08}$. Each regression is weighted by the natural logarithm of housing units in the start year 1999. For the first-stage F-test of two non-stack samples, we report Kleibergen-Paap (2006) robust (clustered) statistics and Montiel Olea-Pflueger (2013) efficient statistics. We report Standard errors are clustered at the CBSA level. ***, **, and * indicate significance at the 1\%, 5\%, and 10\% levels, respectively.
} % end of small font size
} % end of caption
\label{table_GSEM.D99t05vsD05t08.4Reg}

\resizebox{0.85\columnwidth}{!}{%
\begin{tabular}{l*{4}{c}}
\toprule
Dep Var (Panel A, B, and C)                      &\multicolumn{4}{c}{GSEM Growth (99-05 \& 05-08, An)} \\
            \cmidrule{2-5} 
            &\multicolumn{1}{c}{(1)}&\multicolumn{1}{c}{(2)}&\multicolumn{1}{c}{(3)}&\multicolumn{1}{c}{(4)}\\

\midrule
\multicolumn{5}{l}{\textbf{Panel A. OLS estimates}} \\
Net Export Growth (99-05, An) x Dum99t05&    2.928** &    4.552***&    3.800***&    3.286** \\
               &  (1.360)   &  (1.380)   &  (1.368)   &  (1.338)   \\
\addlinespace
Net Export Growth (99-05, An) x Dum05t08&    1.237   &   -0.598   &   -2.144   &   -1.402   \\
               &  (3.075)   &  (2.779)   &  (2.853)   &  (2.922)  \\
\addlinespace
R2-adj         &    0.224   &    0.298   &    0.374   &    0.436   \\
\addlinespace

\midrule
\multicolumn{5}{l}{\textbf{Panel B. Reduced-form estimates}} \\
GIV Net Export Growth (99-05, An) x Dum99t05&    2.173   &    5.288** &    3.373   &    3.522   \\
               &  (2.987)   &  (2.567)   &  (2.633)   &  (2.515)   \\
\addlinespace
GIV Net Export Growth (99-05, An) x Dum05t08&   -4.043   &   -7.309   &   -7.825   &   -5.160   \\
               &  (5.671)   &  (5.218)   &  (5.700)   &  (5.680)   \\
\addlinespace
R2-adj         &    0.224   &    0.298   &    0.374   &    0.436   \\
\addlinespace

\midrule
\multicolumn{5}{l}{\textbf{Panel C . 2SLS estimates}} \\
\addlinespace
Net Export Growth (99-05, An) x Dum99t05&    1.889   &    4.626** &    3.071   &    3.149   \\
               &  (2.539)   &  (2.265)   &  (2.329)   &  (2.307)   \\
\addlinespace
Net Export Growth (99-05, An) x Dum05t08&   -3.514   &   -6.394   &   -7.124   &   -4.614   \\
               &  (4.956)   &  (4.648)   &  (5.245)   &  (5.172)   \\
\addlinespace
CoefEqual\_Chi2 &   11.758   &    9.916   &    9.720   &    9.722   \\
CoefEqual\_PValue&    0.001   &    0.002   &    0.002   &    0.002   \\
\addlinespace
\addlinespace

Dep Var (Panel D): &\multicolumn{4}{c}{Net Export Growth (99-05, An)} \\ 
\midrule 
\multicolumn{5}{l}{\textbf{Panel D . First-stage estimates only for 99-05 (non-stack sample)}} \\
\addlinespace
GIV Net Export Growth (99-05, An) x Dum99t05&    1.150***&    1.143***&    1.098***&    1.118***\\
               &  (0.238)   &  (0.238)   &  (0.271)   &  (0.265)   \\
\addlinespace
KP F-Stat      &    23.36   &    23.11   &    16.44   &    17.78   \\
MOP F-Eff &    22.91   &    22.63   &    16.08   &    17.44   \\

\addlinespace
\addlinespace

\midrule
\multicolumn{5}{l}{\textbf{Controls (for all Panels)}} \\
DumPeriod &  Y     &  Y   &   Y    & Y    \\
Basic Controls x DumPeriod &      &  Y   &   Y    & Y \\
Housing Controls x DumPeriod &       &      & Y       & Y \\
Demographic Controls x DumPeriod &      &      &      &  Y \\
\midrule              
Obs (Panel A, B, and C)          &     1584   &     1584   &     1402   &     1402   \\
Obs (Panel D)           &      792   &      792   &      701   &      701   \\
Cluster SE   &   CBSA   &   CBSA   &   CBSA   &   CBSA   \\
Weight     & Ln(HU99)   & Ln(HU99)   &Ln(HU99)   &Ln(HU99) \\
\bottomrule
\end{tabular}

} % end of resize box

\end{table}

%%%%%%%%%%%%%%%%%%%%%%%%%%%%%%%%%%%%%%%%%%%%%%%%%%%%%%%%%%%%%%%%%%%%%%%%%%%%%%%%%%%%%%%%%%%%%%%%%%%%%%%%%%%%%
%%%%%%%%%%%%%%%%%%%%%%%%%%%%%%%%%%%%%%%%%%%%%%%%%%%%%%%%%%%%%%%%%%%%%%%%%%%%%%%%%%%%%%%%%%%%%%%%%%%%%%%%%%%%%
%%%%%%%%%%%%%%%%%%%%%%%%%%%%%%%%%%%%%%%%%%%%%%%%%%%%%%%%%%%%%%%%%%%%%%%%%%%%%%%%%%%%%%%%%%%%%%%%%%%%%%%%%%%%%
%%%%%%%%%%%%%%%%%%%%%%%%%%%%%%%%%%%%%%%%%%%%%%%%%%%%%%%%%%%%%%%%%%%%%%%%%%%%%%%%%%%%%%%%%%%%%%%%%%%%%%%%%%%%%

%---------------------------------------------------------------
%---------------------------------------------------------------
% Model-based New Prediction
%---------------------------------------------------------------
%---------------------------------------------------------------

\pagebreak 
%---------------------------------------------------------------
%%%%%%%%%%%%%%%%%%%%%%%%%%%%%%%%%%%%%%%%%%%%%%%%
% table_ZIP.Low-Minus-High.PLMNJ.D99t05.4Reg.tex
%%%%%%%%%%%%%%%%%%%%%%%%%%%%%%%%%%%%%%%%%%%%%%%%
%\input{Table_Purch_Full/table_ZIP.Low-Minus-High.PLMNJ.D99t05.4Reg}

\pagebreak 
%---------------------------------------------------------------
%%%%%%%%%%%%%%%%%%%%%%%%%%%%%%%%%%%%%%%%%%%%%%%%
% table_ZIP.LMH.HPI.D00t06vsD07t09.PLMNJ.4Reg
%%%%%%%%%%%%%%%%%%%%%%%%%%%%%%%%%%%%%%%%%%%%%%%%
%\input{Table_Purch_Full/table_ZIP.LMH.HPI.D00t06vsD07t09.PLMNJ.4Reg}

%---------------------------------------------------------------
%---------------------------------------------------------------
% end of this tex file
%---------------------------------------------------------------
%---------------------------------------------------------------

%----------------------------------------------------------------------

%----------------------------------------------------------------------
% section 10: Appendix 

\clearpage
\pagenumbering{arabic}% resets `page` counter to 1
\renewcommand*{\thepage}{A\arabic{page}}
% renew page numbering in the appendix 

\counterwithin{figure}{section}
\counterwithin{table}{section}
% count figures and tables within Appendix section

\appendix

%------------------------------------------------------------
%------------------------------------------------------------
\clearpage
%------------------------------------------------------------

%------------------------------------------------------------
\section{Appendix}

%--------------------------------------------------------------------------------------
%\subsect{Appendix for Data Details}
%--------------------------------------------------------------------------------------

%\input{App_DataDetails}

%--------------------------------------------------------------------------------------
%\subsection{Appendix: GIV for Imports}
%--------------------------------------------------------------------------------------

%-----------------------------------------------------
%-----------------------------------------------------
%-----------------------------------------------------
%-----------------------------------------------------

\subsection{Gravity Model-based IV: US Imports}\label{subsec:GIV_imports}

We have illustrated the central idea of the gravity model-based instrument for US exports in Section \ref{subsec:GIV_exports}. For completeness, we also show how \cite{feenstra2019us} construct IV for US imports here. The gravity-based IV for US imports begins with a simple symmetric constant-elasticity equation in \cite{romalis2007nafta}:
\vspace{-1mm}
\begin{equation}{\label{eq:imp_gravity}}
    \frac{X^{j,US}_{s,v,t}}{X^{j,i}_{s,v,t}} = \Bigg( \frac{w^{j}_{s,t}d^{j,US}\tau^{j,US}_{s,t}}{w^{j}_{s,t}d^{j,i}\tau^{j,i}_{s,t}} \Bigg) ^{1-\sigma} \frac{(P^{US}_{s,t})^{\sigma-1}E^{US}_{s,t}}{(P^{i}_{s,t})^{\sigma-1}E^{i}_{s,t}} = \Bigg(\frac{d^{j,US}\tau^{j,US}_{s,t}}{d^{j,i}\tau^{j,i}_{s,t}} \Bigg) ^{1-\sigma} \frac{(P^{US}_{s,t})^{\sigma-1}E^{US}_{s,t}}{(P^{i}_{s,t})^{\sigma-1}E^{i}_{s,t}}
\end{equation}
$X^{j,US}_{s,v,t}$ is country $j$'s export to US in industry $s$ in product variant $v$ in year $t$. By the similar notation, $X^{j,i}_{s,v,t}$ represents country $j$'s export to country $i$. $w^{j}_{s,t}$ denotes the relative marginal cost of production in industry $s$ in country $j$, which is canceled out in the above equation. $\tau^{j,US}_{s,t}$ and $\tau^{j,i}_{s,t}$ are the \textit{ad valorem} import tariff on country $j$'s export to the US and country $i$, respectively. $d^{j,US}$ and $d^{j,i}$ are the pre-determined bilateral distance and other fixed trade costs from country $j$ to the US and to country $i$, respectively. $P^{US}_{s,t}$ and $P^{i}_{s,t}$ represent the aggregate price index in the US and country $i$. $E^{US}_{s,t}$ and $E^{i}_{s,t}$ denote the total expenditure in the US and country $i$. Lastly, $\sigma$ is the constant elasticity of substitution ($\sigma>1$). 

Like before, the intuition of this gravity-style model is straightforward. The ratio of country $i$'s export to the US relative to country $j$ is decreasing with the ratio of bilateral distance and the ratio of \textit{ad valorem} total import tariff, but increasing with the ratio of aggregate price index and total expenditure.

Suppose that there are $N^{j}_{s,t}$ identical product varieties exported by country $j$ in year $t$ and industry $s$, one can re-arrange the above equation, multiply both sides by $N^{j}_{s,t}$, and sum over countries $i \neq US$:
\vspace{-1mm}
\begin{equation*}
    N^{j}_{s,t}X^{j,US}_{s,v,t}*\sum_{i\neq US} \big[ ( d^{j,i})^{1-\sigma} (P^{i}_{s,t})^{\sigma-1}E^{i}_{s,t} \big] = (d^{j,US}\tau^{j,US}_{s,t})^{1-\sigma} (P^{US}_{s,t})^{\sigma-1}E^{US}_{s,t}* \sum_{i \neq US} \big[ N^{j}_{s,t}X^{j,i}_{s,v,t} (\tau^{j,i}_{s,t})^{\sigma-1} \big ]
\end{equation*}

Since the above equation holds for any countries $i \neq US$, one can choose the set of countries that have similar economic conditions with the US (so that they are market substitution of US market when country $j$ considers its export) to make my prediction more accurate. \cite{feenstra2019us} use the same eight high-income countries by \cite{autor2013china}. 

We denote the sectoral export from country $j$ to the US and to country $i$  as $X^{j,US}_{s,t} \equiv X^{j,US}_{s,v,t}*N^{j}_{s,t}$ and $X^{j,i}_{s,t} \equiv X^{j,i}_{s,v,t}*N^{j}_{s,t}$. Consequently, we can get 
\vspace{-1mm}
\begin{equation*}
    X^{j,US}_{s,t}*\sum_{i\neq US} \big[ ( d^{j,i})^{1-\sigma} (P^{i}_{s,t})^{\sigma-1}E^{i}_{s,t} \big] = (d^{j,US}\tau^{j,US}_{s,t})^{1-\sigma} (P^{US}_{s,t})^{\sigma-1}E^{US}_{s,t}* \sum_{i \neq US} \big[ X^{j,i}_{s,t} (\tau^{j,i}_{s,t})^{\sigma-1} \big ]
\end{equation*}

With a few re-arrangement, we can get the formula for $ X^{j,US}_{s,t}$:
\vspace{-1mm}
\begin{equation}
    X^{j,US}_{s,t} =   \frac{(d^{j,US}\tau^{j,US}_{s,t})^{1-\sigma}(P^{US}_{s,t})^{\sigma-1}E^{US}_{s,t}}{\sum_{i\neq US} \big[ (d^{j,i})^{1-\sigma}(P^{i}_{s,t})^{\sigma-1}E^{i}_{s,t} \big]}  
    * \bigg( \sum_{k\neq US} X^{j,k}_{s,t} \bigg)  * \Bigg\{  \sum_{i\neq US} \bigg[ \frac{ X^{j,i}_{s,t} }{\sum_{k\neq US} X^{j,k}_{s,t}} (\tau^{j,i}_{s,t})^{\sigma -1} \bigg] \Bigg\}
\end{equation}
 
Note in the above formula, we multiply and divide by $\sum_{k\neq US} X^{j,k}_{s,t}$ to prepare for the regression setup in the next step. Now we can take logs of both sides and move the term $\lnb{\sum_{k\neq US} X^{j,k}_{s,t}}$ to the left-hand side of the equation to derive the regression-style formula:
\vspace{-1mm}
\begin{equation} \label{eq:imp_gravityRegression}
\resizebox{0.92\textwidth}{!}{%
\begin{math}
\begin{aligned}
\lnb{X^{j,US}_{s,t}} & = \underbrace{ \lnb{\sum_{k\neq US}X^{j,k}_{s,t}} }_{\text{Term 0}} + \underbrace{ \lnb{(P^{US}_{s,t})^{\sigma-1}E^{US}_{s,t}} }_{\text{Ind-Year FE: } \gamma^{US}_{s,t}} + \underbrace{(1-\sigma)\lnb{d^{j,US}}}_{\text{Exporting-country FE: } \delta^{j,US}} \\
& + \underbrace{(1-\sigma)\lnb{\tau^{j,US}_{s,t}}}_{\text{Term 1}} + \underbrace{(\sigma-1) \lnb{ \Bigg\{  \sum_{i\neq US} \bigg[ \frac{ X^{j,i}_{s,t} }{\sum_{k\neq US} X^{j,k}_{s,t}} (\tau^{j,i}_{s,t})^{\sigma -1} \bigg] \Bigg\}^{\frac{1}{\sigma-1}} }}_{\text{Term 2: } (\sigma-1) \lnb{T^{j}_{s,t}}} + \epsilon^{j}_{s,t} \\
\end{aligned}
\end{math}
} %end of \scalemath \resizebos
\end{equation}
We can see that the US import from country $j$ in the industry $s$ year $t$ can be divided into six terms. ``Term 0'' is the other eight high-income countries' import from country $j$, which represents the world supply. The second term $\gamma^{US}_{s,t}$ is the US demand shocks, which is potentially endogenous. We remove this term by the US industry-by-year fixed effects. The third term $\delta^{j,US}$ is the distance from country $j$ to the US and all other industry- and year-invariant trade costs. Since this term is predetermined rather than a shock, we remove it by the exporting-country fixed effects. ``Term 1" is the tariff on country $j$'s exports imposed by the US, which is by definition out of the control of exporting firms. I keep this term to capture the shock from tariffs. ``Term 2" represents the weighted average tariffs on country $j$'s exports charged by other eight high-income countries. Intuitively, when this weighted average tariffs on country $j$'s exports increase, destination country $j$ will export to the US as a substitution. We keep this term to capture this substitution effect. The last term $\epsilon^{j}_{s,t} = - \lnb{ \sum_{i\neq US} [ (d^{j,i})^{1-\sigma} (P^{i}_{s,t})^{\sigma-1}E^{i}_{s,t} ] } $ is unobserved and only shows up in the regression error term. 

After the above regression, we can construct predicted US imports that are presumably exogenous:
\begin{equation} \label{eq:imp_gravityPreUSImp}
 \lnb{ \widehat{X^{j,US}_{s,t}} } = \lnb{\sum_{k\neq US}X^{j,k}_{s,t}} + \hat{\beta_1} *\lnb{\tau^{j,US}_{s,t}} + \hat{\beta_2}* \lnb{T^{j}_{s,t}}
\end{equation}

%--------------------------------------------------------------------------------------
%\subsection{Empirical.Robustness} This subsection is moved to the main context
%--------------------------------------------------------------------------------------

%\input{Empirical.Robustness}

%--------------------------------------------------------------------------------------
%\subsection{Appendixe: Empirical for Credit Expansion}
%--------------------------------------------------------------------------------------

%------------------------------------------------------------
%------------------------------------------------------------
\subsection{Appendix for the Empirical Results Supporting Credit Expansion}\label{subsec:App_EmpCreditExpansion}

\subsubsection{Broader House Employment Boom (00-06) and Bust (07-10)}
%------------------------------------------------------------
%------------------------------------------------------------
\noindent \textbf{\cite{goukasian2010reaction} House Employment Boom (00-06) and Bust (07-10)}
%------------------------------------------------------------
%------------------------------------------------------------
The relative less relevance and potential data noises in some subcategory industries in the house-related industries (commercial construction and real estate brokerage and management) defined by \cite{goukasian2010reaction} may make the boom and bust trend weaker. This subsection provides evidence for this prediction. 

To test house-related employment boom and bust, I perform the same regressions in Equation  (\ref{eq:RefineHouseEmpShrBoomBustonPLMNJ}) except that the dependent variable is the growth of house-related industries employment share. Table (\ref{table_House.D00t06vsD07t10.PLMNJ.4Reg}) reports OLS, reduced-form, second stage, and the first stage of the stacked regression of house-related employment growth in the boom period (00-06) and the bust (07-10) periods. First, the OLS coefficients in panel A shows that the impact of PLMNJ growth (99-05) are significantly positive in boom period (00-06) and significantly negative in bust period (07-10), both of which are at 1\% level. However, the reduced-form estimates in panel B and 2SLS estimates in panel C are positive but insignificant in the boom period (00-06). These estimates are only significantly negative in the bust period (07-10).  For the coefficient equality test of impact of PLMNJ growth in the boom and bust periods in the Table (\ref{table_RefineHouse.D00t06vsD07t10.PLMNJ.2SLS.wide}), the p-values are almost all below 10\%, meaning the two coefficients are generally statistically different.

The insignificant boom but significant bust in the house-related industries defined by \cite{goukasian2010reaction} are likely result from three reasons. First, the less relevant industries attenuate the boom (00-06). I perform the same test of boom and bust for commercial construction employment in Appendix Table (\ref{table_ComConst.D00t06vsD07t10.PLMNJ.4Reg}) and real estate brokerage and management employment in Table (\ref{table_BrokerageNManagement.D00t06vsD07t10.PLMNJ.4Reg}). The insignificant results show up in both boom and bust for the above two employment industries. Second, data noises are induced by only reporting employment range in the County Business Pattern data by the U.S. Census. Such data noises could stop me from getting significant results via 2SLS. I provide an example for the house-related industries by \cite{goukasian2010reaction}. Appendix Table (\ref{table_House.D00t06.PLMNJ.4RegWin}) shows that reduced-form estimates in column (1) turn to be significant when data are winsoried at 5\%, 10\%, and 15\%. This pattern also shows up in the corresponding 2SLS estimates in column (1). We caution readers that the above data noises mean that the results in our paper shall be viewed as the lower bound of the true impact of credit expansion on the boom and bust cycle in businesses and employment. Third, triggered by massive mortgage defaults the business downturn is amplified by frictions like banking crisis, liquidity crunch, and precautionary saving. These frictions also spill over the negative effects to some less relevant industries. Thus, the employment bust is very strong.

\subsubsection{Exclusion Restriction: Test on PLMNJ Growth}

In section \ref{sec:ExclusionRestriction} Table (\ref{table_PLMNJ.D91t99vsD99t05.NEG.D91t99vsD99t05.4Reg}), I drop four outliers based on the GIV of net export growth due to their extreme impact in the prior period (1991-1999). These four outliers are observations of Howard County (county FIPS: 18067), Tipton County (18159), Durham County (37063), and Orange County (37135). In this appendix, I will compare results with and without dropping outliers to show that a few outliers can make the coefficient of net export growth significantly negative on growth in private-label mortgages (non-jumbo) in the prior period (91-99). 

First, for the prior period (1991-1999), the 2SLS with and without dropping four outliers can be seen from the tiny table below. We can see that 2SLS estimates without dropping is -11.351 and significant at the 1\% level. However, 2SLS estimates with dropping the four outliers is -7.492 but not significant. Comparison between two results show that a few outliers can make the impact of net export growth on private-label mortgages (non-jumbo) growth significantly negative.

%------------------------------------
\begin{table}[h!]
\vspace{-2mm}
\centering 
\caption{
\textbf{2SLS Regression of Private-label Mortgage (non-jumbo) Growth on Net Export Growth in Prior (91-99) With and Without Dropping Outliers}
}
\resizebox{0.9\columnwidth}{!}{%

\begin{tabular}{l*{4}{c}}
\toprule
\textbf{TSLS estimates}            &\multicolumn{4}{c}{Private-label Mortgage (non-jumbo) Growth (07USD, 91-99, annualized)} \\
            \cmidrule{2-5} 
            &\multicolumn{1}{c}{(1)}&\multicolumn{1}{c}{(2)}&\multicolumn{1}{c}{(3)}&\multicolumn{1}{c}{(4)}\\
            
\midrule
\multicolumn{5}{l}{\textbf{Panel A. Sample with Dropping four outliers}} \\
Net Export Growth (91-99, An)&   -1.730   &   -2.067   &   -7.400   &   -7.492   \\
               &  (4.373)   &  (4.312)   &  (5.492)   &  (4.837)   \\
\addlinespace
\multicolumn{5}{l}{\textbf{Panel B. Sample without Dropping four outliers}} \\
Net Export Growth (91-99, An)&   -6.108   &   -6.404   &  -11.185** &  -11.351***\\
               &  (4.297)   &  (4.226)   &  (4.538)   &  (4.055)   \\
\addlinespace
\midrule
\multicolumn{5}{l}{\textbf{Controls}} \\
Basic Controls  &            &  Y   &   Y    & Y        \\
Housing Controls  &           &      & Y       & Y        \\
Demographic Controls &            &      &        &  Y       \\
\midrule          
Obs for Panel A           &      701   &      701   &      623   &      623   \\
Obs for Panel B           &      705   &      705   &      627   &      627   \\
Cluster SE     &     CBSA   &     CBSA   &     CBSA   &     CBSA    \\
Weight         & {\scriptsize Ln(HU-Start)}   & {\scriptsize Ln(HU-Start)}   &{\scriptsize Ln(HU-Start)}   &{\scriptsize Ln(HU-Start)}  \\
\bottomrule

\end{tabular}

} % end of resize box

\end{table}
%------------------------------------

%------------------------------------
\begin{figure}[h!] 
    \centering
    \caption{Reduced-form Regression of Private-label Mortgage (non-jumbo) Growth on Net Export Growth in Prior Period (1991-1999) With and Without Dropping the four outliers}
    \label{fig_RedFormReg_PLMNJG_NEG_with&wtihout_drooping}
    \begin{subfigure}[t]{0.48\textwidth}
        \centering
        \includegraphics[height=4.8cm]{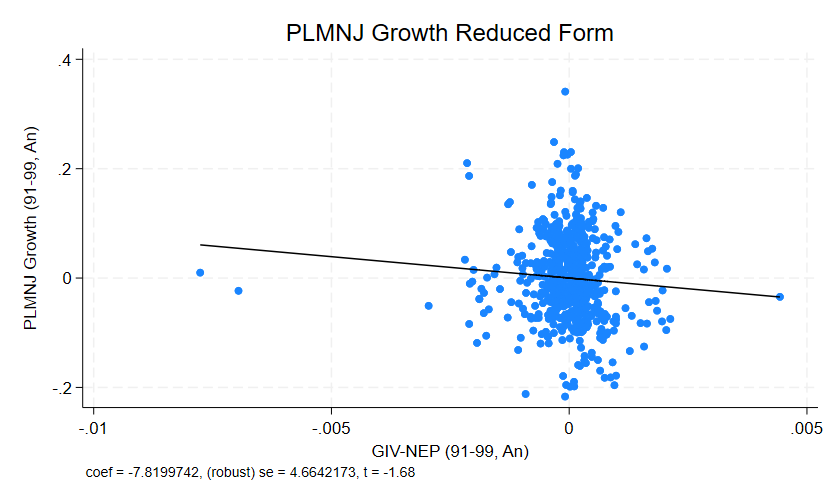}
        \caption{Dropping Outliers}
    \end{subfigure}%
    ~ 
    \begin{subfigure}[t]{0.48\textwidth}
        \centering
        \includegraphics[height=4.8cm]{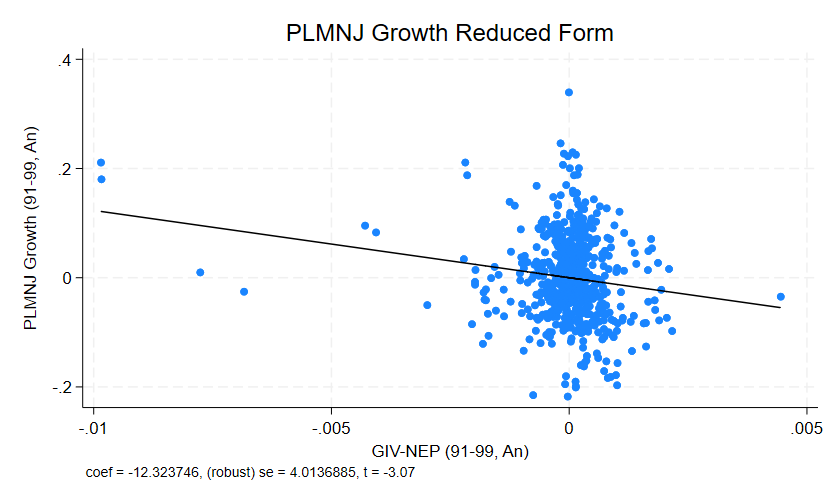}
        \caption{Without Dropping Outliers}
    \end{subfigure}
\end{figure} 
%------------------------------------

Another way to illustrate the above effect of a few outliers is to compare the added-variable plots (AV-plot) of two reduced-form regressions. In Figure (\ref{fig_RedFormReg_PLMNJG_NEG_with&wtihout_drooping}), the left subfigure shows the AV-plot of the reduced-form regression for the sample  dropping the four outliers and the right subfigure is for the sample without dropping outliers. Without dropping, a few outliers make the impact of net export growth on private-label mortgage grow significantly negative in the right subfigure. By dropping the four outliers, the left subfigure shows that, for most observations, such negative impact is insignificant.

\clearpage 
%-----------------------------------------------------
\subsection{Appendix Figures and Tables}

%---------------------------------------------------------------

%%%%%%%%%%%%%%%%%%%%%%%%%%%%%%%%%%%%%%%%%%%%%%%%
% table_RobustTotEmpShr.D00t06.PLMNJ.D99t05.4Reg
%%%%%%%%%%%%%%%%%%%%%%%%%%%%%%%%%%%%%%%%%%%%%%%%

%---------------------------------------------------------------

%%%%%%%%%%%%%%%%%%%%%%%%%%%%%%%%%%%%%%%%%%%%%%%%
% table_RobustTotEmpShr.D00t06.PLMNJ.D99t05.4Reg
%%%%%%%%%%%%%%%%%%%%%%%%%%%%%%%%%%%%%%%%%%%%%%%%

\noindent 

\begin{table}[h!]
\centering
\caption{
\textbf{Four Regressions of Three Alternative Total Employment Growth in Boom Period (00-06) on PLMNJ Growth (99-05)} \smallskip \newline
{\footnotesize 
This table reports OLS, reduced-form, first stage, and second stage results of 2SLS regression $\triangle_{00,06} AlterTotalEmpShr_{c} = \beta * \triangle_{99,05} Ln(PLMNJ_{c}) + \gamma* \bm{Controls_{c}} + \alpha + \epsilon_{c}$. The left-hand-side dependent variable $\triangle_{00,06} AlterTotalEmpShr_{c}$ is the change of the alternative total employment share in working-age population at county $c$ 00-06. Three alternative total employment are wage and salary employment, nonfarm employment, and private nonfarm employment. To prevent excessive influence from outliers, three measures are all winsorized at 0.5\% and 99.5\% level. The key independent variable $\triangle_{99,05} Ln(PLMNJ_{c})$ is the growth rate of the dollar amount (07USD) of private-label mortgage (non-jumbo) (PLMNJ) at county $c$ 99-05. $Controls_{c}$ indicates control variables at county $c$ in 1999. We use the gravity model-based instrumental variable ($\triangle_{99,05}\text{givNetExp}_{m}$) as IV for $\triangle_{99,05} Ln(PLMNJ_{c})$. For the first-stage F-test, we report kleibergen-Paap (2006) robust (clustered) statistics and Montiel Olea-Pflueger (2013) efficient statistics. Each regression is weighted by the natural logarithm of housing units in 1999. Standard errors are clustered at the CBSA level. ***, **, and * indicate significance at the 1\%, 5\%, and 10\% levels, respectively.
} % end of small font size
} % end of caption
\label{table_RobustTotEmpShr.D00t06.PLMNJ.D99t05.4Reg}

\resizebox{0.8\columnwidth}{!}{%

\begin{tabular}{l*{3}{c}}
\toprule
Dep Var (Panel A, B, and C)                     &\multicolumn{1}{c}{(1)}&\multicolumn{1}{c}{(2)}&\multicolumn{1}{c}{(3)}\\

            \cmidrule{2-4}
            
            &\multicolumn{1}{c}{Wage and Salary}&\multicolumn{1}{c}{Nonfarm}&\multicolumn{1}{c}{Private Nonfarm}\\
            
\midrule
\multicolumn{4}{l}{\textbf{Panel A. OLS estimates}} \\
\addlinespace
PLMNJ Growth (99-05, An)&    0.005   &    0.007   &    0.010** \\
               &  (0.004)   &  (0.005)   &  (0.005)   \\
\addlinespace
R2-adj         &    0.102   &   0.0939   &   0.0983   \\
\addlinespace

\midrule
\multicolumn{4}{l}{\textbf{Panel B. Reduced-form estimates}} \\
\addlinespace
GIV Net Export Growth (99-05, An)&    1.023***&    1.248***&    1.398***\\
               &  (0.370)   &  (0.369)   &  (0.351)   \\
\addlinespace
R2-adj         &    0.118   &    0.110   &    0.118   \\
\addlinespace

\midrule
\multicolumn{4}{l}{\textbf{Panel C . 2SLS estimates}} \\
\addlinespace
PLMNJ Growth (99-05, An)&    0.081***&    0.099***&    0.111***\\
               &  (0.027)   &  (0.038)   &  (0.039)   \\
\addlinespace 
\addlinespace

Dep Var (Panel D): &\multicolumn{3}{c}{PLMNJ Growth (99-05, An)} \\ 
\midrule 
\multicolumn{4}{l}{\textbf{Panel D . First-stage estimates}} \\
\addlinespace
GIV Net Export Growth (99-05, An)&   12.555***&   12.555***&   12.555***\\
               &  (3.878)   &  (3.878)   &  (3.878)   \\
\addlinespace
KP F-Stat      &    10.48   &    10.48   &    10.48   \\
MOP F-Stat     &    10.47   &    10.47   &    10.47   \\
\addlinespace
\midrule
\multicolumn{4}{l}{\textbf{Controls (for all Panels)}} \\
Basic Controls  &  Y   &   Y    & Y        \\
Housing Controls &  Y   &   Y    & Y        \\     
Demographic Controls &  Y   &   Y    & Y        \\
\midrule              
Obs            &      684   &      684   &      684   \\
Cluster SE     &     CBSA   &     CBSA   &     CBSA   \\
Weight         & Ln(HU99)   & Ln(HU99)   & Ln(HU99)   \\
\bottomrule

\end{tabular}

} % end of resize box

\end{table}

%---------------------------------------------------------------
%---------------------------------------------------------------
% Empirical: Main Tests 2.1 (2)
% House Employment Growth in Boom (00-06) and Bust (07-10)
%---------------------------------------------------------------
%---------------------------------------------------------------

\pagebreak
%-----------------------------------------------------------------

%%%%%%%%%%%%%%%%%%%%%%%%%%%%%%%%%%%%
% table_House.D00t06vsD07t10.PLMNJ.2SLS.wide
%%%%%%%%%%%%%%%%%%%%%%%%%%%%%%%%%%%%

%-----------------------------------------------------------------------
%%%%%%%%%%%%%%%%%%%%%%%%%%%%%%%%%%%%
% table_House.D00t06vsD07t10.PLMNJ.2SLS.wide
%%%%%%%%%%%%%%%%%%%%%%%%%%%%%%%%%%%%

\noindent 

\begin{table}[h!]
\centering
\caption{
\textbf{2SLS Stacked Regression of House Employment Growth in Boom (00-06) and Bust (07-10) Periods on PLMNJ Growth (99-05)} \smallskip \newline
{\scriptsize
This table reports 2SLS regression $\triangle_{00,06} \& \triangle_{07,10} HouseEmpShr_{c} = \beta_{00,06} * \triangle_{99,05} Ln(PLMNJ_{c}) \times Dum_{00,06} + \beta_{07,10} * \triangle_{99,05} Ln(PLMNJ_{c}) \times Dum_{07,10} + \gamma_{00,06}* \bm{Controls_{c}} \times Dum_{00,06} + \gamma_{07,10}* \bm{Controls_{c}} \times Dum_{07,10} + \epsilon_{period, c}$. The left-hand-side dependent variable $\triangle_{00,06} \& \triangle_{07,10} HouseEmpShr_{c}$ is the change of the house employment share in working-age population at county $c$ 00-06 and 07-10. The key independent variable $\triangle_{99,05} Ln(PLMNJ_{c})$ is the growth rate of the dollar amount (07USD) of private-label mortgages (non-jumbo) at county $c$ 99-05. $Controls_{c}$ indicates control variables at county $c$ in the period start year 1999. We use the gravity model-based instrumental variable $\triangle_{99,05}\text{givNetExp}_{m}$ as the IV for $\triangle_{99,05}Ln(PLMNJ_{c})$. Regression is weighted by the natural logarithm of housing units in 1999.  For the first-stage F-test of two non-stacked samples, we report Kleibergen-Paap (2006) robust (clustered) statistics and Montiel Olea-Pflueger (2013) efficient statistics. Standard errors are clustered at the CBSA level. ***, **, and * indicate significance at the 1\%, 5\%, and 10\% levels, respectively.
\smallskip
} % end of small font size
} % end of caption
\label{table_House.D00t06vsD07t10.PLMNJ.2SLS.wide}

\vspace{-2mm}

\resizebox{\columnwidth}{!}{%
\begin{tabular}{l*{8}{c}}
\toprule
\textbf{TSLS estimates}            &\multicolumn{8}{c}{House Employment Growth (00-06 and 07-10, An)} \\
            \cmidrule{2-9} 
            &\multicolumn{2}{c}{(1)}&\multicolumn{2}{c}{(2)}&\multicolumn{2}{c}{(3)}&\multicolumn{2}{c}{(4)}\\
            
\midrule
PLMNJ Growth (07USD, 99-05, An) x Dum00t06&    0.002   &  (0.003)&    0.001   &  (0.003)&    0.001   &  (0.003)&    0.001   &  (0.003)\\ 
\addlinespace
PLMNJ Growth (07USD, 99-05, An) x Dum07t10&   -0.014***&  (0.005)&   -0.011** &  (0.005)&   -0.013*  &  (0.007)&   -0.015*  &  (0.008)\\ 
\addlinespace
Dum00t06       &    0.001   &  (0.000)&   -0.007***&  (0.002)&   -0.005   &  (0.003)&   -0.010   &  (0.006)\\ 
\addlinespace
Dum07t10       &    0.001   &  (0.001)&    0.023***&  (0.004)&    0.028***&  (0.006)&    0.010   &  (0.010)\\ 
\addlinespace
Ln(Num of HH, 99) x Dum00t06&            &         &    0.000   &  (0.000)&    0.001   &  (0.001)&    0.000   &  (0.001)\\ 
\addlinespace
Ln(Num of HH, 99) x Dum07t10&            &         &   -0.000   &  (0.000)&    0.001   &  (0.001)&    0.001   &  (0.001)\\ 
\addlinespace
Ln(HH Income, 99) x Dum00t06&            &         &    0.001** &  (0.000)&    0.000   &  (0.000)&    0.001   &  (0.001)\\ 
\addlinespace
Ln(HH Income, 99) x Dum07t10&            &         &   -0.002***&  (0.000)&   -0.002***&  (0.001)&   -0.001   &  (0.001)\\ 
\addlinespace
Ratio of Labor Force (1999) x Dum00t06&            &         &   -0.001   &  (0.001)&    0.001   &  (0.001)&    0.001   &  (0.001)\\ 
\addlinespace
Ratio of Labor Force (1999) x Dum07t10&            &         &    0.002   &  (0.002)&    0.000   &  (0.002)&    0.000   &  (0.002)\\ 
\addlinespace
Ln(Num of HU, 99) x Dum00t06&            &         &            &         &   -0.000   &  (0.001)&   -0.000   &  (0.001)\\ 
\addlinespace
Ln(Num of HU, 99) x Dum07t10&            &         &            &         &   -0.001   &  (0.001)&   -0.001   &  (0.001)\\ 
\addlinespace
Housing supply elasticity x Dum00t06&            &         &            &         &   -0.000   &  (0.000)&   -0.000   &  (0.000)\\ 
\addlinespace
Housing supply elasticity x Dum07t10&            &         &            &         &   -0.000   &  (0.000)&   -0.000   &  (0.000)\\ 
\addlinespace
House Vacancy Rate (1999) x Dum00t06&            &         &            &         &    0.004** &  (0.002)&    0.004** &  (0.002)\\ 
\addlinespace
House Vacancy Rate (1999) x Dum07t10&            &         &            &         &   -0.002   &  (0.003)&   -0.001   &  (0.003)\\ 
\addlinespace
Ratio of Renters (1999) x Dum00t06&            &         &            &         &   -0.002   &  (0.001)&   -0.002   &  (0.001)\\ 
\addlinespace
Ratio of Renters (1999) x Dum07t10&            &         &            &         &   -0.001   &  (0.001)&    0.003*  &  (0.001)\\ 
\addlinespace
Ratio of Bachelor Educated (1999) x Dum00t06&            &         &            &         &            &         &   -0.002   &  (0.001)\\ 
\addlinespace
Ratio of Bachelor Educated (1999) x Dum07t10&            &         &            &         &            &         &   -0.005*  &  (0.003)\\ 
\addlinespace
Ratio of White Race (1999) x Dum00t06&            &         &            &         &            &         &   -0.000   &  (0.000)\\ 
\addlinespace
Ratio of White Race (1999) x Dum07t10&            &         &            &         &            &         &    0.001   &  (0.001)\\ 
\addlinespace
Ratio of Immigration (90-00) x Dum00t06&            &         &            &         &            &         &    0.002   &  (0.002)\\ 
\addlinespace
Ratio of Immigration (90-00) x Dum07t10&            &         &            &         &            &         &   -0.010** &  (0.004)\\ 
\addlinespace
\midrule
Obs            &     1582   &         &     1582   &         &     1400   &         &     1400   &         \\
Cluster SE     &     CBSA   &         &     CBSA   &         &     CBSA   &         &     CBSA   &         \\
Weight         & Ln(HU99)   &         & Ln(HU99)   &         & Ln(HU99)   &         & Ln(HU99)   &         \\
KP F-Stat (99-05, non-stack sample)      &    23.45   &         &    21.09   &         &    14.26   &         &    11.92   &         \\
MOP F-Stat (99-05, non-stack sample)  &    22.30   &         &    20.30   &         &    14.35   &         &    11.99   &         \\
CoefEqual\_Chi2 &    5.980   &         &    3.317   &         &    2.648   &         &    2.924   &         \\
CoefEqual\_PValue &    0.014   &         &    0.069   &         &    0.104   &         &    0.087   &         \\
\bottomrule

\end{tabular}

} % end of resize box

\end{table}

\pagebreak 
%---------------------------------------------------------------

%%%%%%%%%%%%%%%%%%%%%%%%%%%%%%%%%%%%%%%%%%%%%%%%
% table_House.D00t06vsD07t10.PLMNJ.4Reg
%%%%%%%%%%%%%%%%%%%%%%%%%%%%%%%%%%%%%%%%%%%%%%%%

%---------------------------------------------------------------

%%%%%%%%%%%%%%%%%%%%%%%%%%%%%%%%%%%%%%%%%%%%%%%%
% table_House.D00t06vsD07t10.PLMNJ.4Reg
%%%%%%%%%%%%%%%%%%%%%%%%%%%%%%%%%%%%%%%%%%%%%%%%

\noindent 

\begin{table}[h!]
\centering
\caption{
\textbf{Four Stacked Regressions of House Employment Growth in Boom (00-06) and Bust (07-10) Periods on PLMNJ Growth (99-05)} \smallskip \newline
{\scriptsize
This table reports OLS, reduced-form, first stage, and second stages of stacked 2SLS regression $\triangle_{00,06} \& \triangle_{07,10} HouseEmpShr_{c} = \beta_{00,06} * \triangle_{99,05} Ln(PLMNJ_{c}) \times Dum_{00,06} + \beta_{07,10} * \triangle_{99,05} Ln(PLMNJ_{c}) \times Dum_{07,10} + \gamma_{00,06}* \bm{Controls_{c}} \times Dum_{00,06} + \gamma_{07,10}* \bm{Controls_{c}} \times Dum_{07,10} + \epsilon_{period, c}$. The left-hand-side dependent variable $\triangle_{00,06} \& \triangle_{07,10} HouseEmpShr_{c}$ is the change of the house employment share in working-age population at county $c$ 00-06 and 07-10. The key independent variable $\triangle_{99,05} Ln(PLMNJ_{c})$ is the growth rate of the dollar amount (07USD) of private-label mortgages (non-jumbo) at county $c$ 99-05. $Controls_{c}$ indicates control variables at county $c$ in the period start year 1999. We use the gravity model-based instrumental variable $\triangle_{99,05}\text{givNetExp}_{m}$ as the IV for $\triangle_{99,05}Ln(PLMNJ_{c})$. Regression is weighted by the natural logarithm of housing units in 1999.  For the first-stage F-test of two non-stacked samples, we report Kleibergen-Paap (2006) robust (clustered) statistics and Montiel Olea-Pflueger (2013) efficient statistics. Standard errors are clustered at the CBSA level. ***, **, and * indicate significance at the 1\%, 5\%, and 10\% levels, respectively.
} % end of small font size
} % end of caption
\label{table_House.D00t06vsD07t10.PLMNJ.4Reg}
\resizebox{0.95\columnwidth}{!}{%
\begin{tabular}{l*{4}{c}}
\toprule
Dep Var (Panel A, B, and C)                      &\multicolumn{4}{c}{House Employment Growth (00-06 \& 07-10, An)} \\
            \cmidrule{2-5} 
            &\multicolumn{1}{c}{(1)}&\multicolumn{1}{c}{(2)}&\multicolumn{1}{c}{(3)}&\multicolumn{1}{c}{(4)}\\

\midrule
\multicolumn{5}{l}{\textbf{Panel A. OLS estimates}} \\
PLMNJ Growth (07USD, 99-05, An) x Dum00t06&    0.003***&    0.002***&    0.002***&    0.002***\\
               &  (0.000)   &  (0.000)   &  (0.001)   &  (0.001)   \\
\addlinespace
PLMNJ Growth (07USD, 99-05, An) x Dum07t10&   -0.005***&   -0.005***&   -0.004***&   -0.004***\\
               &  (0.001)   &  (0.001)   &  (0.001)   &  (0.001)   \\
\addlinespace
R2-adj         &    0.376   &    0.415   &    0.443   &    0.450   \\
\addlinespace

\midrule
\multicolumn{5}{l}{\textbf{Panel B. Reduced-form estimates}} \\
GIV Net Export Growth (99-05, An) x Dum00t06&    0.036   &    0.018   &    0.009   &    0.008   \\
               &  (0.041)   &  (0.041)   &  (0.043)   &  (0.041)   \\
\addlinespace
GIV Net Export Growth (99-05, An) x Dum07t10&   -0.215***&   -0.165** &   -0.161** &   -0.184** \\
               &  (0.077)   &  (0.075)   &  (0.077)   &  (0.080)   \\
\addlinespace
R2-adj         &    0.356   &    0.397   &    0.433   &    0.442   \\
\addlinespace

\midrule
\multicolumn{5}{l}{\textbf{Panel C . 2SLS estimates}} \\
\addlinespace
PLMNJ Growth (07USD, 99-05, An) x Dum00t06&    0.002   &    0.001   &    0.001   &    0.001   \\
               &  (0.003)   &  (0.003)   &  (0.003)   &  (0.003)   \\
\addlinespace
PLMNJ Growth (07USD, 99-05, An) x Dum07t10&   -0.014***&   -0.011** &   -0.013*  &   -0.015*  \\
               &  (0.005)   &  (0.005)   &  (0.007)   &  (0.008)   \\
               
\addlinespace
\addlinespace

Dep Var (Panel D): &\multicolumn{4}{c}{PLMNJ Growth (99-05, An)} \\ 
\midrule 
\multicolumn{5}{l}{\textbf{Panel D . First-stage estimates only for 99-05 (Non-stack sample)}} \\
\addlinespace
GIV NEG (99-05, An)&   15.753***&   15.278***&   12.243***&   12.134***\\
               &  (3.255)   &  (3.331)   &  (3.276)   &  (3.541)   \\
\addlinespace
KP F-Stat      &    23.45   &    21.09   &    14.26   &    11.92   \\
MOP F-Stat     &    22.30   &    20.30   &    14.35   &    11.99   \\
\addlinespace

\midrule
\multicolumn{5}{l}{\textbf{Controls (for all Panels)}} \\
DumPeriod  &    Y        &  Y   &   Y    & Y        \\
Basic Controls x DumPeriod &            &  Y   &   Y    & Y    \\
Housing Controls x DumPeriod &           &      & Y       & Y   \\
Demographic Controls x DumPeriod &            &      &      &  Y \\

\midrule              
Obs (Panel A, B, and C)          &     1582   &     1582   &     1400   &     1400   \\
Obs (Panel D)          &      791   &      791   &      700   &      700   \\
Cluster SE     &     CBSA   &     CBSA   &     CBSA   &     CBSA   \\
Weight         & Ln(HU99)   & Ln(HU99)   & Ln(HU99)   & Ln(HU99)   \\
\bottomrule
\end{tabular}

} % end of resize box

\end{table}

%%%%%%%%%%%%%%%%%%%%%%%%%%%%%%%%%%%%%%%%%%%%%%%%
% table_table_House.D00t06.PLMNJ.4RegWin
%%%%%%%%%%%%%%%%%%%%%%%%%%%%%%%%%%%%%%%%%%%%%%%%

%---------------------------------------------------------------

%%%%%%%%%%%%%%%%%%%%%%%%%%%%%%%%%%%%%%%%%%%%%%%%
% table_House.D00t06.PLMNJ.4RegWin
%%%%%%%%%%%%%%%%%%%%%%%%%%%%%%%%%%%%%%%%%%%%%%%%

\noindent 

\begin{table}[h!]
\centering
\caption{
\textbf{Four Regressions of Winsorized House Employment Growth in Boom Period (00-06) on PLMNJ Growth (99-05)} \smallskip \newline
{\footnotesize 
This table reports OLS, reduced-form, first stage, and second stage results of 2SLS regression $\triangle_{00,06} HouseEmpShr_{c} = \beta * \triangle_{99,05} Ln(PLMNJ_{c}) + \gamma* \bm{Controls_{c}} + \alpha + \epsilon_{c}$. The left-hand-side dependent variable $\triangle_{00,06} HouseEmpShr_{c}$ is the change of the house employment share in working-age population at county $c$ 00-06. I winsored house employment share at various level reported in the table. The key independent variable $\triangle_{99,05} Ln(PLMNJ_{c})$ is the growth rate of the dollar amount (07USD) of private-label mortgage (non-jumbo) (PLMNJ) at county $c$ 99-05. $Controls_{c}$ indicates control variables at county $c$ in 1999. We use the gravity model-based instrumental variable ($\triangle_{99,05}\text{givNetExp}_{m}$) as IV for $\triangle_{99,05} Ln(PLMNJ_{c})$. For the first-stage F-test, we report kleibergen-Paap (2006) robust (clustered) statistics and Montiel Olea-Pflueger (2013) efficient statistics. Each regression is weighted by the natural logarithm of housing units in 1999. Standard errors are clustered at the CBSA level. ***, **, and * indicate significance at the 1\%, 5\%, and 10\% levels, respectively.
} % end of small font size
} % end of caption
\label{table_House.D00t06.PLMNJ.4RegWin}

\resizebox{0.9\columnwidth}{!}{%

\begin{tabular}{l*{4}{c}}
\toprule
Dep Var (All Panels)                     &\multicolumn{4}{c}{Winsored House Employment Growth (00-06, An)} \\
            \cmidrule{2-5} 
            &\multicolumn{1}{c}{(1)}&\multicolumn{1}{c}{(2)}&\multicolumn{1}{c}{(3)}&\multicolumn{1}{c}{(4)}\\

\midrule
\multicolumn{5}{l}{Panel A. Reduced-form Estimates for Winsored House Emp Share (5\%, 95\%)} \\
\addlinespace
GIV Net Export Growth (99-05, An)&    0.053** &    0.038   &    0.032   &    0.031   \\
               &  (0.027)   &  (0.028)   &  (0.028)   &  (0.027)   \\
\addlinespace

\midrule
\multicolumn{5}{l}{Panel B. 2SLS Estimates for Winsored House Emp Share (5\%, 95\%)} \\
\addlinespace
PLMNJ Growth (99-05, An)&    0.003*  &    0.002   &    0.003   &    0.003   \\
               &  (0.002)   &  (0.002)   &  (0.002)   &  (0.002)   \\
\addlinespace

\midrule
\multicolumn{5}{l}{Panel C. Reduced-form Estimates for Winsored House Emp Share (10\%, 90\%)} \\
\addlinespace
GIV Net Export Growth (99-05, An)&    0.046** &    0.032   &    0.027   &    0.026   \\
               &  (0.023)   &  (0.023)   &  (0.023)   &  (0.023)   \\
\addlinespace

\midrule
\multicolumn{5}{l}{Panel D. 2SLS Estimates for Winsored House Emp Share (10\%, 90\%)} \\
\addlinespace
PLMNJ Growth (99-05, An)&    0.003** &    0.002   &    0.002   &    0.002   \\
               &  (0.001)   &  (0.002)   &  (0.002)   &  (0.002)   \\
\addlinespace

\midrule
\multicolumn{5}{l}{Panel E. Reduced-form Estimates for Winsored House Emp Share (15\%, 85\%)} \\
\addlinespace
GIV Net Export Growth (99-05, An)&    0.041** &    0.028   &    0.023   &    0.022   \\
               &  (0.020)   &  (0.020)   &  (0.020)   &  (0.020)   \\
\addlinespace

\midrule
\multicolumn{5}{l}{Panel F. 2SLS Estimates for Winsored House Emp Share (15\%, 85\%)} \\
\addlinespace
PLMNJ Growth (99-05, An)&    0.003** &    0.002   &    0.002   &    0.002   \\
               &  (0.001)   &  (0.001)   &  (0.002)   &  (0.002)   \\
\addlinespace

\addlinespace
\midrule
\multicolumn{5}{l}{\textbf{Controls (for all Panels)}} \\
Basic Controls &            &  Y   &   Y    & Y        \\
Housing Controls &           &      & Y       & Y       \\
Demographic Controls &            &      &        &  Y        \\
\midrule              
Obs            &      791   &      791   &      700   &      700   \\
Cluster SE     &     CBSA   &     CBSA   &     CBSA   &     CBSA   \\
Weight         & Ln(HU99)   & Ln(HU99)   & Ln(HU99)   & Ln(HU99)   \\
\bottomrule

\end{tabular}

} % end of resize box

\end{table}

%---------------------------------------------------------------
%---------------------------------------------------------------
% Empirical: Placebo Tests in the Houing-Related Industries. 
% (1) Commercial Construction (in Figures & Table) and (2) Real Estate Brokerage and Management Employment Growth in Boom (00-06) and Bust (07-10)
%---------------------------------------------------------------
%---------------------------------------------------------------

\pagebreak 
%---------------------------------------------------------------

%%%%%%%%%%%%%%%%%%%%%%%%%%%%%%%%%%%%%%%%%%%%%%%%
% table_BrokerageNManagement.D00t06vsD07t10.PLMNJ.4Reg
%%%%%%%%%%%%%%%%%%%%%%%%%%%%%%%%%%%%%%%%%%%%%%%%

%---------------------------------------------------------------

%%%%%%%%%%%%%%%%%%%%%%%%%%%%%%%%%%%%%%%%%%%%%%%%
% table_BrokerageNManagement.D00t06vsD07t10.PLMNJ.4Reg
%%%%%%%%%%%%%%%%%%%%%%%%%%%%%%%%%%%%%%%%%%%%%%%%

\noindent 

\begin{table}[h!]
\centering
\caption{
\textbf{Four Stacked Regressions of Real Estate Brokerage and Management Employment Growth in Boom (00-06) and Bust (07-10) Periods on PLMNJ Growth (99-05)} \smallskip \newline
{\scriptsize
This table reports OLS, reduced-form, first stage, and second stages of stacked 2SLS regression $\triangle_{00,06} \& \triangle_{07,10} RealEstateEmpShr_{c} = \beta_{00,06} * \triangle_{99,05} Ln(PLMNJ_{c}) \times Dum_{00,06} + \beta_{07,10} * \triangle_{99,05} Ln(PLMNJ_{c}) \times Dum_{07,10} + \gamma_{00,06}* \bm{Controls_{c}} \times Dum_{00,06} + \gamma_{07,10}* \bm{Controls_{c}} \times Dum_{07,10} + \epsilon_{period, c}$. The left-hand-side dependent variable $\triangle_{00,06} \& \triangle_{07,10} RealEstateEmpShr_{c}$ is the change of the real estate brokerage and management employment share in working-age population at county $c$ 00-06 and 07-10. The key independent variable $\triangle_{99,05} Ln(PLMNJ_{c})$ is the growth rate of the dollar amount (07USD) of private-label mortgages (non-jumbo) at county $c$ 99-05. $Controls_{c}$ indicates control variables at county $c$ in the period start year 1999. We use the gravity model-based instrumental variable $\triangle_{99,05}\text{givNetExp}_{m}$ as the IV for $\triangle_{99,05}Ln(PLMNJ_{c})$. Regression is weighted by the natural logarithm of housing units in 1999.  For the first-stage F-test of two non-stacked samples, we report Kleibergen-Paap (2006) robust (clustered) statistics and Montiel Olea-Pflueger (2013) efficient statistics. Standard errors are clustered at the CBSA level. ***, **, and * indicate significance at the 1\%, 5\%, and 10\% levels, respectively.
} % end of small font size
} % end of caption
\label{table_BrokerageNManagement.D00t06vsD07t10.PLMNJ.4Reg}
\resizebox{0.95\columnwidth}{!}{%
\begin{tabular}{l*{4}{c}}
\toprule
Dep Var (Panel A, B, and C)                      &\multicolumn{4}{c}{Brokerage and Management Emp Growth (00-06 \& 07-10, An)} \\
            \cmidrule{2-5} 
            &\multicolumn{1}{c}{(1)}&\multicolumn{1}{c}{(2)}&\multicolumn{1}{c}{(3)}&\multicolumn{1}{c}{(4)}\\

\midrule
\multicolumn{5}{l}{\textbf{Panel A. OLS estimates}} \\
PLMNJ Growth (07USD, 99-05, An) x Dum00t06&    0.035   &    0.032   &    0.055*  &    0.053*  \\
               &  (0.026)   &  (0.026)   &  (0.028)   &  (0.028)   \\
\addlinespace
PLMNJ Growth (07USD, 99-05, An) x Dum07t10&   -0.052   &   -0.038   &   -0.085   &   -0.093   \\
               &  (0.062)   &  (0.060)   &  (0.069)   &  (0.070)   \\
\addlinespace
R2-adj         &   0.0541   &   0.0586   &   0.0621   &   0.0654   \\
\addlinespace

\midrule
\multicolumn{5}{l}{\textbf{Panel B. Reduced-form estimates}} \\
GIV Net Export Growth (99-05, An) x Dum00t06&   -1.224   &   -1.607   &   -0.064   &   -0.681   \\
               &  (1.799)   &  (1.822)   &  (1.654)   &  (1.691)   \\
\addlinespace
GIV Net Export Growth (99-05, An) x Dum07t10&   -6.293   &   -5.623   &   -7.524   &   -8.229   \\
               &  (6.014)   &  (5.847)   &  (6.163)   &  (6.445)   \\
\addlinespace
R2-adj         &   0.0552   &   0.0599   &   0.0623   &   0.0660   \\
\addlinespace

\midrule
\multicolumn{5}{l}{\textbf{Panel C . 2SLS estimates}} \\
\addlinespace
PLMNJ Growth (07USD, 99-05, An) x Dum00t06&   -0.078   &   -0.104   &   -0.005   &   -0.057   \\
               &  (0.114)   &  (0.117)   &  (0.136)   &  (0.145)   \\
\addlinespace
PLMNJ Growth (07USD, 99-05, An) x Dum07t10&   -0.408   &   -0.374   &   -0.626   &   -0.694   \\
               &  (0.389)   &  (0.387)   &  (0.508)   &  (0.549)   \\
               
\addlinespace
\addlinespace

Dep Var (Panel D): &\multicolumn{4}{c}{PLMNJ Growth (99-05, An)} \\ 
\midrule 
\multicolumn{5}{l}{\textbf{Panel D . First-stage estimates only for 99-05 (Non-stack sample)}} \\
\addlinespace
GIV NEG (99-05, An)&   15.753***&   15.278***&   12.243***&   12.134***\\
               &  (3.255)   &  (3.331)   &  (3.276)   &  (3.541)   \\
\addlinespace
KP F-Stat      &    22.29   &    20.53   &    13.35   &    11.22   \\
MOP F-Stat     &    21.10   &    19.69   &    13.30   &    11.18   \\
\addlinespace

\midrule
\multicolumn{5}{l}{\textbf{Controls (for all Panels)}} \\
DumPeriod  &    Y        &  Y   &   Y    & Y        \\
Basic Controls x DumPeriod &            &  Y   &   Y    & Y    \\
Housing Controls x DumPeriod &           &      & Y       & Y   \\
Demographic Controls x DumPeriod &            &      &      &  Y \\

\midrule              
Obs (Panel A, B, and C)          &     1534   &     1534   &     1357   &     1357   \\
Obs (Panel D)         &      761   &      761   &      674   &      674   \\
Cluster SE     &     CBSA   &     CBSA   &     CBSA   &     CBSA   \\
Weight         & Ln(HU99)   & Ln(HU99)   & Ln(HU99)   & Ln(HU99)   \\
\bottomrule
\end{tabular}

} % end of resize box

\end{table}

%%%%%%%%%%%%%%%%%%%%%%%%%%%%%%%%%%%%%%%%%%%%%%%%%%%%%%%%%%%%%%%%%%%%
%%%%%%%%%%%%%%%%%%%%%%%%%%%%%%%%%%%%%%%%%%%%%%%%%%%%%%%%%%%%%%%%%%%%
%%%%%%%%%%%%%%%%%%%%%%%%%%%%%%%%%%%%%%%%%%%%%%%%%%%%%%%%%%%%%%%%%%%%
%%%%%%%%%%%%%%%%%%%%%%%%%%%%%%%%%%%%%%%%%%%%%%%%%%%%%%%%%%%%%%%%%%%%

%---------------------------------------------------------------

%%%%%%%%%%%%%%%%%%%%%%%%%%%%%%%%%%%%%%%%%%%%%%%%
% table_Robust.APLvsNone.RefineHouse.D00t06.D07t10
%%%%%%%%%%%%%%%%%%%%%%%%%%%%%%%%%%%%%%%%%%%%%%%%

%---------------------------------------------------------------

%%%%%%%%%%%%%%%%%%%%%%%%%%%%%%%%%%%%%%%%%%%%%%%%
% table_Robust.APLvsNone.RefineHouse.D00t06.D07t10
%%%%%%%%%%%%%%%%%%%%%%%%%%%%%%%%%%%%%%%%%%%%%%%%

\noindent 

\begin{table}[h!]
\centering
\caption{
\textbf{Robustness Test for Anti-Predatory Lending States vs. Non-Anti-Predatory Lending States.} \\
2SLS Stacked Regression of Refined House Employment Growth in Boom (00-06) and Bust (07-10) Periods on PLMNJ Growth (99-05)  \smallskip \newline
{\scriptsize
This table reports OLS, reduced-form, first stage, and second stages of stacked 2SLS regression $\triangle_{00,06} \& \triangle_{07,10} RefinedHouseEmpShr_{c}  = \beta_{Boom} * \triangle_{99,05} Ln(PLMNJ_{c}) \times Dum_{00,06} + \beta_{Bust} * \triangle_{99,05} Ln(PLMNJ_{c}) \times Dum_{07,10} + \beta_{APL, Boom} * \triangle_{99,05} Ln(PLMNJ_{c}) \times Dum_{00,06} \times Dum_{APL} + \beta_{APL, Bust} * \triangle_{99,05} Ln(PLMNJ_{c}) \times Dum_{07,10} \times Dum_{APL} + \gamma_{Boom} * \bm{Controls_{c}} \times Dum_{00,06} + \gamma_{Bust} * \bm{Controls_{c}} \times Dum_{07,10} + \epsilon_{c}$. The left-hand-side dependent variable $\triangle_{00,06} \& \triangle_{07,10} RefinedHouseEmpShr_{c}$ is the change of the refined house employment share in working-age population at county $c$ 00-06 and 07-10. To reduce the impact of outliers, the dependent variable is winsorized at 2\% and 98\% levels in each period. The key independent variable $\triangle_{99,05} Ln(PLMNJ_{c}$ is the growth rate of the dollar amount (07USD) of private-label mortgages (non-jumbo) at county $c$ 99-05. $Controls_{c}$ indicates control variables at county $c$ in the period start year 1999. In either boom or bust period, we have two endogenous variables here: $\triangle_{99,05} Ln(PLMNJ_{c})$ is instrumented by $\triangle_{99,05}\text{givNetExp}_{m}$ and $\triangle_{99,05} Ln(PLMNJ_{c}) \times Dum_{APL}$ is instrumented by $\triangle_{99,05}\text{givNetExp}_{m} \times Dum_{APL}$. For each of the first-stage F-tests of two endogenous variables, we report Sanderson-Windmeijer (2016) robust (clustered) statistics. To evaluate the overall strength of instruments, we report the p-value of robust (clustered) Kleibergen-Paap test statistics calculated by Windmeijer (2021). Each regression is weighted by the natural logarithm of housing units in 1999. Standard errors are clustered at the CBSA level. ***, **, and * indicate significance at the 1\%, 5\%, and 10\% levels, respectively.
} % end of small font size
} % end of caption
\label{table_Robust.APLvsNone.RefineHouse.D00t06.D07t10}

\resizebox{\columnwidth}{!}{%

\begin{tabular}{l*{4}{c}}
\toprule
\textbf{TSLS Estimates}              &\multicolumn{4}{c}{Refined House Employment Growth x 100 (00-06 or 07-10, An)} \\
            \cmidrule{2-5} 
            &\multicolumn{1}{c}{(1)}&\multicolumn{1}{c}{(2)}&\multicolumn{1}{c}{(3)}&\multicolumn{1}{c}{(4)}\\
            
\midrule
PLMNJ Growth (99-05, An) x Dum00t06&    0.347***&    0.310***&    0.321** &    0.375** \\
               &  (0.111)   &  (0.115)   &  (0.154)   &  (0.157)   \\
\addlinespace
PLMNJ Growth (99-05, An) x Dum07t10&   -0.650***&   -0.523***&   -0.582** &   -0.692** \\
               &  (0.179)   &  (0.181)   &  (0.252)   &  (0.292)   \\
\addlinespace
PLMNJ Growth (99-05, An) x Dum00t06 x DumAPL&    0.078   &    0.061   &    0.047   &    0.029   \\
               &  (0.057)   &  (0.056)   &  (0.072)   &  (0.082)   \\
\addlinespace
PLMNJ Growth (99-05, An) x Dum07t10 x DumAPL&   -0.108   &   -0.034   &   -0.040   &   -0.017   \\
               &  (0.088)   &  (0.082)   &  (0.104)   &  (0.117)   \\
\addlinespace
\midrule
DumPeriod  &    Y        &  Y   &   Y    & Y    \\
Basic Controls x DumPeriod &         &  Y   &   Y   & Y \\
Housing Controls x DumPeriod &         &      & Y     & Y \\
Demographic Controls x DumPeriod &     &     &     &  Y   \\
\midrule
Obs           &     1580   &     1580   &     1398   &     1398   \\
Cluster SE     &     CBSA   &     CBSA   &     CBSA   &     CBSA      \\
Weight         & Ln(HU99)   & Ln(HU99)   & Ln(HU99)   & Ln(HU99)   \\
SW F-Stat: PLMNJ (99 to 05) 1st-Stage &    22.15   &    19.61   &    13.30   &    11.22   \\
SW F-Stat: PLMNJxDumAPL (99 to 05) 1st-Stage &    34.92   &    31.76   &    17.22   &    15.41   \\
KP Robust (99 to 05) UnderID P-Value &   0.0003   &   0.0003   &   0.0038   &   0.0065   \\
CoefEqual\_Chi2 &   13.203   &    8.725   &    5.461   &    6.126   \\
CoefEqual\_PValue &    0.000   &    0.003   &    0.019   &    0.013   \\

\bottomrule

\end{tabular}

} % end of resize box

\end{table}

\clearpage 
%---------------------------------------------------------------

%%%%%%%%%%%%%%%%%%%%%%%%%%%%%%%%%%%%%%%%%%%%%%%%
% table_Robust.NRCvsRC.RefineHouse.D00t06.D07t10
%%%%%%%%%%%%%%%%%%%%%%%%%%%%%%%%%%%%%%%%%%%%%%%%

%---------------------------------------------------------------

%%%%%%%%%%%%%%%%%%%%%%%%%%%%%%%%%%%%%%%%%%%%%%%%
% table_Robust.NRCvsRC.RefineHouse.D00t06.D07t10
%%%%%%%%%%%%%%%%%%%%%%%%%%%%%%%%%%%%%%%%%%%%%%%%

\noindent 

\begin{table}[h!]
\centering
\caption{
\textbf{Robustness Test for Non-Recourse States vs. Recourse States.} \\
2SLS Stacked Regression of Refined House Employment Growth in Boom (00-06) and Bust (07-10) Periods on PLMNJ Growth (99-05)  \smallskip \newline
{\scriptsize
This table reports OLS, reduced-form, first stage, and second stages of stacked 2SLS regression $\triangle_{00,06} \& \triangle_{07,10} RefinedHouseEmpShr_{c}  = \beta_{Boom} * \triangle_{99,05} Ln(PLMNJ_{c}) \times Dum_{00,06} + \beta_{Bust} * \triangle_{99,05} Ln(PLMNJ_{c}) \times Dum_{07,10} + \beta_{NRC, Boom} * \triangle_{99,05} Ln(PLMNJ_{c}) \times Dum_{00,06} \times Dum_{NRC} + \beta_{NRC, Bust} * \triangle_{99,05} Ln(PLMNJ_{c}) \times Dum_{07,10} \times Dum_{NRC} + \gamma_{Boom} * \bm{Controls_{c}} \times Dum_{00,06} + \gamma_{Bust} * \bm{Controls_{c}} \times Dum_{07,10} + \epsilon_{c}$. The left-hand-side dependent variable $\triangle_{00,06} \& \triangle_{07,10} RefinedHouseEmpShr_{c}$ is the change of the refined house employment share in working-age population at county $c$ 00-06 and 07-10. To reduce the impact of outliers, the dependent variable is winsorized at 2\% and 98\% levels in each period. The key independent variable $\triangle_{99,05} Ln(PLMNJ_{c}$ is the growth rate of the dollar amount (07USD) of private-label mortgages (non-jumbo) at county $c$ 99-05. $Dum_{NRC}$ is the dummy variable for counties in non-recourse states. $Controls_{c}$ indicates control variables at county $c$ in the period start year 1999. In either boom or bust period, we have two endogenous variables here: $\triangle_{99,05} Ln(PLMNJ_{c})$ is instrumented by $\triangle_{99,05}\text{givNetExp}_{m}$ and $\triangle_{99,05} Ln(PLMNJ_{c}) \times Dum_{NRC}$ is instrumented by $\triangle_{99,05}\text{givNetExp}_{m} \times Dum_{NRC}$. For each of the first-stage F-tests of two endogenous variables, we report Sanderson-Windmeijer (2016) robust (clustered) statistics. To evaluate the overall strength of instruments, we report the p-value of robust (clustered) Kleibergen-Paap test statistics calculated by Windmeijer (2021). Each regression is weighted by the natural logarithm of housing units in 1999. Standard errors are clustered at the CBSA level. ***, **, and * indicate significance at the 1\%, 5\%, and 10\% levels, respectively.
} % end of small font size
} % end of caption
\label{table_Robust.NRCvsRC.RefineHouse.D00t06.D07t10}

\resizebox{\columnwidth}{!}{%

\begin{tabular}{l*{4}{c}}
\toprule
\textbf{TSLS Estimates}              &\multicolumn{4}{c}{Refined House Employment Growth x 100 (00-06 or 07-10, An)} \\
            \cmidrule{2-5} 
            &\multicolumn{1}{c}{(1)}&\multicolumn{1}{c}{(2)}&\multicolumn{1}{c}{(3)}&\multicolumn{1}{c}{(4)}\\
            
\midrule
PLMNJ Growth (99-05, An) x Dum00t06&    0.369***&    0.327** &    0.333** &    0.380** \\
               &  (0.129)   &  (0.134)   &  (0.169)   &  (0.177)   \\
\addlinespace
PLMNJ Growth (99-05, An) x Dum07t10&   -0.674***&   -0.527** &   -0.594** &   -0.692** \\
               &  (0.206)   &  (0.207)   &  (0.277)   &  (0.327)   \\
\addlinespace
PLMNJ Growth (99-05, An) x Dum00t06 x DumNRC&    0.111*  &    0.098*  &    0.094   &    0.069   \\
               &  (0.058)   &  (0.060)   &  (0.062)   &  (0.066)   \\
\addlinespace
PLMNJ Growth (99-05, An) x Dum07t10 x DumNRC&   -0.234** &   -0.166*  &   -0.171   &   -0.142   \\
               &  (0.102)   &  (0.101)   &  (0.108)   &  (0.125)   \\
\addlinespace
\midrule
DumPeriod  &    Y        &  Y   &   Y    & Y    \\
Basic Controls x DumPeriod &         &  Y   &   Y   & Y \\
Housing Controls x DumPeriod &         &      & Y     & Y \\
Demographic Controls x DumPeriod &     &     &     &  Y   \\
\midrule
Obs           &     1580   &     1580   &     1398   &     1398   \\
Cluster SE     &     CBSA   &     CBSA   &     CBSA   &     CBSA      \\
Weight         & Ln(HU99)   & Ln(HU99)   & Ln(HU99)   & Ln(HU99)   \\
SW F-Stat: PLMNJ (99 to 05) 1st-Stage &    24.93   &    22.53   &    14.96   &    12.80   \\
SW F-Stat: PLMNJxDumAPL (99 to 05) 1st-Stage &    33.12   &    34.51   &    32.73   &    33.87   \\
KP Robust (99 to 05) UnderID P-Value &   0.0006   &   0.0006   &   0.0042   &   0.0072   \\
CoefEqual\_Chi2 &   10.708   &    6.932   &    4.798   &    4.898   \\
CoefEqual\_PValue &    0.001   &    0.008   &    0.028   &    0.027   \\

\bottomrule

\end{tabular}

} % end of resize box

\end{table}

\clearpage 
%---------------------------------------------------------------

%%%%%%%%%%%%%%%%%%%%%%%%%%%%%%%%%%%%%%%%%%%%%%%%
% table_Robust.NJDvsJD.RefineHouse.D00t06.D07t10
%%%%%%%%%%%%%%%%%%%%%%%%%%%%%%%%%%%%%%%%%%%%%%%%

%---------------------------------------------------------------

%%%%%%%%%%%%%%%%%%%%%%%%%%%%%%%%%%%%%%%%%%%%%%%%
% table_Robust.NJDvsJD.RefineHouse.D00t06.D07t10
%%%%%%%%%%%%%%%%%%%%%%%%%%%%%%%%%%%%%%%%%%%%%%%%

\noindent 

\begin{table}[h!]
\centering
\caption{
\textbf{Robustness Test for Non-Judicial States vs. Judicial States.} \\
2SLS Stacked Regression of Refined House Employment Growth in Boom (00-06) and Bust (07-10) Periods on PLMNJ Growth (99-05)  \smallskip \newline
{\scriptsize
This table reports OLS, reduced-form, first stage, and second stages of stacked 2SLS regression $\triangle_{00,06} \& \triangle_{07,10} RefinedHouseEmpShr_{c}  = \beta_{Boom} * \triangle_{99,05} Ln(PLMNJ_{c}) \times Dum_{00,06} + \beta_{Bust} * \triangle_{99,05} Ln(PLMNJ_{c}) \times Dum_{07,10} + \beta_{NJD, Boom} * \triangle_{99,05} Ln(PLMNJ_{c}) \times Dum_{00,06} \times Dum_{NJD} + \beta_{NJD, Bust} * \triangle_{99,05} Ln(PLMNJ_{c}) \times Dum_{07,10} \times Dum_{NJD} + \gamma_{Boom} * \bm{Controls_{c}} \times Dum_{00,06} + \gamma_{Bust} * \bm{Controls_{c}} \times Dum_{07,10} + \epsilon_{c}$. The left-hand-side dependent variable $\triangle_{00,06} \& \triangle_{07,10} RefinedHouseEmpShr_{c}$ is the change of the refined house employment share in working-age population at county $c$ 00-06 and 07-10. To reduce the impact of outliers, the dependent variable is winsorized at 6\% and 94\% levels in each period. The key independent variable $\triangle_{99,05} Ln(PLMNJ_{c}$ is the growth rate of the dollar amount (07USD) of private-label mortgages (non-jumbo) at county $c$ 99-05. $Dum_{NJD}$ is the dummy variable for counties in states where foreclosure of a delinquent property needs judicial judgment. $Controls_{c}$ indicates control variables at county $c$ in the period start year 1999. In either boom or bust period, we have two endogenous variables here: $\triangle_{99,05} Ln(PLMNJ_{c})$ is instrumented by $\triangle_{99,05}\text{givNetExp}_{m}$ and $\triangle_{99,05} Ln(PLMNJ_{c}) \times Dum_{NJD}$ is instrumented by $\triangle_{99,05}\text{givNetExp}_{m} \times Dum_{NJD}$. For each of the first-stage F-tests of two endogenous variables, we report Sanderson-Windmeijer (2016) robust (clustered) statistics. To evaluate the overall strength of instruments, we report the p-value of robust (clustered) Kleibergen-Paap test statistics calculated by Windmeijer (2021). Each regression is weighted by the natural logarithm of housing units in 1999. Standard errors are clustered at the CBSA level. ***, **, and * indicate significance at the 1\%, 5\%, and 10\% levels, respectively.
} % end of small font size
} % end of caption
\label{table_Robust.NJDvsJD.RefineHouse.D00t06.D07t10}

\resizebox{\columnwidth}{!}{%

\begin{tabular}{l*{4}{c}}
\toprule
\textbf{TSLS Estimates}              &\multicolumn{4}{c}{Refined House Employment Growth x 100 (00-06 or 07-10, An)} \\
            \cmidrule{2-5} 
            &\multicolumn{1}{c}{(1)}&\multicolumn{1}{c}{(2)}&\multicolumn{1}{c}{(3)}&\multicolumn{1}{c}{(4)}\\
            
\midrule
PLMNJ Growth (99-05, An) x Dum00t06&    0.289***&    0.240** &    0.272*  &    0.315** \\
               &  (0.112)   &  (0.117)   &  (0.156)   &  (0.159)   \\
\addlinespace
PLMNJ Growth (99-05, An) x Dum07t10&   -0.519***&   -0.366** &   -0.437*  &   -0.530*  \\
               &  (0.172)   &  (0.175)   &  (0.238)   &  (0.279)   \\
\addlinespace
PLMNJ Growth (99-05, An) x Dum00t06 x DumNJD&    0.136***&    0.145***&    0.142***&    0.137***\\
               &  (0.029)   &  (0.027)   &  (0.030)   &  (0.032)   \\
\addlinespace
PLMNJ Growth (99-05, An) x Dum07t10 x DumNJD&   -0.193***&   -0.218***&   -0.224***&   -0.203***\\
               &  (0.054)   &  (0.050)   &  (0.054)   &  (0.057)   \\
\addlinespace
\midrule
DumPeriod  &    Y        &  Y   &   Y    & Y    \\
Basic Controls x DumPeriod &         &  Y   &   Y   & Y \\
Housing Controls x DumPeriod &         &      & Y     & Y \\
Demographic Controls x DumPeriod &     &     &     &  Y   \\
\midrule
Obs           &     1580   &     1580   &     1398   &     1398   \\
Cluster SE     &     CBSA   &     CBSA   &     CBSA   &     CBSA      \\
Weight         & Ln(HU99)   & Ln(HU99)   & Ln(HU99)   & Ln(HU99)   \\
SW F-Stat: PLMNJ (99 to 05) 1st-Stage &    21.48   &    19.53   &    13.58   &    11.60   \\
SW F-Stat: PLMNJxDumAPL (99 to 05) 1st-Stage &    46.17   &    51.23   &    37.93   &    36.49   \\
KP Robust (99 to 05) UnderID P-Value &   0.0009   &   0.0011   &   0.0078   &   0.0117   \\
CoefEqual\_Chi2 &    8.778   &    4.664   &    3.481   &    3.931   \\
CoefEqual\_PValue &    0.003   &    0.031   &    0.062   &    0.047   \\

\bottomrule

\end{tabular}

} % end of resize box

\end{table}

\clearpage 
%---------------------------------------------------------------

%%%%%%%%%%%%%%%%%%%%%%%%%%%%%%%%%%%%%%%%%%%%%%%%
% table_Robust.SandvsNone.RefineHouse.D00t06.D07t10
%%%%%%%%%%%%%%%%%%%%%%%%%%%%%%%%%%%%%%%%%%%%%%%%

%---------------------------------------------------------------

%%%%%%%%%%%%%%%%%%%%%%%%%%%%%%%%%%%%%%%%%%%%%%%%
% table_Robust.SandvsNone.RefineHouse.D00t06.D07t10
%%%%%%%%%%%%%%%%%%%%%%%%%%%%%%%%%%%%%%%%%%%%%%%%

\noindent 

\begin{table}[h!]
\centering
\caption{
\textbf{Robustness Test for Sand States vs. Non-Sand States.} \\
2SLS Stacked Regression of Refined House Employment Growth in Boom (00-06) and Bust (07-10) Periods on PLMNJ Growth (99-05)  \smallskip \newline
{\scriptsize
This table reports OLS, reduced-form, first stage, and second stages of stacked 2SLS regression $\triangle_{00,06} \& \triangle_{07,10} RefinedHouseEmpShr_{c}  = \beta_{Boom} * \triangle_{99,05} Ln(PLMNJ_{c}) \times Dum_{00,06} + \beta_{Bust} * \triangle_{99,05} Ln(PLMNJ_{c}) \times Dum_{07,10} + \beta_{Sand, Boom} \times Dum_{00,06} \times Dum_{Sand} + \beta_{Sand, Bust} * \times Dum_{07,10} \times Dum_{Sand} + \gamma_{Boom} * \bm{Controls_{c}} \times Dum_{00,06} + \gamma_{Bust} * \bm{Controls_{c}} \times Dum_{07,10} + \epsilon_{c}$. The left-hand-side dependent variable $\triangle_{00,06} \& \triangle_{07,10} RefinedHouseEmpShr_{c}$ is the change of the refined house employment share in working-age population at county $c$ 00-06 and 07-10. To reduce the impact of outliers, the dependent variable is winsorized at 5\% and 95\% levels in each period. The key independent variable $\triangle_{99,05} Ln(PLMNJ_{c}$ is the growth rate of the dollar amount (07USD) of private-label mortgages (non-jumbo) at county $c$ 99-05. $Dum_{Sand}$ is the dummy variable for counties in four sand states. $Controls_{c}$ indicates control variables at county $c$ in the period start year 1999. In both boom and bust periods, We use the gravity model-based instrumental variable ($\triangle_{99,05}\text{givNetExp}_{m}$) as IV for $\triangle_{99,05}Ln(PLMNJ_{c})$. For the first-stage F-test of the non-stacked sample (99-05), we report kleibergen-Paap (2006) robust (clustered) statistics and Montiel Olea-Pflueger (2013) efficient statistics. Each regression is weighted by the natural logarithm of housing units in 1999. Standard errors are clustered at the CBSA level. ***, **, and * indicate significance at the 1\%, 5\%, and 10\% levels, respectively.
} % end of small font size
} % end of caption
\label{table_Robust.SandvsNone.RefineHouse.D00t06.D07t10}

\resizebox{\columnwidth}{!}{%

\begin{tabular}{l*{4}{c}}
\toprule
\textbf{TSLS Estimates}              &\multicolumn{4}{c}{Refined House Employment Growth x 100 (00-06 or 07-10, An)} \\
            \cmidrule{2-5} 
            &\multicolumn{1}{c}{(1)}&\multicolumn{1}{c}{(2)}&\multicolumn{1}{c}{(3)}&\multicolumn{1}{c}{(4)}\\
            
\midrule
PLMNJ Growth (99-05, An) x Dum00t06&    0.267** &    0.210*  &    0.230   &    0.289*  \\
               &  (0.113)   &  (0.113)   &  (0.141)   &  (0.160)   \\
\addlinespace
PLMNJ Growth (99-05, An) x Dum07t10&   -0.468***&   -0.308*  &   -0.375*  &   -0.481*  \\
               &  (0.172)   &  (0.169)   &  (0.224)   &  (0.281)   \\
\addlinespace
Dum\_Sand\_xD00t06&    0.034***&    0.043***&    0.043***&    0.034***\\
               &  (0.011)   &  (0.010)   &  (0.012)   &  (0.013)   \\
\addlinespace
Dum\_Sand\_xD07t10&   -0.059***&   -0.072***&   -0.067***&   -0.055** \\
               &  (0.017)   &  (0.016)   &  (0.018)   &  (0.022)   \\
\addlinespace
\midrule
DumPeriod  &    Y        &  Y   &   Y    & Y    \\
Basic Controls x DumPeriod &         &  Y   &   Y   & Y \\
Housing Controls x DumPeriod &         &      & Y     & Y \\
Demographic Controls x DumPeriod &     &     &     &  Y   \\
\midrule
Obs           &     1580   &     1580   &     1398   &     1398   \\
Cluster SE     &     CBSA   &     CBSA   &     CBSA   &     CBSA      \\
Weight         & Ln(HU99)   & Ln(HU99)   & Ln(HU99)   & Ln(HU99)   \\
KP F-Stat      &    16.07   &    14.69   &    9.473   &    7.964   \\
MOP F-Stat     &    15.43   &    14.38   &    9.526   &    7.967   \\
CoefEqual\_Chi2 &    7.697   &    3.946   &    3.180   &    3.358   \\
CoefEqual\_PValue &    0.006   &    0.047   &    0.075   &    0.067   \\

\bottomrule

\end{tabular}

} % end of resize box

\end{table}

\clearpage 
%---------------------------------------------------------------

%%%%%%%%%%%%%%%%%%%%%%%%%%%%%%%%%%%%%%%%%%%%%%%%
% table_Robust.StCapGainTax.RefineHouse.D00t06.D07t10
%%%%%%%%%%%%%%%%%%%%%%%%%%%%%%%%%%%%%%%%%%%%%%%%

%---------------------------------------------------------------

%%%%%%%%%%%%%%%%%%%%%%%%%%%%%%%%%%%%%%%%%%%%%%%%
% table_Robust.StCapGainTax.RefineHouse.D00t06.D07t10
%%%%%%%%%%%%%%%%%%%%%%%%%%%%%%%%%%%%%%%%%%%%%%%%

\noindent 

\begin{table}[h!]
\centering
\caption{
\textbf{Robustness Test for State Capital Gain Tax Rates.} \\
2SLS Stacked Regression of Refined House Employment Growth in Boom (00-06) and Bust (07-10) Periods on PLMNJ Growth (99-05)  \smallskip \newline
{\scriptsize
This table reports OLS, reduced-form, first stage, and second stages of stacked 2SLS regression $\triangle_{00,06} \& \triangle_{07,10} RefinedHouseEmpShr_{c}  = \beta_{Boom} * \triangle_{99,05} Ln(PLMNJ_{c}) \times Dum_{00,06} + \beta_{Bust} * \triangle_{99,05} Ln(PLMNJ_{c}) \times Dum_{07,10} + \beta_{Tax, Boom} \times StateCapGainTax_{s} \times Dum_{00,06}  + \beta_{Tax, Bust} \times StateCapGainTax_{s} * \times Dum_{07,10} + \gamma_{Boom} * \bm{Controls_{c}} \times Dum_{00,06} + \gamma_{Bust} * \bm{Controls_{c}} \times Dum_{07,10} + \epsilon_{c}$. The left-hand-side dependent variable $\triangle_{00,06} \& \triangle_{07,10} RefinedHouseEmpShr_{c}$ is the change of the refined house employment share in working-age population at county $c$ 00-06 and 07-10. To reduce the impact of outliers, the dependent variable is winsorized at 2\% and 98\% levels in each period. The key independent variable $\triangle_{99,05} Ln(PLMNJ_{c}$ is the growth rate of the dollar amount (07USD) of private-label mortgages (non-jumbo) at county $c$ 99-05. $StateCapGainTax_{s}$ is the 2005 state-level capital gain tax rate. $Controls_{c}$ indicates control variables at county $c$ in the period start year 1999. In both boom and bust periods, We use the gravity model-based instrumental variable ($\triangle_{99,05}\text{givNetExp}_{m}$) as IV for $\triangle_{99,05}Ln(PLMNJ_{c})$. For the first-stage F-test of the non-stacked sample (99-05), we report kleibergen-Paap (2006) robust (clustered) statistics and Montiel Olea-Pflueger (2013) efficient statistics. Each regression is weighted by the natural logarithm of housing units in 1999. Standard errors are clustered at the CBSA level. ***, **, and * indicate significance at the 1\%, 5\%, and 10\% levels, respectively.
} % end of small font size
} % end of caption
\label{table_Robust.StCapGainTax.RefineHouse.D00t06.D07t10}

\resizebox{\columnwidth}{!}{%

\begin{tabular}{l*{4}{c}}
\toprule
\textbf{TSLS Estimates}              &\multicolumn{4}{c}{Refined House Employment Growth x 100 (00-06 or 07-10, An)} \\
            \cmidrule{2-5} 
            &\multicolumn{1}{c}{(1)}&\multicolumn{1}{c}{(2)}&\multicolumn{1}{c}{(3)}&\multicolumn{1}{c}{(4)}\\
            
\midrule
PLMNJ Growth (99-05, An) x Dum00t06&    0.385***&    0.339** &    0.339*  &    0.398** \\
               &  (0.131)   &  (0.133)   &  (0.175)   &  (0.184)   \\
\addlinespace
PLMNJ Growth (99-05, An) x Dum07t10&   -0.730***&   -0.570***&   -0.639** &   -0.759** \\
               &  (0.211)   &  (0.207)   &  (0.285)   &  (0.340)   \\
\addlinespace
State Capital Gain Tax x Dum00t06&    0.052   &    0.043   &    0.050   &    0.093   \\
               &  (0.105)   &  (0.100)   &  (0.114)   &  (0.121)   \\
\addlinespace
State Capital Gain Tax x Dum07t10&   -0.292*  &   -0.255   &   -0.317*  &   -0.377*  \\
               &  (0.169)   &  (0.161)   &  (0.189)   &  (0.210)   \\
\addlinespace
\midrule
DumPeriod  &    Y        &  Y   &   Y    & Y    \\
Basic Controls x DumPeriod &         &  Y   &   Y   & Y \\
Housing Controls x DumPeriod &         &      & Y     & Y \\
Demographic Controls x DumPeriod &     &     &     &  Y   \\
\midrule
Obs           &     1580   &     1580   &     1398   &     1398   \\
Cluster SE     &     CBSA   &     CBSA   &     CBSA   &     CBSA      \\
Weight         & Ln(HU99)   & Ln(HU99)   & Ln(HU99)   & Ln(HU99)   \\
KP F-Stat      &    19.90   &    18.22   &    12.54   &    10.55   \\
MOP F-Stat     &    19.09   &    17.68   &    12.64   &    10.61   \\
CoefEqual\_Chi2 &   11.992   &    8.010   &    5.028   &    5.318   \\
CoefEqual\_PValue &    0.001   &    0.005   &    0.025   &    0.021   \\

\bottomrule

\end{tabular}

} % end of resize box

\end{table}

%------------------------------------------------------------
%----------------------------------------------------------------------

%----------------------------------------------------------------------
% end of document
%----------------------------------------------------------------------

\end{document}